\documentclass[10pt]{article}
\newcommand\gothfamily{\usefont{U}{ygoth}{m}{n}}
\DeclareTextFontCommand{\textgoth}{\gothfamily}
\usepackage{amsmath,amssymb,bm,graphicx,authblk}

\begin{document}
\title{Classical Physics: Spacetime and Fields}
\author{{\bf Nikodem Pop{\l}awski}}
\date{}

\maketitle

\section*{Preface}
We present a self-contained introduction to the classical theory of spacetime and fields.
This exposition is based on the most general principles: the principle of general covariance (relativity) and the principle of least action.
The order of the exposition is: 1. Spacetime (principle of general covariance and tensors, affine connection, curvature, metric, space and time, tetrad and spin connection, Lorentz group, spinors); 2. Fields (principle of least action, gravitational field, matter, symmetries and conservation laws, particle limit of field, gravitational field equations, spinor fields, electromagnetic field).
In this order, a particle is a special case of a field existing in spacetime, and classical mechanics can be derived from field theory.
\newline
\newline
I dedicate this book to my Parents: Bo\.{z}enna Pop{\l}awska and Janusz Pop{\l}awski.
I am also grateful to Chris Cox and Francisco Guedes for inspiring this book.
\newline
\newline
{\it The Laws of Physics are simple, beautiful, and universal}.
\newline
\newline

\newpage
\tableofcontents
\renewcommand{\theequation}{\arabic{section}.\arabic{subsection}.\arabic{equation}}

\newpage
\section{Spacetime}
\label{Spacetime}
\subsection{Principle of general covariance and tensors}
\setcounter{equation}{0}
Physical processes are described in coordinate systems in four-dimensional {\em spacetime}, called {\em systems of reference} or {\em frames of reference}.
The {\em Einstein's principle of general covariance} or {\em principle of relativity} states that physical laws do not change their form (are {\em covariant}) under arbitrary, differentiable (and thereby continuous) coordinate transformations.
Equivalently, physical laws have the same form in all admissible frames of reference.

\subsubsection{Vectors}
Let us consider a coordinate transformation from old (unprimed) to new (primed) {\em coordinates} in a four-dimensional manifold:
\begin{equation}
x^i\rightarrow x'^{j}(x^i),
\label{vec1}
\end{equation}
where $x'^{j}$ are differentiable and nondegenerate functions of $x^i$.
The Latin letters $i,j,k,l,m,n,\dots$ are the {\em coordinate indices}, which can be 0, 1, 2, or 3.
Accordingly, $x^i$ are differentiable and nondegenerate functions of $x'^{j}$.
The corresponding transformation matrix is formed from the partial derivatives $\partial x^{i}/\partial x'^j$.
It is four-dimensional and square ($4\times4$), and has a nonzero determinant (the {\em Jacobian} of the transformation): $|\partial x^{i}/\partial x'^j|\neq0$.
The matrix $\partial x'^j/\partial x^{i}$ is inverse to $\partial x^{i}/\partial x'^j$:
\begin{equation}
\sum_i\frac{\partial x'^k}{\partial x^{i}}\frac{\partial x^{i}}{\partial x'^j}=\delta^k_j,
\end{equation}
where
\begin{equation}
\delta^i_k=\left\{ \begin{array}{rr} 1 & i=k\\ 0 & i\neq k \end{array} \right\}.
\label{vec3}
\end{equation}
A {\em scalar} or {\em invariant} is defined as a quantity that does not change under coordinate transformations:
\begin{equation}
\phi=\phi'.
\end{equation}
Accordingly, the differential of a scalar is also a scalar:
\begin{equation}
d\phi=d\phi'.
\end{equation}
If $\phi(x^i)$ is a scalar function of the coordinates $x^i$, then its differential can be written as
\begin{equation}
d\phi(x^i)=\sum_i \frac{\partial\phi}{\partial x^i}dx^i.
\end{equation}

Coordinate differentials $dx^i$ and partial derivatives $\partial_i=\partial/\partial x^i$ transform according to
\begin{eqnarray}
& & dx^{i}=\sum_j \frac{\partial x^{i}}{\partial x'^j}dx'^j, \\
& & \frac{\partial}{\partial x^{i}}=\sum_j \frac{\partial x'^j}{\partial x^{i}}\frac{\partial}{\partial x'^j},\quad \partial_i=\sum_j \frac{\partial x'^j}{\partial x^{i}}\partial'_j.
\end{eqnarray}
A {\em contravariant vector} is defined as a set of quantities that transform under (\ref{vec1}) like coordinate differentials:
\begin{equation}
A^{i}=\sum_j \frac{\partial x^{i}}{\partial x'^j}A'^j.
\label{vec9}
\end{equation}
These quantities are referred to as the {\em components} of the contravariant vector.
A {\em covariant vector} is defined as a set of quantities that transform under (\ref{vec1}) like partial derivatives of a scalar:
\begin{equation}
B_{i}=\sum_j \frac{\partial x'^j}{\partial x^{i}}B'_j.
\label{vec10}
\end{equation}
These quantities are referred to as the components of the covariant vector.
Therefore, coordinate differentials form a contravariant vector and partial derivatives of a scalar form a covariant vector.
The coordinates $x^i$ do not form a vector.
A linear combination of two scalars is a scalar.
A linear combination $aC+bD$ of two contravariant vectors $C$ and $D$, where $a$ and $b$ are scalars, is a contravariant vector $E$ whose components are $E^i=aC^i+bD^i$.
A linear combination $aC+bD$ of two covariant vectors $C$ and $D$ is a covariant vector $E$ whose components are $E_i=aC_i+bD_i$.

An upper index (in a contravariant vector) is called {\em contravariant}, and a lower index is called {\em covariant}.
The derivative with respect to a quantity with a contravariant index $i$ is a quantity with a covariant index $i$.
Conversely, the derivative with respect to a quantity with a covariant index $i$ is a quantity with a contravariant index $i$.
Henceforth, we adopt the following {\em Einstein's summation} notation.
If the same coordinate index $i$ appears in a given quantity twice, as a contravariant index and a covariant index, and we apply the summation $\sum_i$ in this quantity, then we do not need to write the summation sign $\sum_i$.
Accordingly, we can omit the summation signs in the formulae of this section.

\subsubsection{Tensors}
A product of several vectors transforms under differentiable coordinate transformations such that each coordinate index transforms separately:
\begin{equation}
A^{i}B^{j}\dots C_k D_l\dots=\frac{\partial x^{i}}{\partial x'^m}\frac{\partial x^{j}}{\partial x'^n}\frac{\partial x'^p}{\partial x^{k}}\frac{\partial x'^q}{\partial x^{l}}A'^m B'^n \dots C'_p D'_q \dots\,.
\end{equation}
A {\em tensor} is defined as a set of quantities that transform under (\ref{vec1}) like products of the components of vectors, in which each contravariant index transforms according to (\ref{vec9}) and each covariant index transforms according to (\ref{vec10}):
\begin{equation}
T^{ij\dots}_{\phantom{ij}kl\dots}=\frac{\partial x^{i}}{\partial x'^m}\frac{\partial x^{j}}{\partial x'^n}\frac{\partial x'^p}{\partial x^{k}}\frac{\partial x'^q}{\partial x^{l}}T'^{mn\dots}_{\phantom{'mn}pq\dots}.
\label{tens2}
\end{equation}
These quantities are referred to as the components of the tensor.
A tensor is of rank $(p,q)$ if it has $p$ contravariant and $q$ covariant indices.
A scalar is a tensor of rank (0,0), a contravariant vector is a tensor of rank (1,0), and a covariant vector is a tensor of rank (0,1).
A linear combination of two tensors of rank $(p,q)$ is a tensor of rank $(p,q)$ such that its components are the same linear combinations of the corresponding components of the tensors.
The product of two tensors of ranks $(p_1,q_1)$ and $(p_2,q_2)$ is a tensor of rank $(p_1+p_2,q_1+q_2)$.

Tensor indices, which are either all contravariant or all covariant, can be {\em symmetrized}:
\begin{equation}
T_{(ij\dots k)}=\frac{1}{n!}\sum_{\textrm{permutations}}T_{\{ij\dots k\}},
\end{equation}
or {\em antisymmetrized}:
\begin{equation}
T_{[ij\dots k]}=\frac{1}{n!}\sum_{\textrm{permutations}}T_{\{ij\dots k\}}(-1)^N,
\end{equation}
where $n$ is the number of symmetrized or antisymmetrized indices and $N$ is the number of permutations that bring $T_{ij\dots k}$ to $T_{\{ij\dots k\}}$.
For example, for two indices:
\begin{equation}
T_{(ik)}=\frac{1}{2}(T_{ik}+T_{ki}),\quad T_{[ik]}=\frac{1}{2}(T_{ik}-T_{ki}),
\end{equation}
and for three indices: $T_{(ijk)}=(1/6)(T_{ijk}+T_{jki}+T_{kij}+T_{ikj}+T_{jik}+T_{kji})$ and 
$T_{[ijk]}=(1/6)(T_{ijk}+T_{jki}+T_{kij}-T_{ikj}-T_{jik}-T_{kji})$.
If $n>4$ then $T_{[ij\dots k]}=0$.
Symmetrized and antisymmetrized tensors or rank $(k,l)$ are tensors of rank $(k,l)$.

Symmetrization of a tensor $T^{ij\dots}_{\phantom{ij}kl\dots}$ in covariant indices $k,l$ gives a tensor, which is {\em symmetric} in these indices:
\begin{equation}
S^{ij\dots}_{\phantom{ij}kl\dots}=S^{ij\dots}_{\phantom{ij}lk\dots},
\end{equation}
and similarly for contravariant indices.
Antisymmetrization of a tensor $T^{ij\dots}_{\phantom{ij}kl\dots}$ in covariant indices $k,l$ gives a tensor, which is {\em antisymmetric} in these indices:
\begin{equation}
A^{ij\dots}_{\phantom{ij}kl\dots}=-A^{ij\dots}_{\phantom{ij}lk\dots},
\end{equation}
and similarly for contravariant indices.
Any tensor, which has at least two contravariant or two covariant indices, is the sum of its symmetric and antisymmetric parts (with respect to those indices):
\begin{equation}
T^{ij\dots}_{\phantom{ij}kl\dots}=T^{ij\dots}_{\phantom{ij}(kl)\dots}+T^{ij\dots}_{\phantom{ij}[kl]\dots}.
\end{equation}
Symmetrization of an antisymmetric tensor or antisymmetrization of a symmetric tensor bring these tensors to zero.

The number $0$ can be regarded as a tensor of arbitrary rank.
Therefore, all covariant equations of classical physics must have a tensor form:
\begin{equation}
T^{ij\dots}_{\phantom{ij}kl\dots}=0,
\end{equation}
so that the transformation (\ref{vec1}) would bring them to the same tensor form:
\begin{equation}
T'^{ij\dots}_{\phantom{'ij}kl\dots}=0,
\end{equation}
in accordance with the principle of relativity.

\subsubsection{Densities}
The element of volume $d^4 x=dx^0 dx^1 dx^2 dx^3$ in four-dimensional spacetime transforms according to
\begin{equation}
d^4 x=\biggl|\frac{\partial x^{i}}{\partial x'^k}\biggr|d^4 x'.
\end{equation}
A {\em scalar density} is defined as a quantity that transforms such that its product with the element of volume is a scalar, $\mathfrak{s}d^4 x=\mathfrak{s}'d^4 x'$:
\begin{equation}
\mathfrak{s}=\biggl|\frac{\partial x'^i}{\partial x^{k}}\biggr|\mathfrak{s}'.
\end{equation}
A {\em tensor density}, which includes a contravariant and covariant vector density, is defined as a set of quantities that transform like products of the components of a tensor and a scalar density:
\begin{equation}
\mathfrak{T}^{ij\dots}_{\phantom{'ij}kl\dots}=\biggl|\frac{\partial x'^i}{\partial x^{k}}\biggr|\frac{\partial x^{i}}{\partial x'^m}\frac{\partial x^{j}}{\partial x'^n}\frac{\partial x'^p}{\partial x^{k}}\frac{\partial x'^q}{\partial x^{l}}\mathfrak{T}'^{mn\dots}_{\phantom{'mn}pq\dots}.
\end{equation}
These quantities are referred to as the components of the tensor density.
A tensor density is of rank $(p,q)$ if it has $p$ contravariant and $q$ covariant indices.
The above densities are said to be of weight 1.

One can generalize this definition of densities by introducing densitites of weight $w$, which transform according to
\begin{equation}
\mathfrak{s}=\biggl|\frac{\partial x'^i}{\partial x^{k}}\biggr|^w\mathfrak{s}'.
\end{equation}
For example, $d^4 x$ is a scalar density of weight -1.
A linear combination of two densities of rank $(p,q)$ and weight $w$ is a density of rank $(p,q)$ and weight $w$ such that its components are the same linear combinations of the corresponding components of the densities.
The product of two densities of weights $w_1$ and $w_2$ is a density of weight $w_1+w_2$.
Tensor densities can be symmetrized and antisymmetrized analogously to tensors.
Symmetrized and antisymmetrized densities of weight $w$ are densities of weight $w$.
Densities of weight 1 are simply referred to as densities.
Tensors are densities of weight 0.

The square root of the determinant of a tensor of rank $(0,2)$ is a scalar density:
\begin{equation}
\sqrt{|T_{ik}|}=\sqrt{\biggr|\frac{\partial x'^l}{\partial x^{i}}\frac{\partial x'^m}{\partial x^{k}}T'_{lm}\biggr|}=\sqrt{\biggr|\frac{\partial x'^j}{\partial x^{n}}\biggr|^2|T'_{ik}|}=\biggr|\frac{\partial x'^j}{\partial x^{n}}\biggr|\sqrt{|T'_{ik}|}.
\label{dens5}
\end{equation}
Consequently, its product with $d^4 x$ is a scalar:
\begin{equation}
\sqrt{|T_{ik}|}d^4 x=\sqrt{|T'_{ik}|}d^4 x'.
\label{dens6}
\end{equation}
This product can be regarded as an invariant element of volume in four-dimensional spacetime.
Accordingly, such an element requires a tensor of rank $(0,2)$ (the metric tensor in section \ref{metrictensor}).

\subsubsection{Contraction}
Einstein's summation convention also applies within the same tensor or tensor density, if a given coordinate index $i$ appears twice (as a contravariant and covariant index).
Such a tensor or density is said to be {\em contracted} over index $i$.
A contracted tensor of rank $(p,q)$ transforms like a tensor of rank $(p-1,q-1)$:
\begin{equation}
T'^{ij\dots}_{\phantom{'ij}il\dots}=\frac{\partial x'^{i}}{\partial x^m}\frac{\partial x'^{j}}{\partial x^n}\frac{\partial x^p}{\partial x'^{i}}\frac{\partial x^q}{\partial x'^{l}}T^{mn\dots}_{\phantom{mn}pq\dots}=\frac{\partial x'^{j}}{\partial x^n}\frac{\partial x^q}{\partial x'^{l}}\delta^p_m T^{mn\dots}_{\phantom{mn}pq\dots}=\frac{\partial x'^{j}}{\partial x^n}\frac{\partial x^q}{\partial x'^{l}}T^{mn\dots}_{\phantom{mn}mq\dots}\,.
\end{equation}
For example, the contraction of a contravariant and covariant vector $A^i B_i$ is a scalar ({\em scalar product}).
A contracted tensor density of rank $(p,q)$ and weight $w$ transforms like a tensor density of rank $(p-1,q-1)$ and weight $w$:
\begin{eqnarray}
& & \mathfrak{T}'^{ij\dots}_{\phantom{'ij}il\dots}=\biggl|\frac{\partial x^i}{\partial x'^{k}}\biggr|^w\frac{\partial x'^{i}}{\partial x^m}\frac{\partial x'^{j}}{\partial x^n}\frac{\partial x^p}{\partial x'^{i}}\frac{\partial x^q}{\partial x'^{l}}\mathfrak{T}^{mn\dots}_{\phantom{mn}pq\dots}=\biggl|\frac{\partial x^i}{\partial x'^{k}}\biggr|^w\frac{\partial x'^{j}}{\partial x^n}\frac{\partial x^q}{\partial x'^{l}}\delta^p_m \mathfrak{T}^{mn\dots}_{\phantom{mn}pq\dots} \nonumber \\
& & =\biggl|\frac{\partial x^i}{\partial x'^{k}}\biggr|^w\frac{\partial x'^{j}}{\partial x^n}\frac{\partial x^q}{\partial x'^{l}}\mathfrak{T}^{mn\dots}_{\phantom{mn}mq\dots}\,.
\end{eqnarray}
Contraction of a symmetric tensor with an antisymmetric tensor (over indices with respect to which these tensors are respectively symmetric and antisymmetric) gives zero:
\begin{equation}
S_{ik}A^{ik}=0.
\end{equation}
If contraction of two tensors gives zero, these tensors are said to be {\em orthogonal}.
Two orthogonal vectors (one contravariant and one covariant) are said to be {\em perpendicular}.

\subsubsection{Kronecker and Levi-Civita symbols}
The quantity $\delta^i_k$ (\ref{vec3}) is a tensor with constant components:
\begin{equation}
\delta'^{i}_{\phantom{'}k}=\frac{\partial x'^{i}}{\partial x^j}\frac{\partial x^l}{\partial x'^{k}}\delta^j_l=\frac{\partial x'^{i}}{\partial x^j}\frac{\partial x^j}{\partial x'^{k}}=\delta^i_k.
\end{equation}
It is referred to as the {\em Kronecker symbol}.
A completely antisymmetric tensor of rank $(4,0)$, $T^{ijkl}=T^{[ijkl]}$, has 1 independent component $T$.
All its components can be written as $T^{ijkl}=T\epsilon^{ijkl}$, where $\epsilon^{ijkl}$ is the completely antisymmetric, contravariant {\em Levi-Civita permutation symbol}:
\begin{equation}
\epsilon^{0123}=1,\quad \epsilon^{ijkl}=\epsilon^{[ijkl]}=(-1)^N,
\end{equation}
and $N$ is the number of permutations that bring $\epsilon^{ijkl}$ to $\epsilon^{0123}$.
The determinant of a square matrix $S^i_k$, $\mbox{det}(S^i_k)=|S^i_k|$, is defined through the permutation symbol:
\begin{equation}
|S^r_s|\epsilon^{ijkl}=S^i_m S^j_n S^k_p S^l_q\epsilon^{mnpq}.
\label{KLC3}
\end{equation}
Taking $S^i_k=\partial x'^{i}/\partial x^k$ gives
\begin{equation}
\epsilon^{ijkl}=\biggl|\frac{\partial x^r}{\partial x'^{s}}\biggr|\frac{\partial x'^{i}}{\partial x^m}\frac{\partial x'^{j}}{\partial x^n}\frac{\partial x'^{k}}{\partial x^p}\frac{\partial x'^{l}}{\partial x^q}\epsilon^{mnpq}.
\end{equation}
This equation has a form of a transformation law for a tensor density (of weight 1) with constant components: $\epsilon'^{ijkl}=\epsilon^{ijkl}$.
Consequently, the contravariant Levi-Civita symbol is a tensor density of weight 1.
The quantity $T$ is a scalar density of weight -1.

We also introduce the covariant Levi-Civita symbol $\varepsilon_{ijkl}$ through:
\begin{equation}
\epsilon^{ijkl}\varepsilon_{mnpq}=-\left| \begin{array}{rrrr} \delta^i_m & \delta^i_n & \delta^i_p & \delta^i_q \\ \delta^j_m & \delta^j_n & \delta^j_p & \delta^j_q \\ \delta^k_m & \delta^k_n & \delta^k_p & \delta^k_q \\ \delta^l_m & \delta^l_n & \delta^l_p & \delta^l_q \end{array} \right|.
\label{KLC5}
\end{equation}
Therefore, the covariant Levi-Civita symbol is a tensor density of weight -1 and its product with a scalar density is a tensor.
The covariant Levi-Civita symbol is given by
\begin{equation}
\varepsilon_{0123}=-1,\quad \varepsilon_{ijkl}=\varepsilon_{[ijkl]}=(-1)^N,
\end{equation}
where $N$ is the number of permutations that bring $\varepsilon_{ijkl}$ to $\varepsilon_{0123}$, and satisfies
\begin{equation}
|S^r_s|\varepsilon_{ijkl}=S_i^m S_j^n S_k^p S_l^q\varepsilon_{mnpq}.
\end{equation}
Contracting (\ref{KLC5}) gives the following relations:
\begin{eqnarray}
& & \epsilon^{ijkl}\varepsilon_{mnpl}=-\left| \begin{array}{rrr} \delta^i_m & \delta^i_n & \delta^i_p \\ \delta^j_m & \delta^j_n & \delta^j_p \\ \delta^k_m & \delta^k_n & \delta^k_p \end{array} \right|, \quad \epsilon^{ijkl}\varepsilon_{mnkl}=-2(\delta^i_m\delta^j_n-\delta^i_n\delta^j_m), \nonumber \\
& & \epsilon^{ijkl}\varepsilon_{mjkl}=-6\delta^i_m, \quad \epsilon^{ijkl}\varepsilon_{ijkl}=-24.
\label{KLC8}
\end{eqnarray}

\subsubsection{Dual densities}
A contracted product of a covariant tensor and the contravariant Levi-Civita symbol gives a {\em dual} contravariant tensor density of weight 1:
\begin{equation}
\epsilon^{iklm}A_m=\mathfrak{A}^{ikl},\quad \epsilon^{iklm}B_{lm}=\mathfrak{B}^{ik},\quad \epsilon^{iklm}C_{klm}=\mathfrak{C}^i.
\end{equation}
A contracted product of a contravariant tensor and the covariant Levi-Civita symbol gives a dual covariant tensor density of weight -1:
\begin{equation}
\varepsilon_{iklm}A^m=\mathfrak{A}_{ikl},\quad \varepsilon_{iklm}B^{lm}=\mathfrak{B}_{ik},\quad \varepsilon_{iklm}C^{klm}=\mathfrak{C}_i.
\end{equation}
Therefore, there exists an algebraic correspondence between covariant tensors and contravariant densities of weight 1, and between contravariant tensors and covariant densities of weight -1.

\subsubsection{Invariant integrals}
In four-dimensional spacetime, there are four types of integration.

(1) An invariant line integral is an integral of a covariant vector over a one-dimensional curve.
Such a vector is contracted with the line element $dx^i$: $\int A_i dx^i$.

(2) An invariant surface integral is an integral of a tensor of rank (0,2) over a two-dimensional surface.
Such a tensor is contracted with the surface element $df^{ik}$: $\int B_{ik}df^{ik}$, where
\begin{equation}
df^{ik}=\left| \begin{array}{rr} dx^i & dx'^{i} \\ dx^k & dx'^{k} \end{array} \right|=dx^i dx'^{k}-dx^k dx'^{i}
\label{covint31}
\end{equation}
is an antisymmetric tensor.
This infinitesimal element of surface can be geometrically represented as a parallelogram formed by the vectors $dx^i$ and $dx'^{i}$.
The components of $df^{ik}$ are the projections of the area of the parallelogram on the six coordinate planes $x_i x_k$.
The dual density corresponding to the surface element is given by
\begin{equation}
df^\star_{ik}=\frac{1}{2}\varepsilon_{iklm}df^{lm},
\label{covint1}
\end{equation}
which gives
\begin{equation}
df^{lm}=-\frac{1}{2}\epsilon^{lmik}df^\star_{ik},\quad df^{ik}df^\star_{ik}=0.
\label{covint2}
\end{equation}
The element $df^\star_{ik}$ geometrically describes an element of surface equal (in magnitude) to and normal to the element $df^{ik}$: all lines lying in $df^\star_{ik}$ are perpendicular to all lines lying in $df^{ik}$.

(3) An invariant hypersurface (volume) integral is an integral of a tensor of rank (0,3) over a three-dimensional hypersurface.
Such a tensor is contracted with the volume element $dS^{ikl}$: $\int C_{ikl}dS^{ikl}$, where
\begin{equation}
dS^{ikl}=\left| \begin{array}{rrr} dx^i & dx'^{i} & dx^{``i} \\ dx^k & dx'^{k} & dx^{``k} \\ dx^l & dx'^{l} & dx^{``l} \end{array} \right|
\label{covint32}
\end{equation}
is a completely antisymmetric tensor.
This infinitesimal element of hypersurface can be geometrically represented as a parallelepiped formed by the vectors $dx^i$, $dx'^{i}$, and $dx^{``i}$.
The components of $dS^{ikl}$ are the projections of the volume of the parallelepiped on the four coordinate hyperplanes $x_i x_k x_l$.
The dual density corresponding to the hypersurface element is given by
\begin{equation}
dS_i=-\frac{1}{6}\varepsilon_{iklm}dS^{klm},\quad dS^{klm}=-\epsilon^{klmi}dS_i.
\label{covint3}
\end{equation}
The element $dS_i$ geometrically describes a four-vector equal (in magnitude) to the volume of the element $dS^{ikl}$ and normal to this element: perpendicular to all lines lying in $dS^{ikl}$.

(4) An invariant four-volume integral is an integral of a tensor of rank (0,4) over a four-dimensional region of spacetime.
Such a tensor is contracted with the four-volume element $dS^{ijkl}$, defined analogously to $dS^{ikl}$.
The dual density corresponding to the four-volume element is given by
\begin{equation}
d\Omega=\frac{1}{24}\varepsilon_{iklm}dS^{iklm}=dx^0 dx^1 dx^2 dx^3=d^4x.
\label{covint4}
\end{equation}

According to the Gau\ss--Stokes--Green theorems, there exist relations between integrals over different elements:
\begin{eqnarray}
& & dx^i \leftrightarrow df^{ki} \frac{\partial}{\partial x^k}, \label{covint5} \\
& & df^\star_{ik} \leftrightarrow dS_i \frac{\partial}{\partial x^k}-dS_k \frac{\partial}{\partial x^i}, \label{covint6} \\
& & dS_i \leftrightarrow d\Omega\frac{\partial}{\partial x^i}.
\label{covint7}
\end{eqnarray}
To preserve the invariant character of an integral, the product of an integrand and the integration element must be a scalar.
The elements $df^\star_{ik}$, $dS_i$, and $d\Omega$ are densities of weight -1.
Consequently, they must be multiplied by a scalar density $\mathfrak{s}$ (with weight 1), for example, the square root of the determinant of a tensor of rank $(0,2)$ in (\ref{dens5}).

\subsubsection{Antisymmetric derivatives}
To denote a partial derivative of a quantity $T$ with respect to $x^i$, we will use $\partial T/\partial x^i=\partial_i T=T_{,i}$.
A derivative of a covariant vector does not transform like a tensor:
\begin{equation}
\frac{\partial A'_k}{\partial x'^{i}}=\frac{\partial}{\partial x'^{i}}\biggl(\frac{\partial x^m}{\partial x'^{k}}A_m\biggr)=\frac{\partial x^m}{\partial x'^{k}}\frac{\partial A_m}{\partial x'^{i}}+\frac{\partial^2 x^m}{\partial x'^{i}\partial x'^{k}}A_m=\frac{\partial x^l}{\partial x'^{i}}\frac{\partial x^m}{\partial x'^{k}}\frac{\partial A_m}{\partial x^l}+\frac{\partial^2 x^m}{\partial x'^{i}\partial x'^{k}}A_m,
\label{covdev1}
\end{equation}
because of the second term which is linear and homogeneous in $A_i$, unless $x^i$ are linear functions of $x'^{j}$ (for which their second derivatives vanish).
This term is symmetric in the indices $i,k$, thereby the antisymmetric part of $\partial A_k/\partial x^i$ with respect to these indices is a tensor:
\begin{equation}
\partial'_{[i} A'_{k]}=\frac{\partial x^l}{\partial x'^{[i}}\frac{\partial x^m}{\partial x'^{k]}}\partial_l A_m=\frac{\partial x^l}{\partial x'^{i}}\frac{\partial x^m}{\partial x'^{k}}\partial_{[l} A_{m]}.
\end{equation}

The {\em curl} of a covariant vector $A_i$ is defined as twice the antisymmetric part of $\partial_i A_k$: $\partial_i A_k-\partial_k A_i$, and is a tensor.
Similarly, completely antisymmetrized derivatives of tensors of rank $(0,2)$ and $(0,3)$, $\partial_{[i} B_{kl]}$ and $\partial_{[i} C_{klm]}$, are tensors.
If $B_{kl}=A_{[k,l]}$ then $\partial_{[i} B_{kl]}=0$, or conversely, if $\partial_{[i} B_{kl]}=0$ then there exists a vector $A_i$ such that $B_{kl}=A_{[k,l]}$.

The {\em divergence} of a tensor (or density) is a contracted derivative of this tensor (density): $\partial_i T^{il\dots}_{\phantom{il}jk\dots}$.
Because of the correspondence between tensors and dual densities, divergences of (completely antisymmetric if more than 1 index) contravariant densities are densities, dual to completely antisymmetrized derivatives of tensors:
\begin{equation}
\partial_i \mathfrak{C}^i=\epsilon^{iklm}\partial_{[i}C_{klm]},\quad \partial_k \mathfrak{B}^{ik}=\epsilon^{iklm}\partial_{[k}B_{lm]},\quad \partial_l \mathfrak{A}^{ikl}=\epsilon^{iklm}\partial_{[l}A_{m]}.
\end{equation}
For example, the equations $F_{[ik,l]}=0$ and $\mathfrak{F}^{ik}_{\phantom{ik},i}=\mathfrak{j}^k$, that describe Maxwell's electrodynamics (as in (\ref{eft4}) and (\ref{Max3})), are tensorial.
\newline
References: \cite{Schr,LL2}.

\subsection{Affine connection}
\setcounter{equation}{0}
\subsubsection{Covariant differentiation of tensors}
An ordinary derivative of a covariant vector $A_i$ is not a tensor, because its coordinate transformation law (\ref{covdev1}) contains an additional noncovariant term, linear and homogeneous in $A_i$.
Such a term vanishes only if $x^i$ are linear functions of $x'^{j}$, that is, if a coordinate transformation from the old to new coordinates (\ref{vec1}) is linear.
Let us consider a linear combination
\begin{equation}
A_{i;k}=A_{i,k}-\Gamma^{l}_{ik}A_l,
\end{equation}
where the quantity $\Gamma^{l}_{ik}$ (in the second term which is linear and homogeneous in $A_i$) transforms such that $A_{i;k}$ is a tensor:
\begin{equation}
A'_{i;k}=\frac{\partial x^l}{\partial x'^{i}}\frac{\partial x^m}{\partial x'^{k}}A_{l;m}=\frac{\partial x^l}{\partial x'^{i}}\frac{\partial x^m}{\partial x'^{k}}(A_{l,m}-\Gamma^{n}_{lm}A_n).
\end{equation}
Also, (\ref{covdev1}) gives
\begin{equation}
A'_{i;k}=A'_{i,k}-\Gamma'^{l}_{ik}A'_l=\frac{\partial x^m}{\partial x'^{k}}\frac{\partial x^l}{\partial x'^{i}}A_{l,m}+\frac{\partial^2 x^n}{\partial x'^{k}\partial x'^{i}}A_n-\frac{\partial x^n}{\partial x'^{l}}\Gamma'^{l}_{ik}A_n,
\end{equation}
so we obtain
\begin{equation}
\frac{\partial x^n}{\partial x'^{l}}\Gamma'^{l}_{ik}=\frac{\partial x^l}{\partial x'^{i}}\frac{\partial x^m}{\partial x'^{k}}\Gamma^{n}_{lm}+\frac{\partial^2 x^n}{\partial x'^{k}\partial x'^{i}}.
\end{equation}
Multiplying this equation by $\partial x'^{j}/\partial x^n$ gives the transformation law for $\Gamma^{l}_{ik}$:
\begin{equation}
\Gamma'^{j}_{ik}=\frac{\partial x'^{j}}{\partial x^n}\frac{\partial x^l}{\partial x'^{i}}\frac{\partial x^m}{\partial x'^{k}}\Gamma^{n}_{lm}+\frac{\partial x'^{j}}{\partial x^n}\frac{\partial^2 x^n}{\partial x'^{k}\partial x'^{i}}.
\label{affcon5}
\end{equation}

The algebraic object $\Gamma^{l}_{ik}$, which equips spacetime in order to covariantize a derivative of a vector, is referred to as the {\em affine connection}, affinity, or connection.
The connection has generally 64 independent components.
The tensor $A_{i;k}$ is the {\em covariant derivative} of a vector $A_i$ with respect to $x^i$.
The contracted affine connection transforms according to
\begin{equation}
\Gamma'^{i}_{ik}=\frac{\partial x^m}{\partial x'^{k}}\Gamma^{l}_{lm}+\frac{\partial x'^{i}}{\partial x^n}\frac{\partial^2 x^n}{\partial x'^{k}\partial x'^{i}}.
\label{affcon6}
\end{equation}
The affine connection is not a tensor because of the second term on the right-hand side of (\ref{affcon5}).

To denote a covariant derivative of a quantity $T$ with respect to $x^i$, we will use $T_{;i}$ or $\nabla_i T$.
A derivative of a scalar is a covariant vector.
Therefore, a covariant derivative of a scalar is equal to the corresponding partial derivative:
\begin{equation}
\phi_{;i}=\phi_{,i}.
\end{equation}
If we also assume that a covariant derivative of the product of two tensors obeys the same chain rule as a partial derivative:
\begin{equation}
(TU)_{;i}=T_{;i}U+TU_{;i},
\label{affcon9}
\end{equation}
then
\begin{equation}
A_{k,i}B^k+A_k B^k_{\phantom{k},i}=(A_k B^k)_{,i}=(A_k B^k)_{;i}=A_{k;i}B^k+A_k B^k_{\phantom{k};i}=A_{k,i}B^k-\Gamma^{k}_{li}A_k B^l+A_k B^k_{\phantom{k};i}.
\end{equation}
Therefore, we obtain a covariant derivative of a contravariant vector:
\begin{equation}
B^k_{\phantom{k};i}=B^k_{\phantom{k},i}+\Gamma^{k}_{li}B^l.
\end{equation}
Its contraction defines the {\em covariant divergence} of a contravariant vector: $B^i_{\phantom{i};i}$.
The chain rule (\ref{affcon9}) also infers that a covariant derivative of a tensor is equal to the sum of the corresponding partial derivative of this tensor and terms with the affine connection that covariantize each index:
\begin{equation}
T^{ij\dots}_{\phantom{ij}kl\dots;m}=T^{ij\dots}_{\phantom{ij}kl\dots,m}+\Gamma^{i}_{nm}T^{nj\dots}_{\phantom{nj}kl\dots}+\Gamma^{j}_{nm}T^{in\dots}_{\phantom{in}kl\dots}+\dots-\Gamma^{n}_{km}T^{ij\dots}_{\phantom{ij}nl\dots}-\Gamma^{n}_{lm}T^{ij\dots}_{\phantom{ij}kn\dots}-\dots\,.
\label{affcon12}
\end{equation}
A covariant derivative of the Kronecker symbol vanishes:
\begin{equation}
\delta^k_{l;i}=\Gamma^{k}_{ji}\delta^j_l-\Gamma^{j}_{li}\delta^k_j=0.
\end{equation}

The second term on the right of (\ref{affcon5}) does not depend on the affine connection, but only on the coordinate transformation.
Therefore, the difference between two different connections transforms like a tensor of rank (1,2).
Consequently, the variation $\delta\Gamma^{j}_{ik}$, which is an infinitesimal difference between two connections, is a tensor of rank (1,2).

\subsubsection{Parallel transport}
Let us consider two infinitesimally separated points in spacetime, $P(x^i)$ and $Q(x^i+dx^i)$, and a vector field $A$ which takes the value $A^k$ at $P$ and $A^k+dA^k$ at $Q$.
Because $dA^k=A^k_{\phantom{k}{,i}}dx^i$ and $A^k_{\phantom{k}{,i}}$ is not a tensor, the difference $dA^k$ between the vectors $A^k+dA^k$ and $A^k$ is not a vector.
The differential $dA^k$ is not a vector because it arises from subtracting two vectors which are located at two points with different coordinate transformation laws.
The transformation law for $dA^k$ follows from (\ref{covdev1}):
\begin{equation}
dA'_k=d\biggl(\frac{\partial x^m}{\partial x'^{k}}A_m\biggr)=\frac{\partial x^m}{\partial x'^{k}}dA_m+d\biggl(\frac{\partial x^m}{\partial x'^{k}}\biggr)A_m=\frac{\partial x^m}{\partial x'^{k}}dA_m+\frac{\partial^2 x^m}{\partial x'^{i}\partial x'^{k}}A_m dx'^{i}.
\end{equation}

In order to calculate the covariant difference between two vectors at two different points, we must bring these vectors to the same point.
Instead of subtracting from the vector $A^k+dA^k$ at $Q$ the vector $A^k$ at $P$, we must subtract a vector $A^k+\delta A^k$ at $Q$ that corresponds to $A^k$ at $P$, thereby that the resulting difference (covariant differential):
\begin{equation}
DA^k=dA^k-\delta A^k,
\label{partr2}
\end{equation}
is a vector.
The vector $A^k+\delta A^k$ is the {\em parallel-transported} or parallel-translated $A^k$ from $P$ to $Q$.
A parallel-transported linear combination of vectors must be equal to the same linear combination of parallel-transported vectors.
Therefore, $\delta A^k$ is a linear and homogeneous function of $A^k$.
It is also on the order of a differential, thus a linear and homogeneous function of $dx^i$.
The most general form of $\delta A^k$ is
\begin{equation}
\delta A^k=-\Gamma^{k}_{li}A^l dx^i,
\end{equation}
so
\begin{equation}
DA^k=dA^k+\Gamma^{k}_{li}A^l dx^i=A^k_{\phantom{k},i}dx^i+\Gamma^{k}_{li}A^l dx^i=A^k_{\phantom{k};i}dx^i.
\label{partr3}
\end{equation}
Because $\delta A^k$ is not a vector, $\Gamma^{k}_{li}$ is not a tensor.
Because $DA^k$ is a vector, $A^k_{\phantom{k};i}$ is a tensor.
The expressions for covariant derivatives of a covariant vector and tensors result from
\begin{equation}
\delta\phi=0,\quad \delta(TU)=\delta TU+T\delta U.
\end{equation}

\subsubsection{Torsion tensor}
The second term on the right-hand side of (\ref{affcon5}) is symmetric in the indices $i,k$.
Antisymmetrizing (\ref{affcon5}) with respect to these indices eliminates that term, giving
\begin{equation}
S'^j_{\phantom{j}ik}=\frac{\partial x'^{j}}{\partial x^n}\frac{\partial x^l}{\partial x'^{i}}\frac{\partial x^m}{\partial x'^{k}}S^n_{\phantom{n}lm},
\label{torten1}
\end{equation}
where
\begin{equation}
S^j_{\phantom{j}ik}=\Gamma^{j}_{[ik]}=\frac{1}{2}(\Gamma^{j}_{ik}-\Gamma^{j}_{ki})
\label{torten2}
\end{equation}
is the antisymmetric (in the covariant indices) part of the affine connection.
Equation (\ref{torten1}) is a transformation formula for a tensor.
Therefore, (\ref{torten2}) is a tensor, referred to as the Cartan {\em torsion tensor}.
The torsion tensor has generally 24 independent components.
The contracted torsion tensor,
\begin{equation}
S^k_{\phantom{k}ik}=S_i,
\end{equation}
is called the {\em torsion vector}.

\subsubsection{Covariant differentiation of densities}
The differential of the determinant ${\sf S}$ of a square matrix $S$ is given by
\begin{equation}
d{\sf S}=s^k_{\phantom{k}i}dS^i_{\phantom{i}k},
\end{equation}
where $s^k_{\phantom{k}i}$ is the minor corresponding to the component $S^i_{\phantom{i}k}$ of the matrix.
The components of the matrix $S^{-1}$ inverse to $S$,
\begin{equation}
S^i_{\phantom{i}j}(S^{-1})^j_{\phantom{j}k}=(S^{-1})^i_{\phantom{i}j}S^j_{\phantom{j}k}=\delta^i_k,
\end{equation}
are related to the minors of $S$ by
\begin{equation}
(S^{-1})^k_{\phantom{k}i}=\frac{s^k_{\phantom{k}i}}{{\sf S}}.
\end{equation}
The differential $d{\sf S}$ is therefore equal to
\begin{equation}
d{\sf S}={\sf S}(S^{-1})^k_{\phantom{k}i}dS^i_{\phantom{i}k}=-{\sf S}S^k_{\phantom{k}i}d(S^{-1})^i_{\phantom{i}k},
\label{covdifden4}
\end{equation}
which is equivalent to
\begin{equation}
\partial_l{\sf S}={\sf S}(S^{-1})^k_{\phantom{k}i}\partial_l S^i_{\phantom{i}k}=-{\sf S}S^k_{\phantom{k}i}\partial_l(S^{-1})^i_{\phantom{i}k}.
\label{covdifden5}
\end{equation}
Taking $S^i_{\phantom{i}k}=\partial x^i/\partial x'^k$ gives
\begin{equation}
\partial_l \biggl|\frac{\partial x^r}{\partial x'^{s}}\biggr|=\biggl|\frac{\partial x^r}{\partial x'^{s}}\biggr| \frac{\partial x'^{n}}{\partial x^m}\frac{\partial}{\partial x^l}\frac{\partial x^m}{\partial x'^{n}}.
\end{equation}

A derivative of a scalar density $\mathfrak{s}$ of weight $w$ does not transform like a covariant vector density:
\begin{eqnarray}
& & \partial'_i \mathfrak{s}'=\frac{\partial x^l}{\partial x'^{i}}\partial_l \biggl(\biggl|\frac{\partial x^j}{\partial x'^{k}}\biggr|^w \mathfrak{s}\biggr)=\frac{\partial x^l}{\partial x'^{i}}\biggl|\frac{\partial x^j}{\partial x'^{k}}\biggr|^w \partial_l \mathfrak{s}+w\frac{\partial x^l}{\partial x'^{i}}\biggl|\frac{\partial x^j}{\partial x'^{k}}\biggr|^{w-1}\partial_l \biggl|\frac{\partial x^r}{\partial x'^{s}}\biggr|\mathfrak{s} \nonumber \\
& & =\frac{\partial x^l}{\partial x'^{i}}\biggl|\frac{\partial x^j}{\partial x'^{k}}\biggr|^w \partial_l \mathfrak{s}+w\frac{\partial x^l}{\partial x'^{i}}\biggl|\frac{\partial x^j}{\partial x'^{k}}\biggr|^{w-1} \biggl|\frac{\partial x^r}{\partial x'^{s}}\biggr| \frac{\partial x'^{n}}{\partial x^m}\frac{\partial}{\partial x^l}\frac{\partial x^m}{\partial x'^{n}}\mathfrak{s} \nonumber \\
& & =\frac{\partial x^l}{\partial x'^{i}}\biggl|\frac{\partial x^j}{\partial x'^{k}}\biggr|^w \partial_l \mathfrak{s}+w\biggl|\frac{\partial x^j}{\partial x'^{k}}\biggr|^w\frac{\partial x'^{n}}{\partial x^m}\frac{\partial^2 x^m}{\partial x'^{n}\partial x'^{i}}\mathfrak{s}.
\label{covdifden7}
\end{eqnarray}
Let us consider a linear combination
\begin{equation}
\mathfrak{s}_{;i}=\mathfrak{s}_{,i}-w\Gamma_i \mathfrak{s},
\end{equation}
where the quantity $\Gamma_i$ transforms such that $\mathfrak{s}_{;i}$ is a vector density of weight $w$:
\begin{equation}
\mathfrak{s}'_{;i}=\frac{\partial x^l}{\partial x'^{i}}\biggl|\frac{\partial x^j}{\partial x'^{k}}\biggr|^w \mathfrak{s}_{;l}=\frac{\partial x^l}{\partial x'^{i}}\biggl|\frac{\partial x^j}{\partial x'^{k}}\biggr|^w(\mathfrak{s}_{,l}-w\Gamma_l \mathfrak{s}).
\end{equation}
Also, (\ref{covdifden7}) gives
\begin{equation}
\mathfrak{s}'_{;i}=\mathfrak{s}'_{,i}-w\Gamma'_i \mathfrak{s}'=\frac{\partial x^l}{\partial x'^{i}}\biggl|\frac{\partial x^j}{\partial x'^{k}}\biggr|^w \partial_l \mathfrak{s}+w\biggl|\frac{\partial x^j}{\partial x'^{k}}\biggr|^w\frac{\partial x'^{n}}{\partial x^m}\frac{\partial^2 x^m}{\partial x'^{n}\partial x'^{i}}\mathfrak{s}-w\biggl|\frac{\partial x^j}{\partial x'^{k}}\biggr|^w \Gamma'_i \mathfrak{s},
\end{equation}
so we obtain the transformation law for $\Gamma_i$:
\begin{equation}
\Gamma'_i=\frac{\partial x^l}{\partial x'^{i}}\Gamma_l+\frac{\partial x'^{n}}{\partial x^m}\frac{\partial^2 x^m}{\partial x'^{n}\partial x'^{i}},
\end{equation}
which is the same as the transformation law for $\Gamma^{k}_{ki}$ (\ref{affcon6}).
Therefore, the difference $\Gamma_i-\Gamma^{k}_{ki}$ is some covariant vector $V_i$.

If we assume that parallel transport of the product of a scalar density of any weight and a tensor obeys the chain rule:
\begin{equation}
\delta(\mathfrak{s}T)=\delta\mathfrak{s}T+\mathfrak{s}\delta T,
\end{equation}
so the covariant derivative of such product behaves like an ordinary derivative:
\begin{equation}
(\mathfrak{s}T)_{;i}=\mathfrak{s}_{;i}T+\mathfrak{s}T_{;i},
\end{equation}
then the covariant derivative of a tensor density of weight $w$ is equal to the sum of the corresponding ordinary derivative of this tensor, terms with the affine connection that covariantize each index, and the term with $\Gamma_i$:
\begin{eqnarray}
& & \mathfrak{T}^{ij\dots}_{\phantom{ij}kl\dots;m}=\mathfrak{T}^{ij\dots}_{\phantom{ij}kl\dots,m}+\Gamma^{i}_{nm}\mathfrak{T}^{nj\dots}_{\phantom{nj}kl\dots}+\Gamma^{j}_{nm}\mathfrak{T}^{in\dots}_{\phantom{in}kl\dots}+\dots \nonumber \\
& & -\Gamma^{n}_{km}\mathfrak{T}^{ij\dots}_{\phantom{ij}nl\dots}-\Gamma^{n}_{lm}\mathfrak{T}^{ij\dots}_{\phantom{ij}kn\dots}-\dots-w\Gamma_m \mathfrak{T}^{ij\dots}_{\phantom{ij}kl\dots}.
\label{covdifden14}
\end{eqnarray}

The covariant derivative of the contravariant Levi-Civita density is
\begin{equation}
\epsilon^{ijkl}_{\phantom{ijkl};m}=\Gamma^{i}_{nm}\epsilon^{njkl}+\Gamma^{j}_{nm}\epsilon^{inkl}+\Gamma^{k}_{nm}\epsilon^{ijnl}+\Gamma^{l}_{nm}\epsilon^{ijkn}-\Gamma_m \epsilon^{ijkl}.
\label{covdifden15}
\end{equation}
In the summations over $n$ only one term does not vanish for each term on the right-hand side of (\ref{covdifden15}), thereby
\begin{eqnarray}
& & \epsilon^{ijkl}_{\phantom{ijkl};m}=\Gamma^{i}_{n=i|m}\epsilon^{n=i|jkl}+\Gamma^{j}_{n=j|m}\epsilon^{i|n=j|kl}+\Gamma^{k}_{n=k|m}\epsilon^{ij|n=k|l}+\Gamma^{l}_{n=l|m}\epsilon^{ijk|n=l}-\Gamma_m \epsilon^{ijkl} \nonumber \\
& & =(\Gamma^{n}_{nm}-\Gamma_m)\epsilon^{ijkl}=-V_m \epsilon^{ijkl}.
\end{eqnarray}
The Levi-Civita symbol is a tensor density with constant components, thereby it does not change under a parallel transport, $\delta\epsilon=0$.
Therefore, we have
\begin{equation}
\epsilon^{ijkl}_{\phantom{ijkl};m}=0.
\label{covdifden17}
\end{equation}
By means of (\ref{KLC5}), we also have
\begin{equation}
\varepsilon_{ijkl;m}=0.
\label{covdifden18}
\end{equation}
Consequently, we obtain $V_i=0$ and
\begin{equation}
\Gamma_i=\Gamma^{k}_{ki}.
\end{equation}

\subsubsection{Antisymmetric covariant derivatives}
Completely antisymmetrized ordinary derivatives of tensors, $A_{[i,k]}$, $B_{[ik,l]}$ and $C_{[ikl,m]}$, are tensors because of their antisymmetry.
Completely antisymmetrized covariant derivatives of tensors are tensors because $\nabla_i$ is a covariant operation, and are given by direct calculation using the definition of the covariant derivative:
\begin{equation}
A_{[i;k]}=A_{[i,k]}-S^l_{\phantom{l}ik}A_l,\quad B_{[ik;l]}=B_{[ik,l]}-2S^m_{\phantom{m}[ik}B_{l]m}.
\end{equation}
Divergences of contravariant densities, $\mathfrak{C}^i_{\phantom{i},i}$, $\mathfrak{B}^{ik}_{\phantom{ik},i}$ and $\mathfrak{A}^{ikl}_{\phantom{ikl},i}$, which are completely antisymmetric for more than 1 index, are densities because of the correspondence between tensors and dual densities.
Covariant divergences of contravariant densities are densities, and are given by
\begin{eqnarray}
& & \mathfrak{C}^i_{\phantom{i};i}=\mathfrak{C}^i_{\phantom{i},i}+\Gamma^{i}_{ki}\mathfrak{C}^k-\Gamma_i\mathfrak{C}^i=\mathfrak{C}^i_{\phantom{i},i}+2S_i\mathfrak{C}^i,
\label{anticd2} \\
& & \mathfrak{B}^{ik}_{\phantom{ik};i}=\mathfrak{B}^{ik}_{\phantom{ik},i}-S^k_{\phantom{k}il}\mathfrak{B}^{il}+2S_i\mathfrak{B}^{ik},
\label{anticd3}
\end{eqnarray}
and similarly for other tensor ranks.

\subsubsection{Partial integration}
If the product of two quantities (tensors or densities) $TU$ is a contravariant density $\mathfrak{C}^k$ then
\begin{equation}
\int TU_{;k}d\Omega=\int(TU)_{;k}d\Omega-\int T_{;k}Ud\Omega=\int(TU)_{,k}d\Omega+2\int S_k TUd\Omega-\int T_{;k}Ud\Omega.
\end{equation}
The first term on the right-hand side can be transformed into a hypersurface integral $\int TUdS_k$.
If the region of integration extends to infinity and $\mathfrak{C}^k$ corresponds to some physical quantity then the boundary integral $\int TUdS_k$ vanishes, giving
\begin{equation}
\int TU_{;k}d\Omega=2\int S_k TUd\Omega-\int T_{;k}Ud\Omega.
\end{equation}
If $T=\delta^k_i$, then $U=\mathfrak{C}^i$ and
\begin{equation}
\int\mathfrak{C}^i_{\phantom{i};i}d\Omega=2\int S_i \mathfrak{C}^i d\Omega.
\label{parint3}
\end{equation}
Equations (\ref{anticd2}) and (\ref{parint3}) can be written as
\begin{equation}
\nabla_i^\ast\mathfrak{C}^i=\partial_i\mathfrak{C}^i,\quad \int\nabla_i^\ast\mathfrak{C}^i d\Omega=0,
\end{equation}
where
\begin{equation}
\nabla_i^\ast=\nabla_i-2S_i.
\label{parint5}
\end{equation}

\subsubsection{Geodesic frame of reference}
The transformation law (\ref{affcon5}) for the affine connection is equivalent to
\begin{equation}
\Gamma^{j}_{ik}=\frac{\partial x^j}{\partial x'^{n}}\frac{\partial x'^{l}}{\partial x^i}\frac{\partial x'^{m}}{\partial x^k}\Gamma'^{n}_{lm}+\frac{\partial x^j}{\partial x'^{n}}\frac{\partial^2 x'^{n}}{\partial x^i\partial x^k}.
\label{geofr1}
\end{equation}
Let us consider a coordinate transformation
\begin{equation}
x'^{k}=x^k+\frac{1}{2}a^k_{lm}x^l x^m,
\label{geofr2}
\end{equation}
where a constant quantity $a^k_{lm}$ is symmetric in the indices $l,m$.
Accordingly, the partial derivatives are
\begin{equation}
\frac{\partial x'^{k}}{\partial x^i}=\delta^k_i+a^k_{im}x^m,\quad \frac{\partial^2 x'^{k}}{\partial x^i\partial x^j}=a^k_{ij}.
\end{equation}
At the origin of the coordinates, $x^k=x'^{k}=0$, the first relation reduces to
\begin{equation}
\frac{\partial x'^{k}}{\partial x^i}=\delta^k_i.
\label{geofr4}
\end{equation}

Substituting these derivatives into (\ref{geofr1}) gives a relation between the two connections at the origin:
\begin{equation}
\Gamma^{j}_{ik}=\Gamma'^{j}_{ik}+a^j_{ik}.
\end{equation}
Putting
\begin{equation}
a^j_{ik}=\Gamma^{j}_{(ik)}|_{x^l=0}
\label{geofr6}
\end{equation}
gives
\begin{equation}
\Gamma'^{j}_{(ik)}=0.
\label{geofr7}
\end{equation}
Therefore, at a given point, there exists a coordinate frame of reference in which the symmetric part of the connection vanishes.
The relation (\ref{geofr4}) shows that the transformation (\ref{geofr2}) does not change the components of tensors at the origin.
If the affine connection is symmetric in the lower indices, $\Gamma^{j}_{ik}=\Gamma^{j}_{ki}$ (the torsion tensor vanishes), then (\ref{geofr7}) gives
\begin{equation}
\Gamma'^{j}_{ik}=0.
\end{equation}
The coordinate frame of reference in which the torsionless part of the connection vanishes at a given point is referred to as {\em locally geodesic}.

\subsubsection{Affine geodesics and four-velocity}
{\bf Autoparallel curve}.\\
Let us consider a point in spacetime $P(x^k)$ and a vector $dx^k$ at this point.
We construct a point $P'(x^k+dx^k)$ and find the vector $d'x^k$ which is the parallel-transported $dx^k$ from $P$ to $P'$.
Then we construct a point $P''(x^k+dx^k+d'x^k)$ and find the vector $d''x^k$ which is the parallel-transported $d'x^k$ from $P'$ to $P''$.
The next point is $P'''(x^k+dx^k+d'x^k+d''x^k)$, and so on.
Repeating this step constructs a polygonal line which in the limit $dx^k\rightarrow 0$ becomes a curve such that the vector $dx^k/d\lambda$ (where $\lambda$ is a parameter along the curve) tangent to it at any point, when parallely translated to another point on this curve, coincides with the tangent vector there.
Such curve is referred to as an {\em autoparallel curve} or {\em affine geodesic}.
Affine geodesics can be attributed with the concept of length, which, for the polygonal curve, is proportional to the number of parallel-transport steps described above.

The condition that parallel transport of a tangent vector be a tangent vector is
\begin{equation}
\frac{dx^i}{d\lambda}+\delta\biggl(\frac{dx^i}{d\lambda}\biggr)=\frac{dx^i}{d\lambda}-\Gamma^{i}_{kl}\frac{dx^k}{d\lambda}dx^l=M\biggl(\frac{dx^i}{d\lambda}+\frac{d^2x^i}{d\lambda^2}d\lambda\biggr),
\end{equation}
where the proportionality factor $M$ is some function of $\lambda$, or
\begin{equation}
M\frac{d^2x^i}{d\lambda^2}+\Gamma^{i}_{kl}\frac{dx^k}{d\lambda}\frac{dx^l}{d\lambda}=\frac{1-M}{d\lambda}\frac{dx^i}{d\lambda},
\label{affgeo2}
\end{equation}
from which it follows that $M$ must differ from 1 by the order of $d\lambda$.
In the first term on the left-hand side of (\ref{affgeo2}) we can therefore put $M=1$, and we denote $1-M$ by $\phi(\lambda)d\lambda$, thereby
\begin{equation}
\frac{d^2x^i}{d\lambda^2}+\Gamma^{i}_{kl}\frac{dx^k}{d\lambda}\frac{dx^l}{d\lambda}=\phi(\lambda)\frac{dx^i}{d\lambda}.
\label{affgeo3}
\end{equation}

If we replace $\lambda$ by a new variable $s(\lambda)$ then (\ref{affgeo3}) becomes
\begin{equation}
\frac{d^2x^i}{ds^2}+\Gamma^{i}_{kl}\frac{dx^k}{ds}\frac{dx^l}{ds}=\frac{\phi s'-s''}{s'^2}\frac{dx^i}{ds},
\label{affgeo4}
\end{equation}
where the prime denotes differentiation with respect to $\lambda$.
Requiring $\phi s'-s''=0$, which has a general solution $s=\int^\lambda d\lambda\,\exp[\int^\lambda \phi(x)dx]$, brings (\ref{affgeo4}) to the affine geodesic equation:
\begin{equation}
\frac{d^2x^i}{ds^2}+\Gamma^{i}_{kl}\frac{dx^k}{ds}\frac{dx^l}{ds}=0.
\label{affgeo5}
\end{equation} 
The scalar variable $s$ is referred to as the {\em affine parameter}.
The autoparallel equation (\ref{affgeo5}) is invariant under linear transformations $s\rightarrow as+b$ since the two lower limits of integration in the expression for $s(\lambda)$ are arbitrary.\\

\noindent
{\bf Four-velocity vector}.\\
We define the {\em four-velocity} vector:
\begin{equation}
u^i=\frac{dx^i}{ds}.
\label{affgeo6}
\end{equation}
This definition brings (\ref{partr3}) to
\begin{equation}
\frac{DA^k}{ds}=A^k_{\phantom{k};i}u^i,\quad \frac{dA^k}{ds}=A^k_{\phantom{k},i}u^i.
\label{affgeo7}
\end{equation}
Consequently, the autoparallel equation (\ref{affgeo5}) is
\begin{equation}
\frac{Du^i}{ds}=\frac{du^i}{ds}+\Gamma^{i}_{kl}u^k u^l=u^i_{\phantom{i};j}u^j=0.
\label{affgeo8}
\end{equation}
The relations (\ref{affgeo7}) can be generalized to any tensor density $T$:
\begin{equation}
\frac{DT}{ds}=T_{;i}u^i,\quad \frac{dT}{ds}=T_{,i}u^i,
\label{affgeo9}
\end{equation}
The vector $(dx^i/ds)|_Q$ is a parallel translation of $(dx^i/ds)|_P$.
Because $ds$ is a scalar, it is invariant under parallel transport, $ds|_Q=ds|_P$.
Therefore, the vector $dx^i|_Q$ is a parallel translation of $dx^i|_P$, thereby $ds$ measures the length of an infinitesimal section of an affine geodesic.

Only the symmetric part $\Gamma^{i}_{(kl)}$ of the connection enters the autoparallel equation (\ref{affgeo5}) because of the symmetry of $(dx^k/ds)(dx^l/ds)$ with respect to the indices $k,l$; affine geodesics do not depend on torsion.
At any point, a coordinate transformation to the geodesic frame (\ref{geofr1}) brings all the components $\Gamma^{i}_{(kl)}$ to zero, thereby the autoparallel equation becomes $du^i/ds=0$.\\

\noindent
{\bf Projective transformation}.\\
The autoparallel equation is also invariant under a {\em projective transformation}:
\begin{equation}
\Gamma^{i}_{kl}\rightarrow\Gamma^{i}_{kl}+\delta^i_k A_l,
\label{affgeo10}
\end{equation}
where $A_i$ is an arbitrary vector.
Substituting this transformation to (\ref{affgeo8}) gives
\begin{equation}
\frac{du^i}{ds}+\Gamma^{i}_{kl}u^k u^l=-u^i u^k A_k.
\label{affgeo11}
\end{equation}
If we replace $s$ by a new variable $\tilde{s}(s)$ then (\ref{affgeo11}) becomes
\begin{equation}
\frac{dU^i}{d\tilde{s}}+\Gamma^{i}_{kl}U^k U^l=-\frac{u^k A_k \tilde{s}'+\tilde{s}''}{\tilde{s}'^2}\frac{dx^i}{d\tilde{s}},
\label{affgeo12}
\end{equation}
where
\begin{equation}
U^i=\frac{dx^i}{d\tilde{s}}
\end{equation}
and the prime denotes differentiation with respect to $s$.
Requiring $u^k A_k \tilde{s}'+\tilde{s}''=0$, which has a general solution $\tilde{s}=\int^s ds\,\exp[-\int^s A_k u^k(x)dx]$, brings (\ref{affgeo12}) to
\begin{equation}
\frac{dU^i}{d\tilde{s}}+\Gamma^{i}_{kl}U^k U^l=0.
\end{equation}

\subsubsection{Infinitesimal coordinate transformations}
Let us consider a coordinate transformation
\begin{equation}
x'^{i}=x^i+\xi^i,
\label{infcor1}
\end{equation}
where $\xi^i=\delta x^i$ is an infinitesimal vector (a variation of $x^i$).
For a tensor or a tensor density $T$, we define
\begin{eqnarray}
& & \delta T=T'(x'^{i})-T(x^i), \\
& & \bar{\delta} T=T'(x^i)-T(x^i)=\delta T-\xi^k T_{,k}.
\label{infcor3}
\end{eqnarray}
The quantity $\bar{\delta}T$ is a difference of functions at the same coordinate point, so it is respectively a tensor or a tensor density.
For a scalar,
\begin{equation}
\delta\phi=0,\quad \bar{\delta}\phi=-\xi^k \phi_{,k}.
\label{infcor4}
\end{equation}
For a covariant vector,
\begin{eqnarray}
& & \delta A_i=\frac{\partial x^k}{\partial x'^{i}}A_k-A_i\approx -\xi^k_{\phantom{k}_{,i}}A_k, \label{infcor5} \\
& & \bar{\delta} A_i \approx -\xi^k_{\phantom{k}_{,i}}A_k-\xi^k A_{i,k}. \label{infcor6}
\end{eqnarray}
The variation (\ref{infcor5}) is not a tensor, but (\ref{infcor6}) is:
\begin{equation}
\bar{\delta} A_i= -\xi^k_{\phantom{k}_{;i}}A_k-\xi^k A_{i;k}-2S^j_{\phantom{j}ik}\xi^k A_j.
\end{equation}
We refer to $-\bar{\delta} T$ as the {\em Lie derivative} of $T$ along the vector $\xi^i$:
\begin{equation}
{\cal L}_{\xi}T=-\bar{\delta} T.
\label{infcor20}
\end{equation}

For a contravariant vector,
\begin{eqnarray}
& & \delta B^i=\frac{\partial x'^{i}}{\partial x^k}B^k-B^i= \xi^i_{\phantom{i}_{,k}}B^k, \\
& & \bar{\delta} B^i= \xi^i_{\phantom{i}_{,k}}B^k-\xi^k B^i_{\phantom{i},k}=\xi^i_{\phantom{i}_{;k}}B^k-\xi^k B^i_{\phantom{i};k}+2S^i_{\phantom{i}jk}\xi^k B^j.
\label{infcor9}
\end{eqnarray}
For a scalar density, we find
\begin{eqnarray}
& & \delta \mathfrak{s}=\biggl(\biggl|\frac{\partial x^i}{\partial x'^{i}}\biggr|-1\biggr)\mathfrak{s}\approx -\xi^i_{\phantom{i},i}\mathfrak{s}, \label{infcor10} \\
& & \bar{\delta}\mathfrak{s}\approx -\xi^i_{\phantom{i},i}\mathfrak{s}-\xi^k \mathfrak{s}_{,k}=-\xi^i_{\phantom{i};i}\mathfrak{s}-\xi^k \mathfrak{s}_{;k}+2S_i \xi^i \mathfrak{s}.
\end{eqnarray}
The chain rule for $\delta$ infers that for a tensor density of weight $w$ (which includes tensors as densities of weight 0), we have
\begin{eqnarray}
& & \delta \mathfrak{T}^{ij\dots}_{\phantom{ij}kl\dots}\approx \xi^i_{\phantom{i},m}\mathfrak{T}^{mj\dots}_{\phantom{mj}kl\dots}+\xi^j_{\phantom{i},m}\mathfrak{T}^{im\dots}_{\phantom{im}kl\dots}+\dots-\xi^m_{\phantom{m},k}\mathfrak{T}^{ij\dots}_{\phantom{ij}ml\dots}-\xi^m_{\phantom{m},l}\mathfrak{T}^{ij\dots}_{\phantom{ij}km\dots}-\dots \nonumber \\
& & -w\xi^m_{\phantom{m},m}\mathfrak{T}^{ij\dots}_{\phantom{ij}kl\dots}, \\
& & \bar{\delta}\mathfrak{T}^{ij\dots}_{\phantom{ij}kl\dots}\approx \xi^i_{\phantom{i};m}\mathfrak{T}^{mj\dots}_{\phantom{mj}kl\dots}+\xi^j_{\phantom{i};m}\mathfrak{T}^{im\dots}_{\phantom{im}kl\dots}+\dots-\xi^m_{\phantom{m};k}\mathfrak{T}^{ij\dots}_{\phantom{ij}ml\dots}-\xi^m_{\phantom{m};l}\mathfrak{T}^{ij\dots}_{\phantom{ij}km\dots}-\dots \nonumber \\
& & -w\xi^m_{\phantom{m};m}\mathfrak{T}^{ij\dots}_{\phantom{ij}kl\dots}-\xi^m \mathfrak{T}^{ij\dots}_{\phantom{ij}kl\dots;m}+2S^i_{\phantom{i}nm}\xi^m \mathfrak{T}^{nj\dots}_{\phantom{nj}kl\dots}+2S^j_{\phantom{j}nm}\xi^m \mathfrak{T}^{in\dots}_{\phantom{in}kl\dots}+\dots \nonumber \\
& & -2S^n_{\phantom{n}km}\xi^m \mathfrak{T}^{ij\dots}_{\phantom{ij}nl\dots}-2S^n_{\phantom{n}lm}\xi^m \mathfrak{T}^{ij\dots}_{\phantom{ij}kn\dots}-\dots+2wS_m \xi^m \mathfrak{T}^{ij\dots}_{\phantom{ij}kl\dots}.
\label{infcor13}
\end{eqnarray}
A Lie derivative of a tensor density of rank $(k,l)$ and weight $w$ is a tensor density of rank $(k,l)$ and weight $w$.

The formula for the covariant derivative of $T$ can be written as
\begin{equation}
T_{;k}=T_{,k}+\Gamma^{j}_{ik}G^i_j T,
\label{infcor14}
\end{equation}
where $G^i_j$ is a tensor operator acting on tensor densities:
\begin{equation}
G^i_j \phi=0,\quad G^i_j A_k=-\delta^i_k A_j,\quad G^i_j B^k=\delta^k_j B^i,\quad G^i_j \mathfrak{s}=-\delta^i_j \mathfrak{s},
\end{equation}
or generally
\begin{equation}
G^m_n \mathfrak{T}^{ij\dots}_{\phantom{ij}kl\dots}=\delta^i_n \mathfrak{T}^{mj\dots}_{\phantom{mj}kl\dots}+\delta^j_n \mathfrak{T}^{im\dots}_{\phantom{im}kl\dots}+\dots-\delta^m_k \mathfrak{T}^{ij\dots}_{\phantom{ij}nl\dots}-\delta^m_l \mathfrak{T}^{ij\dots}_{\phantom{ij}kn\dots}-\dots-w\delta^m_n \mathfrak{T}^{ij\dots}_{\phantom{ij}kl\dots}.
\end{equation}
This operator also appears in the formula for $\delta T$:
\begin{equation}
\delta T=\xi^i_{\phantom{i},k}G^k_i T.
\label{infcor17}
\end{equation}

\subsubsection{Killing vectors}
A covariant vector $\zeta_i$ that satisfies
\begin{equation}
\zeta_{(i;k)}=0
\label{Kilvec1}
\end{equation}
is referred to as a {\em Killing vector}.
Along an affine geodesic,
\begin{equation}
\frac{D}{ds}(u^i\zeta_i)=u^k(u^i\zeta_i)_{;k}=u^i u^k \zeta_{i;k}+\zeta_i u^k u^i_{\phantom{i};k}=0.
\label{Kilvec2}
\end{equation}
The first term in the sum in (\ref{Kilvec2}) vanishes because of the definition of $\zeta_i$ and the second term vanishes because of the affine geodesic equation.
Therefore, to each Killing vector $\zeta_i$ there corresponds a quantity $u^i \zeta_i$ which does not change along the affine geodesic:
\begin{equation}
u^i \zeta_i=\mbox{const}.
\label{Kilvec3}
\end{equation}
\newline
References: \cite{Schr,LL2,Lord}.

\subsection{Curvature}
\setcounter{equation}{0}
\subsubsection{Curvature tensor}
We define the {\em commutator} $[A,B]$ of two operators $A$ and $B$ as
\begin{equation}
[A,B]=AB-BA=-[B,A].
\end{equation}
The commutator of covariant derivatives is thus
\begin{equation}
[\nabla_i,\nabla_k]=2\nabla_{[i}\nabla_{k]}.
\end{equation}
The commutator of covariant derivatives of a contravariant vector is a tensor:
\begin{eqnarray}
& & [\nabla_j,\nabla_k]B^i=2\nabla_{[j}\nabla_{k]}B^i=2\partial_{[j}\nabla_{k]}B^i-2\Gamma^{l}_{[kj]}\nabla_l B^i+2\Gamma^{i}_{l[j}\nabla_{k]} B^l \nonumber \\
& & =2\partial_{[j}(\Gamma^{i}_{|m|k]}B^m)+2S^l_{\phantom{l}jk}\nabla_l B^i+2\Gamma^{i}_{l[j}\partial_{k]} B^l+2\Gamma^{i}_{l[j}\Gamma^{l}_{|m|k]}B^m \nonumber \\
& & =2(\partial_{[j}\Gamma^{i}_{|m|k]}+\Gamma^{i}_{l[j}\Gamma^{l}_{|m|k]})B^m+2S^l_{\phantom{l}jk}\nabla_l B^i=R^i_{\phantom{i}mjk}B^m+2S^l_{\phantom{l}jk}\nabla_l B^i,
\label{curten3}
\end{eqnarray}
where $||$ embraces indices which are excluded from symmetrization or antisymmetrization.
Therefore, $R^i_{\phantom{i}mjk}$, defined as
\begin{equation}
R^i_{\phantom{i}mjk}=\partial_{j}\Gamma^{i}_{mk}-\partial_{k}\Gamma^{i}_{mj}+\Gamma^{l}_{mk}\Gamma^{i}_{lj}-\Gamma^{l}_{mj}\Gamma^{i}_{lk},
\end{equation}
is a tensor, referred to as the {\em curvature tensor}.\\

\noindent
{\bf Antisymmetry of curvature tensor in last two indices}.\\
The curvature tensor $R^i_{\phantom{i}mjk}$ is antisymmetric in its last two indices:
\begin{equation}
R^i_{\phantom{i}mjk}=-R^i_{\phantom{i}mkj}.
\end{equation}
Therefore, it has generally 96 independent components.
The commutator of covariant derivatives of a covariant vector is
\begin{equation}
[\nabla_j,\nabla_k]A_i=-R^m_{\phantom{m}ijk}A_m+2S^l_{\phantom{l}jk}\nabla_l A_i,
\label{curten5}
\end{equation}
and the commutator of covariant derivatives of a tensor is
\begin{eqnarray}
& & [\nabla_j,\nabla_k]T^{im\dots}_{\phantom{im}lp\dots}=R^i_{\phantom{i}njk}T^{nm\dots}_{\phantom{nm}lp\dots}+R^m_{\phantom{m}njk}T^{in\dots}_{\phantom{in}lp\dots}+\dots-R^n_{\phantom{n}ljk}T^{im\dots}_{\phantom{im}np\dots}-R^n_{\phantom{n}pjk}T^{im\dots}_{\phantom{im}ln\dots} \nonumber \\
& & -\dots+2S^l_{\phantom{l}jk}\nabla_l T^{im\dots}_{\phantom{im}lp\dots}.
\label{curten6}
\end{eqnarray}

\noindent
{\bf Curvature tensor for different connection}.\\
A change in the connection,
\begin{equation}
\tilde{\Gamma}^{i}_{jk}=\Gamma^{i}_{jk}+T^i_{\phantom{i}jk},
\label{curten7}
\end{equation}
where $T^i_{\phantom{i}jk}$ is a tensor, results in the following change of the curvature tensor:
\begin{eqnarray}
& & \tilde{R}^i_{\phantom{i}klm}=\tilde{\Gamma}^{i}_{km,l}-\tilde{\Gamma}^{i}_{kl,m}+\tilde{\Gamma}^{j}_{km}\tilde{\Gamma}^{i}_{jl}-\tilde{\Gamma}^{j}_{kl}\tilde{\Gamma}^{i}_{jm}=\Gamma^{i}_{km,l}-\Gamma^{i}_{kl,m}+\Gamma^{j}_{km}\Gamma^{i}_{jl}-\Gamma^{j}_{kl}\Gamma^{i}_{jm} \nonumber \\
& & +T^i_{\phantom{i}km,l}-T^i_{\phantom{i}kl,m}+\Gamma^{j}_{km}T^i_{\phantom{i}jl}-\Gamma^{j}_{kl}T^i_{\phantom{i}jm}+\Gamma^{i}_{jl}T^j_{\phantom{j}km}-\Gamma^{i}_{jm}T^j_{\phantom{j}kl}+T^j_{\phantom{j}km}T^i_{\phantom{i}jl}-T^j_{\phantom{j}kl}T^i_{\phantom{i}jm} \nonumber \\
& & =R^i_{\phantom{i}klm}+T^i_{\phantom{i}km;l}-T^i_{\phantom{i}kl;m}+T^j_{\phantom{j}km}T^i_{\phantom{i}jl}-T^j_{\phantom{j}kl}T^i_{\phantom{i}jm}.
\label{curten8}
\end{eqnarray}
For a projective transformation (\ref{affgeo10}), $T^i_{\phantom{i}jk}=\delta^i_j A_k$, the curvature tensor changes according to
\begin{equation}
\tilde{R}^i_{\phantom{i}klm}=R^i_{\phantom{i}klm}+\delta^i_k(A_{m;l}-A_{l;m}).
\end{equation}

\noindent
{\bf Variation of curvature tensor}.\\
The variation of the curvature tensor is
\begin{eqnarray}
& & \delta R^i_{\phantom{i}klm}=(\delta\Gamma^{i}_{km})_{,l}-(\delta\Gamma^{i}_{kl})_{,m}+\delta\Gamma^{i}_{jl}\Gamma^{j}_{km}+\Gamma^{i}_{jl}\delta\Gamma^{j}_{km}-\delta\Gamma^{i}_{jm}\Gamma^{j}_{kl}-\Gamma^{i}_{jm}\delta\Gamma^{j}_{kl} \nonumber \\
& & =(\delta\Gamma^{i}_{km})_{;l}-\Gamma^{i}_{jl}\delta\Gamma^{j}_{km}+\Gamma^{j}_{kl}\delta\Gamma^{i}_{jm}+\Gamma^{j}_{ml}\delta\Gamma^{i}_{kj}-(\delta\Gamma^{i}_{kl})_{;m}+\Gamma^{i}_{jm}\delta\Gamma^{j}_{kl}-\Gamma^{j}_{km}\delta\Gamma^{i}_{jl} \nonumber \\
& & -\Gamma^{j}_{lm}\delta\Gamma^{i}_{kj}+\delta\Gamma^{i}_{jl}\Gamma^{j}_{km}+\Gamma^{i}_{jl}\delta\Gamma^{j}_{km}-\delta\Gamma^{i}_{jm}\Gamma^{j}_{kl}-\Gamma^{i}_{jm}\delta\Gamma^{j}_{kl} \nonumber \\
& & =(\delta\Gamma^{i}_{km})_{;l}-(\delta\Gamma^{i}_{kl})_{;m}-2S^n_{\phantom{n}lm}\delta\Gamma^{i}_{kn}.
\label{curten10}
\end{eqnarray}

\subsubsection{Integrability of connection}
The affine connection is {\em integrable} if parallel transport of a vector from point $P$ to point $Q$ is independent of a path along which this vector is parallelly translated, or equivalently, parallel transport of a vector around a closed curve does not change this vector.
For an integrable connection, we can uniquely translate parallelly a given vector $h^i$ at point $P$ to all points in spacetime:
\begin{equation}
\delta h^i=dh^i,
\end{equation}
or
\begin{equation}
h^i_{\phantom{i},k}=-\Gamma^{i}_{jk}h^j.
\label{intcon2}
\end{equation}
Therefore, we have
\begin{equation}
(\Gamma^{i}_{jk}h^j)_{,l}-(\Gamma^{i}_{jl}h^j)_{,k}=\Gamma^{i}_{jk,l}h^j-\Gamma^{i}_{jk}\Gamma^{j}_{ml}h^m-\Gamma^{i}_{jl,k}h^j+\Gamma^{i}_{jl}\Gamma^{j}_{mk}h^m=R^i_{\phantom{i}jlk}h^j=0,
\end{equation}
so, because $h^i$ is arbitrary,
\begin{equation}
R^i_{\phantom{i}klm}=0.
\end{equation}
Spacetime with a vanishing curvature tensor $R^i_{\phantom{i}klm}=0$ is {\em flat}.
Let us consider 4 linearly independent vectors $h^i_a$, where $a$ is 1,2,3,4, and vectors inverse to $h^i_a$:
\begin{equation}
\sum_a h^i_a h_{ka}=\delta^i_k.
\end{equation}
If the affine connection is integrable then (\ref{intcon2}) becomes
\begin{equation}
h^i_{a,k}=-\Gamma^{i}_{lk}h^l_a.
\label{intcon6}
\end{equation}
Multiplying (\ref{intcon6}) by $h_{ja}$ gives
\begin{equation}
\Gamma^{i}_{jk}=-h_{ja}h^i_{a,k}=h_{ja,k}h^i_a.
\label{intcon7}
\end{equation}
An integrable connection has thus 16 independent components.
If the connection is also symmetric, $S^i_{\phantom{i}jk}=0$, then
\begin{equation}
h_{ja,k}-h_{ka,j}=0,
\end{equation}
which is the condition for the independence of the coordinates
\begin{equation}
y_a=\int_P^Q h_{ia}dx^i
\label{intcon9}
\end{equation}
of the path of integration $PQ$.
Adopting $y_a$ as the new coordinates (with point $P=(0,0,0,0)$ in the center) gives
\begin{equation}
\frac{\partial y_a}{\partial x^i}=h_{ia},\quad \frac{\partial x^i}{\partial y_a}=h^i_a,
\label{intcon10}
\end{equation}
so (\ref{intcon7}) becomes
\begin{equation}
\Gamma^{i}_{jk}(x^i)=\frac{\partial x^i}{\partial y_a}\frac{\partial^2 y_a}{\partial x^k\partial x^j}.
\end{equation}
The transformation law for the connection (\ref{affcon5}) gives (with $y_a$ corresponding to $x'^j$)
\begin{equation}
\Gamma^{i}_{jk}(y_a)=0.
\end{equation}
A torsionless integrable connection can be thus transformed to zero; one can always find a system of coordinates which is geodesic everywhere.
If a connection is symmetric but nonintegrable then a geodesic frame of reference can be constructed only at a given point (or along a given world line).

\subsubsection{Parallel transport along closed curve}
Let us consider parallel transport of a covariant vector around an infinitesimal closed curve.
Such a transport changes this vector, according to Stokes' theorem (\ref{covint5}) by
\begin{eqnarray}
& & \Delta A_k=\oint\delta A_k=\oint\Gamma^{i}_{kl}A_i dx^l=\frac{1}{2}\int\biggl(\frac{\partial(\Gamma^{i}_{km}A_i)}{\partial x^l}-\frac{\partial(\Gamma^{i}_{kl}A_i)}{\partial x^m}\biggr)df^{lm} \nonumber \\
& & =\frac{1}{2}\int\biggl(\frac{\partial\Gamma^{i}_{km}}{\partial x^l}A_i-\frac{\partial\Gamma^{i}_{kl}}{\partial x^m}A_i+\Gamma^{i}_{km}\frac{\partial A_i}{\partial x^l}-\Gamma^{i}_{kl}\frac{\partial A_i}{\partial x^m}\biggr)df^{lm},
\label{ptcc1}
\end{eqnarray}
where $df^{lm}$ is the element of an infinitesimal surface bounded by the curve.
This change is a vector because it is a difference between two vectors at the same point.
Along the curve, the change of $A_i$ is caused by the parallel transport: $\delta A_i=\Gamma^{n}_{il}A_n dx^l$, which gives $\partial A_i/\partial x^l=\Gamma^{n}_{il}A_n$ on the curve.
Because the curve is infinitesimal, this derivative is also satisfied on the bounded surface inside the curve.
This derivative can be therefore substituted into (\ref{ptcc1}), giving
\begin{equation}
\Delta A_k\approx\frac{1}{2}\int\biggl[\biggl(\frac{\partial\Gamma^{i}_{km}}{\partial x^l}-\frac{\partial\Gamma^{i}_{kl}}{\partial x^m}\biggr)A_i+(\Gamma^{i}_{km}\Gamma^{n}_{il}-\Gamma^{i}_{kl}\Gamma^{n}_{im})A_n\biggr]df^{lm}\approx\frac{1}{2}R^i_{\phantom{i}klm}A_i\Delta f^{lm},
\end{equation}
where $\Delta f^{lm}=\int df^{lm}$ is the area of the surface and the curvature tensor is evaluated at some point inside the curve.
The change of a contravariant vector in parallel transport around an infinitesimal closed curve results from $\Delta(A_k B^k)=0$:
\begin{equation}
\Delta B^k\approx-\frac{1}{2}R^k_{\phantom{k}ilm}B^i\Delta f^{lm},
\end{equation}
and the corresponding change of a tensor results from the chain rule for parallel transport:
\begin{equation}
\Delta T^{ik\dots}_{\phantom{ik}np\dots}\approx-\frac{1}{2}(R^i_{\phantom{i}jlm}T^{jk\dots}_{\phantom{jk}np\dots}+R^k_{\phantom{k}jlm}T^{ij\dots}_{\phantom{ij}np\dots}+\dots-R^j_{\phantom{j}nlm}T^{ik\dots}_{\phantom{ik}jp\dots}-R^j_{\phantom{j}plm}T^{ik\dots}_{\phantom{ik}nj\dots}-\dots)\Delta f^{lm}.
\end{equation}

\subsubsection{Bianchi identities}
Let us consider
\begin{equation}
\nabla_j\nabla_{[k}\nabla_{l]}B^i=\frac{1}{2}\nabla_j(R^i_{\phantom{i}mkl}B^m)+\nabla_j(S^m_{\phantom{m}kl}\nabla_m B^i)
\label{Biaide1}
\end{equation}
and
\begin{eqnarray}
& & \nabla_{[j}\nabla_{k]}\nabla_l B^i=-\frac{1}{2}R^m_{\phantom{m}ljk}\nabla_m B^i+\frac{1}{2}R^i_{\phantom{i}mjk}\nabla_l B^m+S^m_{\phantom{m}jk}\nabla_m\nabla_l B^i=-\frac{1}{2}R^m_{\phantom{m}ljk}\nabla_m B^i \nonumber \\
& & +\frac{1}{2}R^i_{\phantom{i}mjk}\nabla_l B^m+S^m_{\phantom{m}jk}\nabla_l\nabla_m B^i+S^m_{\phantom{m}jk}R^i_{\phantom{i}nml}B^n+2S^m_{\phantom{m}jk}S^n_{\phantom{n}ml}\nabla_n B^i.
\label{Biaide2}
\end{eqnarray}
Total antisymmetrization of the indices $j,k,l$ in (\ref{Biaide1}) and (\ref{Biaide2}) gives
\begin{equation}
\nabla_{[j}\nabla_k\nabla_{l]}B^i=\frac{1}{2}\nabla_{[j}R^i_{\phantom{i}|m|kl]}B^m+\frac{1}{2}R^i_{\phantom{i}m[kl]}\nabla_{j]}B^m+\nabla_{[j}S^m_{\phantom{m}kl]}\nabla_m B^i+S^m_{\phantom{m}kl}\nabla_{j]}\nabla_m B^i
\end{equation}
and
\begin{eqnarray}
& & \nabla_{[j}\nabla_k\nabla_{l]}B^i=-\frac{1}{2}R^m_{\phantom{m}[ljk]}\nabla_m B^i+\frac{1}{2}R^i_{\phantom{i}m[jk}\nabla_{l]}B^m+S^m_{\phantom{m}[jk}\nabla_{l]}\nabla_m B^i \nonumber \\
& & +S^m_{\phantom{m}[jk}R^i_{\phantom{i}|nm|l]}B^n+2S^m_{\phantom{m}[jk}S^n_{\phantom{n}|m|l]}\nabla_n B^i,
\end{eqnarray}
so
\begin{eqnarray}
& & \frac{1}{2}\nabla_{[j}R^i_{\phantom{i}|m|kl]}B^m+\nabla_{[j}S^m_{\phantom{m}kl]}\nabla_m B^i=-\frac{1}{2}R^m_{\phantom{m}[ljk]}\nabla_m B^i+S^m_{\phantom{m}[jk}R^i_{\phantom{i}|nm|l]}B^n \nonumber \\
& & +2S^m_{\phantom{m}[jk}S^n_{\phantom{n}|m|l]}\nabla_n B^i.
\label{Biaide5}
\end{eqnarray}
Comparing terms in (\ref{Biaide5}) with $B^i$ gives the {\em Bianchi identity}:
\begin{equation}
R^i_{\phantom{i}n[jk;l]}=2R^i_{\phantom{i}nm[j}S^m_{\phantom{m}kl]},
\label{Biaide6}
\end{equation}
while comparing terms with $\nabla_k B^i$ gives the Ricci {\em cyclic identity}:
\begin{equation}
R^m_{\phantom{m}[jkl]}=-2S^m_{\phantom{m}[jk;l]}+4S^m_{\phantom{m}n[j}S^n_{\phantom{n}kl]}.
\label{Biaide7}
\end{equation}
Contracting (\ref{Biaide6}) and (\ref{Biaide7}) with respect to one contravariant and one covariant index gives
\begin{eqnarray}
& & R^i_{\phantom{i}n[ik;l]}=2R^i_{\phantom{i}nm[i}S^m_{\phantom{m}kl]}, \label{Biaide8} \\
& & R^k_{\phantom{k}[jkl]}=-2S^k_{\phantom{k}[jk;l]}+4S^k_{\phantom{k}n[j}S^n_{\phantom{n}kl]}.
\label{Biaide9}
\end{eqnarray}
For a symmetric connection, $S^i_{\phantom{i}jk}=0$, the Bianchi identity and the cyclic identity reduce to
\begin{eqnarray}
& & R^i_{\phantom{i}n[jk;l]}=0, \label{Biaide10} \\
& & R^m_{\phantom{m}[jkl]}=0.
\label{Biaide11}
\end{eqnarray}
The cyclic identity (\ref{Biaide11}) imposes 16 constraints on the curvature tensor, thereby the curvature tensor with a vanishing torsion has 80 independent components.

\subsubsection{Ricci tensor}
Contraction of the curvature tensor with respect to the contravariant index and the second covariant index gives the {\em Ricci tensor}:
\begin{equation}
R_{ik}=R^j_{\phantom{j}ijk}=\Gamma^{j}_{ik,j}-\Gamma^{j}_{ij,k}+\Gamma^{l}_{ik}\Gamma^{j}_{lj}-\Gamma^{l}_{ij}\Gamma^{j}_{lk}.
\end{equation}
Contraction of the curvature tensor with respect to the contravariant index and the third covariant index gives the Ricci tensor with the opposite sign because of the antisymmetry of the curvature tensor with respect to its last indices.
Contraction of the curvature tensor with respect to the contravariant index and the first covariant index gives the homothetic or {\em segmental curvature} tensor:
\begin{equation}
Q_{ik}=R^j_{\phantom{j}jik}=\Gamma^{j}_{jk,i}-\Gamma^{j}_{ji,k},
\label{Ric2}
\end{equation}
which is a curl.
A change in the connection (\ref{curten7}) results in the following changes of the Ricci tensor and segmental curvature tensor:
\begin{eqnarray}
& & R_{ik}\rightarrow R_{ik}+T^l_{\phantom{l}ik;l}-T^l_{\phantom{i}il;k}+T^j_{\phantom{j}ik}T^l_{\phantom{l}jl}-T^j_{\phantom{j}il}T^l_{\phantom{l}jk}, \\
& & Q_{ik}\rightarrow Q_{ik}+T^j_{\phantom{j}jk,i}-T^j_{\phantom{j}ji,k}.
\end{eqnarray}
For a projective transformation (\ref{affgeo10})
\begin{eqnarray}
& & R_{ik}\rightarrow R_{ik}+A_{k;i}-A_{i;k}, \label{Ric5} \\
& & Q_{ik}\rightarrow Q_{ik}+4(A_{k,i}-A_{i,k}).
\end{eqnarray}
Therefore, the symmetric part of the Ricci tensor is invariant under projective transformations.
The variation of the Ricci tensor is
\begin{equation}
\delta R_{ik}=(\delta\Gamma^{l}_{ik})_{;l}-(\delta\Gamma^{l}_{il})_{;k}-2S^j_{\phantom{j}lk}\delta\Gamma^{l}_{ij},
\label{Ric7}
\end{equation}
which follows from (\ref{curten10}), whereas the variation of the segmental curvature tensor is
\begin{equation}
\delta Q_{ik}=(\delta\Gamma^{j}_{jk})_{,i}-(\delta\Gamma^{j}_{ji})_{,k}.
\end{equation}

\subsubsection{Geodesic deviation}
Let us consider a family of affine geodesics characterized by the affine parameter $s$, measured along each curve from its point of intersection with a given hypersurface, and distinguished by a scalar parameter $t$: $x^i=x^i(s,t)$.
We define
\begin{equation}
v^i=\frac{\partial x^i}{\partial t},
\end{equation}
which gives
\begin{equation}
v^i_{\phantom{i};k}u^k-u^i_{\phantom{i};k}v^k=v^i_{\phantom{i},k}u^k-u^i_{\phantom{i},k}v^k-2S^i_{\phantom{i}kl}u^k v^l=\frac{du^i}{dt}-\frac{dv^i}{ds}-2S^i_{\phantom{i}kl}u^k v^l=-2S^i_{\phantom{i}kl}u^k v^l,
\end{equation}
where $u^i=\partial x^i/\partial s$ is the four-velocity along each curve.
We therefore have
\begin{eqnarray}
& & \frac{D^2 v^i}{ds^2}=(v^i_{\phantom{i};j}u^j)_{;k}u^k=(u^i_{\phantom{i};j}v^j)_{;k}u^k-2(S^i_{\phantom{i}kl}u^k v^l)_{;j}u^j \nonumber \\
& & =u^i_{\phantom{i};jk}v^j u^k+u^i_{\phantom{i};j}v^j_{\phantom{j};k}u^k-2(S^i_{\phantom{i}kl}u^k v^l)_{;j}u^j \nonumber \\
& & =u^i_{\phantom{i};kj}v^j u^k-R^i_{\phantom{i}ljk}u^l v^j u^k-2S^l_{\phantom{l}jk}u^i_{\phantom{i};l}v^j u^k+u^i_{\phantom{i};j}v^j_{\phantom{j};k}u^k-2(S^i_{\phantom{i}kl}u^k v^l)_{;j}u^j \nonumber \\
& & =u^i_{\phantom{i};kj}v^j u^k-R^i_{\phantom{i}ljk}u^l v^j u^k-2S^l_{\phantom{l}jk}u^i_{\phantom{i};l}v^j u^k+u^i_{\phantom{i};j}(u^j_{\phantom{j};k}v^k-2S^j_{\phantom{i}kl}u^k v^l) \nonumber \\
& & -2(S^i_{\phantom{i}kl}u^k v^l)_{;j}u^j=(u^i_{\phantom{i};k}u^k)_{;j}v^j+R^i_{\phantom{i}jkl}u^j u^k v^l-2(S^i_{\phantom{i}kl}u^k v^l)_{;j}u^j \nonumber \\
& & =R^i_{\phantom{i}jkl}u^j u^k v^l-2\frac{D}{ds}(S^i_{\phantom{i}kl}u^k v^l),
\label{geodev3}
\end{eqnarray}
which can be written as
\begin{equation}
\frac{D}{ds}\biggl(\frac{Dv^i}{ds}+2S^i_{\phantom{i}kl}u^k v^l \biggr)=R^i_{\phantom{i}jkl}u^j u^k v^l.
\label{geodev4}
\end{equation}
This is the equation of {\em geodesic deviation}.
If we replace affine geodesics by arbitrary curves then $u^i_{\phantom{i};k}u^k\neq0$ and (\ref{geodev4}) becomes
\begin{equation}
\frac{D}{ds}\biggl(\frac{Dv^i}{ds}+2S^i_{\phantom{i}kl}u^k v^l \biggr)=R^i_{\phantom{i}jkl}u^j u^k v^l+(u^i_{\phantom{i};k}u^k)_{;j}v^j.
\label{geodev5}
\end{equation}
The separation vector
\begin{equation}
\xi^i=v^i dt
\end{equation}
connects points on two infinitely close affine geodesics with $t$ and $t+dt$ for the same $s$.
Multiplying (\ref{geodev3}) by $dt$ gives another form of the equation of geodesic deviation,
\begin{equation}
\frac{D^2 \xi^i}{ds^2}=R^i_{\phantom{i}jkl}u^j u^k \xi^l-2\frac{D}{ds}(S^i_{\phantom{i}kl}u^k \xi^l).
\label{geodev7}
\end{equation}
\newline
References: \cite{Schr,LL2,Lord,Hehl1}.

\subsection{Metric}
\setcounter{equation}{0}
\subsubsection{Metric tensor}
\label{metrictensor}
The affine parameter $s$ is a measure of the length only along an affine geodesic.
In order to extend the concept of length to all points in spacetime, the spacetime is equipped with an algebraic object $g_{ik}$, referred to as the  {\em metric tensor} and defined as
\begin{equation}
ds^2=g_{ik}dx^i dx^k.
\label{metten1}
\end{equation}
The quantity $ds$ in (\ref{metten1}) is called the {\em line element}.
The metric tensor is a symmetric covariant tensor of rank (0,2):
\begin{equation}
g_{ik}=g_{ki}.
\end{equation}
The {\em inverse metric tensor} $g^{ik}$ is defined as the tensor inverse to $g_{ik}$:
\begin{equation}
g_{ij}g^{ik}=\delta_j^k.
\label{metten7}
\end{equation}
The inverse metric tensor is a symmetric contravariant tensor of rank (2,0):
\begin{equation}
g^{ik}=g^{ki}.
\end{equation}
The affine parameter $s$, whose differential is given by (\ref{metten1}), is referred to as the {\em interval}.\\

\noindent
{\bf Metricity}.\\
Because the differential $ds$ of the affine parameter does not change under parallel transport along an affine geodesic from point $P(x^i)$ to point $Q(x^i+dx^i)$, $ds|_Q=ds|_P$, and $dx^i|_Q$ is a parallel translation of $dx^i|_P$, the relation (\ref{metten1}) imposes that $g_{ik}|_Q=g_{ik}|_P+g_{ik,j}dx^j$ is a parallel translation of $g_{ik}|_P$:
\begin{equation}
g_{ik}|_Q=g_{ik}|_P+\delta g_{ik}.
\end{equation}
Consequently, the covariant differential of the metric tensor is zero:
\begin{equation}
Dg_{ik}=g_{ik;j}dx^j=dg_{ik}-\delta g_{ik}=g_{ik,j}dx^j-\delta g_{ik}=0.
\end{equation}
Therefore, the covariant derivative of the metric tensor is zero:
\begin{equation}
g_{ik;j}=0.
\label{metten5}
\end{equation}
This relation is equivalent to
\begin{equation}
g_{ik,j}-\Gamma^{l}_{ij}g_{lk}-\Gamma^{l}_{kj}g_{il}=0.
\end{equation}
Because the inverse metric tensor is a function of the metric tensor only, its covariant derivative is zero:
\begin{equation}
g^{ik}_{\phantom{ik};j}=0.
\end{equation}
This relation also follows from the covariant derivative of (\ref{metten7}):
\begin{equation}
g_{ij;l}g^{ik}+g_{ij}g^{ik}_{\phantom{ik};l}=0.
\end{equation}
The relation (\ref{metten5}) is referred to as {\em metricity} or {\em metric compatibility} of the affine connection, and relates the connection to the metric tensor.\\

\begin{figure}[th]
\centering
\includegraphics[width=2.3in]{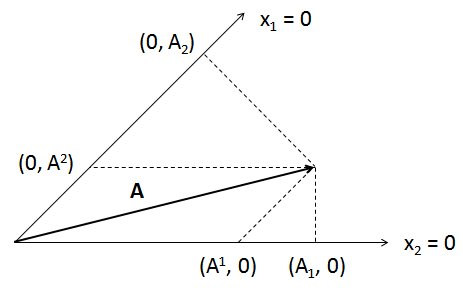}
\caption{Contravariant and covariant components of a vector.}
\label{vector}
\end{figure}

\noindent
{\bf Raising and lowering indices}.\\
The metric tensor allows to associate covariant and contravariant vectors:
\begin{eqnarray}
& & A^i=g^{ik}A_k, \label{metten9} \\
& & B_i=g_{ik}B^k,
\label{metten10}
\end{eqnarray}
because such an association is also consistent with the covariant differentials of these vectors which are vectors:
\begin{equation}
DA^i=D(g^{ik}A_k)=g^{ik}DA_i,\quad DB_i=D(g_{ik}B^k)=g_{ik}DB^k.
\end{equation}
The operation in (\ref{metten9}) {\em raises} a coordinate index and the operation in (\ref{metten10}) {\em lowers} a coordinate index.
The operations of raising and lowering of indices commute with covariant differentiation with respect to $\Gamma^{\rho}_{\mu\,\nu}$.
For covariant and contravariant indices of tensors and densities this association is
\begin{eqnarray}
& & g_{im}\mathfrak{T}^{ij\dots}_{\phantom{ij}kl\dots}=\mathfrak{T}^{\phantom{m}j\dots}_{m\phantom{j}kl\dots}, \\
& & g^{km}\mathfrak{T}^{ij\dots}_{\phantom{ij}kl\dots}=\mathfrak{T}^{ijm\dots}_{\phantom{ijm}l\dots}.
\end{eqnarray}
The operations of raising and lowering of indices can be applied several times, for example:
\begin{equation}
T^{ik}=g^{ij}g^{kl}T_{jl}.
\end{equation}
The contravariant and covariant components of a two-dimensional vector are shown in Figure \ref{vector}.\\

\noindent
{\bf Change of metric tensor under coordinate transformation}.\\
Under a coordinate transformation $x'^{i}=x^i+\xi^i(x^k)$ (\ref{infcor1}), where $\xi^i$ are infinitesimal functions of the coordinates, the inverse transformation law (\ref{tens2}) for the contravariant metric tensor gives
\begin{equation}
g'^{ik}(x'^j)=g^{lm}(x^j)\frac{\partial x'^i}{\partial x^l}\frac{\partial x'^k}{\partial x^m}=g^{lm}(\delta^i_l+\xi^i_{\phantom{i},l})(\delta^k_m+\xi^k_{\phantom{k},m})\approx g^{ik}(x^j)+g^{im}\xi^k_{\phantom{k},m}+g^{lk}\xi^i_{\phantom{i},l},
\end{equation}
omitting terms quadratic in $\xi^i$.
To represent all terms as functions of the same coordinate point, we expand $g'^{ik}(x^j+\xi^j)$ in powers of $\xi^j$, neglecting terms of higher order in $\xi^j$: $g'^{ik}(x'^j)=g'^{ik}(x^j)+\xi^j g'^{ik}_{\phantom{ik},j}\approx g'^{ik}(x^j)+\xi^j g^{ik}_{\phantom{ik},j}$.
The last equality in this relation, replacing $g'^{ik}$ with $g^{ik}$, can be applied because the difference between these two tensors multiplied by $\xi^j$ is of higher order in $\xi^j$.
Consequently, we obtain
\begin{equation}
g'^{ik}(x^j)=g^{ik}(x^j)+g^{im}\xi^k_{\phantom{k},m}+g^{lk}\xi^i_{\phantom{i},l}-\xi^j g^{ik}_{\phantom{ik},j}.
\label{metten603}
\end{equation}
This relation is equivalent to
\begin{equation}
-{\cal L}_{\xi}g^{ik}=\bar{\delta}g^{ik}=g'^{ik}(x^j)-g^{ik}(x^j)=2\xi^{(i;k)}+4S^{(ik)}_{\phantom{(ik)}l}\xi^l,
\label{metten21}
\end{equation}
where $\bar{\delta}$ is given by (\ref{infcor3}), ${\cal L}_{\xi}$ is the Lie derivative along the vector $\xi^i$ (\ref{infcor20}), and we denote $^{;i}= _{;k}g^{ik}$.
It is a special case of (\ref{infcor13}).\\

\noindent
{\bf Normalization of four-velocity}.\\
The four-velocity vector (\ref{affgeo6}) is normalized because of (\ref{metten1}):
\begin{equation}
u^i u_i=g_{ik}u^i u^k=\frac{g_{ik}dx^i dx^k}{ds^2}=1.
\label{metten22}
\end{equation}
This vector thus has 3 independent components.\\

\noindent
{\bf Determinant of metric tensor}.\\
Let us consider the determinant of the matrix composed from the components of the covariant metric tensor $g_{ik}$:
\begin{equation}
\mathfrak{g}=|g_{ik}|.
\end{equation}
The differential and derivatives of the determinant of the metric tensor are given, following (\ref{covdifden4}) and (\ref{covdifden5}), by
\begin{eqnarray}
& & d\mathfrak{g}=\mathfrak{g}g^{ik}dg_{ik}=-\mathfrak{g}g_{ik}dg^{ik}, \label{metten16} \\
& & \mathfrak{g}_{,l}=\mathfrak{g}g^{ik}g_{ik,l}=-\mathfrak{g}g_{ik}g^{ik}_{\phantom{ik},l}.
\label{metten17}
\end{eqnarray}
The variation of the determinant of the metric tensor is thus
\begin{equation}
\delta\mathfrak{g}=\mathfrak{g}g^{ik}\delta g_{ik}=-\mathfrak{g}g_{ik}\delta g^{ik}.
\label{metten18}
\end{equation}
The covariant derivative of the determinant of the metric tensor vanishes:
\begin{equation}
\mathfrak{g}_{;j}=0.
\label{metten19}
\end{equation}

\noindent
{\bf Completely antisymmetric unit pseudotensor}.\\
The square root of the absolute value of the determinant of the metric tensor, $\sqrt{|\mathfrak{g}|}$, is a scalar density of weight 1, according to (\ref{dens5}).
We can use it to construct from the Levi-Civita symbols a quantity which behaves like a tensor with respect to continuous coordinate transformations:
\begin{eqnarray}
& & E_{iklm}=\sqrt{|\mathfrak{g}|}\varepsilon_{iklm}, \label{metten14} \\
& & E^{iklm}=\frac{1}{\sqrt{|\mathfrak{g}|}}\epsilon^{iklm}=g^{in}g^{kp}g^{lq}g^{mr}E_{npqr}.
\label{metten15}
\end{eqnarray}
If we change the sign of one or three of the coordinates, then the components of $E^{iklm}$ do not change because $\epsilon^{iklm}$ and $\varepsilon_{iklm}$ have the same components in all coordinate systems, whereas some of the components of a tensor change sign.
The components (\ref{metten14}) and (\ref{metten15}) are thus referred to as those of the {\em completely antisymmetric unit pseudotensor}.
The relations (\ref{KLC8}) are also valid if we replace $\epsilon$ and $\varepsilon$ by $E$.
The contractions of $E$ with tensors have the same property under changing the sign of the coordinates and therefore are pseudotensors, including pseudovectors and pseudoscalars.
The relations (\ref{covdifden17}) and (\ref{covdifden18}) give
\begin{equation}
E^{ijkl}_{\phantom{ijkl};m}=0,\quad E_{ijkl;m}=0.
\label{metten20}
\end{equation}

\noindent
{\bf Invariant integrals}.\\
To preserve the invariant character of integrals, the elements of integration $df^\star_{ik}$ (\ref{covint1}), $dS_i$ (\ref{covint3}), and $d\Omega$ (\ref{covint4}) must be multiplied by a scalar density, which can be taken as $\sqrt{|\mathfrak{g}|}$.
A pseudotensor element
\begin{equation}
\sqrt{|\mathfrak{g}|}df^\star_{ik}=\frac{1}{2}E_{iklm}df^{lm}
\label{covint21}
\end{equation}
geometrically describes an element of surface normal to $df^{ik}$ and equal in magnitude to its area.
A pseudotensor element
\begin{equation}
\sqrt{|\mathfrak{g}|}dS_i=-\frac{1}{6}E_{iklm}dS^{klm}
\label{covint22}
\end{equation}
geometrically describes an element of hypersurface normal to $dS^{ikl}$ and equal in magnitude to its volume.
A pseudoscalar element $\sqrt{|\mathfrak{g}|}d\Omega$ is an invariant element of four-volume.
Its invariance follows from the transformation law (\ref{dens6}).
The relation (\ref{covint5}) for a vector $A_i$ gives
\begin{equation}
\oint A_i dx^i=\int df^{ki}\frac{\partial A_i}{\partial x^k}=\frac{1}{2}\int\Bigl(\frac{\partial A_k}{\partial x^i}-\frac{\partial A_i}{\partial x^k}\Bigr)df^{ik}. \label{covint23}
\end{equation}
The relation (\ref{covint6}) for an antisymmetric tensor density $\sqrt{|\mathfrak{g}|}F^{ik}=-\sqrt{|\mathfrak{g}|}F^{ki}$ gives
\begin{equation}
\frac{1}{2}\oint F^{ik}\sqrt{|\mathfrak{g}|}df^\star_{ik}=\frac{1}{2}\int\Bigl(dS_i \frac{\partial(\sqrt{|\mathfrak{g}|} F^{ik})}{\partial x^k}-dS_k \frac{\partial(\sqrt{|\mathfrak{g}|} F^{ik})}{\partial x^i}\Bigr)=\int\frac{\partial(\sqrt{|\mathfrak{g}|}F^{ik})}{\partial x^k}dS_i. \label{covint24}
\end{equation}
The relation (\ref{covint7}) for a vector density $\sqrt{|\mathfrak{g}|}B^i$ gives
\begin{equation}
\oint B^i\sqrt{|\mathfrak{g}|}dS_i=\int \frac{\partial(\sqrt{|\mathfrak{g}|}B^i)}{\partial x^i}d\Omega.
\label{covint25}
\end{equation}

\subsubsection{Christoffel symbols}
The metricity condition (\ref{metten5}) imposes 40 constraints on the affine connection:
\begin{eqnarray}
& & g_{ik;j}+g_{kj;i}-g_{ji;k}=g_{ik,j}-\Gamma^{l}_{ij}g_{lk}-\Gamma^{l}_{kj}g_{il}+g_{kj,i}-\Gamma^{l}_{ki}g_{lj}-\Gamma^{l}_{ji}g_{kl}-g_{ji,k}+\Gamma^{l}_{jk}g_{li} \nonumber \\
& & +\Gamma^{l}_{ik}g_{jl}=g_{ik,j}+g_{kj,i}-g_{ji,k}-2\Gamma^{l}_{(ij)}g_{kl}-2S^l_{\phantom{l}kj}g_{il}-2S^l_{\phantom{l}ki}g_{jl}=0.
\label{Chrsym1}
\end{eqnarray}
Multiplying (\ref{Chrsym1}) by $g^{km}$ and contracting gives
\begin{equation}
\Gamma^{m}_{(ij)}=\mathring{\Gamma}^{m}_{ij}+2S_{(ij)}^{\phantom{(ij)}m},
\end{equation}
where
\begin{equation}
\mathring{\Gamma}^{m}_{ij}=\frac{1}{2}g^{mk}(g_{kj,i}+g_{ki,j}-g_{ij,k})
\label{Chrsym3}
\end{equation}
are referred to as the {\em Christoffel symbols}.
Using (\ref{metten7}), they can be written as
\begin{equation}
\mathring{\Gamma}^{m}_{ij}=-\frac{1}{2}(g_{kj}g^{mk}_{\phantom{mk},i}+g_{ki}g^{mk}_{\phantom{mk},j}-g^{mk}g_{il}g_{jn}g^{ln}_{\phantom{ln},k}).
\label{Chrsym4}
\end{equation}
The Christoffel symbols are symmetric in their covariant indices:
\begin{equation}
\mathring{\Gamma}^{k}_{ij}=\mathring{\Gamma}^{k}_{ji}.
\end{equation}
Multiplying (\ref{Chrsym3}) by $g_{mk}$ and contracting gives
\begin{equation}
\mathring{\Gamma}^{m}_{ij}g_{mk}=\frac{1}{2}(g_{kj,i}+g_{ki,j}-g_{ij,k}).
\label{Chrsym69}
\end{equation}

\noindent
{\bf Contortion tensor}.\\
Because $\Gamma^{k}_{ij}=\Gamma^{k}_{(ij)}+S^k_{\phantom{k}ij}$, the metric-compatible affine connection is equal to
\begin{equation}
\Gamma^{k}_{ij}=\mathring{\Gamma}^{k}_{ij}+C^k_{\phantom{k}ij},
\label{Chrsym6}
\end{equation}
where
\begin{equation}
C^k_{\phantom{k}ij}=S_{ij}^{\phantom{ij}k}+S_{ji}^{\phantom{ji}k}+S^k_{\phantom{k}ij}
\label{Chrsym7}
\end{equation}
is the {\em contortion tensor}.
This tensor is antisymmetric in its first two indices:
\begin{equation}
C_{ijk}=-C_{jik}.
\end{equation}
The inverse relation between the torsion and contortion tensors is
\begin{equation}
S^i_{\phantom{i}jk}=C^i_{\phantom{i}[jk]}.
\end{equation}
The Christoffel symbols are the torsionless part of the connection.\\

\noindent
{\bf Levi-Civita connection}.\\
The difference between two affine connections is a tensor, thereby the sum of a connection and a tensor of rank (1,2) is a connection.
Therefore, the Christoffel symbols form a connection, referred to as the {\em Levi-Civita connection}.
We define the covariant derivative with respect to the Levi-Civita connection analogously to (\ref{affcon12}) and (\ref{covdifden14}), with $\Gamma^{k}_{ij}$ replaced by $\mathring{\Gamma}^{k}_{ij}$, and denote it $_{:i}$ instead of $_{;i}$, or $\mathring{\nabla}_i$ instead of $\nabla_i$.
The covariant derivative with respect to the Levi-Civita connection of the metric tensor vanishes, as for that with respect to any connection:
\begin{equation}
g_{ik:j}=g_{ik,j}-\mathring{\Gamma}^{l}_{ij}g_{lk}-\mathring{\Gamma}^{l}_{kj}g_{il}=0.
\label{Chrsym10}
\end{equation}
This equation agrees with (\ref{Chrsym3}) and gives the relation between derivatives of the metric tensor and the Christoffel symbols:
\begin{equation}
g_{ik,j}=\mathring{\Gamma}^{l}_{ij}g_{lk}+\mathring{\Gamma}^{l}_{kj}g_{il}.
\label{Chrsym100}
\end{equation}
Similarly, we have
\begin{equation}
g^{ik}_{\phantom{ik}:j}=g^{ik}_{\phantom{ik},j}+\mathring{\Gamma}^{i}_{lj}g^{lk}+\mathring{\Gamma}^{k}_{lj}g^{il}=0.
\label{Chrsym11}
\end{equation}
The Levi-Civita covariant derivative of a tensor density with respect to the interval $s$ is, analogously to (\ref{affgeo9}),
\begin{equation}
\frac{\mathring{D}T}{ds}=T_{:i}u^i.
\end{equation}

Since the covariant derivatives of the Levi-Civita symbols are equal to zero, according to (\ref{covdifden17}) and (\ref{covdifden18}), their covariant derivatives with respect to the Levi-Civita connection are zero:
\begin{equation}
\epsilon^{ijkl}_{\phantom{ijkl}:m}=0,\quad \varepsilon_{ijkl:m}=0.
\label{Chrsym20}
\end{equation}
The following Levi-Civita covariant derivatives are also zero:
\begin{equation}
\mathfrak{g}_{:j}=0,\quad E^{ijkl}_{\phantom{ijkl}:m}=0,\quad E_{ijkl:m}=0.
\end{equation}

\noindent
{\bf Lie derivative of metric tensor}.\\
The Lie derivative of the contravariant metric tensor along an infinitesimal vector $\xi^i$ (\ref{metten21}) can be written in terms of the covariant derivative with respect to the Levi-Civita connection:
\begin{equation}
-{\cal L}_{\xi}g^{ik}=\bar{\delta}g^{ik}=\xi^{i:k}+\xi^{k:i},
\end{equation}
where $^{:i}= _{:k}g^{ik}$.
For the covariant metric tensor, we have
\begin{equation}
-{\cal L}_{\xi}g_{ik}=\bar{\delta}g_{ik}=-\xi_{i:k}-\xi_{k:i},
\label{Chrsym22}
\end{equation}
so that the condition $g_{ij}g'^{kj}=\delta_i^k$ is satisfied to terms of first order. 
A Killing vector (\ref{Kilvec1}) for the Levi-Civita connection satisfies
\begin{equation}
\zeta_{(i:k)}=0.
\label{Chrsym23}
\end{equation}
This vector therefore becomes a generator of a transformation
\begin{equation}
x'^{i}=x^i+\epsilon\zeta^i,
\label{Chrsym24}
\end{equation}
where $\epsilon$ is an infinitesimal scalar, which coincides with (\ref{infcor1}) for $\xi^i=\epsilon\zeta^i$.
Such transformations are {\em isometries}: they do not change the metric tensor.\\

\noindent
{\bf Identities with Levi-Civita connection}.\\
The following formulae are satisfied:
\begin{eqnarray}
& & \mathring{\Gamma}^{k}_{ki}=\frac{1}{2}g^{jk}g_{jk,i}=-\frac{1}{2}g_{jk}g^{jk}_{\phantom{jk},i}=\frac{1}{2}\frac{\mathfrak{g},i}{\mathfrak{g}}=(\textrm{ln}\sqrt{|\mathfrak{g}|})_{,i}, \label{Chrsym14} \\
& & \mathring{\Gamma}^{k}_{ij}g^{ij}=-\frac{1}{\sqrt{|\mathfrak{g}|}}(\sqrt{|\mathfrak{g}|}g^{ik})_{,i}, \\
& & B^i_{\phantom{i}:i}=\frac{1}{\sqrt{|\mathfrak{g}|}}(\sqrt{|\mathfrak{g}|}B^i)_{,i}, \label{Chrsym16} \\
& & F^{ik}_{\phantom{ik}:i}=\frac{1}{\sqrt{|\mathfrak{g}|}}(\sqrt{|\mathfrak{g}|}F^{ik})_{,i},
\label{Chrsym17} \\
& & A_{i:k}-A_{k:i}=A_{i,k}-A_{k,i},
\label{Chrsym18}
\end{eqnarray}
where $F^{ik}=-F^{ki}$.
The quantities (\ref{Chrsym16}) and (\ref{Chrsym17}) are the covariant divergences of a vector $B^i$ and a tensor $F^{ik}$.
The quantity (\ref{Chrsym18}) is the covariant curl of a vector $A_i$.
The covariant divergences of contravariant densities (\ref{anticd2}) and (\ref{anticd3}) with respect to the Levi-Civita connection reduce to
\begin{equation}
\mathfrak{C}^i_{\phantom{i}:i}=\mathfrak{C}^i_{\phantom{i},i},\quad \mathfrak{B}^{ik}_{\phantom{ik}:i}=\mathfrak{B}^{ik}_{\phantom{ik},i}.
\label{Chrsym234}
\end{equation}
The covariant divergence of a contravariant vector density with respect to the Levi-Civita connection is thus equal to the covariant derivative using $\nabla_i^\ast$ (\ref{parint5}).

The Christoffel symbols satisfy all formulae that are satisfied by $\Gamma^{k}_{ij}$ in which $S^i_{\phantom{i}jk}=0$.
Because the Levi-Civita connection is a symmetric connection, it can be brought to zero by transforming the coordinates to a geodesic frame of reference.
In a geodesic frame, the covariant derivative $\mathring{\nabla}_i$ with respect to the Levi-Civita connection coincides with the ordinary derivative $\partial_i$.
The integral relations (\ref{covint23}), (\ref{covint24}), and (\ref{covint25}) can be written in a covariant form:
\begin{eqnarray}
& & \oint A_i dx^i=\frac{1}{2}\int(A_{k:i}-A_{i:k})df^{ik}, \label{Chrsym26} \\
& & \frac{1}{2}\oint F^{ik}\sqrt{|\mathfrak{g}|}df^\star_{ik}=\int F^{ik}_{\phantom{ik}:k}\sqrt{|\mathfrak{g}|}dS_i, \\
& & \oint B^i\sqrt{|\mathfrak{g}|}dS_i=\int B^i_{\phantom{i}:i}\sqrt{|\mathfrak{g}|}d\Omega.
\label{Chrsym19}
\end{eqnarray}

\noindent
{\bf Variation of Christoffel symbols}.\\
The variation of the Levi-Civita connection is, as for any connection, a tensor:
\begin{eqnarray}
& & \delta\mathring{\Gamma}^{k}_{ij}=\frac{1}{2}g^{kl}\bigl((\delta g_{lj})_{,i}+(\delta g_{li})_{,j}-(\delta g_{ij})_{,l}\bigr)+\frac{1}{2}\delta g^{kl}(g_{lj,i}+g_{li,j}-g_{ij,l}) \nonumber \\
& & =\frac{1}{2}g^{kl}\bigl((\delta g_{lj})_{:i}+(\delta g_{li})_{:j}-(\delta g_{ij})_{:l}\bigr)+\frac{1}{2}g^{kl}(\mathring{\Gamma}^{m}_{li}\delta g_{mj}+\mathring{\Gamma}^{m}_{ji}\delta g_{lm}+\mathring{\Gamma}^{m}_{lj}\delta g_{mi}+\mathring{\Gamma}^{m}_{ij}\delta g_{lm} \nonumber \\
& & -\mathring{\Gamma}^{m}_{il}\delta g_{mj}-\mathring{\Gamma}^{m}_{jl}\delta g_{im})+\delta g^{kl}\mathring{\Gamma}^{m}_{ij}g_{lm}=\frac{1}{2}g^{kl}\bigl((\delta g_{lj})_{:i}+(\delta g_{li})_{:j}-(\delta g_{ij})_{:l}\bigr) \nonumber \\
& & +g^{kl}\mathring{\Gamma}^{m}_{ij}\delta g_{lm}+\delta g^{kl}\mathring{\Gamma}^{m}_{ij}g_{lm}=\frac{1}{2}g^{kl}\bigl((\delta g_{lj})_{:i}+(\delta g_{li})_{:j}-(\delta g_{ij})_{:l}\bigr)+\mathring{\Gamma}^{m}_{ij}\delta\delta^k_m \nonumber \\
& & =\frac{1}{2}g^{kl}\bigl((\delta g_{lj})_{:i}+(\delta g_{li})_{:j}-(\delta g_{ij})_{:l}\bigr).
\end{eqnarray}

\noindent
{\bf Antisymmetry of curvature tensor in first two indices}.\\
The commutator of covariant derivatives (\ref{curten6}) of the metric tensor gives
\begin{equation}
[\nabla_j,\nabla_k]g_{im}=-R^n_{\phantom{n}ijk}g_{nm}-R^n_{\phantom{n}mjk}g_{in}+2S^l_{\phantom{l}jk}\nabla_l g_{im}=
-R_{mijk}-R_{imjk}=0.
\label{Chrsym199}
\end{equation}
Consequently, the curvature tensor is antisymmetric in its first two indices:
\begin{equation}
R_{ijkl}=-R_{jikl},
\end{equation}
and the segmental curvature tensor (\ref{Ric2}) vanishes.
Consequently, there is only one independent way to contract the curvature tensor, which gives the Ricci tensor up to a sign.

\subsubsection{Riemann tensor}
The commutator of covariant derivatives with respect to the Levi-Civita connection of a covariant vector is
\begin{equation}
[\mathring{\nabla}_j,\mathring{\nabla}_k]A_i=-\mathring{R}^m_{\phantom{m}ijk}A_m,
\label{Riem0}
\end{equation}
analogously to (\ref{curten5}) and without the torsion tensor of this connection that vanishes.
The curvature tensor constructed from the Levi-Civita connection is referred to as the Riemannian curvature tensor or the {\em Riemann tensor}:
\begin{equation}
\mathring{R}^i_{\phantom{i}mjk}=\partial_{j}\mathring{\Gamma}^{i}_{mk}-\partial_{k}\mathring{\Gamma}^{i}_{mj}+\mathring{\Gamma}^{i}_{lj}\mathring{\Gamma}^{l}_{mk}-\mathring{\Gamma}^{i}_{lk}\mathring{\Gamma}^{l}_{mj}.
\label{Riem1}
\end{equation}
Similarly, the commutators of covariant derivatives of a contravariant vector and of a tensor are respectively given by (\ref{curten3}) and (\ref{curten6}), in which $R^i_{\phantom{i}jkl}$ is replaced with $\mathring{R}^i_{\phantom{i}jkl}$ and $S^i_{\phantom{i}jk}=0$.
The commutator of covariant derivatives of the metric tensor vanishes:
\begin{equation}
[\mathring{\nabla}_j,\mathring{\nabla}_k]g_{lp}=-\mathring{R}^m_{\phantom{m}ljk}g_{mp}-\mathring{R}^m_{\phantom{m}pjk}g_{lm}=0,
\end{equation}
so the covariant Riemann tensor $\mathring{R}_{imjk}$ is also antisymmetric in the indices $i,m$.\\

\noindent
{\bf Symmetry and antisymmetry properties of Riemann tensor}.\\
Substituting (\ref{Chrsym3}) in (\ref{Riem1}) gives
\begin{equation}
\mathring{R}_{iklm}=\frac{1}{2}(g_{im,kl}+g_{kl,im}-g_{il,km}-g_{km,il})+g_{jn}(\mathring{\Gamma}^{j}_{im}\mathring{\Gamma}^{n}_{kl}-\mathring{\Gamma}^{j}_{il}\mathring{\Gamma}^{n}_{km}),
\label{Riem3}
\end{equation}
which explicitly shows the following symmetry and antisymmetry properties:
\begin{eqnarray}
& & \mathring{R}_{iklm}=-\mathring{R}_{ikml}, \\
& & \mathring{R}_{iklm}=-\mathring{R}_{kilm}, \label{Riem5} \\
& & \mathring{R}_{iklm}=\mathring{R}_{lmik}.
\end{eqnarray}
Accordingly, the {\em Riemannian Ricci tensor} is symmetric:
\begin{equation}
\mathring{R}_{ik}=\mathring{R}^j_{\phantom{j}ijk}=\mathring{R}_{ki}.
\label{Riem7}
\end{equation}

\noindent
{\bf Relation between curvature and Riemann tensors}.\\
Substituting (\ref{Chrsym6}) in (\ref{curten7}) and (\ref{curten8}) gives the relation between the curvature and Riemann tensors:
\begin{equation}
R^i_{\phantom{i}klm}=\mathring{R}^i_{\phantom{i}klm}+C^i_{\phantom{i}km:l}-C^i_{\phantom{i}kl:m}+C^j_{\phantom{j}km}C^i_{\phantom{i}jl}-C^j_{\phantom{j}kl}C^i_{\phantom{i}jm}.
\label{Riem8}
\end{equation}
Contracting (\ref{Riem8}) with respect to in the indices $i,l$ gives
\begin{equation}
R_{km}=\mathring{R}_{km}+C^i_{\phantom{i}km:i}-C^i_{\phantom{i}ki:m}+C^j_{\phantom{j}km}C^i_{\phantom{i}ji}-C^j_{\phantom{j}ki}C^i_{\phantom{i}jm}.
\label{Riem9}
\end{equation}
Consequently, the Ricci scalar or the {\em curvature scalar},
\begin{equation}
R=R_{ik}g^{ik},
\end{equation}
is given by
\begin{equation}
R=\mathring{R}-g^{ik}(2C^l_{\phantom{l}il:k}+C^j_{\phantom{j}ij}C^l_{\phantom{l}kl}-C^l_{\phantom{l}im}C^m_{\phantom{m}kl}),
\label{Riem11}
\end{equation}
where $\mathring{R}$ is the Riemannian curvature scalar or the {\em Riemann scalar}:
\begin{equation}
\mathring{R}=\mathring{R}_{ik}g^{ik}.
\end{equation}
The variation of the Riemann tensor is, analogously to (\ref{curten10}),
\begin{equation}
\delta \mathring{R}^i_{\phantom{i}klm}=(\delta\mathring{\Gamma}^{i}_{km})_{:l}-(\delta\mathring{\Gamma}^{i}_{kl})_{:m},
\end{equation}
and the variation of the Riemannian Ricci tensor is
\begin{equation}
\delta \mathring{R}_{ik}=(\delta\mathring{\Gamma}^{l}_{ik})_{:l}-(\delta\mathring{\Gamma}^{l}_{il})_{:k}.
\label{Riem14}
\end{equation}

\noindent
{\bf Contracted Bianchi identities}.\\
Contracting the identities (\ref{Biaide8}) and (\ref{Biaide9}) with the metric tensor gives
\begin{equation}
R_{nk;l}-R_{nl;k}+R^i_{\phantom{i}nkl;i}=-2R_{nm}S^m_{\phantom{m}kl}-2R^i_{\phantom{i}nmk}S^m_{\phantom{m}il}+2R^i_{\phantom{i}nml}S^m_{\phantom{m}ik}
\label{Riem15}
\end{equation}
and the {\em contracted cyclic identity}:
\begin{equation}
R_{jl}-R_{lj}=-2S_{j;l}+2S_{l;j}-2S^k_{\phantom{k}lj;k}+4S_n S^n_{\phantom{n}lj}.
\label{Riem16}
\end{equation}
Further contraction of (\ref{Riem15}) with the metric tensor gives the {\em contracted Bianchi identity}:
\begin{equation}
R^i_{\phantom{i}l;i}-\frac{1}{2}R_{;l}=2R_{km}S^{mk}_{\phantom{mk}l}-R^{ik}_{\phantom{ik}ml}S^m_{\phantom{m}ik}.
\label{Riem17}
\end{equation}
The Bianchi identity (\ref{Biaide10}) and the cyclic identity (\ref{Biaide11}) for the Riemann tensor are
\begin{eqnarray}
& & \mathring{R}^i_{\phantom{i}n[jk:l]}=0, \\
& & \mathring{R}^m_{\phantom{m}[jkl]}=0.
\label{Riem19}
\end{eqnarray}
Contracting these equations with the metric tensor gives
\begin{eqnarray}
& & \mathring{R}_{nk:l}+\mathring{R}^i_{\phantom{i}nkl:i}-\mathring{R}_{nl:k}=0, \label{Riem20} \\
& & \mathring{R}_{jl}-\mathring{R}_{lj}=0,
\end{eqnarray}
in agreement with (\ref{Riem7}).
Further contraction of (\ref{Riem20}) with the metric tensor gives the contracted Bianchi identity:
\begin{equation}
G^i_{\phantom{i}k:i}=0,
\label{Riem22}
\end{equation}
for the symmetric {\em Einstein tensor}, defined as
\begin{equation}
G_{ik}=\mathring{R}_{ik}-\frac{1}{2}\mathring{R}g_{ik}=G_{ki}.
\label{Riem23}
\end{equation}
This identity is a covariant conservation of the Einstein tensor.

\subsubsection{Properties of Riemann tensor}
\noindent
{\bf Two dimensions}.\\
In two dimensions there is only 1 independent component of the Riemann tensor, $P_{1212}$.
The Riemann scalar is
\begin{equation}
P=\frac{2P_{1212}}{\mathfrak{s}},
\label{prop1}
\end{equation}
where $\mathfrak{s}$ is the determinant of the two-dimensional metric tensor $\gamma_{ik}$:
\begin{equation}
\mathfrak{s}=|\gamma_{ik}|=\gamma_{11}\gamma_{22}-\gamma_{12}^2.
\end{equation}
In the Cartesian coordinates (\ref{intprop00}), a surface near the point $x=y=0$ is given by
\begin{equation}
z=\frac{x^2}{2\rho_1}+\frac{y^2}{2\rho_2},
\label{prop3}
\end{equation}
where $\rho_1$ and $\rho_2$ are the radii of curvature.
Substituting (\ref{prop3}) into the line element $dl^2=dx^2+dy^2+dz^2$ gives
\begin{equation}
dl^2=\gamma_{ik}dx^i dx^k=\biggl(1+\frac{x^2}{\rho_1^2}\biggr)dx^2+\biggl(1+\frac{y^2}{\rho_2^2}\biggr)dy^2+2\frac{xy}{\rho_1\rho_2}dx\,dy.
\end{equation}
The second derivatives of the corresponding components of $\gamma_{ik}(x,y)$ at the point $x=y=0$ give $P_{1212}$ and thus
\begin{equation}
\frac{1}{2}P|_{x=y=0}=K=\frac{1}{\rho_1\rho_2},
\label{prop5}
\end{equation}
where $K$ is the Gau\ss\,\,curvature.\\

\noindent
{\bf Three dimensions}.\\
In three dimensions there are 3 independent pairs, 12, 23, and 31, thereby the Riemann tensor has 6 independent components: 3 with identical pairs and $3\cdot2/2=3$ with different pairs (the cyclic identity does not reduce the number of independent components).
The Ricci tensor has also 6 components, which are related to the components of the Riemann tensor by
\begin{equation}
P_{\alpha\beta\gamma\delta}=P_{\alpha\gamma}\gamma_{\beta\delta}-P_{\alpha\delta}\gamma_{\beta\gamma}+P_{\beta\delta}\gamma_{\alpha\gamma}-P_{\beta\gamma}\gamma_{\alpha\delta}+\frac{P}{2}(\gamma_{\alpha\delta}\gamma_{\beta\gamma}-\gamma_{\alpha\gamma}\gamma_{\beta\delta}).
\label{prop6}
\end{equation}
Choosing the Cartesian coordinates at a given point, for which
\begin{equation}
g_{\alpha\beta}=\mbox{diag}(1,1,1),
\label{prop7}
\end{equation}
and diagonalizing $P_{\alpha\beta}$, which is equivalent to 3 rotations, brings $P_{\alpha\beta}$ to the canonical form with $6-3=3$ independent components.
Consequently, the Riemann tensor in three dimensions has 3 physically independent components.
The Gau\ss\,\,curvature of a surface formed by geodesic lines and perpendicular to the $x^3$ axis is given by
\begin{equation}
K=\frac{P_{1212}}{\gamma_{11}\gamma_{22}-\gamma_{12}^2}.
\label{prop8}
\end{equation}

\noindent
{\bf Four dimensions}.\\
In four dimensions there are 6 independent pairs, 01, 02, 03, 12, 23, and 31, thereby there are 6 components with identical pairs and $6\cdot5/2=15$ with different pairs.
The cyclic identity reduces the number of independent components by 1, thereby the Riemann tensor in four dimensions has generally 20 independent components.
Choosing the Cartesian coordinates at a given point and applying 6 rotations brings $\mathring{R}_{ijkl}$ to the canonical form with $20-6=14$ physically independent components.

The {\em Weyl tensor} is defined as
\begin{equation}
W_{iklm}=\mathring{R}_{iklm}-\frac{1}{2}(\mathring{R}_{il}g_{km}+\mathring{R}_{km}g_{il}-\mathring{R}_{im}g_{kl}-\mathring{R}_{kl}g_{im})+\frac{1}{6}\mathring{R}(g_{il}g_{km}-g_{im}g_{kl}).
\end{equation}
This tensor has all the symmetry and antisymmetry properties of the Riemann tensor, and is also traceless (any contraction of the Weyl tensor vanishes).

\subsubsection{Metric geodesics}
Let us consider two points in spacetime, $P$ and $Q$.
Among curves that connect these points, one curve has the minimal value of the interval $s=\int ds$, and is referred to as a {\em metric geodesic}.
The equation of a metric geodesic is given by the condition that $\int ds$ be an extremum with the endpoints of the curve fixed:
\begin{eqnarray}
& & \delta\int ds=\delta\int(g_{ik}dx^i dx^k)^{1/2}=\int\frac{\delta dx^i g_{ij}dx^j}{ds}+\frac{1}{2}\int\frac{\delta g_{ij}dx^i dx^j}{ds}=\int g_{ij}u^j\delta dx^i \nonumber \\
& & +\frac{1}{2}\int g_{ij,k}\delta x^k u^i u^j ds=\int d(u_i\delta x^i)-\int du_i\delta x^i+\frac{1}{2}\int g_{ij,k}\delta x^k u^i u^j ds \nonumber \\
& & =-\int\frac{du_i}{ds}\delta x^i ds+\frac{1}{2}\int g_{jk,i}\delta x^i u^j u^k ds+\int d(u_i\delta x^i)=0.
\label{metgeo1}
\end{eqnarray}
The first two terms in the last line can be written as
\begin{eqnarray}
& & -\int\biggl[\frac{d}{ds}(g_{ij}u^j)-\frac{1}{2}g_{jk,i}u^j u^k\biggr]\delta x^i ds=-\int\biggl[g_{ij}\frac{du^j}{ds}+u^k g_{ij,k}u^j-\frac{1}{2}g_{jk,i}u^j u^k\biggr]\delta x^i ds \nonumber \\
& & =-\int\biggl[g_{ij}\frac{du^j}{ds}+\frac{1}{2}(g_{ij,k}+g_{ik,j}-g_{jk,i})u^j u^k\biggr]\delta x^i ds=
-\int\biggl[g_{im}\frac{du^m}{ds}+\mathring{\Gamma}^{m}_{jk}g_{im}u^j u^k\biggr]\delta x^i ds \nonumber \\
& & =-\int\biggl[g_{im}\frac{\mathring{D}u^m}{ds}\biggr]\delta x^i ds.
\end{eqnarray}
The variation (\ref{metgeo1}) is therefore
\begin{equation}
\delta\int_1^2 ds=-\int_1^2\biggl[g_{im}\frac{\mathring{D}u^m}{ds}\biggr]\delta x^i ds+(u_i\delta x^i)\Big|_1^2=0,
\label{metgeo2}
\end{equation}
where $1$ and $2$ denote the world points corresponding to the initial and final spacetime position of the particle.

The last term in (\ref{metgeo2}) vanishes because $\delta x^i=0$ at the endpoints.
Since $\delta x^i$ is arbitrary, we obtain the metric geodesic equation:
\begin{equation}
\frac{\mathring{D}u^m}{ds}=\frac{du^m}{ds}+\mathring{\Gamma}^{m}_{jk}u^j u^k=u^i u^m_{\phantom{m}:i}=0.
\label{metgeo3}
\end{equation}
The metric geodesic equation (\ref{metgeo3}) can be written as
\begin{equation}
\frac{d^2x^i}{ds^2}+\mathring{\Gamma}^{i}_{kl}\frac{dx^k}{ds}\frac{dx^l}{ds}=0.
\label{metgeo4}
\end{equation}
Using (\ref{Chrsym6}) and (\ref{Chrsym7}), the affine geodesic equation (\ref{affgeo5}) can be written as
\begin{equation}
\frac{d^2x^i}{ds^2}+\mathring{\Gamma}^{i}_{kl}\frac{dx^k}{ds}\frac{dx^l}{ds}+2S_{kl}^{\phantom{kl}i}\frac{dx^k}{ds}\frac{dx^l}{ds}=0.
\label{metgeo5}
\end{equation}
If the torsion tensor is completely antisymmetric then the last term in (\ref{metgeo5}) vanishes and the affine geodesic equation coincides with the metric geodesic equation.
The equation of geodesic deviation with respect to the Levi-Civita connection is, analogously to (\ref{geodev4}),
\begin{equation}
\frac{\mathring{D}^2v^i}{ds^2}=\mathring{R}^i_{\phantom{i}jkl}u^j u^k v^l.
\end{equation}

\noindent
{\bf Geodesic equation and normalization of four-velocity}.\\
The metric geodesic equation (\ref{metgeo3}) is consistent with the normalization of the four-velocity vector $u^l u^i g_{li}=1$ (\ref{metten22}).
Differentiating this relation with respect to $s$ gives
\begin{equation}
2\frac{du^l}{ds}u^i g_{li}+u^l u^i\frac{dg_{li}}{ds}=2\frac{du^l}{ds}u_l+u^l u^i g_{li,k}u^k=0.
\label{mass129}
\end{equation}
Using
\begin{equation}
\mathring{\Gamma}^{j}_{ik}u^i u^k u_j=\frac{1}{2}(g_{li,k}+g_{lk,i}-g_{ik,l})u^i u^k u^l=\frac{1}{2}g_{li,k}u^i u^k u^l
\end{equation}
turns (\ref{mass129}) into
\begin{equation}
\frac{du^j}{ds}u_j+\mathring{\Gamma}^{j}_{ik}u^i u^k u_j=0.
\label{mass6}
\end{equation}
Consequently, $du^j/ds$ satisfies
\begin{equation}
\frac{du^j}{ds}+\mathring{\Gamma}^{j}_{ik}u^i u^k=f^{jk}u_k,
\label{metgeo9}
\end{equation}
where $f^{jk}$ is an arbitrary antisymmetric tensor.
If this tensor is zero then (\ref{metgeo9}) reduces to (\ref{metgeo3}).\\

\noindent
{\bf Conservation of contracted Killing vector}.\\
If $\zeta_i$ is a Killing vector of the Levi-Civita connection then along a metric geodesic,
\begin{equation}
\frac{\mathring{D}}{ds}(u^i\zeta_i)=u^k(u^i\zeta_i)_{:k}=u^i u^k \zeta_{i:k}+\zeta_i u^k u^i_{\phantom{i}:k}=0.
\label{metgeo7}
\end{equation}
The first term in the sum in (\ref{metgeo7}) vanishes because of (\ref{Chrsym23}) and the second term vanishes because of the metric geodesic equation.
Therefore, to each Killing vector of the Levi-Civita connection there corresponds a quantity $u^i \zeta_i$ which does not change along the metric geodesic, analogously to (\ref{Kilvec3}):
\begin{equation}
u^i \zeta_i=g_{ik}u^i \zeta^k=\mbox{const}.
\label{metgeo8}
\end{equation}

\subsubsection{Galilean frame of reference and Minkowski tensor}
At a given point, the nondegenerate ($\mathfrak{g}\neq0$) metric tensor can be brought to a diagonal (canonical) form $g_{ik}=\textrm{diag}(\pm1,\pm1,\pm1,\pm1)$.
{\em Physical} spacetime is described by the metric tensor with $\mathfrak{g}<0$.
Without loss of generality, we assume that the canonical form of the metric tensor is
\begin{equation}
g_{ik}=\eta_{ik}=\textrm{diag}(1,-1,-1,-1),\quad g^{ik}=\eta^{ik}=\textrm{diag}(1,-1,-1,-1).
\label{Galfr1}
\end{equation}
A frame of reference in which $g_{ik}$ has the canonical form is referred to as {\em Galilean}.
The transformation (\ref{geofr2}) with (\ref{geofr6}) brings a symmetric affine connection, thus the Christoffel symbols, to zero at a given point without changing the components of the metric tensor because of (\ref{geofr4}).
Therefore, a frame of reference can be locally both geodesic and Galilean.
Such a frame is called {\em inertial}.
In this frame, the first derivatives of the metric tensor vanish because of (\ref{Chrsym100}).
The corresponding metric tensor (\ref{Galfr1}) is referred to as the {\em Minkowski tensor}.
In a locally inertial frame the coordinates $x^i$, not only the differentials $dx^i$, are components of a contravariant vector.

In the absence of torsion, spacetime with a vanishing Riemann tensor $\mathring{R}^i_{\phantom{i}klm}=0$ is flat.
In the new coordinates $y_a$ (\ref{intcon9}), (\ref{intcon10}) gives
\begin{equation}
g^{ab}(y)=g_{ik}(x)\frac{\partial x^i}{\partial y_a}\frac{\partial x^k}{\partial y_b}=g_{ik}(x)h^{ia}h^{kb}=\eta^{ab}.
\end{equation}
Therefore, in a flat spacetime without torsion one can always find a system of coordinates which is Galilean everywhere.

\subsubsection{Riemann normal coordinates}
If the frame of reference is locally geodesic and Galilean at a given point, taken as the origin of the coordinates, then the metric tensor at a point near the origin depends on the derivatives of the metric at the origin.
In this frame, the Christoffel symbols at the origin vanish.
We expand the metric tensor up to quadratic terms:
\begin{equation}
g_{ij}(x^k)=g_{ij}(0)+g_{ij,k}(0)x^k+\frac{1}{2}g_{ij,kl}(0)x^k x^l=\eta_{ij}+\frac{1}{2}g_{ij,kl}(0)x^k x^l,
\label{normal1}
\end{equation}
where the metric tensor at the origin is equal to the Minkowski tensor and the first derivatives of the metric tensor at the origin vanish because of (\ref{Chrsym100}).
We choose the coordinates such that
\begin{equation}
x^i=a^i s
\end{equation}
for every metric geodesic curve passing through the origin and parameterized with the interval $s$, where $a^i$ is a constant four-vector and $s=0$ at the origin.
Such coordinates are referred to as the {\em Riemann normal coordinates}.
Accordingly, the derivatives of $x^i$ with respect to $s$ are
\begin{equation}
\frac{dx^i}{ds}(0)=a^i,\quad  \frac{d^2 x^i}{ds^2}(0)=\frac{d^3 x^i}{ds^3}(0)=0.
\label{normal3}
\end{equation}
Consequently, the metric geodesic equation (\ref{metgeo4}) gives
\begin{equation}
\mathring{\Gamma}^{i}_{jk}(0)a^j a^k=0,
\end{equation}
therefore the condition for the geodesic frame of reference (\ref{geofr7}) is satisfied:
\begin{equation}
\mathring{\Gamma}^{i}_{jk}(0)=0.
\end{equation}

Differentiating (\ref{metgeo4}) with respect to $s$ gives
\begin{equation}
\frac{d^3x^i}{ds^3}+\frac{d\mathring{\Gamma}^{i}_{jk}}{ds}\frac{dx^j}{ds}\frac{dx^k}{ds}+2\mathring{\Gamma}^{i}_{jk}\frac{d^2 x^j}{ds^2}\frac{dx^k}{ds}=0.
\end{equation}
At the origin, the relations (\ref{normal3}) reduce this equation to
\begin{equation}
\frac{d\mathring{\Gamma}^{i}_{jk}}{ds}\frac{dx^j}{ds}\frac{dx^k}{ds}=\mathring{\Gamma}^{i}_{jk,l}\frac{dx^l}{ds}\frac{dx^j}{ds}\frac{dx^k}{ds}=\mathring{\Gamma}^{i}_{jk,l}(0)a^l a^j a^k=0.
\end{equation}
Therefore, the Christoffel symbols satisfy
\begin{equation}
\mathring{\Gamma}^{i}_{(jk,l)}(0)=0.
\label{normal8}
\end{equation}

In the geodesic frame of reference, the Riemann tensor (\ref{Riem1}) reduces to
\begin{equation}
\mathring{R}^i_{\phantom{i}jkl}=\mathring{\Gamma}^{i}_{jl,k}-\mathring{\Gamma}^{i}_{jk,l}.
\end{equation}
Consequently, using (\ref{normal8}) gives
\begin{equation}
\mathring{R}^i_{\phantom{i}jkl}+\mathring{R}^i_{\phantom{i}kjl}=\mathring{\Gamma}^{i}_{jl,k}-\mathring{\Gamma}^{i}_{jk,l}+\mathring{\Gamma}^{i}_{kl,j}-\mathring{\Gamma}^{i}_{kj,l}=-3\mathring{\Gamma}^{i}_{jk,l},
\end{equation}
which gives
\begin{equation}
\mathring{\Gamma}^{i}_{jk,l}(0)=-\frac{1}{3}(\mathring{R}^i_{\phantom{i}jkl}+\mathring{R}^i_{\phantom{i}kjl})(0).
\label{normal11}
\end{equation}
Differentiating (\ref{Chrsym100}) with respect to the coordinates and using vanishing of the first derivatives of the metric tensor at the origin gives
\begin{equation}
g_{ij,kl}=\mathring{\Gamma}^{m}_{kj,l}g_{mi}+\mathring{\Gamma}^{m}_{ki,l}g_{mj}.
\end{equation}
Substituting (\ref{normal11}) into this equation gives
\begin{eqnarray}
& & g_{ij,kl}(0)=-\frac{1}{3}(\mathring{R}_{ikjl}+\mathring{R}_{ijkl}+\mathring{R}_{jkil}+\mathring{R}_{jikl})=-\frac{1}{3}(\mathring{R}_{ikjl}-\mathring{R}_{kijl}+\mathring{R}_{jikl})=-\frac{1}{3}(\mathring{R}_{ikjl}-\mathring{R}_{kjil}) \nonumber \\
& & =-\frac{1}{3}(\mathring{R}_{ikjl}+\mathring{R}_{iljk}).
\end{eqnarray}
Consequently, the covariant metric tensor (\ref{normal1}) in the Riemann normal coordinates at a point near the origin, in quadratic approximation, is given by
\begin{equation}
g_{ij}(x^k)=\eta_{ij}-\frac{1}{6}(\mathring{R}_{ikjl}+\mathring{R}_{iljk})(0)x^k x^l=\eta_{ij}-\frac{1}{3}\mathring{R}_{ikjl}(0)x^k x^l.
\label{normal14}
\end{equation}
The deviation of the metric tensor from the Minkowski tensor is proportional to the curvature.
The corresponding contravariant metric tensor is given by
\begin{equation}
g^{ij}(x^k)=\eta^{ij}+\frac{1}{3}\mathring{R}^{i\phantom{k}j}_{\phantom{i}k\phantom{j}l}(0)x^k x^l.
\end{equation}
Similar calculations lead to the expansion of the covariant metric tensor in quartic approximation:
\begin{equation}
g_{ij}(x^k)=\eta_{ij}-\frac{1}{3}\mathring{R}_{ikjl}(0)x^k x^l-\frac{1}{6}\mathring{R}_{ikjl:m}(0)x^k x^l x^m-\Bigl(\frac{1}{20}\mathring{R}_{ikjl:mn}-\frac{2}{45}\mathring{R}_{ikl}^{\phantom{ikl}p}\mathring{R}_{jmnp}\Bigr)(0)x^k x^l x^m x^n.
\end{equation}
\newline
References: \cite{Schr,LL2,Lord,Hehl1}.

\subsection{Space and time}
\setcounter{equation}{0}
\subsubsection{Intervals, velocity, and proper time}
\label{timespacedistances}
The form of the Minkowski tensor (\ref{Galfr1}) distinguishes the coordinate $x^0$ from the rest of the coordinates $x^\alpha$, where the index $\alpha$ can be 1,2,3.
The {\em temporal} coordinate $x^0$ can be written as $x^0=ct$, where $t$ is referred to as {\em time} and $c$ is called the {\em velocity of propagation of interaction} or the {\em speed of light}.
The coordinates $x^\alpha$ are {\em spatial} and span {\em space}.
Henceforth, they are denoted by the first Greek letters $\alpha,\beta,\gamma,\delta,\dots$.
The coordinates $x^i$ describe a {\em world point} or {\em event}, and span {\em spacetime}.
An infinitesimal interval $ds$ is {\em timelike} if $ds^2=dx^i dx_i>0$, {\em spacelike} if $ds^2<0$, and {\em null} if $ds^2=0$.
Similarly, a vector $V^i$ is timelike if $V^i V_i>0$, spacelike if $V^i V_i<0$, and null if $V^i V_i=0$.
If two non-null vectors are orthogonal to each other, then one of them is timelike and the other is spacelike.\\

\noindent
{\bf Cartesian coordinates}.\\
In a Galilean coordinate system, the spatial coordinates are {\em Cartesian}, denoted by
\begin{equation}
x^1=x,\quad x^2=y,\quad x^3=z,
\label{intprop00}
\end{equation}
and the metric tensor is given by (\ref{Galfr1}).
In this frame, the square of the line element (interval) (\ref{metten1}) between two infinitesimally separated points (events) is
\begin{equation}
ds^2=\eta_{ik}dx^i dx^k=c^2 dt^2-\sum_\alpha dx^\alpha dx^\alpha=c^2 dt^2-dl^2=c^2 dt^2-dx^2-dy^2-dz^2,
\label{intprop2}
\end{equation}
where $dx^i$ are infinitesimal coordinate differences between the two points.
The square of the interval between two finitely separated points is
\begin{equation}
\Delta s^2=\eta_{ik}\Delta x^i \Delta x^k=c^2 \Delta t^2-\sum_\alpha \Delta x^\alpha\Delta x^\alpha=c^2 \Delta t^2-\Delta x^2-\Delta y^2-\Delta z^2,
\label{intprop3}
\end{equation}
where $\Delta x^i$ are finite coordinate differences between the two points.\\

\noindent
{\bf Velocity}.\\
A curve $x^i(\lambda)$, where $\lambda$ is a parameter, is referred to as a {\em world line} of a given moving point.
The points of this line determine the space coordinates of the moving point at different moments of time.
The quantities
\begin{equation}
v^\alpha=\frac{dx^\alpha}{dt}
\label{intprop1}
\end{equation}
are the components of a three-dimensional vector, the {\em velocity} of this point.\\

\noindent
{\bf Rest frame and proper time}.\\
If $\Delta s$ is timelike, one can always find a frame of reference in which the two events occur at the same place, $\Delta x^\alpha=0$.
A frame of reference in which $dx^\alpha=0$, thereby $v^\alpha=0$, describes a point at {\em rest} and is referred to as the {\em rest frame} or the {\em comoving frame}.
In this frame $t=\tau$,
\begin{equation}
ds^2=c^2 d\tau^2,
\label{intprop4}
\end{equation}
where $\tau$ is the {\em proper time}.
If $dx^\alpha\neq0$, thereby $v^\alpha\neq0$, along a world line then the point {\em moves} or is in {\em motion}.
The proper time for a moving point is equal to the time measured by a clock moving with this point.
The proper time always runs forward (as in section \ref{eventhorizon}).

If $\Delta s$ is spacelike, one can always find a frame of reference in which the two events occur at the same time (are {\em synchronous}), $\Delta x^0=0$.
If $ds=0$ along a world line, this world line describes the propagation of a signal ({\em interaction}), with $(\sum_\alpha v^\alpha v^\alpha)^{1/2}=c$.
Equations (\ref{intprop2}) and (\ref{intprop4}) give
\begin{equation}
d\tau^2=dt^2-\frac{1}{c^2}\sum_\alpha dx^\alpha dx^\alpha,
\label{intprop5}
\end{equation}
so the proper time $\tau$ goes more slowly than the coordinate time $t$.
If $\Delta s$ is timelike, the two events occur at different times: $t_1\neq t_2$.
If $t_2>t_1$ then $t_2$ is in the {\em future} with respect to $t_1$ and $t_1$ is in the {\em past} with respect to $t_2$.\\

\begin{figure}[th]
\centering
\includegraphics[width=2in]{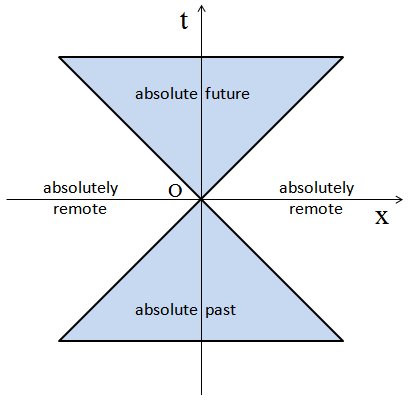}
\caption{Light cone.}
\label{cone}
\end{figure}

\noindent
{\bf Light cone}.\\
The time $t_0$ of an event is the {\em present} time relative to that event.
All events for which $t<t_0$ form the {\em absolute past} relative to the event $O$ at the present (events in this region occur {\em before} $O$ in all systems of reference).
All events for which $t>t_0$ form the {\em absolute future} relative to the event $O$ at the present (events in this region occur {\em after} $O$ in all systems of reference).
Such a division into the absolute past and the absolute future with respect to $O$ is possible only for events for which their intervals with respect to $O$ are timelike, as shown in Figure \ref{cone}.
For $O=(0,0,0,0)$, these events $(ct,x,y,z)$ lie within a cone $(ct)^2-x^2-y^2-z^2=0$ which is called the null cone or {\em light cone}.
All events for which their intervals with respect to $O$ are spacelike are {\em absolutely remote} relative to $O$.
The {\em principle of causality} states that any event $O$ can be affected only by events in the absolute past relative to $O$.

\subsubsection{Distances}
In the rest frame, $dx^\alpha=0$ gives $u^\alpha=0$.
At each point in space, the condition $dx^\alpha=0$ gives the relation between the proper time and the coordinate time:
\begin{equation}
d\tau=\frac{1}{c}\sqrt{g_{00}}dx^0,
\label{dist1}
\end{equation}
which requires
\begin{equation}
g_{00}>0.
\label{dist2}
\end{equation}
The relation (\ref{metten22}) gives
\begin{equation}
u^0=(g_{00})^{-1/2}.
\end{equation}
The metric tensor that does not satisfy $\mathfrak{g}<0$ cannot be the metric of a real spacetime.
The metric tensor that does not satisfy (\ref{dist2}) corresponds to a system of reference which cannot be realized with real bodies.
In this case, however, a suitable transformation of the coordinates can bring $g_{00}$ to a positive value.\\

\begin{figure}[th]
\centering
\includegraphics[width=1.5in]{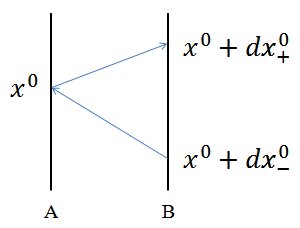}
\caption{Distance.}
\label{distance}
\end{figure}

\noindent
{\bf Spatial metric tensor}.\\
The distance between two infinitesimally separated points cannot be obtained by imposing $dx^0$ because $x^0$ transforms differently at these points.
Instead, we consider a signal that leaves point $B(x^\alpha+dx^\alpha)$ at $x^0+dx^0_-$, reaching point $A(x^\alpha)$ at $x^0$ and coming back to point $B$ at $x^0+dx^0_+$, as shown in Figure \ref{distance}.
Accordingly, we have
\begin{equation}
ds^2=g_{00}(dx^0)^2+2g_{0\alpha}dx^0 dx^\alpha+g_{\alpha\beta}dx^\alpha dx^\beta=0
\end{equation}
gives
\begin{equation}
dx^0_{\pm}=\frac{1}{g_{00}}(-g_{0\alpha}dx^\alpha\pm\sqrt{(g_{0\alpha}g_{0\beta}-g_{00}g_{\alpha\beta})dx^\alpha dx^\beta}).
\end{equation}
The difference in the time coordinate between emitting and receiving the signal at point $B$ is equal to the difference between $dx^0_+$ and $dx^0_-$ times $\sqrt{g_{00}}/c$, and the {\em distance} $dl$ between points $A$ and $B$ is equal to this difference times $c/2$:
\begin{equation}
dl^2=\gamma_{\alpha\beta}dx^\alpha dx^\beta,
\label{intprop11}
\end{equation}
where
\begin{equation}
\gamma_{\alpha\beta}=-g_{\alpha\beta}+\frac{g_{0\alpha}g_{0\beta}}{g_{00}}
\label{intprop12}
\end{equation}
is the symmetric {\em spatial metric tensor} of spacetime, that is, the metric tensor of space.
In general, $g_{ik}$ and thus $\gamma_{\alpha\beta}$ depend on the time coordinate $x^0$.
Consequently, the integral of $dl$ depends on the world line between the two given space points, thereby a definite distance between two points is not uniquely defined.\\

\noindent
{\bf Synchronization}.\\
The event at point $A$ at $x^0$ is {\em synchronized} with the event at point $B$ at the arithmetic mean of the time coordinates of emitting and receiving the signal:
\begin{equation}
x^0+\frac{1}{2}(dx^0_- +dx^0_+)=x^0+g_\alpha dx^\alpha,
\end{equation}
where
\begin{equation}
g_\alpha=-\frac{g_{0\alpha}}{g_{00}}.
\label{intprop14}
\end{equation}
Therefore, we have
\begin{equation}
\delta x^0=g_\alpha\delta x^\alpha,
\label{intprop15}
\end{equation}
which is equivalent to $\delta x_0=0$, is the difference in $x^0$ between two synchronized infinitesimally separated points.\\

\noindent
{\bf Relation between four-velocity and velocity}.\\
In terms of (\ref{intprop11}) and (\ref{intprop14}), the square of the line element is equal to
\begin{equation}
ds^2=g_{00}(dx^0-g_\alpha dx^\alpha)^2-dl^2,
\label{intprop16}
\end{equation}
The three-dimensional velocity (\ref{intprop1}):
\begin{equation}
v^\alpha=\frac{dx^\alpha}{d\tau}=\frac{c\,dx^\alpha}{\sqrt{g_{00}}(dx^0-g_\beta dx^\beta)}
\label{intprop17}
\end{equation}
is defined in terms of the synchronized proper time, corresponding to the difference in $x^0$ between two synchronized infinitesimally separated points (\ref{intprop15}):
\begin{equation}
d\tau=\frac{1}{c}\sqrt{g_{00}}(dx^0-\delta x^0)=\frac{1}{c}\sqrt{g_{00}}(dx^0-g_\alpha dx^\alpha).
\end{equation}
The {\em speed} $v$, defined through
\begin{equation}
v^2=\gamma_{\alpha\beta}v^\alpha v^\beta,
\label{intprop20}
\end{equation}
is therefore equal to the rate 
 of the distance, given by (\ref{intprop11}), with respect to the proper time:
\begin{equation}
v=\frac{dl}{d\tau}.
\end{equation}

The metric (\ref{intprop16}) becomes
\begin{equation}
ds^2=g_{00}(dx^0-g_\alpha dx^\alpha)^2-v^2 d\tau^2=g_{00}(dx^0-g_\alpha dx^\alpha)^2\biggl(1-\frac{v^2}{c^2}\biggr),
\label{intprop19}
\end{equation}
Consequently, the spatial components of the four-velocity (\ref{affgeo6}) are proportional to those of the velocity (\ref{intprop17}):
\begin{equation}
u^\alpha=\frac{dx^\alpha}{\sqrt{g_{00}}(dx^0-g_\beta dx^\beta)\sqrt{1-v^2/c^2}}=\frac{v^\alpha}{c\sqrt{1-v^2/c^2}}.
\label{intprop21}
\end{equation}
Dividing the square root of (\ref{intprop19}) by $ds$ and using (\ref{affgeo6}) gives the temporal component:
\begin{equation}
\quad u^0=\frac{1}{\sqrt{g_{00}}\sqrt{1-v^2/c^2}}+g_\alpha u^\alpha,
\end{equation}
from which it follows
\begin{equation}
u_0=g_{00}u^0+g_{0\alpha}u^\alpha=\frac{\sqrt{g_{00}}}{\sqrt{1-v^2/c^2}}.
\label{intprop22}
\end{equation}

\noindent
{\bf Properties of spatial metric tensor}.\\
The spatial metric tensor (\ref{intprop12}) and the quantity (\ref{intprop14}) satisfy the following formulae:
\begin{eqnarray}
& & \gamma^{\alpha\beta}=-g^{\alpha\beta}, \label{spvec5} \\
& & g^\alpha=\gamma^{\alpha\beta}g_\beta=-g^{0\alpha},
\label{spvec7} \\
& & g^{00}=\frac{1}{g_{00}}-g_\alpha g^\alpha, \\
& & \mathfrak{g}=-g_{00}\mathfrak{s},
\label{spvec6}
\end{eqnarray}
where $\gamma^{\alpha\beta}$ is the inverse of $\gamma_{\alpha\beta}$:
\begin{equation}
\gamma^{\alpha\delta}\gamma_{\beta\delta}=\delta^\alpha_\beta,
\label{spvec4}
\end{equation}
and
\begin{equation}
\mathfrak{s}=\mbox{det}\,\gamma_{\alpha\beta}.
\label{spvec9}
\end{equation}
For example, contracting (\ref{intprop12}) with (\ref{spvec5}) gives
\begin{eqnarray}
& & \gamma^{\alpha\delta}\gamma_{\beta\delta}=g^{\alpha\delta}g_{\beta\delta}-g^{\alpha\delta}g_{0\delta}\frac{g_{0\beta}}{g_{00}}=g^{\alpha i}g_{\beta i}-g^{\alpha 0}g_{\beta 0}-(g^{\alpha i}g_{0i}-g^{\alpha 0}g_{00})\frac{g_{0\beta}}{g_{00}} \nonumber \\
& & =\delta^\alpha_\beta-\delta^\alpha_0\frac{g_{0\beta}}{g_{00}}=\delta^\alpha_\beta,
\end{eqnarray}
in accordance with (\ref{spvec4}).\\

\noindent
{\bf Spatial unit pseudotensors}.\\
In three-dimensional space, the completely antisymmetric permutation symbols are defined as
\begin{equation}
e^{123}=\epsilon^{0123}=1,\quad e^{\alpha\beta\gamma}=e^{[\alpha\beta\gamma]},\quad e_{123}=-\varepsilon_{0123}=1,\quad e_{\alpha\beta\gamma}=e_{[\alpha\beta\gamma]}.
\label{spvec101}
\end{equation}
The spatial unit antisymmetric pseudotensor is analogous to (\ref{metten14}) and (\ref{metten15}):
\begin{equation}
\eta_{\alpha\beta\gamma}=\sqrt{\mathfrak{s}}e_{\alpha\beta\gamma},\quad \eta^{\alpha\beta\gamma}=\frac{1}{\sqrt{\mathfrak{s}}}e^{\alpha\beta\gamma}.
\label{spvec106}
\end{equation}
If we change the sign of one or three of the spatial coordinates, then the components of $\eta_{\alpha\beta\gamma}$ do not change.
The following formulae, analogous to (\ref{KLC8}), are satisfied:
\begin{eqnarray}
& & e_{\alpha\beta\gamma}e_{\lambda\mu\nu}=\left| \begin{array}{rrr} \delta_{\alpha\lambda} & \delta_{\alpha\mu} & \delta_{\alpha\nu} \\ \delta_{\beta\lambda} & \delta_{\beta\mu} & \delta_{\beta\nu} \\ \delta_{\gamma\lambda} & \delta_{\gamma\mu} & \delta_{\gamma\nu} \end{array} \right|, \quad e_{\alpha\beta\gamma}e_{\lambda\mu\gamma}=\delta_{\alpha\lambda}\delta_{\beta\mu}-\delta_{\alpha\mu}\delta_{\beta\lambda}, \nonumber \\
& & e_{\alpha\beta\gamma}e_{\lambda\beta\gamma}=2\delta_{\alpha\lambda}, \quad e_{\alpha\beta\gamma}e_{\alpha\beta\gamma}=6,
\end{eqnarray}
where $\delta_{\alpha\beta}$ is the {\em Cartesian metric tensor},
\begin{equation}
\delta_{\alpha\beta}=\delta^{\alpha\beta}=\textrm{diag}(1,1,1).
\label{spvec44}
\end{equation}

\subsubsection{Spatial vectors}
Let us consider a coordinate transformation (\ref{vec1}) of the spatial coordinates:
\begin{equation}
x^\alpha\rightarrow x'^{\beta}(x^\alpha),
\label{spvec101}
\end{equation}
whereas the time coordinate $x^0$ is unchanged.
A {\em spatial scalar} is a quantity that is invariant under (\ref{spvec101}).
Analogously to (\ref{vec9}), a {\em contravariant spatial vector} is defined as a set of quantities (components) that transform under (\ref{spvec101}) like spatial coordinate differentials:
\begin{equation}
A^{\alpha}=\frac{\partial x^{\alpha}}{\partial x'^\beta}A'^\beta.
\end{equation}
Analogously to (\ref{vec10}), a {\em covariant spatial vector} is defined as a set of quantities that transform under (\ref{spvec101}) like spatial partial derivatives of a scalar:
\begin{equation}
B_{\alpha}=\frac{\partial x'^\beta}{\partial x^{\alpha}}B'_\beta.
\end{equation}
A linear combination $aC+bD$ of two spatial vectors $C$ and $D$ (either both contravariant or both covariant), where $a$ and $b$ are scalars, is a  spatial vector $E$ whose components are
\begin{equation}
E^\alpha=aC^\alpha+bD^\alpha,\quad  E_\alpha=aC_\alpha+bD_\alpha.
\label{spvec41}
\end{equation}
If $a=b=1$, then these relations represent the addition of vectors.
Analogously to (\ref{tens2}), a {\em spatial tensor} is defined as a set of quantities that transform under (\ref{spvec101}) like products of the components of spatial vectors:
\begin{equation}
T^{\alpha\beta\dots}_{\phantom{\alpha\beta}\gamma\delta\dots}=\frac{\partial x^{\alpha}}{\partial x'^\mu}\frac{\partial x^{\beta}}{\partial x'^\nu}\frac{\partial x'^\rho}{\partial x^{\gamma}}\frac{\partial x'^\sigma}{\partial x^{\delta}}T'^{\mu\nu\dots}_{\phantom{'\mu\nu}\rho\sigma\dots}.
\end{equation}

The spatial components $A^\alpha$ of a contravariant four-vector $A^i$ form a contravariant spatial vector, whereas the temporal component $A^0$ is a spatial scalar:
\begin{equation}
A^i=(A^0,A^\alpha).
\end{equation}
The spatial components $B_\alpha$ of a covariant four-vector $B_i$ form a covariant spatial vector, whereas the temporal component $B_0$ is a spatial scalar:
\begin{equation}
B_i=(B_0,B_\alpha).
\end{equation}
The component $g_{00}$ of the metric tensor is a spatial scalar, the quantities $g_\alpha$ (\ref{intprop14}) form a covariant spatial vector (like the components $g_{0\alpha}$), and the quantities $\gamma_{\alpha\beta}$ (\ref{intprop12}) form a covariant spatial tensor.
All the tensor operations (covariant differentiation, raising and lowering of indices) are carried out in a three-dimensional space with $\gamma_{\alpha\beta}$, $g_\alpha$, and $g_{00}$.
The covariant components of a spatial vector are related to the contravariant components by the spatial metric tensors (\ref{intprop12}) and (\ref{spvec4}), which raise and lower the indices of spatial vectors analogously to the metric tensor acting on four-vectors:
\begin{eqnarray}
& & A_\alpha=\gamma_{\alpha\beta}A^\beta, \\
& & B^\alpha=\gamma^{\alpha\beta}B_\beta.
\end{eqnarray}

\noindent
{\bf Cartesian vectors}.\\
In a locally Galilean system of coordinates, the coordinates are Cartesian (\ref{intprop00}) and the spatial metric tensor is
\begin{equation}
\gamma_{\alpha\beta}=\delta_{\alpha\beta},
\end{equation}
where $\delta_{\alpha\beta}$ is given by (\ref{spvec44}).
Consequently, the contravariant $A^\alpha$ and covariant $A_\alpha$ components of a spatial vector are identical, forming a spatial vector ${\bf A}$.
Henceforth, spatial vectors are denoted by the bold letters.
The contravariant and covariant spatial components of a four-vector differ from each other by the sign.
For four-vectors defined as contravariant, we associate the spatial components with the contravariant components:
\begin{equation}
A^i=(A^0,A^\alpha)=(A^0,{\bf A}),\quad A_i=(A^0,-{\bf A}).
\end{equation}
For four-vectors defined as covariant, we associate the spatial components with the covariant components:
\begin{equation}
B_i=(B_0,B_\alpha)=(B_0,{\bf B}),\quad B^i=(B_0,-{\bf B}).
\end{equation}
A linear combination $a{\bf A}+b{\bf B}$ of two spatial vectors ${\bf A}$ and ${\bf B}$ is a spatial vector whose components form the same linear combination, analogously to (\ref{spvec41}).\\

\noindent
{\bf Shift vector and lapse}.\\
We define the {\em shift vector} ${\bf g}$, whose covariant components are given by (\ref{intprop14}) and contravariant components are given by (\ref{spvec7}):
\begin{equation}
g_\alpha=-\frac{g_{0\alpha}}{g_{00}},\quad g^\alpha=-g^{0\alpha}.
\label{spvec121}
\end{equation}
The shift vector is a spatial vector.
We define the {\em lapse} as $1/\sqrt{h}$, where
\begin{equation}
h=g_{00}.
\label{spvec120}
\end{equation}
The lapse is a spatial scalar.\\

\noindent
{\bf Radius and velocity vectors}.\\
In a Galilean coordinate system, the spatial coordinates $x^\alpha$ form a spatial vector ${\bf x}$, referred to as the {\em radius vector}.
The components $v^\alpha$ (\ref{intprop1}) form the time derivative of the radius vector, which defines the spatial vector of {\em velocity}:
\begin{equation}
{\bf v}=\frac{d{\bf x}}{dt}.
\label{relkin1}
\end{equation}
The magnitude of the velocity is equal to the speed (\ref{intprop20}):
\begin{equation}
|{\bf v}|=v.
\label{relkin2}
\end{equation}

\noindent
{\bf Products of vectors}.\\
Henceforth, we give formulae in the spatial-tensor notation and the Cartesian-vector notation.
The {\em scalar product} (dot product) of two spatial vectors is
\begin{equation}
{\bf A}\cdot{\bf B}=\gamma_{\alpha\beta}A^\alpha B^\beta.
\end{equation}
The {\em square} of a spatial vector ${\bf A}$ is
\begin{equation}
A^2={\bf A}\cdot{\bf A}=\gamma_{\alpha\beta}A^\alpha A^\beta.
\end{equation}
and its {\em magnitude} (also length or norm) is
\begin{equation}
A=|{\bf A}|=\sqrt{A^2}.
\end{equation}
The {\em angle} between two spatial vectors $\theta$ is defined through
\begin{equation}
{\bf A}\cdot{\bf B}=AB\,\cos\theta.
\label{spvec13}
\end{equation}
Two vectors are perpendicular if their scalar product is equal to zero.
Any two vectors satisfy the {\em Cauchy--Schwarz inequality}:
\begin{equation}
({\bf A}\cdot{\bf B})^2\le A^2 B^2,
\end{equation}
which gives $|\cos\theta|\le 1$.

The {\em vector product} (cross product) of two spatial vectors ${\bf A}$ and ${\bf B}$:
\begin{equation}
{\bf C}={\bf A}\times{\bf B},
\end{equation}
is defined as the spatial pseudovector that is dual to the antisymmetric tensor
\begin{equation}
C_{\alpha\beta}=A_\alpha B_\beta-A_\beta B_\alpha.
\end{equation}
Consequently, its components $C^{\alpha}$ satisfy
\begin{eqnarray}
& & C^{\alpha}=\frac{1}{2}\eta^{\alpha\beta\gamma}C_{\beta\gamma}=\eta^{\alpha\beta\gamma}A_\beta B_\gamma,\quad C_{\alpha}=\frac{1}{2}\eta_{\alpha\beta\gamma}C^{\beta\gamma}=\eta_{\alpha\beta\gamma}A^\beta B^\gamma, \label{spvec17} \\
& & C_{\alpha\beta}=\eta_{\alpha\beta\gamma}C^\gamma,\quad C^{\alpha\beta}=\eta^{\alpha\beta\gamma}C_\gamma.
\end{eqnarray}
The vector product of two spatial vectors ${\bf A}$ and ${\bf B}$ satisfies
\begin{equation}
{\bf A}\times{\bf B}=AB\,\sin\theta\,{\bf n},
\label{spvec60}
\end{equation}
where ${\bf n}$ is a unit vector perpendicular to both ${\bf A}$ and ${\bf B}$, in the direction given by the right-handed corkscrew rule.\\

\noindent
{\bf Spatial derivatives}.\\
The {\em spatial covariant derivative} $\nabla_\alpha$ acts on spatial vectors analogously to the metric covariant derivative acting on four-vectors:
\begin{eqnarray}
& & \nabla_\alpha A^\beta=\partial_\alpha A^\beta+\lambda^{\beta}_{\gamma\alpha} A^\gamma, \label{spvec23} \\
& & \nabla_\alpha A_\beta=\partial_\alpha A_\beta-\lambda^{\gamma}_{\beta\alpha} A_\gamma,
\label{spvec24}
\end{eqnarray}
where $\lambda^{\delta}_{\alpha\beta}$ are the three-dimensional, {\em spatial Christoffel symbols}, constructed from $\gamma_{\alpha\beta}$ in the same way as the Christoffel symbols $\mathring{\Gamma}^i_{jk}$ (\ref{Chrsym3}) are constructed from $g_{ik}$:
\begin{equation}
\lambda^{\delta}_{\alpha\beta}=\frac{1}{2}\gamma^{\delta\gamma}(\gamma_{\gamma\alpha,\beta}+\gamma_{\gamma\beta,\alpha}-\gamma_{\alpha\beta,\gamma}).
\label{spvec25}
\end{equation}
The spatial covariant derivative acts on spatial tensors and densities analogously to (\ref{affcon12}) and (\ref{covdifden14}), in which $\Gamma^i_{jk}$ is replaced with $\lambda^{\delta}_{\alpha\beta}$.\\

\noindent
{\bf Gradient, divergence, and curl}.\\
The spatial components of a covariant-vector operator $\partial_i$ acting on a scalar $\phi$ form the {\em gradient} of $\phi$:
\begin{equation}
\partial_i\phi=\biggl(\frac{\partial\phi}{c\partial t},\frac{\partial\phi}{\partial x^\alpha}\biggr)=\biggl(\frac{\partial\phi}{c\partial t},\mbox{{\bf grad}}\,\phi\biggr)=\biggl(\frac{\partial\phi}{c\partial t},{\bm \nabla}\phi\biggr).
\end{equation}
The gradient operator is thus given by
\begin{equation}
(\mbox{{\bf grad}})_\alpha=({\bm \nabla})_\alpha=\partial_\alpha,\quad (\mbox{{\bf grad}})^\alpha=({\bm \nabla})^\alpha=\gamma^{\alpha\beta}\partial_\beta.
\label{spvec26}
\end{equation}
The divergence of a spatial vector ${\bf A}$ is, analogously to (\ref{Chrsym16}), equal to
\begin{equation}
\mbox{div}\,{\bf A}={\bm \nabla}\cdot{\bf A}=\nabla_\alpha A^\alpha=\frac{1}{\sqrt{\mathfrak{s}}}\partial_\alpha(\sqrt{\mathfrak{s}}A^\alpha).
\label{spvec28}
\end{equation}
The curl of a spatial vector ${\bf A}$ is defined as the spatial pseudovector that is dual to the antisymmetric tensor $\nabla_\alpha A_\beta-\nabla_\beta A_\alpha=\partial_\alpha A_\beta-\partial_\beta A_\alpha$:
\begin{equation}
(\mbox{{\bf curl}}\,{\bf A})^\alpha=({\bm \nabla}\times{\bf A})^\alpha=\frac{1}{2}\eta^{\alpha\beta\gamma}(\partial_\beta A_\gamma-\partial_\gamma A_\beta)=\eta^{\alpha\beta\gamma}\partial_\beta A_\gamma.
\label{spvec29}
\end{equation}
The {\em Laplace--Beltrami operator} or {\em Laplacian} is the divergence of the gradient:
\begin{equation}
\triangle=\mbox{div}\,\mbox{{\bf grad}}=\nabla^2={\bm \nabla}\cdot{\bm \nabla}=\frac{1}{\sqrt{\mathfrak{s}}}\partial_\alpha(\sqrt{\mathfrak{s}}\gamma^{\alpha\beta}\partial_\beta).
\label{spvec30}
\end{equation}
In a system of coordinates, in which $g_{00}=1$ and $g_{0\alpha}=0$, the four-divergence operator (\ref{Chrsym16}) reduces to the {\em d'Alembert operator} or {\em d'Alembertian}:
\begin{equation}
\Box=\frac{1}{c^2}\frac{\partial^2}{\partial t^2}-\triangle.
\label{relkin100}
\end{equation}

\noindent
{\bf Properties of vectors and derivatives}.\\
We denote
\begin{equation}
({\bf A}\cdot{\bm \nabla}){\bf B}=
({\bf A}\cdot{{\bf grad}}){\bf B}=A^\alpha\nabla_\alpha{\bf B}.
\end{equation}
The following formulae are satisfied:
\begin{eqnarray}
& & {\bf A}\times{\bf B}=-{\bf B}\times{\bf A}, \\
& & {\bf A}\cdot({\bf B}\times{\bf C})={\bf B}\cdot({\bf C}\times{\bf A})={\bf C}\cdot({\bf A}\times{\bf B}), \\
& & {\bf A}\times({\bf B}\times{\bf C})={\bf B}({\bf A}\cdot{\bf C})-{\bf C}({\bf A}\cdot{\bf B}), \\
& & ({\bf A}\cdot{\bf B})^2+({\bf A}\times{\bf B})^2=A^2 B^2, \\
& & \mbox{{\bf curl}}\,\mbox{{\bf grad}}\,\phi=0, \\
& & \mbox{div}\,\mbox{{\bf curl}}\,{\bf A}=0, \\
& & \mbox{{\bf grad}}(\phi\psi)=\mbox{{\bf grad}}\,\phi\,\psi+\phi\,\mbox{{\bf grad}}\,\psi, \\
    & & \mbox{\bf grad}({\bf A}\cdot{\bf B})=({\bf A}\cdot\mbox{{\bf grad}}){\bf B}+({\bf B}\cdot\mbox{{\bf grad}}){\bf A}+{\bf A}\times\mbox{{\bf curl}}\,{\bf B}+{\bf B}\times\mbox{{\bf curl}}\,{\bf A}, \label{spvec555} \\
& & \mbox{div}(\phi{\bf A})=\mbox{{\bf grad}}\,\phi\cdot{\bf A}+\phi\,\mbox{div}\,{\bf A}, \label{spvec888} \\ 
& & \mbox{{\bf curl}}(\phi{\bf A})=\mbox{{\bf grad}}\,\phi\times{\bf A}+\phi\,\mbox{{\bf curl}}\,{\bf A}, \\
& & \mbox{div}({\bf A}\times{\bf B})={\bf B}\cdot\mbox{{\bf curl}}\,{\bf A}-{\bf A}\cdot\mbox{{\bf curl}}\,{\bf B}, \label{spvec444} \\
& & \mbox{{\bf curl}}({\bf A}\times{\bf B})=({\bf B}\cdot\mbox{{\bf grad}}){\bf A}-({\bf A}\cdot\mbox{{\bf grad}}){\bf B}+{\bf A}\,\mbox{div}\,{\bf B}-{\bf B}\,\mbox{div}\,{\bf A}, \\
& & \mbox{{\bf curl}}\,\mbox{{\bf curl}}\,{\bf A}=\mbox{{\bf grad}}\,\mbox{div}\,{\bf A}-\triangle{\bf A}.
\end{eqnarray}

\subsubsection{Spatial integrals}
In three-dimensional space, there are three types of integration.
A line integral is an integral of a covariant spatial vector over a curve: $\int A_\alpha dx^\alpha$.
A surface integral is an integral of a spatial tensor of rank (0,2) over a surface: $\int B_{\alpha\beta}df^{\alpha\beta}$, where
\begin{equation}
df^{\alpha\beta}=\left| \begin{array}{rr} dx^\alpha & dx'^{\alpha} \\ dx^\beta & dx'^{\beta} \end{array} \right|=dx^\alpha dx'^{\beta}-dx^\beta dx'^{\alpha}
\end{equation}
is an antisymmetric tensor analogous to (\ref{covint31}).
This infinitesimal element of surface can be geometrically represented as a parallelogram formed by the vectors $dx^\alpha$ and $dx'^{\alpha}$.
The components of $df^{\alpha\beta}$ are the projections of the area of the parallelogram on the coordinate planes $x^\alpha x^\beta$.
A volume integral is an integral of a spatial tensor of rank (0,3) over a volume: $\int C_{\alpha\beta\gamma}dS^{\alpha\beta\gamma}$, where $dS^{\alpha\beta\gamma}$ is defined analogously to (\ref{covint32}).

The dual density corresponding to the surface element is given by
\begin{equation}
df_\alpha=\frac{1}{2}\varepsilon_{\alpha\beta\gamma}df^{\beta\gamma},\quad df^{\alpha\beta}=\epsilon^{\alpha\beta\gamma}df_\gamma,\quad df^{\alpha\beta}df_\beta=0.
\end{equation}
The element $df_\alpha$ is orthogonal to $df^{\alpha\beta}$ and geometrically describes a vector normal to the surface.
Because of (\ref{spvec101}), this element is related to (\ref{covint1}):
\begin{equation}
df_\alpha=df^\star_{0\alpha}.
\end{equation}
The dual density corresponding to the volume element is given by
\begin{equation}
dV=\frac{1}{6}\varepsilon_{\alpha\beta\gamma}dS^{\alpha\beta\gamma}=dx^1 dx^2 dx^3.
\end{equation}
Because of (\ref{spvec101}), this element is related to (\ref{covint3}):
\begin{equation}
dV=dS_0,
\end{equation}
and geometrically describes the projection of the hypersurface element on the hyperplane of constant $x^0$.

The spatial analogue of (\ref{covint5}) is
\begin{equation}
dx^\alpha \leftrightarrow df^{\beta\alpha} \frac{\partial}{\partial x^\beta}.
\label{spvec33}
\end{equation}
The spatial analogue of (\ref{covint7}) is
\begin{equation}
df_\alpha \leftrightarrow dV\frac{\partial}{\partial x^\alpha}.
\label{spvec34}
\end{equation}
To preserve the invariant character of an integral in space, the product of an integrand and the integration element must be a three-dimensional scalar.
Analogously to invariant integrals in spacetime, the elements $df_\alpha$ and $dV$ must be multiplied by a three-dimensional scalar density, for example, the square root of the determinant $\mathfrak{s}$ of the spatial metric tensor $\gamma_{\alpha\beta}$.
Accordingly, the product $\sqrt{\mathfrak{s}}df_\alpha$ geometrically describes a vector $\sqrt{\mathfrak{s}}d{\bf f}$ normal to the surface element and equal in magnitude to the area of the element, analogously to (\ref{covint21}).
The product $\sqrt{\mathfrak{s}}dV$ is the element of geometrical spatial volume, analogously to (\ref{covint22}).
Using the curl (\ref{spvec29}), the relation (\ref{spvec33}) for a spatial vector $A_\alpha$ gives {\em Stokes' theorem}: 
\begin{eqnarray}
& & \oint{\bf A}\cdot d{\bf l}=\oint A_\alpha dx^\alpha=\int\frac{\partial A_\alpha}{\partial x^\beta}df^{\beta\alpha}=\int\partial_\beta A_\alpha e^{\beta\alpha\gamma}df_\gamma=\int\partial_\beta A_\alpha \eta^{\gamma\beta\alpha}\sqrt{\mathfrak{s}}df_\gamma \nonumber \\
& & =\int\mbox{{\bf curl}}\,{\bf A}\cdot\sqrt{\mathfrak{s}}d{\bf f}.
\label{covint28}
\end{eqnarray}
This relation is the spatial analogue of (\ref{Chrsym26}).
If the surface of integration is a plane, then Stokes' theorem reduces to {\em Green's theorem}.
Using the divergence (\ref{spvec28}), the relation (\ref{spvec34}) for a spatial vector density $\sqrt{\mathfrak{s}}A^\alpha$ gives {\em Gau\ss' theorem}: 
\begin{eqnarray}
& & \oint{\bf A}\cdot\sqrt{\mathfrak{s}}d{\bf f}=\oint A^\alpha\sqrt{\mathfrak{s}}df_\alpha=\int \frac{\partial(\sqrt{\mathfrak{s}}A^\alpha)}{\partial x^\alpha}dV=\int\mbox{div}\,{\bf A}\sqrt{\mathfrak{s}}dV.
\label{covint29}
\end{eqnarray}
This relation is the spatial analogue of (\ref{Chrsym19}).

The line integral $\oint{\bf A}\cdot d{\bf l}$ is the {\em circulation} of a vector ${\bf A}$ along a closed contour ${\bf l}$.
The surface integral $\int{\bf A}\cdot\sqrt{\mathfrak{s}}d{\bf f}$ is the {\em flux} of ${\bf A}$ through a surface ${\bf f}$.
If a surface is closed, the flux is $\oint{\bf A}\cdot\sqrt{\mathfrak{s}}d{\bf f}$.
In a locally Galilean frame of reference, the coordinates are Cartesian and $\mathfrak{s}=1$.
Stokes' theorem (\ref{covint28}) equals the flux of the curl of a vector through a surface to the circulation of this vector along the closed contour that is the boundary of the surface.
Gau\ss' theorem (\ref{covint29}) equals the volume integral of the divergence of a vector over a region of space to the flux of this vector through the closed surface that is the boundary of the region.

\subsubsection{Velocity of propagation of interaction}
According to the principle of relativity, physical laws have the same form in all admissible frames of reference.
Consequently, the velocity of propagation of interaction $c$ is constant in all these frames.
This constancy provides another, physical argument for the invariance of an infinitesimal interval $ds$ (an infinitesimal change of the affine parameter $s$).
We consider an interval, whose value in one frame of reference $K$ is $ds$ and in another frame of reference $K'$ is $ds'$.
If $ds=0$ (propagation of interaction), then $ds'=0$ because of the principle of relativity.
Furthermore, the differentials $ds$ and $ds'$ are infinitesimals of the same order.
Consequently, their squares must be proportional to one another:
\begin{equation}
ds^2=f\,ds'^2.
\label{velprop1}
\end{equation}
The function $f$ cannot depend on the coordinates because different points in space and different moments in time must be equivalent in order to be consistent with the homogeneity of space and time.
Also, it cannot depend on the direction of the velocity of $K'$ relative to $K$ because different directions must be equivalent in order to be consistent with the isotropy of space.
Therefore, $f$ is a function of the speed of $K'$ relative to $K$.

Let us consider three frames of reference $K$, $K_1$, and $K_2$.
If the speed of $K_1$ relative to $K$ is $v_1$ and the speed of $K_2$ relative to $K$ is $v_2$, then
\begin{equation}
ds^2=f(v_1)ds_1^2,\quad ds^2=f(v_2)ds_2^2.
\label{velprop2}
\end{equation}
Similarly, the intervals in $K_1$ and $K_2$ are related to one another by
\begin{equation}
ds_1^2=f(v_{12})ds_2^2.
\label{velprop3}
\end{equation}
The relations (\ref{velprop2}) and (\ref{velprop3}) give
\begin{equation}
\frac{f(v_2)}{f(v_1)}=f(v_{12}).
\label{velprop4}
\end{equation}
In this relation, $v_{12}$ depends on the angle between the vectors ${\bf v}_1$ and ${\bf v}_2$ of $K_1$ and $K_2$ relative to $K$, and the left-hand side does not depend on this angle ($v_1$ and $v_2$ are the magnitudes of these vectors).
Therefore, the function $f(v)$ reduces to a constant and (\ref{velprop4}) determines this constant to be 1.
Consequently, (\ref{velprop1}) reduces to
\begin{equation}
ds^2=ds'^2,
\end{equation}
showing that $ds$ is a scalar.

\subsubsection{Event horizon}
\label{eventhorizon}
A {\em hypersurface} in a four-dimensional spacetime consists of points whose coordinates satisfy an equation of constraint:
\begin{equation}
f(x^i)=0,
\label{emb31}
\end{equation}
where $f$ is a function of the coordinates.
The {\em normal vector} to this hypersurface is given by
\begin{equation}
n_i=\frac{\partial f}{\partial x^i}.
\label{emb32}
\end{equation}
All infinitesimal displacements $dx^i$ along such a hypersurface satisfy, according to (\ref{emb32}),
\begin{equation}
df=n_i dx^i=0.
\label{emb34}
\end{equation}

If the normal vector is a null vector:
\begin{equation}
n_i n^i=0,
\label{evhor1}
\end{equation}
then this hypersurface is a null hypersurface.
Equations (\ref{emb34}) and (\ref{evhor1}) indicate that $n^i$ lies itself on the null hypersurface to which it is normal:
\begin{equation}
dx^i \propto n^i,
\end{equation}
which also gives
\begin{equation}
ds^2=dx_i dx^i \propto n_i n^i=0.
\end{equation}
Therefore, all world lines on a null hypersurface are null.
The light cones at the points of such a hypersurface are tangent to this hypersurface.
Since all physical world lines must lie within the local light cones, the forward-time motion through a null hypersurface can occur in only one direction.
To avoid any discontinuities, this direction is the same for all points on such a hypersurface.
A null hypersurface is therefore an {\em event horizon}: a boundary in spacetime beyond which events cannot affect events on the other side.
All laws of classical physics are known to be {\em time-symmetric}, that is, symmetric under the transformation $t\rightarrow-t$.
However, the existence of event horizons, which are solutions to these laws and provide boundary conditions for spacetime, violates this symmetry.
The unidirectional character of the motion through an event horizon can be used to define the past and future: the {\em arrow of time}.

\begin{footnotesize}
\subsubsection{Embedded surfaces}
{\bf Intrinsic metric}.\\
A {\em surface} embedded in a three-dimensional space consists of points whose radius vectors are vector functions of two parameters $\xi^\alpha$, where the index $\alpha$ can be 1 or 2: ${\bf x}={\bf x}(\xi^1,\xi^2)$.
A vector
\begin{equation}
\partial_\alpha{\bf x}=\frac{\partial{\bf x}}{\partial \xi^\alpha}
\end{equation}
is {\em tangent} to the surface.
We define the {\em induced} or {\em intrinsic metric tensor} on the surface as
\begin{equation}
\gamma_{\alpha\beta}=\partial_\alpha{\bf x}\cdot\partial_\beta{\bf x}.
\label{emb2}
\end{equation}
The length element $dl$ on the surface is given by the {\em first fundamental form}:
\begin{equation}
dl^2=d{\bf x}\cdot d{\bf x}=\frac{\partial{\bf x}}{\partial \xi^\alpha}\cdot\frac{\partial{\bf x}}{\partial \xi^\beta}d\xi^\alpha d\xi^\beta=\gamma_{\alpha\beta}d\xi^\alpha d\xi^\beta,
\end{equation}
and the area element is given by
\begin{equation}
dS=\sqrt{\mbox{det}\gamma_{\alpha\beta}}d\xi^1 d\xi^2.
\end{equation}
The inverse intrinsic metric tensor $\gamma^{\alpha\beta}$ is defined according to
\begin{equation}
\gamma^{\alpha\delta}\gamma_{\beta\delta}=\delta^\alpha_\beta.
\label{emb5}
\end{equation}

\noindent
{\bf Extrinsic curvature}.\\
We define the unit {\em normal vector} to a surface as
\begin{equation}
{\bf n}=\frac{\partial_1{\bf x}\times\partial_2{\bf x}}{|\partial_1{\bf x}\times\partial_2{\bf x}|},\quad  {\bf n}\cdot{\bf n}=1.
\end{equation}
This vector is perpendicular to a tangent vector:
\begin{equation}
\partial_\alpha{\bf x}\cdot{\bf n}=0.
\label{emb7}
\end{equation}
If the surface is curved, then the normal vectors at two close points on the surface are not parallel.
The change of the normal vector is given by the {\em extrinsic curvature tensor}:
\begin{equation}
K_{\alpha\beta}=\partial_\alpha\partial_\beta{\bf x}\cdot{\bf n}.
\label{emb8}
\end{equation}
The extrinsic curvature is symmetric,
\begin{equation}
K_{\alpha\beta}=K_{\beta\alpha}.
\label{emb9}
\end{equation}
Differentiating the relation (\ref{emb7}) with respect to $\xi^\beta$ and using (\ref{emb8}) gives
\begin{equation}
K_{\alpha\beta}=-\partial_\alpha{\bf x}\cdot\partial_\beta{\bf n}.
\label{emb10}
\end{equation}
The quantity $K_{\alpha\beta}d\xi^\alpha d\xi^\beta$ is the {\em second fundamental form}.
The {\em intrinsic Christoffel symbols} $\mathring{\Gamma}^{\alpha}_{\beta\gamma}$, symmetric in the lower indices, are constructed from the intrinsic metric tensor analogously to the spatial Christoffel symbols (\ref{spvec25}) constructed from the spatial metric tensor.
They are used to construct the covariant derivative $\nabla_\alpha$ acting on the vectors tangent to the surface, analogously to (\ref{spvec23}) and (\ref{spvec24}).\\

\noindent
{\bf Gau\ss\, and Weingarten equations}.\\
The covariant derivatives acting on ${\bf x}$ and ${\bf n}$ are equal to the partial derivatives:
\begin{equation}
\nabla_\alpha{\bf x}=\partial_\alpha{\bf x},\quad \nabla_\alpha{\bf n}=\partial_\alpha{\bf n}.
\end{equation}
The second derivatives of ${\bf x}$, which are the first derivatives of the tangent vectors, satisfy the {\em Gau\ss\, equation}:
\begin{equation}
\partial_\alpha\partial_\beta{\bf x}=\mathring{\Gamma}^{\gamma}_{\alpha\beta}\partial_\gamma{\bf x}+K_{\alpha\beta}{\bf n},
\end{equation}
which can be written in a covariant form:
\begin{equation}
\nabla_\alpha\nabla_\beta{\bf x}=K_{\alpha\beta}{\bf n}.
\end{equation}
Multiplying this equation by ${\bf n}$ gives (\ref{emb8}).
The first derivatives of the normal vector satisfy the {\em Weingarten equation}:
\begin{equation}
\partial_\alpha{\bf n}=-K_\alpha^{\phantom{\alpha}\beta}\partial_\beta{\bf x}=-K_{\alpha\gamma}\gamma^{\beta\gamma}\partial_\beta{\bf x},
\end{equation}
which can be written in a covariant form:
\begin{equation}
\nabla_\alpha{\bf n}=-K_\alpha^{\phantom{\alpha}\beta}\nabla_\beta{\bf x}.
\end{equation}
Multiplying this equation by $\partial_\gamma{\bf x}$ and using (\ref{emb2}) gives (\ref{emb10}).
The intrinsic metric tensor and the extrinsic curvature can also be written in a covariant form:
\begin{eqnarray}
& & \gamma_{\alpha\beta}=\nabla_\alpha{\bf x}\cdot\nabla_\beta{\bf x}, \\
& & K_{\alpha\beta}=\nabla_\alpha\nabla_\beta{\bf x}\cdot{\bf n}.
\label{emb17}
\end{eqnarray}

\noindent
{\bf Codazzi--Mainardi--Peterson equation}.\\
Using the Gau\ss\, equation, the relation
\begin{equation}
\partial_\alpha\partial_\beta\partial_\gamma{\bf x}=\partial_\beta\partial_\alpha\partial_\gamma{\bf x}
\end{equation}
can be written as
\begin{equation}
\partial_\alpha(\mathring{\Gamma}^{\delta}_{\beta\gamma}\partial_\delta{\bf x}+K_{\beta\gamma}{\bf n})=\partial_\beta(\mathring{\Gamma}^{\delta}_{\alpha\gamma}\partial_\delta{\bf x}+K_{\alpha\gamma}{\bf n}).
\end{equation}
Effecting the differentiation and using again the Gau\ss\, equation gives
\begin{eqnarray}
& & \partial_\delta{\bf x}\partial_\alpha\mathring{\Gamma}^{\delta}_{\beta\gamma}+\mathring{\Gamma}^{\delta}_{\beta\gamma}\mathring{\Gamma}^{\epsilon}_{\alpha\delta}\partial_\epsilon{\bf x}+\mathring{\Gamma}^{\delta}_{\beta\gamma}K_{\alpha\delta}{\bf n}+\partial_\alpha K_{\beta\gamma}{\bf n}+K_{\beta\gamma}\partial_\alpha{\bf n} \nonumber \\
& & =\partial_\delta{\bf x}\partial_\beta\mathring{\Gamma}^{\delta}_{\alpha\gamma}+\mathring{\Gamma}^{\delta}_{\alpha\gamma}\mathring{\Gamma}^{\epsilon}_{\beta\delta}\partial_\epsilon{\bf x}+\mathring{\Gamma}^{\delta}_{\alpha\gamma}K_{\beta\delta}{\bf n}+\partial_\beta K_{\alpha\gamma}{\bf n}+K_{\alpha\gamma}\partial_j{\bf n}.
\label{emb20}
\end{eqnarray}
Multiplying this equation by $\partial_\zeta{\bf x}$ and using (\ref{emb2}), (\ref{emb7}), and (\ref{emb10}) gives
\begin{equation}
\gamma_{\delta\zeta}\partial_\alpha\mathring{\Gamma}^{\delta}_{\beta\gamma}+\gamma_{\epsilon\zeta}\mathring{\Gamma}^{\delta}_{\beta\gamma}\mathring{\Gamma}^{\epsilon}_{\alpha\delta}-K_{\alpha\zeta}K_{\beta\gamma}-\gamma_{\delta\zeta}\partial_\beta\mathring{\Gamma}^{\delta}_{\alpha\gamma}-\gamma_{\epsilon\zeta}\mathring{\Gamma}^{\delta}_{\alpha\gamma}\mathring{\Gamma}^{\epsilon}_{\beta\delta}+K_{\beta\zeta}K_{\alpha\gamma}=0,
\end{equation}
which is equivalent to the {\em Gau\ss\, equation}:
\begin{equation}
r^\epsilon_{\phantom{\epsilon}\gamma\alpha\beta}=K^\epsilon_{\phantom{\epsilon}\alpha} K_{\gamma\beta}-K^\epsilon_{\phantom{\epsilon}\beta} K_{\gamma\alpha},
\label{emb22}
\end{equation}
where $r^\epsilon_{\phantom{\epsilon}\gamma\alpha\beta}$ is the {\em intrinsic curvature tensor} constructed from the intrinsic Christoffel symbols analogously to the Riemann tensor (\ref{Riem1}) constructed from the Levi-Civita connection.
Multiplying (\ref{emb20}) by ${\bf n}$ and using (\ref{emb7}) and $\partial_\alpha{\bf n}\cdot{\bf n}=0$ gives the {\em Codazzi--Mainardi--Peterson equation}:
\begin{equation}
\mathring{\Gamma}^{\delta}_{\beta\gamma}K_{\alpha\delta}+\partial_\alpha K_{\beta\gamma}=\mathring{\Gamma}^{\delta}_{\alpha\gamma}K_{\beta\delta}+\partial_\beta K_{\alpha\gamma},
\end{equation}
which can be written in a covariant form:
\begin{equation}
\nabla_\alpha K_{\beta\gamma}=\nabla_\beta K_{\alpha\gamma}.
\label{emb23}
\end{equation}

\noindent
{\bf Gau\ss\, curvature}.\\
The {\em Gau\ss\,\,curvature} is defined as
\begin{equation}
K=\frac{\mbox{det}K_{\alpha\beta}}{\mbox{det}\gamma_{\alpha\beta}}=\frac{K_{11}K_{22}-K_{12}K_{21}}{\gamma_{11}\gamma_{22}-\gamma_{12}\gamma_{21}}.
\label{emb24}
\end{equation}
Using the Gau\ss\, equation, it leads to the {\em Gau\ss\, theorem}:
\begin{equation}
K=\frac{r_{1212}}{\mbox{det}\gamma_{\alpha\beta}},
\end{equation}
which is consistent with (\ref{prop1}) and (\ref{prop5}).

A {\em curve} on a surface consists of points whose radius vectors depend on a parameter $t$: ${\bf x}={\bf x}(\xi^1(t),\xi^2(t))$.
Such a curve is geodesic if it satisfies the metric geodesic equation analogous to (\ref{metgeo4}):
\begin{equation}
\frac{d^2\xi^\alpha}{dt^2}+\mathring{\Gamma}^{\alpha}_{\beta\gamma}\frac{d\xi^\beta}{dt}\frac{d\xi^\gamma}{dt}=0.
\end{equation}
A geodesic curve also satisfies
\begin{equation}
\frac{d^2{\bf x}}{dt^2}\sim{\bf n},\quad \frac{d}{dt}\Bigl(\frac{d{\bf x}}{dt}\cdot\frac{d{\bf x}}{dt}\Bigr)=0.
\end{equation}
\end{footnotesize}

\begin{footnotesize}
\subsubsection{Embedded hypersurfaces}
A hypersurface embedded in a four-dimensional spacetime consists of points whose coordinates are functions of three parameters $\xi^\alpha$, where the index $\alpha$ can be 1, 2, or 3: $x^i=x^i(\xi^1,\xi^2,\xi^3)$.
Equivalently, these coordinates satisfy an equation of constraint (\ref{emb31}).
The normal vector (\ref{emb32}) is orthogonal to the hypersurface:
\begin{equation}
n_i n_{j:k}\epsilon^{ijkl}=0,
\end{equation}
where $\epsilon^{ijkl}$ is the completely antisymmetric permutation symbol.
This condition is equivalent to
\begin{equation}
n_{[i}n_{j:k]}=\frac{1}{6}(n_i n_{j:k}+n_j n_{k:i}+n_k n_{i:j}-n_k n_{j:i}-n_i n_{k:j}-n_j n_{i:k})=0.
\label{hyp1}
\end{equation}
If the normal vector to a hypersurface is timelike, then the hypersurface is spacelike.
Such a normal vector can be normalized:
\begin{equation}
n^i n_i=1,
\end{equation}
which gives
\begin{equation}
n^i n_{i:k}=0.
\label{hyp2}
\end{equation}
In this case, the four-velocity of a point in spacetime can be taken as the normal vector:
\begin{equation}
n^i=u^i.
\end{equation}
If the only nonzero component of the four-velocity is the time component, then the hypersurface is a hypersurface of constant time and represents a volume in space, in which the point exists at this time.
A division of spacetime into such hypersurfaces is referred to as a {\em foliation} of spacetime.\\

\noindent
{\bf Intrinsic metric}.\\
We consider a spacelike hypersurface.
We define the {\em projection tensor} onto the hypersurface:
\begin{equation}
h^i_{\phantom{i}j}=\delta^i_j-n^i n_j,
\label{hyp3}
\end{equation}
which is orthogonal to $n^i$:
\begin{equation}
h^i_{\phantom{i}j}n^j=0.
\end{equation}
The projection tensor satisfies
\begin{equation}
h^i_{\phantom{i}j}h^j_{\phantom{j}k}=h^i_{\phantom{i}k}.
\end{equation}
The indices in the projection tensor can be raised or lowered by the metric tensor:
\begin{equation}
h_{ik}=h^j_{\phantom{j}k}g_{ij}= g_{ik}-n_i n_k=h_{ki}, \quad  h^{ik}=h^i_{\phantom{i}j}g^{jk}=g^{ik}-n^i n^k=h^{ki}.
\label{hyp5}
\end{equation}
The tensors $h_{ik}$ and $h^{ik}$ are symmetric and not inverse to one another.

The projection $\perp$ of a tensor $T$ onto a hypersurface is defined as the contraction of the tensor $T$ with the projection tensor through all indices.
For example, the projections of vectors are
\begin{equation}
\perp V^i=h^i_{\phantom{i}k}V^k, \quad  \perp V_i=h^k_{\phantom{k}i}V_k.
\end{equation}
These projections are tangent vectors to the hypersurface.
The projection of the metric tensor gives
\begin{equation}
\perp g_{ij}=h^k_{\phantom{k}i}h^l_{\phantom{l}j}g_{kl}=h_{ij}, \quad  \perp g^{ij}=h^i_{\phantom{i}k}h^j_{\phantom{j}l}g^{kl}=h^{ij}.
\end{equation}
Consequently, the relation
\begin{equation}
h_{ij}\perp V^i\perp V^j=g_{ij}\perp V^i\perp V^j
\end{equation}
shows that the tensor $h_{ij}$ is the intrinsic metric tensor $\gamma_{ij}$ on the hypersurface, analogously to (\ref{emb2}):
\begin{equation}
\gamma_{ij}=h_{ij}.
\end{equation}
The inverse intrinsic metric tensor $\gamma^{ij}$ is defined as in (\ref{emb5}).
The projection of the normal vector vanishes:
\begin{equation}
\perp n^i=0.
\label{hyp6}
\end{equation}
If the normal vector to the hypersurface is timelike, then the projections of tensors have only spatial components.
Using the tensor $n^i n_j$ instead of $h^i_{\phantom{i}j}$ projects a tensor onto the direction of the normal vector.

The projection of the covariant derivative (with respect to a torsionless affine connection) of a vector defines the intrinsic covariant derivative of a vector on the hypersurface:
\begin{equation}
D_k V^l=\perp \nabla_k V^l=h^i_{\phantom{i}k}h^l_{\phantom{l}j}\nabla_i V^j.
\end{equation}
The intrinsic covariant derivative of the intrinsic metric tensor vanishes:
\begin{equation}
D_k \gamma_{ij}=\perp \nabla_k \gamma_{ij}=\perp \nabla_k(g_{ij}-n_i n_j)=-\perp (n_i\nabla_k n_j+n_j\nabla_k n_i)=0,
\end{equation}
which is a consequence of the metric compatibility of the affine connection (\ref{metten5}).
Accordingly, the intrinsic covariant derivative is constructed from the intrinsic Christoffel symbols, which are constructed from the intrinsic metric tensor.
If $\nabla_k$ is related to the Levi-Civita connection of the metric $g_{ij}$, then $D_k$ is related to the Levi-Civita connection of the intrinsic metric $\gamma_{ij}$.\\

\noindent
{\bf Extrinsic curvature}.\\
If the parallel transport of the normal vector to a hypersurface along a vector $W^i=\perp V^i$ on the hypersurface does not vanish,
\begin{equation}
W^i\nabla_i n^j\neq0,
\end{equation}
then the hypersurface is curved.
Such a hypersurface has a nonzero extrinsic curvature tensor, defined as
\begin{equation}
K_{ij}=-\perp \nabla_i n_j=-h^k_{\phantom{k}i}h^l_{\phantom{l}j}\nabla_k n_l.
\label{hyp7}
\end{equation}
Using (\ref{hyp2}) and (\ref{hyp3}), the extrinsic curvature is equal to
\begin{equation}
K_{ij}=-(\delta^k_i-n^k n_i)(\delta^l_j-n^l n_j)n_{l:k}=-n_{j:i}+n_i n^k n_{j:k}.
\label{hyp8}
\end{equation}
The extrinsic curvature is a tensor with only spatial components:
\begin{equation}
K_{ij}n^j=0.
\end{equation}
Antisymmetrizing the indices in the extrinsic curvature and using (\ref{hyp8}) gives
\begin{equation}
K_{ij}-K_{ji}=n_{i:j}-n_{j:i}+n^k n_i n_{j:k}-n^k n_j n_{i:k}.
\end{equation}
The term on the right-hand side is equal to the term in (\ref{hyp1}) contracted with $n^k$, which vanishes.
Consequently, the extrinsic curvature is symmetric, as in (\ref{emb9}).
This symmetry also results from (\ref{emb32}):
\begin{equation}
K_{ij}=-\perp \nabla_i \partial_j f=-\perp \nabla_j \partial_i f=K_{ji}.
\end{equation}
For the Levi-Civita connection, (\ref{Chrsym22}) gives
\begin{equation}
K_{ij}=-\perp n_{j:i}=-\perp n_{(i:j)}=-\frac{1}{2}\perp {\cal L}_{n}g_{ij},
\end{equation}
where ${\cal L}_{n}$ is the Lie derivative of the metric tensor along the vector $n^i$.
If the normal vector is the four-velocity of a point in spacetime, then (\ref{hyp8}) gives
\begin{equation}
K_{ij}=-u_{j;i}+u_i\frac{Du_j}{ds}.
\end{equation}
The contraction of the extrinsic curvature tensor gives the {\em extrinsic curvature scalar}:
\begin{equation}
K=K_{ij}\gamma^{ij}.
\end{equation}

For a spacelike hypersurface, the spatial coordinates on this hypersurface can be taken as the parameters $\xi^\alpha$.
Differentiating the equation of constraint for a hypersurface $f(x^i(\xi^\alpha))=0$ with respect to $\xi^\alpha$ gives
\begin{equation}
\frac{\partial f}{\partial \xi^\alpha}=\frac{\partial f}{\partial x^i}\frac{\partial x^i}{\partial \xi^\alpha}=n_i \frac{\partial x^i}{\partial \xi^\alpha}=0.
\end{equation}
Differentiating covariantly this equation with respect to $\xi^\beta$ gives
\begin{equation}
\frac{\nabla n_i}{\partial \xi^\beta}\frac{\partial x^i}{\partial \xi^\alpha}+n_i\frac{\nabla^2 x^i}{\partial \xi^\beta \partial \xi^\alpha}=\frac{\nabla}{\partial x^i}\frac{\partial f}{\partial \xi^\beta}\frac{\partial x^i}{\partial \xi^\alpha}+n_i\nabla_\beta\nabla_\alpha x^i=\frac{\nabla^2 f}{\partial \xi^\alpha \partial \xi^\beta}+n_i\nabla_\beta\nabla_\alpha x^i=\frac{\nabla n_\beta}{\partial \xi^\alpha}+n_i\nabla_\beta\nabla_\alpha x^i=0,
\end{equation}
where the covariant derivatives $\nabla_\alpha$ are constructed from the metric tensor $\gamma_{\alpha\beta}$ and the corresponding Levi-Civita connection.
Accordingly, using (\ref{hyp7}), we obtain
\begin{equation}
K_{\alpha\beta}=-\nabla_\alpha n_\beta=n_i\nabla_\beta\nabla_\alpha x^i,
\end{equation}
which is consistent with the extrinsic curvature tensor for a surface (\ref{emb17}).\\

\noindent
{\bf Gau\ss--Codazzi equations}.\\
The intrinsic covariant derivative of a vector $W^i=\perp V^i$ on a hypersurface is
\begin{eqnarray}
& & D_j W_k=\perp \nabla_j W_k=(\delta^m_j-n^m n_j)(\delta^n_k-n^n n_k)\nabla_m W_n \nonumber \\
& & =\nabla_j W_k-n^m n_j \nabla_m W_k-n^n n_k \nabla_j W_n+n^m n_j n^n n_k \nabla_m W_n \nonumber \\
& & =\nabla_j W_k-n^m n_j \nabla_m W_k+n_k W^n \nabla_j n_n+n^m n_j n^n n_k \nabla_m W_n,
\end{eqnarray}
where we used $n_i W^i=0$, which gives $n^i\nabla_j W_i=-W^i\nabla_j n_i$.
Consequently, the second derivative is
\begin{eqnarray}
& & D_i D_j W_k=\perp(\nabla_i D_j W_k)=\perp(\nabla_i \perp \nabla_j W_k) \nonumber \\
& & =\perp \nabla_i \nabla_j W_k+\perp \nabla_i(-n^m n_j \nabla_m W_k+n_k W^n \nabla_j n_n+n^m n_j n^n n_k \nabla_m W_n) \nonumber \\
& & =\perp \nabla_i \nabla_j W_k+\perp \nabla_i n_k W^n \nabla_j n_n=\perp \nabla_i \nabla_j W_k+K_{ik}K_{jn}W^n,
\label{hyp41}
\end{eqnarray}
where we used (\ref{hyp6}) and (\ref{hyp7}).
The commutator of intrinsic covariant derivatives gives the intrinsic curvature tensor:
\begin{equation}
[D_i,D_j]W_k=-r^l_{\phantom{l}kij}W_l,
\end{equation}
whereas the commutator of covariant derivatives gives the Riemann tensor, according to (\ref{Riem0}).
Therefore, antisymmetrizing the indices $i,j$ in (\ref{hyp41}) gives
\begin{equation}
r^l_{\phantom{l}kij}W_l=\perp \mathring{R}^l_{\phantom{l}kij}W_l-K_{ik}K_{jl}W^l+K_{jk}K_{il}W^l,
\end{equation}
which leads to
\begin{equation}
\perp \mathring{R}_{lkij}=r_{lkij}+K_{ik}K_{jl}-K_{jk}K_{il}.
\label{hyp42}
\end{equation}
This equation is consistent with (\ref{emb22}) for $\mathring{R}_{lkij}=0$, satisfied for a curved surface in a flat space.

The projection of $n^i \mathring{R}_{ijkl}$ is given by
\begin{eqnarray}
& & \perp(n^i \mathring{R}_{ijkl})=\perp(\nabla_l \nabla_k n_j-\nabla_k \nabla_l n_j)=\perp(\nabla_l(K_{kj}-n_k n^i n_{j:i})-\nabla_k(K_{lj}-n_l n^i n_{j:i})) \nonumber \\
& & =\perp(\nabla_l K_{kj}-\nabla_k K_{lj}+(\nabla_k n_l-\nabla_l n_k)n^i n_{j:i})=D_l K_{kj}-D_k K_{lj},
\label{hyp43}
\end{eqnarray}
where we used (\ref{emb32}) and (\ref{hyp6}).
This equation is consistent with (\ref{emb23}) for $\mathring{R}_{lkij}=0$, satisfied for a curved surface in a flat space.
Equations (\ref{hyp42}) and (\ref{hyp43}) are referred to as the {\em Gau\ss--Codazzi equations}.

If the normal vector to a hypersurface is spacelike, then the hypersurface is timelike.
An example of such a hypersurface is a hypersurface on which a given spatial coordinate is constant.
The normal vector can be normalized:
\begin{equation}
n^i n_i=-1.
\end{equation}
The projection tensor onto a timelike hypersurface differs from (\ref{hyp3}) by a sign:
\begin{equation}
h^i_{\phantom{i}j}=\delta^i_j+n^i n_j,
\end{equation}
whereas all other definitions are the same as for spacelike hypersurfaces.\\

\noindent
{\bf Continuity at hypersurface}.\\
If a hypersurface is spacelike or timelike, and forms a boundary between two submanifolds in spacetime, then the intrinsic metric tensor $\gamma_{ij}$ and the extrinsic curvature tensor $K_{ij}$ are continuous across the hypersurface.
Consequently, the first and second fundamental forms are continuous across the hypersurface.
These two covariant conditions are referred to as the {\em Darmois--Israel junction conditions}.
Equivalently, the metric tensor $g_{ij}$ and its derivatives $g_{ij,k}$ are continuous across the hypersurface.
These two conditions are referred to as the {\em Lichnerowicz junction conditions}.\\
\end{footnotesize}
\newline
References: \cite{Schr,LL2,Lord,Hehl1,Wald}.

\subsection{Tetrad and spin connection}
\setcounter{equation}{0}
\subsubsection{Tetrad}
At each point in spacetime, in addition to a coordinate system, it is possible to set up four linearly independent vectors $e_a^i$ such that
\begin{equation}
e^i_a e_{ib}=\eta_{ab},
\label{tet1}
\end{equation}
where $\eta_{ab}=\mbox{diag}(1,-1,-1,-1)$ is the coordinate-invariant Minkowski tensor (\ref{Galfr1}), coinciding with the metric tensor in a Galilean form.
The first Latin letters $a,b,c,d,\dots$, which can be 0, 1, 2, or 3, are the coordinate-invariant {\em Lorentz indices}.
This set of four vectors is referred to as a {\em tetrad}.
The inverse tetrad $e^{ai}$ satisfies
\begin{eqnarray}
& & e^i_a e_i^b=\delta^b_a, \\
& & e^i_a e_k^a=\delta^i_k.
\end{eqnarray}
The coordinate metric tensors $g_{ik}$ and $g^{ik}$ are related to the Minkowski metric tensor through the tetrad:
\begin{eqnarray}
& & g_{ik}=e^a_i e^b_k \eta_{ab}, \label{tet4} \\
& & g^{ik}=e_a^i e_b^k \eta^{ab},
\label{tet5}
\end{eqnarray}
where $\eta^{ab}$ satisfies
\begin{equation}
\eta_{ac}\eta^{bc}=\delta_a^b.
\end{equation}

\noindent
{\bf Invariant indices}.\\
Any vector $V$ can be specified by its components $V^i$ with respect to the coordinate system or by the coordinate-invariant projections $V^a$ of the vector onto the tetrad field:
\begin{eqnarray}
& & V^a=e^a_i V^i,\quad V_a=e_a^i V_i, \\
& & V^i=e^i_a V^a,\quad V_i=e_i^a V_a,
\end{eqnarray}
and similarly for tensors and densities with more indices.
The tensor $\eta_{ab}$ and its inverse $\eta^{ab}$ are used to lower and raise Lorentz indices, as the tensor $g_{ik}$ and its inverse $g^{ik}$ are used to lower and raise coordinate indices.

The line element can be written in the Galilean form:
\begin{equation}
ds^2=\eta_{ab}e^a_i dx^i e^b_k dx^k,
\label{tet15}
\end{equation}
in which the linear forms $e^a_i dx^i$ are not (in general) exact differentials of any function of the coordinates.
The line element (\ref{intprop16}) corresponds to the tetrad $e^a_i$ ($\alpha=1,2,3$):
\begin{equation}
e^0_i=(\sqrt{h},-\sqrt{h}{\bf g}),\quad e^\alpha_i=(0,{\bf e}^\alpha),
\end{equation}
where ${\bf g}$ and $h$ are given by (\ref{spvec121}) and (\ref{spvec120}), and the choice of the vectors ${\bf e}^\alpha$ depends on the spatial form $dl^2$.\\

\noindent
{\bf Determinant of tetrad}.\\
The determinant of the matrix composed from the components of the tetrad,
\begin{equation}
\mathfrak{e}=|e^a_i|,
\label{tet9}
\end{equation}
is related to the determinant $\mathfrak{g}$ of the metric tensor $g_{ik}$, using (\ref{tet4}), by
\begin{equation}
\mathfrak{e}=\sqrt{|\mathfrak{g}|}.
\label{tet10}
\end{equation}
The differential and derivatives of the determinant (\ref{tet9}) are given, analogously to (\ref{metten16}) and (\ref{metten17}), by
\begin{eqnarray}
& & d\mathfrak{e}=\mathfrak{e}e^i_a de^a_i=-\mathfrak{e}e^a_i de^i_a, \\
& & \mathfrak{e}_{,k}=\mathfrak{e}e^i_a e^a_{i,k}=-\mathfrak{e}e^a_i e^i_{a,k}.
\end{eqnarray}
The variation of (\ref{tet9}) is thus, analogously to (\ref{metten18}), equal to
\begin{equation}
\delta\mathfrak{e}=\mathfrak{e}e^i_a\delta e^a_i=-\mathfrak{e}e^a_i\delta e^i_a.
\label{tet13}
\end{equation}
Similarly to (\ref{metten19}), the covariant derivative of (\ref{tet9}) vanishes:
\begin{equation}
\mathfrak{e}_{;j}=0.
\end{equation}

\subsubsection{Lorentz transformation}
The relation (\ref{tet4}) imposes 10 constraints on the 16 components of the tetrad, leaving 6 components arbitrary.
If we change from one tetrad $e^i_a$ to another, $\tilde{e}^i_b$, then the vectors of the new tetrad are linear combinations of the vectors of the old tetrad:
\begin{equation}
\tilde{e}^i_a=\Lambda^b_{\phantom{b}a}e^i_b.
\label{Lortr1}
\end{equation}
The relation (\ref{tet4}) applied to the tetrad field $\tilde{e}^i_b$,
\begin{equation}
g_{ik}=\tilde{e}^a_i\tilde{e}^b_k \eta_{ab},
\end{equation}
imposes on the matrix $\Lambda^b_{\phantom{b}a}$ the orthogonality condition:
\begin{equation}
\Lambda^c_{\phantom{c}a}\Lambda^d_{\phantom{d}b}\eta_{cd}=\eta_{ab}.
\label{Lortr3}
\end{equation}
We refer to $\Lambda^b_{\phantom{b}a}$ as a {\em Lorentz matrix}, and to a transformation of form (\ref{Lortr1}) as the {\em Lorentz transformation}.

\subsubsection{Tetrad transport}
A natural choice for the zeroth component of a tetrad at a given point is
\begin{equation}
e^i_0=u^i.
\label{FW1}
\end{equation}
Along a world line this tetrad should be transported such that the zeroth component always coincides with the four-velocity.
The {\em Fermi--Walker transport} of a tetrad is defined as
\begin{equation}
\frac{D^\ast e^i_a}{ds}=\frac{Du^i}{ds}u_j e^j_a-u^i\frac{Du_j}{ds}e^j_a.
\label{FW2}
\end{equation}
Putting $a=0$ in (\ref{FW2}) gives
\begin{equation}
\frac{D^\ast u^i}{ds}=\frac{Du^i}{ds},
\end{equation}
so the Fermi--Walker transport of the four-velocity is equivalent to its covariant change and thus (\ref{FW1}) is valid at all points.
This transport preserves the orthogonality relation for tetrads (\ref{tet1}) because (\ref{FW2}) gives
\begin{equation}
\frac{D^\ast}{ds}(e^i_a e_{ib})=0.
\end{equation}

\subsubsection{Spin connection}
We define
\begin{equation}
\omega^i_{\phantom{i}ak}=e^i_{a;k}=e^i_{a,k}+\Gamma^{i}_{jk}e^j_a.
\label{spcon1}
\end{equation}
The quantities
\begin{equation}
\omega^a_{\phantom{a}bi}=e^a_k\omega^k_{\phantom{k}bi}=e^a_k(e^k_{b,i}+\Gamma^{k}_{ji}e^j_b)
\label{spcon2}
\end{equation}
transform like vectors under coordinate transformations.
We can extend the notion of covariant differentiation to quantities with Lorentz coordinate-invariant indices by regarding $\omega^{ab}_{\phantom{ab}i}$ as a connection, referred to as Lorentz or {\em spin connection}.
For a contravariant Lorentz vector
\begin{equation}
V^a_{\phantom{a}|i}=V^a_{\phantom{a},i}+\omega^a_{\phantom{a}bi}V^b,
\end{equation}
where $_{|i}$ is the covariant derivative of such a quantity with respect to $x^i$.
The covariant derivative of a scalar $V^a W_a$ coincides with its ordinary derivative:
\begin{equation}
(V^a W_a)_{|i}=(V^a W_a)_{,i},
\end{equation}
which gives the covariant derivative of a covariant Lorentz vector:
\begin{equation}
W_{a|i}=W_{a,i}-\omega^b_{\phantom{b}ai}W_b.
\end{equation}
The chain rule infers that the covariant derivative of a Lorentz tensor is equal to the sum of the corresponding ordinary derivative of this tensor and terms with spin connection corresponding to each Lorentz index:
\begin{equation}
T^{ab\dots}_{\phantom{ab}cd\dots|i}=T^{ab\dots}_{\phantom{ab}cd\dots,i}+\omega^a_{\phantom{a}ei}T^{eb\dots}_{\phantom{eb}cd\dots}+\omega^b_{\phantom{b}ei}T^{ae\dots}_{\phantom{ae}cd\dots}+\dots-\omega^e_{\phantom{e}ci}T^{ab\dots}_{\phantom{ab}ed\dots}-\omega^e_{\phantom{e}di}T^{ab\dots}_{\phantom{ab}ce\dots}-\dots\,.
\end{equation}
We assume that the covariant derivative $_{|i}$ is total, that is, also recognizes coordinate indices, acting on them like $_{;i}$.
For a tensor with both coordinate and Lorentz indices
\begin{equation}
T^{aj\dots}_{\phantom{aj}bk\dots|i}=T^{aj\dots}_{\phantom{aj}bk\dots,i}+\omega^a_{\phantom{a}ei}T^{ej\dots}_{\phantom{ej}bk\dots}+\Gamma^{j}_{li}T^{al\dots}_{\phantom{al}bk\dots}+\dots-\omega^e_{\phantom{e}bi}T^{aj\dots}_{\phantom{aj}ek\dots}-\Gamma^{l}_{ki}T^{aj\dots}_{\phantom{aj}bl\dots}-\dots\,.
\end{equation}

A total covariant derivative of a tetrad is
\begin{equation}
e^i_{a|k}=e^i_{a,k}+\Gamma^{i}_{jk}e^j_a-\omega^b_{\phantom{b}ak}e^i_b=0,
\label{spcon7}
\end{equation}
because of (\ref{spcon1}).
Therefore, total covariant differentiation commutes with converting between coordinate and Lorentz indices.
Equation (\ref{spcon7}) determines the spin connection $\omega^a_{\phantom{a}bi}$ in terms of the affine connection, the tetrad, and its partial derivatives, in accordance with (\ref{spcon2}).
Conversely, the affine connection is determined by the spin connection, the tetrad, and its partial derivatives:
\begin{equation}
\Gamma^{j}_{ik}=\omega^j_{\phantom{j}ik}+e^a_{i,k}e^j_a.
\end{equation}
The torsion tensor is then related to these quantities by
\begin{equation}
S^j_{\phantom{j}ik}=\omega^j_{\phantom{j}[ik]}+e^a_{[i,k]}e^j_a,
\label{spcon9}
\end{equation}
and the torsion vector is
\begin{equation}
S_i=\omega^k_{\phantom{k}[ik]}+e^a_{[i,k]}e^k_a.
\end{equation}

Metric compatibility of the affine connection leads to
\begin{equation}
g_{ik;j}=g_{ik|j}=e^a_i e^b_k\eta_{ab|j}=-e^a_i e^b_k(\omega^c_{\phantom{c}aj}\eta_{cb}+\omega^c_{\phantom{c}bj}\eta_{ac})=-(\omega_{kij}+\omega_{ikj})=0,
\end{equation}
so the spin connection is antisymmetric in its first two indices:
\begin{equation}
\omega^a_{\phantom{a}bi}=-\omega_{b\phantom{a}i}^{\phantom{b}a}.
\end{equation}
Accordingly, the spin connection has 24 independent components.
The contortion tensor is related to the spin connection by
\begin{equation}
C_{ijk}=\omega_{ijk}+\Delta_{ijk},
\label{spcon13}
\end{equation}
where
\begin{equation}
\Delta_{ijk}=e_{ia}e^a_{[j,k]}-e_{ja}e^a_{[i,k]}-e_{ka}e^a_{[i,j]}
\end{equation}
are the {\em Ricci rotation coefficients}.
The first term on the right-hand side in (\ref{spcon13}) is expected because both the contortion tensor and spin connection are antisymmetric in their first two indices.
The quantities
\begin{equation}
\mathring{\omega}^i_{\phantom{i}ak}=e^i_{a:k}=e^i_{a,k}+\mathring{\Gamma}^{i}_{jk}e^j_a
\label{spcon15}
\end{equation}
form the {\em Levi-Civita spin connection} and are related to the Ricci rotation coefficients by (\ref{spcon13}) with $C_{ijk}=0$,
\begin{equation}
\mathring{\omega}_{ijk}=-\Delta_{ijk},
\end{equation}
so
\begin{equation}
C_{ijk}=\omega_{ijk}-\mathring{\omega}_{ijk}.
\label{spcon17}
\end{equation}

\subsubsection{Tetrad representation of curvature tensor}
The commutator of the covariant derivatives of a tetrad with respect to the affine connection is
\begin{equation}
2e^k_{a;[ji]}=R^k_{\phantom{k}lij}e^\sigma_a+2S^l_{\phantom{l}ij}e^k_{a;l}.
\end{equation}
This commutator can also be expressed in terms of the spin connection:
\begin{eqnarray}
& & e^k_{a;[ji]}=\omega^k_{\phantom{k}a[j;i]}=(e^k_b \omega^b_{\phantom{b}a[j})_{;i]}=\omega_{ba[j}\omega^{k b}_{\phantom{k b}i]}+\omega^b_{\phantom{b}a[j;i]}e^k_b \nonumber \\
& & =\omega_{ba[j}\omega^{k b}_{\phantom{k b}i]}+\omega^b_{\phantom{b}a[j,i]}e^k_b+S^l_{\phantom{l}ij}\omega^k_{\phantom{k}al}.
\end{eqnarray}
Consequently, the curvature tensor with two Lorentz and two coordinate indices depends only on the spin connection and its ordinary derivatives:
\begin{equation}
R^a_{\phantom{a}bij}=\omega^a_{\phantom{a}bj,i}-\omega^a_{\phantom{a}bi,j}+\omega^a_{\phantom{a}ci}\omega^c_{\phantom{c}bj}-\omega^a_{\phantom{a}cj}\omega^c_{\phantom{c}bi}.
\label{tetrep3}
\end{equation}
Because the spin connection is antisymmetric in its first two indices, the tensor (\ref{tetrep3}) is antisymmetric in its first two (Lorentz) indices, like the Riemann tensor.
The contraction of the curvature tensor (\ref{tetrep3}) with a tetrad gives the Ricci tensor with one Lorentz and one coordinate index:
\begin{equation}
R_{bj}=R^a_{\phantom{a}bij}e^i_a.
\end{equation}
The contraction of the tensor $R^a_{\phantom{a}i}$ with a tetrad gives the Ricci scalar,
\begin{equation}
R=R^a_{\phantom{a}i}e^i_a=R^{ab}_{\phantom{ab}ij}e^i_a e^j_b.
\label{tetrep5}
\end{equation}

The Riemann tensor with two Lorentz and two coordinate indices depends on the Levi-Civita connection (\ref{spcon15}) the same way the curvature tensor depends on the affine connection:
\begin{equation}
\mathring{R}^a_{\phantom{a}bij}=\mathring{\omega}^a_{\phantom{a}bj,i}-\mathring{\omega}^a_{\phantom{a}bi,j}+\mathring{\omega}^a_{\phantom{a}ci}\mathring{\omega}^c_{\phantom{c}bj}-\mathring{\omega}^a_{\phantom{a}cj}\mathring{\omega}^c_{\phantom{c}bi}.
\label{tetrep6}
\end{equation}
The contraction of (\ref{tetrep6}) with a tetrad gives the Riemannian Ricci tensor with one Lorentz and one coordinate index:
\begin{equation}
\mathring{R}_{bj}=\mathring{R}^a_{\phantom{a}bij}e^i_a.
\end{equation}
The contraction of the tensor $\mathring{R}^a_{\phantom{a}i}$ with a tetrad gives the Riemann scalar,
\begin{equation}
\mathring{R}=\mathring{R}^a_{\phantom{a}i}e^i_a=\mathring{R}^{ab}_{\phantom{ab}ij}e^i_a e^j_b.
\label{tetrep8}
\end{equation}
\newline
References: \cite{Lord,Hehl1,Uti,KS,MTW}.

\subsection{Lorentz group}
\setcounter{equation}{0}
 \subsubsection{Subgroups of Lorentz group and Einstein principle of relativity}
 Lorentz transformations (\ref{Lortr1}) relate different tetrads at a given point in spacetime, where the metric tensor can be brought to the Galilean form: $g_{ik}=\eta_{ik}$.
The completely antisymmetric pseudotensors (\ref{metten14}), (\ref{metten15}), and (\ref{spvec106}) satisfy $E_{ijkl}=\varepsilon_{ijkl}$, $E^{ijkl}=\epsilon^{ijkl}$, $\eta_{\alpha\beta\gamma}=e_{\alpha\beta\gamma}$, and $\eta^{\alpha\beta\gamma}=e^{\alpha\beta\gamma}$.
The first two pseudotensors we denote in the Galilean frame of reference with $e$: $e_{ijkl}$ and $e^{ijkl}$.
In this section, the Greek letters $\mu,\nu,\rho,\sigma,\dots$ denote the spacetime Lorentz indices (like $a,b,c,d,\dots$) and the first Greek letters $\alpha,\beta,\gamma,\dots$ denote the space Lorentz indices.

A composition of two Lorentz transformations $\Lambda_1$ and $\Lambda_2$,
\begin{equation}
\Lambda^a_{\phantom{a}b}=\Lambda^a_{(1)c}\Lambda^c_{(2)b},
\end{equation}
satisfies the relation (\ref{Lortr3}), thereby it is a Lorentz transformation.
The Kronecker symbol $\delta^a_b$ also satisfies (\ref{Lortr3}), thereby it can be regarded as the identity Lorentz transformation.
Therefore, Lorentz transformations form a group, referred to as the {\em Lorentz group}.
Taking the determinant of (\ref{Lortr3}) gives
\begin{equation}
|\Lambda^a_{\phantom{a}b}|=\pm1.
\end{equation}
A Lorentz transformation with $|\Lambda^a_{\phantom{a}b}|=1$ is {\em proper} and with $|\Lambda^a_{\phantom{a}b}|=-1$ is {\em improper}.
Proper Lorentz transformations form a group because the determinant of the product of two proper Lorentz transformations is 1.
Improper Lorentz transformations include the {\em parity} transformation $P$:
\begin{equation}
\Lambda^a_{\phantom{a}b}(P)=\textrm{diag}(1,-1,-1,-1),\quad t\rightarrow t,\,\,{\bf x}\rightarrow-{\bf x},
\label{sub3}
\end{equation}
and the {\em time reversal} $T$:
\begin{equation}
\Lambda^a_{\phantom{a}b}(T)=\textrm{diag}(-1,1,1,1),\quad t\rightarrow-t,\,\,{\bf x}\rightarrow{\bf x}.
\label{sub4}
\end{equation}
The relation (\ref{Lortr3}) gives $\Lambda^0_{\phantom{0}0}\Lambda^0_{\phantom{0}0}-\Lambda^0_{\phantom{0}\alpha}\Lambda^0_{\phantom{0}\alpha}=1$, thereby
\begin{equation}
|\Lambda^0_{\phantom{0}0}|\geq1.
\end{equation}

Lorentz transformations with $\Lambda^0_{\phantom{0}0}\geq1$ are {\em orthochronous} and form a group.
If $x^i$ is a timelike vector, $x^i x_i>0$, then for an orthochronous transformation $x'^0=\Lambda^0_{\phantom{0}0}x^0+\Lambda^0_{\phantom{0}\alpha}x^\alpha$,
\begin{equation}
|\Lambda^0_{\phantom{0}\alpha}x^\alpha|\leq\sqrt{\Lambda^0_{\phantom{0}\alpha}\Lambda^0_{\phantom{0}\alpha}x^\beta x^\beta}<\sqrt{(\Lambda^0_{\phantom{0}0})^2(x^0)^2}=|\Lambda^0_{\phantom{0}0}x^0|.
\end{equation}
Therefore, the time component of a timelike vector does not change the sign under orthochronous transformations.
Einstein's {\em special principle of relativity} states that physical laws do not change their form under transformations within the orthochronous proper subgroup of the Lorentz group.
Equivalently, physical laws have the same form in all admissible inertial frames of reference.
The special principle of relativity is a special case of the general principle of relativity, in which arbitrary differentiable coordinate transformations are restricted to linear transformations (orthochronous proper Lorentz transformations) between inertial frames of reference.

Under the parity transformation, the spatial components of contravariant and covariant vectors, which form spatial vectors, change the sign.
The permutation symbols do not change under this transformation.
Accordingly, the spatial components of dual vector densities, such as the components of a vector product (\ref{spvec17}) or a curl (\ref{spvec29}), do not change the sign.
Such quantities, that transform under proper Lorentz transformations like vectors and do not change the sign in their spatial components under the parity transformation, are referred to as axial vectors or {\em pseudovectors}.
Similarly, the scalar contraction of the Levi-Civita symbol and a tensor changes the sign, while a scalar does not.
Quantities that transform under proper Lorentz transformations like scalars and change the sign under the parity transformation are referred to as {\em pseudoscalars}.

\subsubsection{Infinitesimal Lorentz transformations}
Let us consider an infinitesimal Lorentz transformation
\begin{equation}
\Lambda^\mu_{\phantom{\mu}\nu}=\delta^\mu_\nu+\epsilon^\mu_{\phantom{\mu}\nu},
\label{infLor1}
\end{equation}
where $\epsilon^\mu_{\phantom{\mu}\nu}$ are infinitesimal quantities.
The relation (\ref{Lortr3}) gives
\begin{equation}
\epsilon_{\mu\nu}=-\epsilon_{\nu\mu},
\label{infLor2}
\end{equation}
where the indices are raised and lowered using the Minkowski metric tensor.
Therefore, Lorentz transformations are given by 6 independent antisymmetric parameters $\epsilon_{\mu\nu}$.
The corresponding transformation of a contravariant vector $A^\mu$ is
\begin{equation}
A'^\mu=A^\mu+\epsilon^\mu_{\phantom{\mu}\nu}A^\nu=A^\mu+\frac{1}{2}\epsilon^{\rho\sigma}(\delta^\mu_\rho \eta_{\sigma\nu}-\delta^\mu_\sigma \eta_{\rho\nu})A^\nu=A^\mu+\frac{1}{2}\epsilon^{\rho\sigma}J^\mu_{\nu\rho\sigma}A^\nu,
\end{equation}
where
\begin{equation}
J^\mu_{\nu\rho\sigma}=\delta^\mu_\rho \eta_{\sigma\nu}-\delta^\mu_\sigma \eta_{\rho\nu}.
\label{infLor4}
\end{equation}
We define matrices $J_{\rho\sigma}$ such that
\begin{equation}
(J_{\rho\sigma})^\mu_\nu=J^\mu_{\nu\rho\sigma}.
\label{infLor5}
\end{equation}
Therefore, in the matrix notation (with $A^\mu$ treated as a column),
\begin{equation}
A'=\Bigl(1+\frac{1}{2}\epsilon^{\rho\sigma}J_{\rho\sigma}\Bigr)A.
\label{infLor6}
\end{equation}
The 6 matrices $J_{\rho\sigma}$ are the infinitesimal {\em generators} of the {\em vector representation} of the Lorentz group.
The explicit form of the generators of the Lorentz group in the vector representation is
\begin{eqnarray}
& & J_{01}=\left( \begin{array}{cccc}
0 & -1 & 0 & 0 \\
-1 & 0 & 0 & 0 \\
0 & 0 & 0 & 0 \\
0 & 0 & 0 & 0 \end{array} \right),\quad 
J_{02}=\left( \begin{array}{cccc}
0 & 0 & -1 & 0 \\
0 & 0 & 0 & 0 \\
-1 & 0 & 0 & 0 \\
0 & 0 & 0 & 0 \end{array} \right), \nonumber \\
& & J_{03}=\left( \begin{array}{cccc}
0 & 0 & 0 & -1 \\
0 & 0 & 0 & 0 \\
0 & 0 & 0 & 0 \\
-1 & 0 & 0 & 0 \end{array} \right),\quad 
J_{12}=\left( \begin{array}{cccc}
0 & 0 & 0 & 0 \\
0 & 0 & -1 & 0 \\
0 & 1 & 0 & 0 \\
0 & 0 & 0 & 0 \end{array} \right), \nonumber \\
& & J_{23}=\left( \begin{array}{cccc}
0 & 0 & 0 & 0 \\
0 & 0 & 0 & 0 \\
0 & 0 & 0 & -1 \\
0 & 0 & 1 & 0 \end{array} \right),\quad 
J_{31}=\left( \begin{array}{cccc}
0 & 0 & 0 & 0 \\
0 & 0 & 0 & 1 \\
0 & 0 & 0 & 0 \\
0 & -1 & 0 & 0 \end{array} \right).
\end{eqnarray}

\subsubsection{Generators and Lie algebra of Lorentz group}
The commutator of the generators of the Lorentz group in the vector representation is given, using (\ref{infLor4}) and (\ref{infLor5}), by
\begin{equation}
[J_{\kappa\tau},J_{\rho\sigma}]^\mu_\nu=(J_{\kappa\tau})^\mu_\lambda(J_{\rho\sigma})^\lambda_\nu-(J_{\rho\sigma})^\mu_\lambda(J_{\kappa\tau})^\lambda_\nu=(-J_{\kappa\rho}\eta_{\tau\sigma}-J_{\tau\sigma}\eta_{\kappa\rho}+J_{\kappa\sigma}\eta_{\tau\rho}+J_{\tau\rho}\eta_{\kappa\sigma})^\mu_\nu,
\end{equation}
so
\begin{equation}
[J_{\kappa\tau},J_{\rho\sigma}]=-J_{\kappa\rho}\eta_{\tau\sigma}-J_{\tau\sigma}\eta_{\kappa\rho}+J_{\kappa\sigma}\eta_{\tau\rho}+J_{\tau\rho}\eta_{\kappa\sigma}.
\label{genLie2}
\end{equation}
The relation (\ref{genLie2}) constitutes the {\em Lie algebra} of the {\em Lorentz group}.
If a set of quantities $\phi$ transforms under a Lorentz transformation $\Lambda$ with a matrix $D(\Lambda)$
\begin{equation}
\phi\rightarrow D(\lambda)\phi,
\end{equation}
then $D$ is a representation of the Lorentz group if
\begin{equation}
D(I)=I,\quad D(\Lambda_1\Lambda_2)=D(\Lambda_1)D(\Lambda_2),
\end{equation}
where $I$ denotes the identity transformation, and $\Lambda_1$ and $\Lambda_2$ are two Lorentz transformations.
Therefore, we have
\begin{equation}
D(\Lambda^{-1})=D^{-1}(\Lambda),
\end{equation}
where $\Lambda^{-1}$ is the Lorentz transformation to $\Lambda$: $\Lambda\Lambda^{-1}=I$.
For an infinitesimal Lorentz transformation in any representation,
\begin{equation}
D(\Lambda)=I+\frac{1}{2}\epsilon^{\rho\sigma}J_{\rho\sigma},
\label{genLie6}
\end{equation}
according to (\ref{infLor6}).
The relation
\begin{equation}
D(\Lambda_1\Lambda_2\Lambda_1^{-1})=D(\Lambda_1)D(\Lambda_2)D^{-1}(\Lambda_1)
\end{equation}
gives (\ref{genLie2}), valid for any representation of the Lorentz group.\\

\begin{footnotesize}
If $\Lambda_1$ and $\Lambda_2$ are two group transformations then $\Lambda_3=\Lambda_1\Lambda_2\Lambda_1^{-1}$ is a group transformation.
If $\Lambda_2=I+\epsilon_2 G_2$ is an infinitesimal group transformation with generator $G_2$ then $\Lambda_3=I+\epsilon_2\Lambda_1 G_2 \Lambda_1^{-1}$ is an infinitesimal group transformation with generator $G_3=\Lambda_1 G_2 \Lambda_1^{-1}$.
If $\Lambda_1=I+\epsilon_1 G_1$ is an infinitesimal group transformation with generator $G_1$ then, neglecting terms in $\epsilon_1$ of higher order, $G_3=G_2+\epsilon_1[G_1,G_2]$, thereby $[G_1,G_2]$ is a generator.
For a finite number $N$ of linearly independent generators, a general infinitesimal group transformation is $\Lambda=I+\Sigma_{a=1}^N\epsilon_a G_a$.
Because $[G_a,G_b]$ is a generator, it is a linear combination of the $N$ generators: $[G_a,G_b]=\Sigma_{c=1}^N f_{abc}G_c$, where $f_{abc}$ are the structure constants of the Lie algebra of the given group.
For the Lorentz group, $\epsilon_a G_a=D(\Lambda)-I$, where $D(\Lambda)$ is given by (\ref{genLie6}).    
\end{footnotesize}

\subsubsection{Rotations and boosts}
{\em Rotations} are proper orthochronous Lorentz transformations with 
\begin{equation}
\Lambda^0_{\phantom{0}\alpha}=\Lambda^\alpha_{\phantom{\alpha}0}=0,\quad \Lambda^0_{\phantom{0}0}=1.
\end{equation}
Rotations act only on the spatial coordinates $x^\alpha$ and form a group, referred to as the {\em rotation group}.
{\em Boosts} are proper orthochronous Lorentz transformations with 
\begin{equation}
\Lambda^\alpha_{\phantom{\alpha}\beta}=0.
\end{equation}
We define
\begin{eqnarray}
& & J_\alpha=\frac{1}{2}e_{\alpha\beta\gamma}J^{\beta\gamma}, \\
& & K_\alpha=J_{0\alpha},
\end{eqnarray}
and
\begin{eqnarray}
& & \vartheta_\alpha=\frac{1}{2}e_{\alpha\beta\gamma}\epsilon^{\beta\gamma}, \label{rotbo5} \\
& & \eta_\alpha=\epsilon_{0\alpha}.
\label{rotbo6}
\end{eqnarray}
The explicit form of the generators of the rotation group $J_\alpha$ in the vector representation is
\begin{equation}
J_1=\left( \begin{array}{ccc}
0 & 0 & 0 \\
0 & 0 & -1 \\
0 & 1 & 0 \end{array} \right),\quad 
J_2=\left( \begin{array}{ccc}
0 & 0 & 1 \\
0 & 0 & 0 \\
-1 & 0 & 0 \end{array} \right),\quad 
J_3=\left( \begin{array}{ccc}
0 & -1 & 0 \\
1 & 0 & 0 \\
0 & 0 & 0 \end{array} \right).
\end{equation}
For an infinitesimal Lorentz transformation (\ref{genLie6})
\begin{equation}
D=I+{\bm \vartheta}\cdot{\bf J}+{\bm \eta}\cdot{\bf K}.
\end{equation}

\noindent
{\bf Finite rotations and boosts}.\\
A finite Lorentz transformation can be regarded as a composition of successive identical infinitesimal Lorentz transformations:
\begin{equation}
D=\mbox{lim}_{n\rightarrow\infty}(I+{\bm \theta}\cdot{\bf J}/n+{\bm \eta}\cdot{\bf K}/n)^n=e^{{\bm \theta}\cdot{\bf J}+{\bm \eta}\cdot{\bf K}}.
\end{equation}
The finite parameters ${\bm \theta}$, ${\bm \eta}$ are the {\em canonical parameters} for a given Lorentz transformations.
For a finite Lorentz transformation, (\ref{genLie6}) gives
\begin{equation}
D(\Lambda)=\exp\Bigl(\frac{1}{2}\epsilon^{\rho\sigma}J_{\rho\sigma}\Bigr),
\end{equation}
so
\begin{equation}
J_{\mu\nu}=\frac{\partial D(\Lambda)}{\partial\epsilon^{\mu\nu}}\Bigl|_{\Lambda=I}.
\label{rotbo11}
\end{equation}
The explicit form of a finite Lorentz transformation in the vector representation is
\begin{eqnarray}
& & R_1=e^{\theta J_1}=\left( \begin{array}{cccc}
1 & 0 & 0 & 0 \\
0 & 1 & 0 & 0 \\
0 & 0 & \cos\theta & -\sin\theta \\
0 & 0 & \sin\theta & \cos\theta \end{array} \right),\quad 
R_2=e^{\theta J_2}=\left( \begin{array}{cccc}
1 & 0 & 0 & 0 \\
0 & \cos\theta & 0 & \sin\theta \\
0 & 0 & 1 & 0 \\
0 & -\sin\theta & 0 & \cos\theta \end{array} \right), \nonumber \\
& & R_3=e^{\theta J_3}=\left( \begin{array}{cccc}
1 & 0 & 0 & 0 \\
0 & \cos\theta & -\sin\theta & 0 \\
0 & \sin\theta & \cos\theta & 0 \\
0 & 0 & 0 & 1 \end{array} \right),\quad 
B_1=e^{\eta K_1}=\left( \begin{array}{cccc}
\cosh\eta & \sinh\eta & 0 & 0 \\
\sinh\eta & \cosh\eta & 0 & 0 \\
0 & 0 & 1 & 0 \\
0 & 0 & 0 & 1 \end{array} \right), \nonumber \\
& & B_2=e^{\eta K_2}=\left( \begin{array}{cccc}
\cosh\eta & 0 & \sinh\eta & 0 \\
0 & 1 & 0 & 0 \\
\sinh\eta & 0 & \cosh\eta & 0 \\
0 & 0 & 0 & 1 \end{array} \right),\quad 
B_3=e^{\eta K_3}=\left( \begin{array}{cccc}
\cosh\eta & 0 & 0 & \sinh\eta \\
0 & 1 & 0 & 0 \\
0 & 0 & 1 & 0 \\
\sinh\eta & 0 & 0 & \cosh\eta \end{array} \right), \nonumber \\
\label{rotbo12}
\end{eqnarray}
where $R_\alpha$ denotes a rotation about the $x^\alpha$ axis and $B_\alpha$ denotes a boost along this axis.
The canonical parameters ${\bm \theta}$ and ${\bm \eta}$ are respectively referred to as the {\em angle of rotation} and {\em rapidity}.
The parameters ${\bm \vartheta}$ and ${\bm \eta}$ in (\ref{rotbo5}) and (\ref{rotbo6}) are thus respectively infinitesimal values of the angle of rotation and rapidity.
A rotation about any axis, say $z$, by an angle $\theta$ turns the two other axes, $x$ and $y$, into new axes, $x'$ and $y'$, such that the angle between $x$ and $x'$ (or $y$ and $y'$) (\ref{spvec13}) is $\theta$.
The rotation group is {\em compact}: $\theta_\alpha\in[0,2\pi]$ and $\theta=2\pi\Leftrightarrow\theta=0$.
The explicit form of a finite rotation in the three-dimensional vector representation is
\begin{eqnarray}
& & R_1(\theta)=\left( \begin{array}{ccc}
1 & 0 & 0 \\
0 & \cos\theta & -\sin\theta \\
0 & \sin\theta & \cos\theta \end{array} \right),\quad 
R_2(\theta)=\left( \begin{array}{ccc}
\cos\theta & 0 & \sin\theta \\
0 & 1 & 0 \\
-\sin\theta & 0 & \cos\theta \end{array} \right), \nonumber \\
& & R_3(\theta)=\left( \begin{array}{ccc}
\cos\theta & -\sin\theta & 0 \\
\sin\theta & \cos\theta & 0 \\
0 & 0 & 1 \end{array} \right).
\label{rotbo13}
\end{eqnarray}
For instance,
\begin{equation}
\left( \begin{array}{c}
V_x \\
V_y \\
V_z \end{array} \right)
\rightarrow\left( \begin{array}{c}
V'_x \\
V'_y \\
V'_z \end{array} \right)
=R_3\left( \begin{array}{c}
V_x \\
V_y \\
V_z \end{array} \right)
=\left( \begin{array}{c}
V_x\cos\theta-V_y\sin\theta \\
V_x\sin\theta+V_y\cos\theta \\
V_z \end{array} \right).
\end{equation}
The relation (\ref{rotbo11}) gives
\begin{equation}
J_\alpha=\frac{\partial R_\alpha(\theta)}{\partial\theta}\Bigl|_{\theta=0}.
\label{rotbo15}
\end{equation}
The orthogonality relation (\ref{Lortr3}) applied to any of the rotation matrices (\ref{rotbo13}) shows that a rotation matrix $R$ is orthogonal, that is, its transpose $R^T$ is equal to its inverse $R^{-1}$:
\begin{equation}
R^T_\alpha=R^{-1}_\alpha,\quad R_\alpha R^T_\alpha=R^T_\alpha R_\alpha=I,
\end{equation}
where $I$ is the identity matrix.\\

\noindent
{\bf Commutation relations for rotations and boosts}.\\
The commutation relation (\ref{genLie2}) gives
\begin{eqnarray}
& & [J_\alpha,J_\beta]=e_{\alpha\beta\gamma}J_\gamma, \label{rotbo17} \\
& & [J_\alpha,K_\beta]=e_{\alpha\beta\gamma}K_\gamma, \label{rotbo18} \\
& & [K_\alpha,K_\beta]=-e_{\alpha\beta\gamma}J_\gamma.
\label{rotbo19}
\end{eqnarray}
Therefore, rotations do not commute and form a nonabelian group, rotations and boosts do not commute, and boosts do not commute.
Changing the order of two nonparallel boosts is equivalent to applying a rotation, referred to as the {\em Thomas--Wigner rotation}.
The structure constants of the Lie algebra of the rotation group are $f_{abc}=e_{abc}$.
Moreover, the square of the generators of rotation,
\begin{equation}
J^2=J_\alpha J_\alpha,
\end{equation}
commutes with $J_\alpha$:
\begin{equation}
[J^2,J_\beta]=[J_\alpha,J_\beta]J_\alpha+J_\alpha[J_\alpha,J_\beta]=e_{\alpha\beta\gamma}(J_\gamma J_\alpha+J_\alpha J_\gamma)=0.
\label{rotbo21}
\end{equation}
Definining
\begin{eqnarray}
{\bf L}=\frac{1}{2}({\bf J}+i{\bf K}), \\
{\bf Q}=\frac{1}{2}({\bf J}-i{\bf K}),
\end{eqnarray}
gives
\begin{eqnarray}
& & [L_\alpha,L_\beta]=e_{\alpha\beta\gamma}L_\gamma, \\
& & [Q_\alpha,Q_\beta]=e_{\alpha\beta\gamma}Q_\gamma, \\
& & [L_\alpha,Q_\beta]=0,
\end{eqnarray}
so the Lorentz group is isomorphic with the product of two complex rotation groups.
Accordingly, the Lorentz group can be regarded as the group of four-dimensional rotations in the Minkowski space, or the group of {\em tetrad rotations}.

\subsubsection{Poincar\'{e} group}
\noindent
{\bf Translations}.\\
Under an infinitesimal coordinate transformation (\ref{infcor1}) in a locally flat spacetime, (\ref{Chrsym22}) gives
\begin{equation}
\eta_{ik}\rightarrow\eta_{ik}-\xi_{i,k}-\xi_{k,i}.
\label{Poin1}
\end{equation}
Therefore, the tensor $\eta_{ik}$ is invariant under (\ref{infcor1}) (isometric) if $\xi^i$ is a Killing vector,
\begin{equation}
\xi_{(i,k)}=0,
\end{equation}
which has the solution
\begin{equation}
\xi^i=\epsilon^{ik}x_k+\epsilon^i,
\label{Poin3}
\end{equation}
where $\epsilon^{ik}$ and $\epsilon^i$ are constant.
The first term on the right-hand side of (\ref{Poin3}) corresponds to a Lorentz rotation described by 6 parameters $\epsilon^{ik}$ satisfying (\ref{infLor2}).
The second term on the right-hand side of (\ref{Poin3}) corresponds to a {\em translation}.
A combination of two translations does not change if their order is reversed, thereby translations commute:
\begin{equation}
[T_\mu,T_\nu]=0,
\label{Poin4}
\end{equation}
where $T_\mu$ is the generator of translation.
The relations (\ref{rotbo17}) and (\ref{rotbo18}) mean that $J^\alpha$ and $K^\alpha$ are spatial vectors under rotations.
Spatial translations are spatial vectors under rotations, while a time translation is a scalar:
\begin{eqnarray}
& & [J_\alpha,T_\beta]=e_{\alpha\beta\gamma}T_\gamma, \label{Poin5} \\
& & [J_\alpha,T_0]=0.
\label{Poin6}
\end{eqnarray}
The last relation indicates that the generators of rotations, like the generators of spatial translations, correspond to {\em conserved} quantities, which are quantities that do not change in time.
The covariant generalization of (\ref{Poin5}) and (\ref{Poin6}) is
\begin{equation}
[J_{\mu\nu},T_\rho]=T_\mu\eta_{\nu\rho}-T_\nu\eta_{\mu\rho}.
\label{Poin7}
\end{equation}
The relations (\ref{genLie2}), (\ref{Poin4}) and (\ref{Poin7}) constitute the Lie algebra of the inhomogeneous Lorentz or {\em Poincar\'{e} group}.
In particular,
\begin{eqnarray}
& & [K_\alpha,T_\beta]=-T_0\delta_{\alpha\beta}, \\
& & [K_\alpha,T_0]=-T_\alpha.
\end{eqnarray}
The last relation indicates that the generators of boosts do not correspond to conserved quantities.\\

\noindent
{\bf Differential representation}.\\
For an infinitesimal rotation about the $z$ axis,
\begin{eqnarray}
& & (I+\vartheta J_z)f(ct,{\bf x})=D(R_z(\vartheta))f({ct,\bf x})=f(ct,R_z(\vartheta){\bf x})\approx f(ct,x-\vartheta y,\vartheta x+y,z) \nonumber \\
& & =f(ct,{\bf x})-\vartheta y\frac{\partial f}{\partial x}+\vartheta x\frac{\partial f}{\partial y},
\end{eqnarray}
or
\begin{equation}
J_z=x\frac{\partial}{\partial y}-y\frac{\partial}{\partial x},
\end{equation}
which gives the differential representation of rotations:
\begin{equation}
J_\alpha=e_{\alpha\beta\gamma}x_\beta\partial_\gamma.
\label{Poin14}
\end{equation}
For an infinitesimal boost along the $z$ axis,
\begin{eqnarray}
& & (I+\eta K_z)f(ct,{\bf x})=D(B_z(\eta))f({ct,{\bf x}})=f(B_z(\eta)(ct,{\bf x}))\approx f(ct+\eta z,y,z+\eta ct) \nonumber \\
& & =f(ct,{\bf x})+\eta z\frac{\partial f}{c\partial t}+\eta ct\frac{\partial f}{\partial z},
\end{eqnarray}
or
\begin{equation}
K_z=z\frac{\partial}{c\partial t}+ct\frac{\partial}{\partial z},
\end{equation}
which gives the differential representation of boosts:
\begin{equation}
K_\alpha=x_\alpha\frac{\partial}{c\partial t}+ct\frac{\partial}{\partial x^\alpha}.
\end{equation}

The relation for an infinitesimal translation, analogous to (\ref{genLie6}), is
\begin{equation}
D(t)=I+\epsilon^\mu T_\mu,
\end{equation}
so a finite translation is given by
\begin{equation}
D(t)=e^{\epsilon^\mu T_\mu}.
\end{equation}
Translation in (\ref{Poin3}) can also be written as
\begin{equation}
t_\mu(\epsilon)x^\nu=x^\nu+\epsilon\delta^\nu_\mu.
\end{equation}
The relation analogous to (\ref{rotbo15}) is
\begin{equation}
T_\mu=\frac{\partial t_\mu(\epsilon)}{\partial\epsilon}\Bigl|_{\epsilon=0}.
\end{equation}
The differential representation of a translation is thus
\begin{equation}
T_\mu=\frac{\partial}{\partial x^\mu}.
\label{Poin22}
\end{equation}

\subsubsection{Invariants of Lorentz and Poincar\'{e} group}
Analogously to (\ref{rotbo21}),
\begin{eqnarray}
& & [L^2,L_\beta]=0, \\
& & [Q^2,Q_\beta]=0,
\end{eqnarray}
so $L^2$ and $Q^2$ commute with all 6 generators of the Lorentz group.
Consequently, $J^2+K^2$ and ${\bf J}\cdot{\bf K}$ commute with all generators of the Lorentz group, that is, are the invariants or {\em Casimir operators} of the Lorentz group.
The Casimir operators of the Lorentz group do not commute with the generators of translation $T_\mu$, thereby they are not the invariants of the Poincar\'{e} group.
Instead, the {\em mass operator}
\begin{equation}
m^2=-T^\mu T_\mu
\label{Cas3}
\end{equation}
and
\begin{equation}
W^2=W^\mu W_\mu,
\end{equation}
where $W^\mu$ is the {\em Pauli--Luba\'{n}ski pseudovector}
\begin{equation}
W^\mu=\frac{1}{2}e^{\mu\nu\rho\sigma}J_{\rho\sigma}T_\nu,
\label{Cas5}
\end{equation}
commute with all generators of the Poincar\'{e} group, thereby they are the Casimir operators of the Poincar\'{e} group.
The Pauli--Luba\'{n}ski pseudovector obeys the commutation relations
\begin{eqnarray}
& & [T_\mu,W_\nu]=0, \\
& & [J_{\mu\nu},W_\rho]=W_\mu\eta_{\nu\rho}-W_\nu\eta_{\mu\rho}, \label{Cas7} \\
& & [W^\mu,W^\nu]=e^{\mu\nu\rho\sigma}W_\rho T_\sigma.
\end{eqnarray}
The relation (\ref{Cas7}) is analogous to (\ref{Poin7}) because $W^\mu$ behaves like a vector under proper Lorentz transformations.

We define the {\em four-momentum operator}
\begin{equation}
P_\mu=iT_\mu,
\label{Cas9}
\end{equation}
whose time component is the {\em energy operator} $P_0=iT_0$ and spatial components form the {\em momentum operator} $P_\alpha=iT_\alpha$.
We define the {\em angular four-momentum operator}
\begin{equation}
M_{\mu\nu}=iJ_{\mu\nu},
\label{Cas10}
\end{equation}
whose spatial components form the {\em angular momentum operator}
\begin{equation}
M_\alpha=iJ_\alpha.
\label{Cas11}
\end{equation}
Therefore, the following relations are satisfied:
\begin{eqnarray}
& & [M_{\mu\nu},M_{\rho\sigma}]=-i(M_{\mu\rho}\eta_{\nu\sigma}+M_{\nu\sigma}\eta_{\mu\rho}-M_{\mu\sigma}\eta_{\nu\rho}-M_{\nu\rho}\eta_{\mu\sigma}), \\
& & [P_\mu,P_\nu]=0, \\
& & [M_{\mu\nu},P_\rho]=i(P_\mu\eta_{\nu\rho}-P_\nu\eta_{\mu\rho}), \\
& & m^2=P^\mu P_\mu, \label{Cas15} \\
& & W^\mu=-\frac{1}{2}e^{\mu\nu\rho\sigma}M_{\rho\sigma}P_\nu, \\
& & [P_\mu,W_\nu]=0, \\
& & [M_{\mu\nu},W_\rho]=i(W_\mu\eta_{\nu\rho}-W_\nu\eta_{\mu\rho}), \\
& & [W^\mu,W^\nu]=-ie^{\mu\nu\rho\sigma}W_\rho P_\sigma, \\
& & [M_\alpha,M_\beta]=ie_{\alpha\beta\gamma}M_\gamma.
\label{Cas20}
\end{eqnarray}

\subsubsection{Relativistic kinematics}
Lorentz transformation matrices $\Lambda$ act on tetrads, which are constructed in a locally Minkowski (flat) spacetime.
They include rotations $R_\alpha$ and boosts $B_\alpha$ (\ref{rotbo12}), which are linear transformations.
Therefore, Lorentz matrices rotate Cartesian coordinates in that spacetime:
\begin{equation}
x^i=\Lambda^i_{\phantom{i}j}x'^j.
\end{equation}
They have constant coordinate transformation matrices:
\begin{equation}
\frac{\partial x^i}{\partial x'^{j}}=\Lambda^i_{\phantom{i}j}.
\end{equation}
which transform a contravariant vector (\ref{vec9}).
The inverse coordinate transformation matrices, which are also constant, transform a covariant vector (\ref{vec10}):
\begin{equation}
\frac{\partial x'^i}{\partial x^{j}}=(\Lambda^{-1})^i_{\phantom{i}j}.
\end{equation}
Consequently, Lorentz matrices rotate vectors and tensors in a locally flat spacetime:
\begin{eqnarray}
& & A^i=\Lambda^i_{\phantom{i}j}A'^j,\quad B_i=(\Lambda^{-1})^j_{\phantom{j}i}B'_j,
\label{relkin2} \\
& & T^{ij\dots}_{\phantom{ij}kl\dots}=\Lambda^i_{\phantom{i}m}\Lambda^j_{\phantom{j}n}(\Lambda^{-1})^p_{\phantom{p}k}(\Lambda^{-1})^q_{\phantom{q}l}T^{'mn\dots}_{\phantom{'mn}pq\dots},
\label{relkin3}
\end{eqnarray}
where the last equation is a special case of (\ref{tens2}).\\

\noindent
{\bf Special Lorentz transformation}.\\
Let us consider a boost in the direction of the $X$ axis:
\begin{equation}
x^i=(B_1)^i_{\phantom{i}j}x'^j=(e^{\eta K_1})^i_{\phantom{i}j}x'^j,
\end{equation}
where $x^i$ and $x'^i$ have a form of a column (4$\times$1 matrix), and the boost matrix $B_1=e^{\eta K_1}$ is given by (\ref{rotbo12}).
It is equivalent to defining the $X$ axis as the axis parallel to this boost.
Therefore, the coordinates in an inertial $K$-system (unprimed) are related to the coordinates in an inertial $K'$-system (primed) by
\begin{eqnarray}
& & ct=ct'\cosh\eta+x'\sinh\eta, \nonumber \\
& & x=x'\cosh\eta+ct'\sinh\eta, \nonumber \\
& & y=y',\quad  z=z'.
\label{relkin4}
\end{eqnarray}
These relations also follow from the invariance of the finite interval (\ref{intprop2}):
\begin{equation}
s^2=(ct)^2-x^2-y^2-z^2={s'}^2=(ct')^2-{x'}^2-{y'}^2-{z'}^2.
\end{equation}
The transformation (\ref{relkin4}) relates the coordinates of a point in two inertial frames $K$ and $K'$, in which the $X,X'$ axes are coincident, and the $Y,Z$ axes are parallel to the $Y',Z'$ axes.

Let us consider the motion of the origin of the $K'$-system in the $K$-system.
Therefore, $x'=0$ and (\ref{relkin4}) gives
\begin{equation}
ct=ct'\cosh\eta,\quad  x=ct'\sinh\eta,
\end{equation}
which relates the rapidity $\eta$ to the speed $V=dx/dt$ of $K'$ relative to $K$:
\begin{equation}
\tanh\eta=\beta,
\label{relkin6}
\end{equation}
where
\begin{equation}
\beta=\frac{V}{c}.
\label{relkin7}
\end{equation}
Accordingly, $\cosh\eta=\gamma$ and $\sinh\eta=\beta\gamma$, where
\begin{equation}
\gamma=(1-\beta^2)^{-1/2}=\Bigl(1-\frac{V^2}{c^2}\Bigr)^{-1/2}.
\label{relkin8}
\end{equation}
This relation indicates that $V$ must be lesser than $c$ and therefore $\beta$ must be lesser than 1.
The relations (\ref{relkin4}) become
\begin{eqnarray}
& & ct=\gamma(ct'+\beta x'), \nonumber \\
& & x=\gamma(x'+\beta ct'), \nonumber \\
& & y=y',\quad z=z',
\label{relkin9}
\end{eqnarray}
and are referred to as a {\em special Lorentz transformation} in the $X$ direction.
The inverse transformation is given by
\begin{eqnarray}
& & ct'=\gamma(ct-\beta x), \nonumber \\
& & x'=\gamma(x-\beta ct), \nonumber \\
& & y'=y,\quad z'=z.
\label{relkin10}
\end{eqnarray}
This transformation differs from (\ref{relkin9}) by the opposite sign next to $\beta$, which is a consequence of symmetry: if $K'$ moves relative to $K$ with velocity ${\bf V}$, then $K$ moves relative to $K'$ with velocity $-{\bf V}$.\\

\noindent
{\bf Transformation of vectors}.\\
In a locally flat spacetime, contravariant vectors transform like $dx^i$ and thus like $x^i$ because the Lorentz transformation is linear.
Consequently, the Cartesian components of a contravariant vector $W^i$ transform according to (\ref{relkin2}) with the matrix $\Lambda$, analogously to (\ref{relkin9}):
\begin{eqnarray}
& & W^0=\gamma(W^{0'}+\beta W^{1'}), \nonumber \\
& & W^1=\gamma(W^{1'}+\beta W^{0'}), \nonumber \\
& & W^2=W^{2'},\quad  W^3=W^{3'}.
\label{relkin33}
\end{eqnarray}
The Cartesian components of a covariant vector $W_i$ transform according to (\ref{relkin2}) with the matrix $\Lambda^{-1}$, analogously to (\ref{relkin10}):
\begin{eqnarray}
& & W_0=\gamma(W_{0'}-\beta W_{1'}), \nonumber \\
& & W_1=\gamma(W_{1'}-\beta W_{0'}), \nonumber \\
& & W_2=W_{2'},\quad  W_3=W_{3'}.
\end{eqnarray}
Tensors transform like products of vectors, following (\ref{relkin3}).
For example, a tensor of rank (0,2) transforms according to
\begin{eqnarray}
& & T_{00}=\gamma(T_{00'}-\beta T_{01'})=\gamma^2(T_{0'0'}-\beta T_{1'0'}-\beta T_{0'1'}+\beta^2 T_{1'1'}), \nonumber \\
& & T_{01}=\gamma(T_{01'}-\beta T_{00'})=\gamma^2(T_{0'1'}-\beta T_{1'1'}-\beta T_{0'0'}+\beta^2 T_{1'0'}), \nonumber \\
& & T_{0\perp}=\gamma(T_{0'\perp'}-\beta T_{1'\perp'}), \nonumber \\
& & T_{11}=\gamma(T_{11'}-\beta T_{10'})=\gamma^2(T_{1'1'}-\beta T_{0'1'}-\beta T_{1'0'}+\beta^2 T_{0'0'}), \nonumber \\
& & T_{1\perp}=\gamma(T_{1'\perp'}-\beta T_{0'\perp'}), \nonumber \\
& & T_{\perp\perp}=T_{\perp'\perp'},
\label{relkin16}
\end{eqnarray}
where the index $\perp$ denotes either 2 or 3, and the transposed components $T_{ik}^T=T_{ki}$ transform like the transpositions of the right-hand sides in (\ref{relkin16}).
If the tensor $T_{ik}$ is antisymmetric, then 
$T_{01}=T_{0'1'}$.\\

\noindent
{\bf Tetrad for special Lorentz transformation}.\\
For a special Lorentz transformation along the $X$ axis, the differentials of the relations (\ref{relkin9}) are
\begin{eqnarray}
& & c\,dt=\gamma(c\,dt'+\beta\,dx'), \nonumber \\
& & dx=\gamma(dx'+\beta c\,dt'), \nonumber \\
& & dy=dy',\quad dz=dz'.
\label{relkin17}
\end{eqnarray}
Substituting them into the square of the line element (\ref{intprop2}) gives
\begin{equation}
ds^2=[\gamma(c\,dt'+\beta\,dx')]^2-[\gamma(dx'+\beta c\,dt')]^2-dy'^2-dz'^2.
\end{equation}
Comparing it with $ds^2=\eta_{ab}e^a_i dx'^i e^b_k d'x^k$ (\ref{tet15}) gives the tetrad $e^a_i$ for a special Lorentz transformation along the $X$ axis:
\begin{eqnarray}
& & e^0_i=(\gamma,\gamma\beta,0,0), \nonumber \\
& & e^1_i=(\gamma\beta,\gamma,0,0), \nonumber \\
& & e^2_i=(0,0,1,0),\quad e^3_i=(0,0,0,1).
\end{eqnarray}
The forms $e^a_i dx'^i$ are exact differentials because the tetrad coefficients are functions of a constant speed $V$.
The tetrad rows form a matrix:
\begin{equation}
e^a_i=\left( \begin{array}{cccc}
\gamma & \gamma\beta & 0 & 0 \\
\gamma\beta & \gamma & 0 & 0 \\
0 & 0 & 1 & 0 \\
0 & 0 & 0 & 1 \end{array} \right), 
\end{equation}
which is equal to the boost matrix $B_1$ in (\ref{rotbo12}).
Its determinant $\mathfrak{e}$ (\ref{tet9}) is therefore equal to the determinant of a proper Lorentz matrix, which is 1, in agreement with (\ref{tet10}) for $\mathfrak{g}=-1$ in a flat spacetime.\\

\noindent
{\bf Composition of velocities}.\\
For a special Lorentz transformation along the $X$ axis, the relations (\ref{relkin17}) can be written as
\begin{eqnarray}
& & dt=\gamma\Bigl(dt'+\frac{V}{c^2}dx'\Bigr), \nonumber \\
& & dx=\gamma(dx'+Vdt'), \nonumber \\
& & dy=dy',\quad  dz=dz'.
\end{eqnarray}
Using the components of the velocity (\ref{relkin1}) in $K$ and $K'$:
\begin{equation}
v_x=\frac{dx}{dt},\quad v_y=\frac{dy}{dt},\quad v_z=\frac{dz}{dt}, \quad v'_x=\frac{dx'}{dt'},\quad v'_y=\frac{dy'}{dt'},\quad v'_z=\frac{dz'}{dt'},
\end{equation}
gives the transformation of velocities:
\begin{eqnarray}
& & v_x=\frac{dx'+Vdt'}{dt'+V\,dx'/c^2}=\frac{v'_x+V}{1+Vv'_x/c^2}, \nonumber \\
& & v_y=\frac{dy'}{\gamma(dt'+V\,dx'/c^2)}=\frac{v'_y}{\gamma(1+Vv'_x/c^2)},\quad  v_z=\frac{v'_z}{\gamma(1+Vv'_x/c^2)}.
\label{relkin18}
\end{eqnarray}
If all the velocities are in the $X$ direction, then the first relation in (\ref{relkin18}) is
\begin{equation}
v=\frac{v'+V}{1+Vv'/c^2}.
\end{equation}
This relation is equivalent to the addition of rapidities:
\begin{equation}
\frac{v}{c}=\tanh\eta=\tanh(\eta'+\eta_0)=\frac{\tanh\eta'+\tanh\eta_0}{1+\tanh\eta'\tanh\eta_0}=\frac{v'/c+V/c}{1+(v'/c)(V/c)}.
\end{equation}

Two special Lorentz transformations in the same direction commute because of (\ref{rotbo19}): changing their order does not change the total Lorentz transformation.
If a Lorentz transformation from $K'$ to $K$ has parameters $\beta_1$ and $\gamma_1$, and a Lorentz transformation from $K''$ to $K'$ has parameters $\beta_2$ and $\gamma_2$, then a Lorentz transformation from $K''$ to $K$ has parameters $\beta_3$ and $\gamma_3$ such that
\begin{equation}
\beta_3=\frac{\beta_1+\beta_2}{1+\beta_1\beta_2},\quad \gamma_3=\gamma_1\gamma_2(1+\beta_1\beta_2).
\end{equation}
Two special Lorentz transformations in different directions do not commute because of the Thomas--Wigner rotation.\\

\noindent{\bf Aberration}.\\
If the velocity ${\bf v}$ lies in the $XY$ plane then the velocity ${\bf v}'$ lies in the $X'Y'$ plane.
If $\theta$ is the angle between ${\bf v}$ and the $X$ axis, and $\theta'$ is the angle between ${\bf v'}$ and the $X'$ axis, then ${\bf v}$ has components $v_x=v\cos\theta$, $v_y=v\sin\theta$ and ${\bf v}'$ has components $v'_x=v'\cos\theta'$, $v'_y=v'\sin\theta'$.
The formulas (\ref{relkin18}) give
\begin{equation}
\tan\theta=\frac{v'\sin\theta'}{\gamma(v'\cos\theta'+V)},
\label{relkin101}
\end{equation}
which relates the directions of the velocity in the two frames.
For $v=v'=c$, they give the {\em aberration} of a signal:
\begin{equation}
\sin\theta=\frac{\sqrt{1-V^2/c^2}\,\sin\theta'}{1+(V/c)\cos\theta'},\quad \cos\theta=\frac{\cos\theta'+V/c}{1+(V/c)\cos\theta'}.
\label{relkin28}
\end{equation}
If $V\ll c$, then $\sin\theta-\sin\theta'\approx-(V/c)\sin\theta'\cos\theta'$ and also $\Delta\theta=\theta'-\theta\ll\theta$.
Accordingly, $\sin\theta-\sin\theta'\approx\cos\theta'(\theta-\theta')$, giving $\Delta\theta=(V/c)\sin\theta'$.\\

\noindent
{\bf Lorentz--FitzGerald contraction}.\\
Let us consider two points at rest in an inertial frame of reference $K$ with positions $x_1$ and $x_2$, thereby the distance between them is $\Delta x=x_2-x_1$.
In the inertial frame $K'$, moving relative to $K$ in the $X$ direction with speed $V$, $x_1=\gamma(x'_1+Vt'_1)$ and $x_2=\gamma(x'_2+Vt'_2)$, thereby if $t'_1=t'_2$ is the time at which we measure (simultaneously) the positions of the two points then $\Delta x=\gamma(x'_2-x'_1)=\gamma\Delta x'$.
Therefore, the length of an object in $K'$, whose length in the rest frame $K$ is $l$ ({\em proper length}), is
\begin{equation}
l'=\frac{l}{\gamma}<l,
\end{equation}
which is referred to as the {\em Lorentz--FitzGerald contraction}.
The volume of an object in $K'$, whose volume in the rest frame $K$ is $V$ ({\em proper volume}), is
\begin{equation}
V'=\frac{V}{\gamma}.
\label{relkin23}
\end{equation}
Let us suppose that there are two rods of equal lengths, moving parallel relative to each other.
From the point of view of an observer moving with the first rod, the second one is shorter, and from the point of view of an observer moving with the second rod, the first one is shorter.
There is no contradiction in this statement because the positions of both ends of a rod must be measured simultaneously and the simultaneity is not invariant: from the transformation law (\ref{relkin9}) it follows that if $\delta t=0$ then $\delta t'\neq0$ and if $\delta t'=0$ then $\delta t\neq0$.\\

\noindent
{\bf Time dilation}.\\
Let us consider a clock (any mechanism with a periodic or evolutionary behavior) at rest in $K'$ with position $x'$; the time difference between two events with $t'_1$ and $t'_2$, as measured by this clock, is $\Delta t'=t'_2-t'_1$.
In the frame $K$, $t_1=\gamma(t'_1+Vx'/c^2)$ and $t_2=\gamma(t'_2+Vx'/c^2)$, thereby
\begin{equation}
\Delta t=t_2-t_1=\gamma\Delta t'>\Delta t'.
\end{equation}
Therefore, the rate of time is slower for moving clocks than those at rest ({\em time dilation}), in agreement with (\ref{intprop5}) and (\ref{intprop11}), from which $c^2 d\tau^2=c^2 dt^2-dl^2$ and
\begin{equation}
d\tau=\frac{1}{\gamma}dt.
\label{relkin25}
\end{equation}
Let us suppose that there are two clocks linked to the inertial frames $K$ and $K'$, and that when the clock in $K$ passes by the clock in $K'$ the readings of the two clocks coincide.
From the point of view of an observer in $K$ clocks in $K'$ go more slowly, and from the point of view of an observer in $K'$ clocks in $K$ go more slowly.
There is no contradiction in this statement because to compare the rates of the two clocks in $K$ and $K'$ we must compare the readings of the same moving clock in $K'$ with different clocks in $K$; we require several clocks in one frame and one in the other, thus the measurement process is not symmetric with respect to the two frames of reference.
The clock that goes more slowly is the one which is being compared with different clocks in the other frame.
The time interval measured by a clock is equal to the integral
\begin{equation}
\Delta t=\frac{1}{c}\int ds
\label{relkin26}
\end{equation}
along its world line.
Since the world line is a straight line for a clock at rest and a curved line for a clock moving such that it returns to the starting point, the integral $\int ds$ taken between two world points has its maximum value if it is taken along the straight line connecting these two points.\\

\noindent
{\bf Doppler effect}.\\
Let us suppose that an observer in frame $K$ measures a periodic signal with period $T$, frequency $\nu=1/T$ and wavelength $\lambda=c/\nu$, propagating in the $-X$ direction; the number of pulses in time $dt$ is $n=\nu dt$.
A second observer in frame $K'$, moving in the $X$ direction with speed $V$ relative to the first one, travels a distance $Vdt$ and measures $Vdt/\lambda$ more pulses: $n'=\nu(1+V/c)dt$.
Because the time interval $dt$ with respect to $K'$ is $dt'=dt/\gamma$, the frequency of the signal in $K'$ is $\nu'=\gamma\nu(1+V/c)$ or
\begin{equation}
\nu'=e^\eta\nu.
\label{relkin29}
\end{equation}
This dependence of the frequency of a signal on a frame of reference is referred to as the {\em Doppler effect}.\\

\noindent
{\bf Vector form of Lorentz transformation}.\\
A boost along an arbitrary direction is represented by
\begin{equation}
{\bm\beta}=\frac{{\bf V}}{c},
\end{equation}
whose magnitude is (\ref{relkin7}).
For this boost, the spatial vector ${\bf x}=(x,y,z)$ transforms such that its component parallel to the velocity ${\bf V}=c{\bm\beta}$ of $K'$ relative to $K$, ${\bf x}_\parallel=({\bf x}\cdot{\bf V}){\bf V}/V^2$, behaves like $x$ in (\ref{relkin9}) and its component perpendicular to ${\bf V}$, ${\bf x}_\perp={\bf x}-{\bf x}_\parallel$, behaves like $y$ or $z$ in (\ref{relkin9}):
\begin{eqnarray}
& & t=\gamma\Bigl(t'+\frac{{\bf V}\cdot{\bf x}'}{c^2}\Bigr), \nonumber \\
& & {\bf x}_\perp={\bf x}'_\perp, \nonumber \\
& & {\bf x}_\parallel=\gamma({\bf x}'_\parallel+{\bf V}t').
\label{relkin11}
\end{eqnarray}
Combining these relations gives
\begin{equation}
{\bf x}={\bf x}'_\perp+\gamma({\bf x}'_\parallel+{\bf V}t')={\bf x}'+(\gamma-1){\bf x}'_\parallel+\gamma{\bf V}t'={\bf x}'+\gamma{\bf V}t'+\frac{(\gamma-1)({\bf x}'\cdot{\bf V}){\bf V}}{V^2},
\label{relkin12}
\end{equation}
where $V=|{\bf V}|$.
Therefore, the transformation law for the coordinates in two inertial frames of reference is
\begin{equation}
\left( \begin{array}{c}
ct \\
{\bf x} \end{array} \right)
=\left( \begin{array}{cc}
\gamma & \gamma{\bm\beta} \\
\gamma{\bm\beta} & 1+[(\gamma-1){\bm\beta}/\beta^2]{\bm\beta} \end{array} \right)
\left( \begin{array}{cc}
ct' \\
{\bf x}' \end{array} \right),
\label{relkin13}
\end{equation}
or equivalently
\begin{equation}
\left( \begin{array}{c}
ct' \\
{\bf x}' \end{array} \right)
=\left( \begin{array}{cc}
\gamma & -\gamma{\bm\beta} \\
-\gamma{\bm\beta} & 1+[(\gamma-1){\bm\beta}/\beta^2]{\bm\beta} \end{array} \right)
\left( \begin{array}{cc}
ct \\
{\bf x} \end{array} \right).
\label{relkin14}
\end{equation}
The matrix in (\ref{relkin14}) is called a {\em boost matrix}.

For a boost along an arbitrary direction, contravariant vectors in the local Minkowski spacetime transform like $dx^i$ and thus like $x^i$, according to (\ref{relkin11}) and (\ref{relkin13}), giving
\begin{equation}
\left( \begin{array}{c}
W^0 \\
{\bf W} \end{array} \right)
=\left( \begin{array}{cc}
\gamma & \gamma{\bm\beta} \\
\gamma{\bm\beta} & 1+[(\gamma-1){\bm\beta}/\beta^2]{\bm\beta} \end{array} \right)
\left( \begin{array}{cc}
W'^0 \\
{\bf W}' \end{array} \right).
\label{relkin15}
\end{equation}
The components of the velocity (\ref{relkin1}) in $K$ and $K'$ are
\begin{equation}
{\bf v}=\frac{d{\bf x}}{dt},\quad {\bf v'}=\frac{d{\bf x'}}{dt'}.
\label{relkin19}
\end{equation}
Using the differentials of (\ref{relkin11}) and (\ref{relkin12}) in (\ref{relkin19}) give the Lorentz transformation of velocities:
\begin{equation}
{\bf v}=\frac{d{\bf x}'+\gamma{\bf V}dt'+(\gamma-1)(d{\bf x}'\cdot{\bf V}){\bf V}/V^2}{\gamma(dt'+{\bf V}\cdot d{\bf x}'/c^2)}=\frac{{\bf v'}+\gamma{\bf V}+(\gamma-1)({\bf v'}\cdot{\bf V}){\bf V}/V^2}{\gamma(1+{\bf v'}\cdot{\bf V}/c^2)}.
\label{relkin21}
\end{equation}
This transformation is not symmetric under interchanging ${\bf v}'$ and ${\bf V}$ unless these vectors are parallel to one another.
If $v'=c$ then $v=c$, in agreement with the constancy of the speed of propagation of interaction.\\

\noindent
{\bf Nonrelativistc limit}.\\
The above formulae refer to {\em relativistic kinematics}.
When $v\ll c$, at which $\gamma\rightarrow1$, these formulae reduce to their {\em nonrelativistic} limit.
This limit is equivalent to $c\rightarrow\infty$.
The Lorentz transformation (\ref{relkin9}) is equivalent to
\begin{equation}
t=\gamma\Bigl(t'+\frac{V}{c^2}x'\Bigr),\quad  x=\gamma(x'+Vt'),\quad  y=y',\quad  z=z'.
\end{equation}
In the nonrelativistic limit, it reduces to
\begin{equation}
t=t',\quad  x=x'+Vt',\quad  y=y',\quad  z=z'.
\label{relkin32}
\end{equation}
The time is therefore an absolute (invariant) quantity in nonrelativistic ({\em Newtonian}) physics.
The relations (\ref{relkin32}) do not depend on $c$ and therefore do not have a symmetry in $ct'$ and $x'$ like (\ref{relkin9}).
For a boost along an arbitrary direction, the Lorentz transformation (\ref{relkin13}) reduces to
\begin{equation}
t=t',\quad  {\bf x}={\bf x}'+{\bf V}t',
\label{relkin30}
\end{equation}
which is equivalent to (\ref{relkin32}) if ${\bf V}$ is along the $X$ axis.
These formulas constitute the {\em Galilei transformation}.
The transformation law for velocities (\ref{relkin21}) reduces to the simple addition of vectors:
\begin{equation}
{\bf v}={\bf v'}+{\bf V}.
\label{relkin31}
\end{equation}
Consequently, any two Galilei transformations commute because vector addition is a commuting operation.

\subsubsection{Four-acceleration}
\noindent
{\bf Components of four-velocity}.\\
In the Galilean system of coordinates, the line element in (\ref{intprop2}) is equal to
\begin{equation}
ds=cdt\Bigl(1-\frac{\sum_\alpha v^\alpha v^\alpha}{c^2}\Bigr)^{1/2}=cdt\Bigl(1-\frac{v^2}{c^2}\Bigr)^{1/2}=\frac{cdt}{\gamma}.
\label{foac1}
\end{equation}
This line element is a special case of that in (\ref{intprop19}) for $g_{00}=1$ and $g_{0\alpha}=0$.
The corresponding differential of the proper time is equal to (\ref{relkin25}).
In a locally inertial frame of reference, the components of the four-velocity in the Cartesian coordinates are
\begin{equation}
u^0=\frac{dx^0}{ds}=\frac{cdt}{ds}=\gamma,\quad u^\alpha=\frac{dx^\alpha}{ds}=\frac{dx^\alpha}{cdt/\gamma}=\frac{\gamma}{c}v^\alpha,
\label{foac2}
\end{equation}
which can be written as
\begin{equation}
u^i=\Bigl(\gamma,\frac{\gamma}{c}{\bf v}\Bigr),\quad u_i=\Bigl(\gamma,-\frac{\gamma}{c}{\bf v}\Bigr),
\label{foac3}
\end{equation}
where ${\bf v}$ is the velocity (\ref{relkin1}) and $\gamma=1/\sqrt{1-v^2/c^2}$ (as in (\ref{relkin8})).
The spatial components $u^\alpha$ in (\ref{foac2}) coincide with those in (\ref{intprop21}), whereas $u^0$ in (\ref{foac2}) is a special case of that in (\ref{intprop21}) for $g_{00}=1$ and $g_{0\alpha}=0$.\\

\noindent
{\bf Components of four-acceleration}.\\
We define the {\em four-acceleration}:
\begin{equation}
w^i=\frac{Du^i}{ds}=\frac{D^2 x^i}{ds^2}=u^k u^i_{\phantom{i};k}.
\label{foac4}
\end{equation}
This vector is orthogonal to $u^i$ because of (\ref{metten22}):
\begin{equation}
w^i u_i=\frac{1}{2}\frac{D}{ds}(u^i u_i)=0,
\label{foac5}
\end{equation}
thus having 3 independent components.
In a locally flat spacetime, a covariant differential $D$ reduces to an ordinary differential $d$, so the four-acceleration is given by
\begin{equation}
w^i=\frac{du^i}{ds}=\frac{d^2 x^i}{ds^2}=u^k u^i_{\phantom{i},k}.
\label{foac6}
\end{equation}
Its components in the Cartesian coordinates are
\begin{eqnarray}
& & w^0=\frac{du^0}{ds}=\frac{d\gamma}{cdt}\frac{dx^0}{ds}=\frac{\gamma}{c}\frac{d\gamma}{dt}=\frac{1}{2c}\frac{d}{dt}(\gamma^2)=\frac{1}{2c}\frac{d}{dt}\Bigl(\frac{1}{1-{\bf v}^2/c^2}\Bigr)=\Bigl(1-\frac{{\bf v}^2}{c^2}\Bigr)^{-2}\Bigl(\frac{{\bf v}}{c^3}\cdot\frac{d{\bf v}}{dt}\Bigr) \nonumber \\
& & =\frac{\gamma^4}{c^3}{\bf v}\cdot{\bf a}, \nonumber \\
& & w^\alpha=\frac{du^\alpha}{ds}=\frac{du^\alpha}{cdt}\frac{dx^0}{ds}=\frac{\gamma}{c^2}\frac{d}{dt}(\gamma v^\alpha)=\frac{\gamma^2}{c^2}a^\alpha+\frac{v^\alpha}{c^2}\gamma\frac{d\gamma}{dt}=\frac{\gamma^2}{c^2}a^\alpha+\frac{\gamma^4}{c^4}({\bf v}\cdot{\bf a})v^\alpha,
\end{eqnarray}
which can be written as
\begin{equation}
w^i=\Bigl(\frac{\gamma^4}{c^3}{\bf v}\cdot{\bf a},\frac{\gamma^2}{c^2}{\bf a}+\frac{\gamma^4}{c^4}({\bf v}\cdot{\bf a}){\bf v}\Bigr),\quad w_i=\Bigl(\frac{\gamma^4}{c^3}{\bf v}\cdot{\bf a},-\frac{\gamma^2}{c^2}{\bf a}-\frac{\gamma^4}{c^4}({\bf v}\cdot{\bf a}){\bf v}\Bigr),
\label{foac8}
\end{equation}
where ${\bf a}$ is the three-dimensional {\em acceleration} vector:
\begin{equation}
a^\alpha=\frac{dv^\alpha}{dt}=\frac{d^2 x^\alpha}{dt^2},\quad {\bf a}=\frac{d{\bf v}}{dt}=\frac{d^2{\bf x}}{dt^2}.
\label{foac9}
\end{equation}
The invariant square of the four-acceleration is thus
\begin{equation}
w^i w_i=\frac{\gamma^8}{c^6}({\bf v}\cdot{\bf a})^2-\Bigl(\frac{\gamma^2}{c^2}{\bf a}+\frac{\gamma^4}{c^4}({\bf v}\cdot{\bf a}){\bf v}\Bigr)^2=-\frac{\gamma^4}{c^4}\Bigl({\bf a}^2+\frac{\gamma^2}{c^2}({\bf v}\cdot{\bf a})^2\Bigr).
\label{foac10}
\end{equation}

\noindent
{\bf Instantaneous rest frame}.\\
If ${\bf v}=0$ at a given instant of time, the corresponding frame of reference is referred to as the {\em instantaneous rest frame}.
In this frame
\begin{equation}
w^i w_i=-\frac{a^2}{c^4},
\label{foac11}
\end{equation}
so
\begin{equation}
a_0=c^2\sqrt{-w^i w_i}
\label{foac12}
\end{equation}
is the magnitude of the acceleration in the instantaneous rest frame, called the {\em proper acceleration}.
Along an affine geodesic, the four-acceleration with respect to the affine connection (\ref{foac4}) vanishes because of (\ref{affgeo8}).
Along a metric geodesic, the four-acceleration with respect to the Levi-Civita connection (defined by (\ref{foac4}) with colon instead of semicolon) vanishes because of (\ref{metgeo3}).
The equation of geodesic deviation (\ref{geodev7}) determines the relative four-acceleration of two bodies moving along two infinitely close affine geodesics.\\

\noindent
{\bf Thomas precession}.\\
Let us suppose that a noninertial frame $K'$ moves with velocity ${\bf v}$ relative to an inertial frame of reference $K$.
If the velocity of $K'$ changes by $d{\bf v}'$ relative to the initial frame $K'$, then it changes by $d{\bf v}$ relative to $K$.
In the nonrelativistic limit, the two changes are equal, $d{\bf v}=d{\bf v}'$, and $K'$ does not rotate with respect to $K$.
In relativistic kinematics, these changes are different because of the Thomas--Wigner rotation.
The velocity of $K'$ relative to $K$ after the change, ${\bf v}+d{\bf v}$, is equal to $d{\bf v}'$ boosted by ${\bf v}$.
Using the Lorentz transformation (\ref{relkin21}), in which ${\bf v}'$ is replaced with $d{\bf v}'$ and ${\bf V}$ is replaced with ${\bf v}$, we obtain
\begin{equation}
{\bf v}+d{\bf v}=\frac{d{\bf v'}+\gamma{\bf v}+(\gamma-1)(d{\bf v'}\cdot{\bf v}){\bf v}/v^2}{\gamma(1+d{\bf v'}\cdot{\bf v}/c^2)},
\end{equation}
where $\gamma=(1-v^2/c^2)^{-1/2}$.
Keeping only terms linear in $d{\bf v'}$ leads to
\begin{equation}
{\bf v}+d{\bf v}=\frac{d{\bf v'}}{\gamma}+{\bf v}(1-d{\bf v'}\cdot{\bf v}/c^2)+\frac{\gamma-1}{\gamma}(d{\bf v'}\cdot{\bf v})\frac{{\bf v}}{v^2}.
\end{equation}
Taking the vector product of this relation with ${\bf v}$ gives
\begin{equation}
{\bf v}\times d{\bf v}=\frac{{\bf v}\times d{\bf v}'}{\gamma}.
\end{equation}
The angle of infinitesimal rotation from the nonrelativistic sum ${\bf v}+d{\bf v}'$ to the relativistic ${\bf v}+d{\bf v}$ determines the relativistic rotation of $K'$ with respect to $K$.
Using (\ref{spvec60}) with $\sin(d\theta)\approx d\theta$ and $d{\bm \theta}={\bf n}\,d\theta$, where ${\bf n}$ is a unit vector parallel to the axis of rotation, gives
\begin{equation}
d{\bm \theta}=\frac{({\bf v}+d{\bf v}')\times({\bf v}+d{\bf v})}{v^2}\approx\frac{{\bf v}\times(d{\bf v}-d{\bf v}')}{v^2}=\frac{1-\gamma}{v^2}({\bf v}\times d{\bf v}).
\end{equation}
Consequently, we obtain the angular velocity of the {\em Thomas precession}:
\begin{equation}
{\bm\Omega}=\frac{d{\bm \theta}}{dt}=\frac{\gamma-1}{v^2}({\bf a}\times{\bf v})=\frac{\gamma^2}{\gamma+1}\frac{{\bf a}\times{\bf v}}{c^2},
\end{equation}
where ${\bf a}=d{\bf v}/dt$ is the acceleration of $K'$ relative to $K$.
When $v\ll c$,
\begin{equation}
{\bm\Omega}\approx\frac{{\bf a}\times{\bf v}}{2c^2}.
\end{equation}
\newline
References: \cite{LL2,Lord}.

\subsection{Spinors}
\setcounter{equation}{0}
\subsubsection{Spinor representation of Lorentz group}
Let $\gamma^a$ be the coordinate-invariant 4$\times$4 {\em Dirac matrices} defined as
\begin{equation}
\gamma^a\gamma^b+\gamma^b\gamma^a=2\eta^{ab}I_4,
\label{spin1}
\end{equation}
where $I_4$ is the four-dimensional unit matrix (4 is the lowest dimension for which (\ref{spin1}) has solutions).
Accordingly, the spacetime-dependent Dirac matrices, $\gamma^i=e^i_a \gamma^a$, satisfy
\begin{equation}
\gamma^i\gamma^j+\gamma^j\gamma^i=2g^{ij}I_4.
\label{spin2}
\end{equation}
Under a tetrad rotation, (\ref{Lortr1}) gives
\begin{equation}
\tilde{\gamma}^a=\Lambda^a_{\phantom{a}b}\gamma^b.
\end{equation}
Let $L$ be a 4$\times$4 matrix such that
\begin{equation}
\gamma^a=\Lambda^a_{\phantom{a}b}L\gamma^b L^{-1}=L\tilde{\gamma}^a L^{-1},
\label{spin4}
\end{equation}
where $L^{-1}$ is the matrix inverse to $L$: $LL^{-1}=L^{-1}L=I_4$.
The condition (\ref{spin4}) represents the constancy of the Dirac matrices $\gamma^a$ under the combined tetrad rotation and transformation $\gamma\rightarrow L\gamma L^{-1}$.
We refer to $L$ as the {\em spinor representation} of the Lorentz group.
The relation (\ref{spin4}) gives the matrix $L$ as a function of the Lorentz matrix $\Lambda^a_{\phantom{a}b}$.
For an infinitesimal Lorentz transformation (\ref{infLor1}), the solution for $L$ is
\begin{equation}
L=I_4+\frac{1}{2}\epsilon_{ab}G^{ab},\quad L^{-1}=I_4-\frac{1}{2}\epsilon_{ab}G^{ab},
\label{spin5}
\end{equation}
where $G^{ab}$ are the {\em generators} of the spinor representation of the Lorentz group:
\begin{equation}
G^{ab}=\frac{1}{4}(\gamma^a\gamma^b-\gamma^b\gamma^a).
\label{spin6}
\end{equation}
The relation (\ref{spin5}) for the spinor representation is analogous to the relation (\ref{infLor6}) for the vector representation.
The generators $G^{ab}$ are analogous to the generators $J_{\mu\nu}$.
Therefore, they also satisfy the relation (\ref{genLie2}).\\

\noindent
{\bf Spinor and adjoint spinor}.\\
A {\em spinor} $\psi$ is defined as a quantity that, under tetrad rotations, transforms according to
\begin{equation}
\tilde{\psi}=L\psi.
\label{spin7}
\end{equation}
It is represented by a column with four components.
An {\em adjoint spinor} $\bar{\psi}$ is defined as a quantity that transforms according to
\begin{equation}
\tilde{\bar{\psi}}=\bar{\psi}L^{-1}.
\label{spin8}
\end{equation}
It is represented by a row with four components.
The product $\bar{\psi}\psi$ is a scalar:
\begin{equation}
\tilde{\bar{\psi}}\tilde{\psi}=\bar{\psi}\psi,
\label{spin9}
\end{equation}
which follows from the above transformation laws and also from the matrix multiplication of a row and a column. 
The indices of the $\gamma^a$ and $L$ that are implicit in the 4$\times$4 matrix multiplication in (\ref{spin1}), (\ref{spin2}), and (\ref{spin4}) are spinor indices.
The relation (\ref{spin4}) shows that the Dirac matrices $\gamma^a$ can be regarded as quantities that have, in addition to the invariant index $a$, one spinor index and one adjoint-spinor index.
The product $\psi\bar{\psi}$ transforms like the Dirac matrices:
\begin{equation}
\tilde{\psi}\tilde{\bar{\psi}}=L\psi\bar{\psi}L^{-1}.
\end{equation}

\noindent
{\bf Bilinear forms}.\\
The spinors $\psi$ and $\bar{\psi}$ can be used to construct tensors in combinations that are linear both in $\psi$ and $\bar{\psi}$ and referred to as {\em bilinear forms}.
For example, $\bar{\psi}\gamma^a\psi$ transforms like a contravariant Lorentz vector:
\begin{equation}
\bar{\psi}\gamma^a\psi\rightarrow\bar{\psi}L^{-1}\Lambda^a_{\phantom{a}b}L\gamma^b L^{-1}L\psi=\Lambda^a_{\phantom{a}b}\bar{\psi}\gamma^b\psi,
\label{spin11}
\end{equation}
and $\bar{\psi}\gamma^{[a}\gamma^{c]}\psi$ transforms like an antisymmetric contravariant Lorentz tensor:
\begin{equation}
\bar{\psi}\gamma^{[a}\gamma^{c]}\psi\rightarrow\bar{\psi}L^{-1}\Lambda^a_{\phantom{a}b}\Lambda^c_{\phantom{c}d}L\gamma^{[b} L^{-1}L\gamma^{d]} L^{-1}L\psi=\Lambda^a_{\phantom{a}b}\Lambda^c_{\phantom{c}d}\bar{\psi}\gamma^{[b}\gamma^{d]}\psi.
\label{spin12}
\end{equation}

\subsubsection{Spinor connection}
The derivative of a spinor does not transform like a spinor:
\begin{equation}
\tilde{\psi}_{,i}=L\psi_{,i}+L_{,i}\psi.
\end{equation}
If we introduce the {\em spinor connection} $\Gamma_i$ that transforms according to
\begin{equation}
\tilde{\Gamma}_i=L\Gamma_i L^{-1}+L_{,i}L^{-1},
\label{spinco2}
\end{equation}
then a {\em covariant derivative} of a spinor,
\begin{equation}
\psi_{;i}=\psi_{,i}-\Gamma_i \psi,
\label{spinco3}
\end{equation}
is a spinor:
\begin{equation}
\tilde{\psi}_{;i}=\tilde{\psi}_{,i}-\tilde{\Gamma}_i\tilde{\psi}=L\psi_{,i}+L_{,i}\psi-(L\Gamma_i L^{-1}+L_{,i}L^{-1})L\psi=L\psi_{;i}.
\end{equation}
Because $\bar{\psi}\psi$ is a scalar,
\begin{equation}
(\bar{\psi}\psi)_{;i}=(\bar{\psi}\psi)_{,i},
\end{equation}
the chain rule for covariant differentiation gives the covariant derivative of an adjoint spinor,
\begin{equation}
\bar{\psi}_{;i}=\bar{\psi}_{,i}+\bar{\psi}\Gamma_i.
\label{spinco6}
\end{equation}
We also have
\begin{equation}
\psi_{|i}=\psi_{;i},\quad \bar{\psi}_{|i}=\bar{\psi}_{;i}.
\label{spinco7}
\end{equation}

The Dirac matrices $\gamma^a$ transform like $\psi\bar{\psi}$, whose covariant derivative is
\begin{equation}
(\psi\bar{\psi})_{;i}=\psi_{;i}\bar{\psi}+\psi\bar{\psi}_{;i}=(\psi\bar{\psi})_{,i}-\Gamma_i \psi\bar{\psi}+\psi\bar{\psi}\Gamma_i=(\psi\bar{\psi})_{,i}-[\Gamma_i,\psi\bar{\psi}].
\end{equation}
Therefore, the covariant derivative of a Dirac matrix is
\begin{equation}
\gamma^a_{\phantom{a};i}=\gamma^a_{\phantom{a},i}-[\Gamma_i,\gamma^a]=-[\Gamma_i,\gamma^a],
\end{equation}
which gives
\begin{equation}
\gamma^j_{\phantom{j};i}=\gamma^j_{\phantom{j}|i}=\gamma^j_{\phantom{j},i}+\Gamma^{j}_{ki}\gamma^k-[\Gamma_i,\gamma^j].
\end{equation}
Accordingly, we obtain
\begin{equation}
\gamma^a_{\phantom{a}|i}=\omega^{a}_{\phantom{a}bi}\gamma^b-[\Gamma_i,\gamma^a].
\label{spinco11}
\end{equation}
The quantity $\bar{\psi}\gamma^i\psi_{|i}$ transforms under Lorentz rotations like a scalar:
\begin{equation}
\bar{\psi}\gamma^i\psi_{|i}\rightarrow\bar{\psi}L^{-1}L\gamma^i L^{-1}L\psi_{|i}=\bar{\psi}\gamma^i\psi_{|i}.
\end{equation}

\noindent
{\bf Fock--Ivanenko coefficients}.\\
The relation $\eta_{ab|i}=0$ infers that
\begin{equation}
\gamma^a_{\phantom{a}|i}=0,
\label{spinco13}
\end{equation}
because the Dirac matrices $\gamma^a$ only depend on $\eta_{ab}$.
Multiplying both sides of (\ref{spinco11}) by $\gamma_a$ from the left gives
\begin{equation}
\omega_{abi}\gamma^a \gamma^b-\gamma_a \Gamma_i \gamma^a+4\Gamma_i=0.
\label{spinco14}
\end{equation}
We seek the solution of (\ref{spinco14}) in the form
\begin{equation}
\Gamma_i=-\frac{1}{4}\omega_{abi}\gamma^a \gamma^b-A_i,
\label{spinco15}
\end{equation}
where $A_i$ is a spinor-tensor quantity with one vector index.
Substituting (\ref{spinco15}) to (\ref{spinco14}), together with the identity $\gamma_c \gamma^a \gamma^b \gamma^c=4\eta^{ab}$, gives
\begin{equation}
-\gamma_a A_i \gamma^a+4A_i=0,
\label{spinco16}
\end{equation}
so $A_i$ is an arbitrary vector multiple of $I_4$.
Therefore, the spinor connection $\Gamma_i$ is given, up to the addition of an arbitrary vector multiple of $I_4$, by the {\em Fock--Ivanenko coefficients}:
\begin{equation}
\Gamma_i=-\frac{1}{4}\omega_{abi}\gamma^a \gamma^b=-\frac{1}{2}\omega_{abi}G^{ab}.
\label{spinco17}
\end{equation}
Using the definition (\ref{spcon1}), we can also write (\ref{spinco17}) as
\begin{equation}
\Gamma_i=-\frac{1}{8}e^j_{c;i}[\gamma_j,\gamma^c]=\frac{1}{8}[\gamma^j_{\phantom{j};i},\gamma_j].
\end{equation}

For the Levi-Civita connection, the covariant derivative of a spinor (\ref{spinco3}) becomes
\begin{equation}
\psi_{:i}=\psi_{,i}-\mathring{\Gamma}_i \psi,
\label{spinco19}
\end{equation}
and the covariant derivative of an adjoint spinor (\ref{spinco6}) becomes
\begin{equation}
\bar{\psi}_{:i}=\bar{\psi}_{,i}+\bar{\psi}\mathring{\Gamma}_i,
\label{spinco20}
\end{equation}
where the {\em Levi-Civita spinor connection} $\mathring{\Gamma}_i$ is given, similarly to (\ref{spinco17}), by
\begin{equation}
\mathring{\Gamma}_i=-\frac{1}{4}\mathring{\omega}_{abi}\gamma^a \gamma^b=-\frac{1}{2}\mathring{\omega}_{abi}G^{ab}.
\label{spinco21}
\end{equation}
Substituting (\ref{spcon17}) into (\ref{spinco21}) gives
\begin{equation}
\Gamma_i=\mathring{\Gamma}_i-\frac{1}{4}C_{jki}\gamma^j \gamma^k=\mathring{\Gamma}_i-\frac{1}{4}C_{jki}\gamma^{[j} \gamma^{k]}.
\end{equation}
Accordingly, (\ref{spinco19}) and (\ref{spinco20}) yield
\begin{eqnarray}
& & \psi_{;k}=\psi_{:k}+\frac{1}{4}C_{ijk}\gamma^i \gamma^j \psi, \label{spinco23} \\
& & \bar{\psi}_{;k}=\bar{\psi}_{:k}-\frac{1}{4}C_{ijk}\bar{\psi}\gamma^i \gamma^j.
\label{spinco24}
\end{eqnarray}

\subsubsection{Curvature spinor}
The commutator of total covariant derivatives of a spinor is
\begin{eqnarray}
& & \psi_{|ji}-\psi_{|ij}=(\psi_{|j})_{,i}-\Gamma_i\psi_{|j}-\Gamma^{k}_{ji}\psi_{|k}-(\psi_{|i})_{,j}+\Gamma_j\psi_{|i}+\Gamma^{k}_{ij}\psi_{|k} \nonumber \\
& & =-\Gamma_{j,i}\psi+\Gamma_i\Gamma_j\psi+\Gamma_{i,j}\psi-\Gamma_j\Gamma_i\psi+2S^k_{\phantom{k}ij}\psi_{|k}=K_{ij}\psi+2S^k_{\phantom{k}ij}\psi_{|k},
\label{cursp1}
\end{eqnarray}
where $K_{ij}=-K_{ji}$ is defined as
\begin{equation}
K_{ij}=\Gamma_{i,j}-\Gamma_{j,i}+[\Gamma_i,\Gamma_j].
\label{cursp2}
\end{equation}
Substituting (\ref{spinco2}) to (\ref{cursp2}) gives
\begin{equation}
\tilde{K}_{ij}=\tilde{\Gamma}_{i,j}-\tilde{\Gamma}_{j,i}+[\tilde{\Gamma}_i,\tilde{\Gamma}_j]=L(\Gamma_{i,j}-\Gamma_{j,i}+[\Gamma_i,\Gamma_j])L^{-1}=LK_{ij}L^{-1},
\end{equation}
so $K_{ij}$ transforms under tetrad rotations like the Dirac matrices $\gamma^a$, that is, $K_{ij}$ is a spinor with one spinor index and one adjoint-spinor index.
We refer to $K_{ij}$ as the {\em curvature spinor}.

The relation (\ref{spinco13}) leads to
\begin{equation}
\gamma^k_{\phantom{k}|i}=0.
\label{cursp4}
\end{equation}
Therefore, the commutator of covariant derivatives of the spacetime-dependent Dirac matrices vanishes:
\begin{equation}
2\gamma^k_{\phantom{k}|[ji]}=R^k_{\phantom{k}lij}\gamma^l+2S^l_{\phantom{l}ij}\gamma^k_{\phantom{k}|l}+[K_{ij},\gamma^k]=R^k_{\phantom{k}lij}\gamma^l+[K_{ij},\gamma^k]=0.
\label{cursp5}
\end{equation}
Multiplying both sides of (\ref{cursp5}) by $\gamma_k$ from the left gives
\begin{equation}
R_{klij}\gamma^k \gamma^l+\gamma_k K_{ij}\gamma^k-4K_{ij}=0.
\label{cursp6}
\end{equation}
We seek the solution of (\ref{cursp6}) in the form
\begin{equation}
K_{ij}=\frac{1}{4}R_{klij}\gamma^k \gamma^l+B_{ij},
\label{cursp7}
\end{equation}
where $B_{ij}$ is a spinor-tensor quantity with two vector indices.
Substituting (\ref{cursp7}) to (\ref{cursp6}) gives
\begin{equation}
\gamma_k B_{ij}\gamma^k-4B_{ij}=0,
\end{equation}
so $B_{ij}$ is an antisymmetric-tensor multiple of $I_4$.
The tensor $B_{ij}$ is related to the vector $A_i$ in (\ref{spinco15}) by
\begin{equation}
B_{ij}=A_{j,i}-A_{i,j}+[A_i,A_j].
\label{cursp9}
\end{equation}
Because $\psi$ has no indices other than spinor indices, $A_i$ is a vector and $[A_i,A_j]=0$.
The invariance of (\ref{cursp6}) under the addition of an antisymmetric-tensor multiple $B_{ij}$ of the unit matrix to the curvature spinor is related to the invariance of (\ref{spinco14}) under the addition of a vector multiple $A_i$ of the unit matrix to the spinor connection.
Setting $A_i=0$, which corresponds to the Fock--Ivanenko spinor connection, gives $B_{ij}=0$.
Therefore, the curvature spinor $K_{ij}$ is given, up to the addition of an arbitrary antisymmetric-tensor multiple of $I_4$, by
\begin{equation}
K_{ij}=\frac{1}{4}R_{klij}\gamma^k \gamma^l=\frac{1}{2}R_{klij}G^{kl}.
\label{cursp10}
\end{equation}

For the Levi-Civita connection, the spinor connection is equal to (\ref{spinco21}) instead of (\ref{spinco17}).
Consequently, the Riemannian curvature spinor (\ref{cursp2}) is equal to
\begin{equation}
\mathring{K}_{ij}=\mathring{\Gamma}_{i,j}-\mathring{\Gamma}_{j,i}+[\mathring{\Gamma}_i,\mathring{\Gamma}_j],
\end{equation}
and the relation (\ref{cursp10}) gives
\begin{equation}
\mathring{K}_{ij}=\frac{1}{4}\mathring{R}_{klij}\gamma^k \gamma^l=\frac{1}{2}\mathring{R}_{klij}G^{kl}.
\end{equation}
\newline
References: \cite{Lord,Hehl1}.
\newline
\newline
Spacetime is a fabric, in which various fields representing matter exist.
These fields can be described by vectors, tensors, and spinors.
They satisfy the equations derived from two fundamental principles: the principle of relativity and the principle of least action.
The physics of fields is referred to as {\em field theory} and constitutes Chapter \ref{Fields} (Fields).

\newpage
\section{Fields}
\label{Fields}
\subsection{Principle of least action}
\setcounter{equation}{0}
{\bf Lagrangian density}.\\
The most general formulation of the law that governs the dynamics of classical systems is Hamilton's {\em principle of least action}, according to which every classical system is characterized by a definite scalar-density function $\mathfrak{L}$, and the dynamics of the system is such that a certain condition is satisfied.
Let $\phi_A(x^i)$ be a set of differentiable tensor functions of the coordinates, indexed by $A$.
Such functions are referred to as physical {\em fields}.
Let $\mathfrak{L}$ be a scalar density constructed from the fields $\phi_A$ and their derivatives.
Consequently, a quantity
\begin{equation}
S=\frac{1}{c}\int\mathfrak{L}\,d\Omega,
\label{Lageq1}
\end{equation}
where the integration is over some four-dimensional region in spacetime, is a scalar.
Let $\delta\phi_A$ be arbitrary and independent, small changes of $\phi_A$ (regarded as a dynamical variable) over the region of integration, which vanish at the boundary of this region.
Then the change in $S$ can be written as
\begin{equation}
\delta S=\sum_A \delta_A S,
\end{equation}
where
\begin{equation}
\delta_A S=\frac{1}{c}\int F_A\delta\phi_A d\Omega.
\end{equation}

The principle of least action states that the dynamics of a physical system is given by the condition the scalar $S$ be a local minimum.
Therefore, any infinitesimal change in the dynamics of the system does not alter the value of $S$:
\begin{equation}
\delta S=0
\label{Lageq4}
\end{equation}
($S$ is a local extremum).
The condition (\ref{Lageq4}) is referred to as the {\em principle of stationary action}, which is the necessary part of the principle of least action.
If the fields $\phi_A$ transform covariantly under coordinate transformations, then the variational condition (\ref{Lageq4}) gives covariant equations
\begin{equation}
F_A=0.
\label{Lageq5}
\end{equation}
These equations are also invariant for other transformations (such as tetrad rotations or internal symmetries), for which $\mathfrak{L}$ is invariant.

The density $\mathfrak{L}$ is referred to as the {\em Lagrangian density}, $S$ is the {\em action} functional, $\delta S=0$ is the principle of least action, and (\ref{Lageq5}) are the {\em field equations}.
The field equations of a physical system are the result of the action for this system being a local extremum among all possible configurations.
The condition that the action be a local minimum imposes additional restrictions on possible choices for $S$.
The number of independent field equations for a given system is referred to as the number of the {\em degrees of freedom} representing this system.\\

\noindent
{\bf Lagrange equations for fields}.\\
In most physical cases $\mathfrak{L}$ contains only $\phi_A$ and their first derivatives.
A Lagrangian density containing higher derivatives can always be written in terms of first derivatives by increasing the number of the components $\phi_A$.
If $\mathfrak{L}$ depends only on $\phi$ and $\partial_i\phi$: $\mathfrak{L}=\mathfrak{L}(\phi,\phi_{,i})$, then
\begin{eqnarray}
& & \delta S=\frac{1}{c}\int\biggl(\frac{\partial\mathfrak{L}}{\partial\phi}\delta\phi+\frac{\partial\mathfrak{L}}{\partial(\phi_{,i})}\delta(\phi_{,i})\biggr)d\Omega=\frac{1}{c}\int\biggl(\frac{\partial\mathfrak{L}}{\partial\phi}\delta\phi+\frac{\partial\mathfrak{L}}{\partial(\phi_{,i})}(\delta\phi)_{,i}\biggr)d\Omega \nonumber \\
& & =\frac{1}{c}\int\biggl[\frac{\partial\mathfrak{L}}{\partial\phi}\delta\phi-\partial_i\biggl(\frac{\partial\mathfrak{L}}{\partial(\phi_{,i})}\biggr)\delta\phi+\partial_i\biggl(\frac{\partial\mathfrak{L}}{\partial(\phi_{,i})}\delta\phi\biggr)\biggr]d\Omega.
\label{Lageq6}
\end{eqnarray}
The last term in the integrand in the second line of (\ref{Lageq6}) is a divergence.
Its four-volume integral can be transformed, using the Gau\ss--Stokes theorem (\ref{covint7}), into a hypersurface integral over the boundary of the integration region.
Since $\delta\phi=0$ on the boundary, this term does not contribute to the variation of the action:
\begin{equation}
\delta S=\frac{1}{c}\int\biggl[\frac{\partial\mathfrak{L}}{\partial\phi}-\partial_i\biggl(\frac{\partial\mathfrak{L}}{\partial(\phi_{,i})}\biggr)\biggr]\delta\phi\, d\Omega+\int\frac{\partial\mathfrak{L}}{\partial(\phi_{,i})}\delta\phi\,dS_i=\frac{1}{c}\int\biggl[\frac{\partial\mathfrak{L}}{\partial\phi}-\partial_i\biggl(\frac{\partial\mathfrak{L}}{\partial(\phi_{,i})}\biggr)\biggr]\delta\phi\,d\Omega.
\end{equation}
If $\delta S=0$ for arbitrary variations $\delta\phi$ that vanish on the boundary, then
\begin{equation}
\frac{\partial\mathfrak{L}}{\partial\phi}-\partial_i\biggl(\frac{\partial\mathfrak{L}}{\partial(\phi_{,i})}\biggr)=0,\quad 
\frac{\partial\mathfrak{L}}{\partial\phi}-\frac{\partial}{\partial x^i}\biggl(\frac{\partial\mathfrak{L}}{\partial(\phi_{,i})}\biggr)=0.
\label{Lageq8}
\end{equation}
Defining the {\em variational derivative} of $\mathfrak{L}$ with respect to $\phi$,
\begin{equation}
\frac{\delta\mathfrak{L}}{\delta\phi}=\frac{\partial\mathfrak{L}}{\partial\phi}-\partial_i\biggl(\frac{\partial\mathfrak{L}}{\partial(\phi_{,i})}\biggr),
\end{equation}
we can write (\ref{Lageq8}) as
\begin{equation}
\frac{\delta\mathfrak{L}}{\delta\phi}=0.
\end{equation}
The set of equations (\ref{Lageq8}), for each field $\phi_A$, is referred to as the {\em Lagrange equations}.

There is some arbitrariness in the choice of $\mathfrak{L}$; adding to it the divergence of an arbitrary vector density or multiplying it by a constant produces the same field equations.
If a system consists of two noninteracting parts $A$ and $B$, with corresponding Lagrangian densitites $\mathfrak{L}_A(\phi_A,\partial\phi_A)$ and $\mathfrak{L}_B(\phi_B,\partial\phi_B)$, then the Lagrangian density for this system is the sum $\mathfrak{L}_A+\mathfrak{L}_B$.
This additivity of the Lagrangian density means that the field equations for either of the two parts do not involve quantities pertaining to the other part.
If $\mathfrak{L}_A$ also depends on $\phi_B,\partial\phi_B$, or $\mathfrak{L}_B$ depends on $\phi_A,\partial\phi_A$, or both, then the subsystems $A$ and $B$ interact.\\

\noindent
{\bf Covariant form of Lagrange equations}.\\
The Lagrange equations are covariant and can be written in an explicitly covariant form.
A covariant derivative with respect to the Levi-Civita connection $\phi_{:i}$ is equal to the sum of the corresponding partial derivative $\phi_{,i}$ and terms that are linear in $\phi$.
Consequently, the partial derivatives of $\mathfrak{L}$ with respect to both derivatives of $\phi$ are equal:
\begin{equation}
\frac{\partial\mathfrak{L}}{\partial(\phi_{:i})}=\frac{\partial\mathfrak{L}}{\partial(\phi_{,i})},
\label{Lageq11}
\end{equation}
turning (\ref{Lageq8}) into
\begin{equation}
\frac{\partial\mathfrak{L}}{\partial\phi}-\partial_i\biggl(\frac{\partial\mathfrak{L}}{\partial(\phi_{:i})}\biggr)=0.
\label{Lageq12}
\end{equation}
Because $\mathfrak{L}$ is a scalar density, $\partial\mathfrak{L}/\partial(\phi_{:i})$ is a contravariant vector density, satisfying (\ref{Chrsym234}):
\begin{equation}
\mathring{\nabla}_i\biggl(\frac{\partial\mathfrak{L}}{\partial(\phi_{:i})}\biggr)=\partial_i\biggl(\frac{\partial\mathfrak{L}}{\partial(\phi_{:i})}\biggr).
\end{equation}
Therefore, the Lagrange equations (\ref{Lageq12}) have a covariant form:
\begin{equation}
\frac{\partial\mathfrak{L}}{\partial\phi}-\mathring{\nabla}_i\biggl(\frac{\partial\mathfrak{L}}{\partial(\phi_{:i})}\biggr)=0.
\label{Lageq14}
\end{equation}
The covariant derivative $\mathring{\nabla}_i$ with respect to the Levi-Civita connection can be replaced with $\nabla_i^\ast$ (\ref{parint5}).
\newline
References: \cite{Schr,LL2,Lord}.

\subsection{Gravitational field}
\setcounter{equation}{0}
\subsubsection{Action for gravitational field}
Let us consider a Lagrangian density $\mathfrak{L}$ that depends on the affine (or spin) connection and its first derivatives.
Such Lagrangian density can be decomposed into the covariant part $\mathfrak{L}_\textrm{g}$ that contains derivatives of the affine/spin connection, which is referred to as the Lagrangian density for the {\em gravitational field}, and the covariant part $\mathfrak{L}_\textrm{m}$ that does not contain these derivatives, which is referred to as the Lagrangian density for {\em matter}:
\begin{equation}
\mathfrak{L}=\mathfrak{L}_\textrm{g}+\mathfrak{L}_\textrm{m}.
\label{Laggr1}
\end{equation}
The simplest covariant scalar that can be constructed from the affine/spin connection and its first derivatives is the Ricci scalar $R$.
The corresponding Lagrangian density for the gravitational field is proportional to the product of $R$ and the scalar density $\sqrt{-\mathfrak{g}}$:
\begin{equation}
\mathfrak{L}_\textrm{g}=-\frac{1}{2\kappa}\sqrt{-\mathfrak{g}}R=-\frac{1}{2\kappa}\sqrt{-\mathfrak{g}}\Bigl(\mathring{R}-g^{ik}(2C^l_{\phantom{l}il:k}+C^j_{\phantom{j}ij}C^l_{\phantom{l}kl}-C^l_{\phantom{l}im}C^m_{\phantom{m}kl})\Bigr),
\label{Laggr2}
\end{equation}
where $\kappa$ is {\em Einstein's gravitational constant} and we used (\ref{Riem11}).
The action for the gravitational field is thus
\begin{equation}
S_\textrm{g}=\frac{1}{c}\int\mathfrak{L}_\textrm{g} d\Omega=-\frac{1}{2\kappa c}\int R\sqrt{-\mathfrak{g}}d\Omega.
\label{Laggr3}
\end{equation}

The metric tensor and the affine connection are two fundamental quantities describing a gravitational field.
Since the affine connection is metric-compatible, given by (\ref{Chrsym6}), it is a function of the metric tensor, its derivatives and the torsion tensor.
Accordingly, the metric and torsion tensors are dynamical variables in varying the action.
Equivalently, the tetrad and spin connection can be taken as dynamical variables.\\

\noindent
{\bf Lagrangian of gravitational field without derivatives of connection}.\\
Let us consider the Riemannian part of the Lagrangian density for the gravitational field (\ref{Laggr2}), which is proportional to the Riemann scalar $\mathring{R}$:
\begin{equation}
\mathring{\mathfrak{L}}_\textrm{g}=-\frac{1}{2\kappa}\sqrt{-\mathfrak{g}}\mathring{R}.
\label{Laggr4}
\end{equation}
The scalar density $\sqrt{-\mathfrak{g}}\mathring{R}$ is linear in first derivatives of the Christoffel symbols $\mathring{\Gamma}^{i}_{kl}$:
\begin{eqnarray}
& & \sqrt{-\mathfrak{g}}\mathring{R}=\sqrt{-\mathfrak{g}}g^{ik}(\mathring{\Gamma}^{l}_{ik,l}-\mathring{\Gamma}^{l}_{il,k}+\mathring{\Gamma}^{m}_{ik}\mathring{\Gamma}^{l}_{ml}-\mathring{\Gamma}^{m}_{il}\mathring{\Gamma}^{l}_{mk}) \nonumber \\
& & =(\sqrt{-\mathfrak{g}}g^{ik}\mathring{\Gamma}^{l}_{ik})_{,l}-\mathring{\Gamma}^{l}_{ik}(\sqrt{-\mathfrak{g}}g^{ik})_{,l}-(\sqrt{-\mathfrak{g}}g^{ik}\mathring{\Gamma}^{l}_{il})_{,k}+\mathring{\Gamma}^{l}_{il}(\sqrt{-\mathfrak{g}}g^{ik})_{,k} \nonumber \\
& & +\sqrt{-\mathfrak{g}}g^{ik}(\mathring{\Gamma}^{m}_{ik}\mathring{\Gamma}^{l}_{ml}-\mathring{\Gamma}^{m}_{il}\mathring{\Gamma}^{l}_{mk}).
\label{Laggr5}
\end{eqnarray}
We can therefore subtract from $\sqrt{-\mathfrak{g}}\mathring{R}$ total derivatives without altering the field equations, replacing it by a noncovariant quantity $\mathcal{G}$ that does not contain first derivatives of the Christoffel symbols:
\begin{eqnarray}
& & \mathcal{G}=\sqrt{-\mathfrak{g}}\mathring{R}-(\sqrt{-\mathfrak{g}}g^{ik}\mathring{\Gamma}^{l}_{ik})_{,l}+(\sqrt{-\mathfrak{g}}g^{ik}\mathring{\Gamma}^{l}_{il})_{,k} \nonumber \\
& & =\mathring{\Gamma}^{l}_{il}(\sqrt{-\mathfrak{g}}g^{ik})_{,k}-\mathring{\Gamma}^{l}_{ik}(\sqrt{-\mathfrak{g}}g^{ik})_{,l}+\sqrt{-\mathfrak{g}}g^{ik}(\mathring{\Gamma}^{m}_{ik}\mathring{\Gamma}^{l}_{ml}-\mathring{\Gamma}^{m}_{il}\mathring{\Gamma}^{l}_{mk}) \nonumber \\
& & =\mathring{\Gamma}^{l}_{il}\bigl((\sqrt{-\mathfrak{g}}g^{ik})_{:k}+\mathring{\Gamma}^{j}_{jk}\sqrt{-\mathfrak{g}}g^{ik}-\sqrt{-\mathfrak{g}}\mathring{\Gamma}^{i}_{jk}g^{jk}-\sqrt{-\mathfrak{g}}\mathring{\Gamma}^{k}_{jk}g^{ij}\bigr) \nonumber \\
& & -\mathring{\Gamma}^{l}_{ik}\bigl((\sqrt{-\mathfrak{g}}g^{ik})_{:l}+\mathring{\Gamma}^{j}_{jl}\sqrt{-\mathfrak{g}}g^{ik}-\sqrt{-\mathfrak{g}}\mathring{\Gamma}^{i}_{jl}g^{jk}-\sqrt{-\mathfrak{g}}\mathring{\Gamma}^{k}_{jl}g^{ij}\bigr) \nonumber \\
& & +\sqrt{-\mathfrak{g}}g^{ik}(\mathring{\Gamma}^{m}_{ik}\mathring{\Gamma}^{l}_{ml}-\mathring{\Gamma}^{m}_{il}\mathring{\Gamma}^{l}_{mk})=\mathring{\Gamma}^{l}_{il}\bigl(\mathring{\Gamma}^{j}_{jk}\sqrt{-\mathfrak{g}}g^{ik}-\sqrt{-\mathfrak{g}}\mathring{\Gamma}^{i}_{jk}g^{jk} \nonumber \\
& & -\sqrt{-\mathfrak{g}}\mathring{\Gamma}^{k}_{jk}g^{ij}\bigr)-\mathring{\Gamma}^{l}_{ik}\bigl(\mathring{\Gamma}^{j}_{jl}\sqrt{-\mathfrak{g}}g^{ik}-\sqrt{-\mathfrak{g}}\mathring{\Gamma}^{i}_{jl}g^{jk}-\sqrt{-\mathfrak{g}}\mathring{\Gamma}^{k}_{jl}g^{ij}\bigr) \nonumber \\
& & +\sqrt{-\mathfrak{g}}g^{ik}(\mathring{\Gamma}^{m}_{ik}\mathring{\Gamma}^{l}_{ml}-\mathring{\Gamma}^{m}_{il}\mathring{\Gamma}^{l}_{mk})=\sqrt{-\mathfrak{g}}g^{ik}(\mathring{\Gamma}^{m}_{il}\mathring{\Gamma}^{l}_{mk}-\mathring{\Gamma}^{m}_{ik}\mathring{\Gamma}^{l}_{ml}).
\label{Laggr6}
\end{eqnarray}
This quantity coincides (with the opposite sign) with the last two terms in the first line of (\ref{Laggr5}).
We also define
\begin{equation}
{\sf G}=\frac{\mathcal{G}}{\sqrt{-\mathfrak{g}}}=g^{ik}(\mathring{\Gamma}^{m}_{il}\mathring{\Gamma}^{l}_{mk}-\mathring{\Gamma}^{m}_{ik}\mathring{\Gamma}^{l}_{ml}).
\label{Laggr7}
\end{equation}
The Riemannian part (\ref{Laggr4}) of the Lagrangian density for the gravitational field reduces accordingly to
\begin{equation}
\mathring{\mathfrak{L}}_\textrm{g}=-\frac{1}{2\kappa}\mathcal{G}=-\frac{1}{2\kappa}\sqrt{-\mathfrak{g}}{\sf G}.
\label{Laggr8}
\end{equation}

\noindent
{\bf Positivity of gravitational constant}.\\
Any coordinate transformation results in variations of $g^{ik}$, thereby
\begin{equation}
\mathring{S}_\textrm{g}=\frac{1}{c}\int\mathring{\mathfrak{L}}_\textrm{g}d\Omega=-\frac{1}{2\kappa c}\int \mathring{R}\sqrt{-\mathfrak{g}}d\Omega
\end{equation}
is not necessarily a minimum with respect to these variations (only an extremum) because not all $\delta g^{ik}$ correspond to actual variations of the gravitational field.
In order to exclude the variations $\delta g^{ik}$ resulting from changing the coordinates, we must impose on the metric tensor 4 arbitrary constraints.
If we choose
\begin{equation}
g_{0\alpha}=0,\quad |g_{\alpha\beta}|=\textrm{const},
\end{equation}
then ${\sf G}$ becomes
\begin{equation}
{\sf G}=-\frac{1}{4}g^{00}g^{\alpha\beta}g^{\gamma\delta}g_{\alpha\gamma,0}g_{\beta\delta,0}.
\end{equation}
In the locally Galilean frame of reference, $g_{\alpha\beta}=-\delta_{\alpha\beta}$, thereby
\begin{equation}
{\sf G}=-\frac{1}{4}g^{00}(g_{\alpha\beta,0})^2.
\end{equation}
For physical systems, $g^{00}>0$.
Therefore, in order for $\mathring{S}_\textrm{g}$ to have a minimum, $\kappa$ must be positive, otherwise an arbitrarily rapid change of $g_{\alpha\beta}$ in time would result in an arbitrarily low value of $\mathring{S}_\textrm{g}$ and there would be no minimum of $S$.

\subsubsection{Gravitational potential}
If the metric tensor $g_{ij}$ is approximately equal to the Minkowski metric tensor $\eta_{ij}$, then the corresponding gravitational field is {\em weak}.
The component $g_{00}$ is approximately equal to 1.
We can therefore write
\begin{equation}
g_{00}\approx1+\frac{2\phi}{c^2},
\label{grapot1}
\end{equation}
where $\phi$ is referred to as the {\em gravitational potential}.
Therefore, nonrelativistic gravitational fields, corresponding to the limit $c\rightarrow\infty$, are weak.
Also $u^0\approx1$ and $u^\alpha\approx0$.
In this limit, the leading component of the Levi-Civita connection is
\begin{equation}
\mathring{\Gamma}^{\alpha}_{00}\approx-\frac{1}{2}g^{\alpha\beta}\frac{\partial g_{00}}{\partial x^\beta}=\frac{1}{c^2}\frac{\partial\phi}{\partial x^\alpha},
\label{grapot2}
\end{equation}
thereby the metric geodesic equation (\ref{metgeo3}) reduces to
\begin{equation}
\frac{d{\bf v}}{dt}={\bf g},\quad {\bf g}=-\mbox{{\bf grad}}\,\phi,
\label{grapot3}
\end{equation}
where ${\bf g}$ is the {\em gravitational acceleration}.
The quantity ${\sf G}$ in (\ref{Laggr7}) reduces to
\begin{equation}
{\sf G}=\frac{2}{c^4}(\mbox{{\bf grad}}\,\phi)^2.
\label{grapot4}
\end{equation}
The leading component of the Riemannian Ricci tensor is
\begin{equation}
\mathring{R}_{00}\approx\frac{\partial\mathring{\Gamma}^{\alpha}_{00}}{\partial x^\alpha}=\frac{1}{c^2}\frac{\partial^2\phi}{\partial x^{\alpha2}}=\frac{1}{c^2}\triangle\phi.
\label{grapot5}
\end{equation}
\newline
References: \cite{LL2,Lord}.

\subsection{Matter}
\setcounter{equation}{0}
\subsubsection{Tetrad energy--momentum tensor}
The variation of the action for matter,
\begin{equation}
S_\textrm{m}=\frac{1}{c}\int\mathfrak{L}_\textrm{m} d\Omega,
\label{dmemd1}
\end{equation}
with respect to the tetrad:
\begin{equation}
\delta S_\textrm{m}=\frac{1}{c}\int\mathfrak{T}^{\phantom{i}a}_i\delta e^i_a d\Omega,
\label{dtemd1}
\end{equation}
defines the {\em tetrad energy--momentum density} $\mathfrak{T}^{\phantom{i}a}_i$.
Equivalently, we have
\begin{equation}
\delta\mathfrak{L}_\textrm{m}=\mathfrak{T}^{\phantom{i}a}_i\delta e^i_a
\label{dtemd2}
\end{equation}
or
\begin{equation}
\mathfrak{T}^{\phantom{i}a}_i=\frac{\delta\mathfrak{L}_\textrm{m}}{\delta e^i_a}=\frac{\partial\mathfrak{L}_\textrm{m}}{\partial e^i_a}-\partial_l\biggl(\frac{\partial\mathfrak{L}_\textrm{m}}{\partial(e^i_{a,l})}\biggr).
\label{dtemd3}
\end{equation}
The corresponding tensor density with two coordinate indices is
\begin{equation}
\mathfrak{T}_{ij}=e_{aj}\mathfrak{T}^{\phantom{i}a}_{i}.
\end{equation}
The {\em tetrad energy--momentum tensor} is defined as
\begin{equation}
t_{ij}=\frac{\mathfrak{T}_{ij}}{\mathfrak{e}}.
\label{dtemd5}
\end{equation}
This tensor is generally not symmetric.

\subsubsection{Canonical energy--momentum tensor}
A matter Lagrangian density $\mathfrak{L}_\textrm{m}$ can be written as $\mathfrak{L}_\textrm{m}=\mathfrak{e}L$, where $L$ is a scalar.
If $\mathfrak{L}$ depends on matter fields $\phi$ and their covariant derivatives $\phi_{|i}$, then such fields are said to be {\em minimally coupled} to the affine connection.
Let the matter field $\phi$ have all its indices Lorentz indices (no vector indices).
Consequently, the tetrad appears in $L$ only through a covariant combination $e^i_a \phi_{|i}$.
Varying $\mathfrak{L}$ with respect to the tetrad gives, using (\ref{tet13}),
\begin{equation}
\delta\mathfrak{L}_\textrm{m}=\mathfrak{e}\delta L-\mathfrak{e}e^a_i L\delta e^i_a=\mathfrak{e}\frac{\partial L}{\partial\phi_{|a}}\phi_{|i}\delta e^i_a-\mathfrak{L}_\textrm{m} e^a_i \delta e^i_a=\biggl(\frac{\partial\mathfrak{L}_\textrm{m}}{\partial\phi_{|a}}\phi_{|i}-e^a_i \mathfrak{L}_\textrm{m}\biggr)\delta e^i_a.
\label{cemd1}
\end{equation}
The last term in (\ref{cemd1}),
\begin{equation}
\Theta^{\phantom{i}a}_i=\frac{\partial\mathfrak{L}_\textrm{m}}{\partial\phi_{|a}}\phi_{|i}-e^a_i \mathfrak{L}_\textrm{m},
\label{cemd2}
\end{equation}
is referred to as the {\em canonical energy--momentum density}.
The corresponding tensor density with two coordinate indices is
\begin{equation}
\Theta^{\phantom{j}i}_j=\frac{\partial\mathfrak{L}_\textrm{m}}{\partial\phi_{|i}}\phi_{|j}-\delta^i_j \mathfrak{L}_\textrm{m}=\frac{\partial\mathfrak{L}_\textrm{m}}{\partial\phi_{,i}}\phi_{|j}-\delta^i_j \mathfrak{L}_\textrm{m}.
\label{cemd3}
\end{equation}
Comparing (\ref{cemd1}) with (\ref{dtemd2}) shows that the canonical energy--momentum density is identical with the tetrad energy--momentum density:
\begin{equation}
\Theta^{\phantom{i}a}_i=\mathfrak{T}^{\phantom{i}a}_i.
\label{cemd4}
\end{equation}
The {\em canonical energy--momentum tensor} is defined as
\begin{equation}
\theta^{\phantom{i}j}_i=\frac{\Theta^{\phantom{i}j}_i}{\mathfrak{e}}.
\label{cemd5}
\end{equation}
This tensor is therefore identical with the tetrad energy--momentum tensor (\ref{dtemd5}):
\begin{equation}
\theta_{ij}=t_{ij}.
\end{equation}

\subsubsection{Metric energy--momentum tensor}
The variation of the matter action (\ref{dmemd1}) with respect to the metric tensor:
\begin{equation}
\delta S_\textrm{m}=\frac{1}{2c}\int{\cal T}_{ik}\delta g^{ik}d\Omega=-\frac{1}{2c}\int{\cal T}^{ik}\delta g_{ik}d\Omega,
\label{dmemd2}
\end{equation}
defines the {\em metric energy--momentum density} ${\cal T}_{ik}$.
The last equality follows from $g_{ik}\delta g^{ik}=-g^{ik}\delta g_{ik}$.
The metric energy--momentum density is symmetric:
\begin{equation}
{\cal T}_{ik}={\cal T}_{ki}.
\end{equation}
Equivalently, we have
\begin{equation}
\frac{1}{2}{\cal T}_{ik}=\frac{\delta\mathfrak{L}_\textrm{m}}{\delta g^{ik}}=\frac{\partial\mathfrak{L}_\textrm{m}}{\partial g^{ik}}-\partial_l\biggl(\frac{\partial\mathfrak{L}_\textrm{m}}{\partial(g^{ik}_{\phantom{ij},l})}\biggr).
\label{dmemd4}
\end{equation}
The symmetric {\em metric energy--momentum tensor} is defined as
\begin{equation}
T_{ik}=\frac{{\cal T}_{ik}}{\sqrt{-\mathfrak{g}}}=T_{ki}.
\label{dmemd5}
\end{equation}

\subsubsection{Spin tensor}
The variation of the matter action (\ref{dmemd1}) with respect to the spin connection,
\begin{equation}
\delta S_\textrm{m}=\frac{1}{2c}\int\mathfrak{S}_{ab}^{\phantom{ab}i}\delta\omega^{ab}_{\phantom{ab}i}d\Omega,
\label{spden1}
\end{equation}
defines the {\em spin density} $\mathfrak{S}_{ab}^{\phantom{ab}i}$:
\begin{equation}
\mathfrak{S}_{ab}^{\phantom{ab}i}=2\frac{\delta\mathfrak{L}_\textrm{m}}{\delta\omega^{ab}_{\phantom{ab}i}}=2\frac{\partial\mathfrak{L}_\textrm{m}}{\partial\omega^{ab}_{\phantom{ab}i}},
\label{spden2}
\end{equation}
which is antisymmetric in the Lorentz indices because of the antisymmetry of $\omega^{ab}_{\phantom{ab}i}$:
\begin{equation}
\mathfrak{S}_{ab}^{\phantom{ab}i}=-\mathfrak{S}_{ba}^{\phantom{ba}i}.
\end{equation}
The second equality in (\ref{spden2}) is satisfied because a matter Lagrangian density $\mathfrak{L}_\textrm{m}$ may depend on the spin connection but not on its derivatives; a scalar density depending on derivatives of $\omega^{ab}_{\phantom{ab}i}$ is a Lagrangian density for the gravitational field.
The variations $\delta\omega^{ab}_{\phantom{ab}i}$ are independent of $\delta e^i_a$, thereby the spin density is independent of the energy--momentum density.
The relation (\ref{spcon13}) indicates that the spin density with three coordinate indices, which is antisymmetric in the first two indices, is generated by the contortion tensor:
\begin{equation}
\mathfrak{S}_{ij}^{\phantom{ij}k}=-\mathfrak{S}_{ji}^{\phantom{ji}k}=2\frac{\delta\mathfrak{L}_\textrm{m}}{\delta C^{ij}_{\phantom{ij}k}}.
\label{spden4}
\end{equation}
Accordingly, the variation of $\mathfrak{L}_\textrm{m}$ with respect to the torsion tensor,
\begin{equation}
\tau_{i}^{\phantom{i}jk}=2\frac{\delta\mathfrak{L}_\textrm{m}}{\delta S^{i}_{\phantom{i}jk}},
\end{equation}
is a homogeneous linear function of the spin connection because of (\ref{Chrsym7}):
\begin{eqnarray}
& & \tau_{ijk}=2\frac{\delta\mathfrak{L}_\textrm{m}}{\delta S^{ijk}}=2\frac{\delta\mathfrak{L}_\textrm{m}}{\delta C^{lmn}}\frac{\partial C^{lmn}}{\partial S^{ijk}}=\mathfrak{S}_{lmn}(\delta^l_i\delta^m_{[j}\delta^n_{k]}+\delta^m_i\delta^n_{[j}\delta^l_{k]}+\delta^n_i\delta^m_{[j}\delta^l_{k]}) \nonumber \\
& & =\mathfrak{S}_{ijk}-\mathfrak{S}_{jki}+\mathfrak{S}_{kij}, \\
& & \mathfrak{S}_{ijk}=\tau_{[ij]k}, \label{spden7}
\end{eqnarray}
antisymmetric in the last two indices:
\begin{equation}
\tau_{ijk}=-\tau_{ikj}.
\end{equation}
The variation of $\mathfrak{L}_\textrm{m}$ with respect to the metric-compatible affine connection in the metric-affine variational formulation of gravity is equivalent to the variation with respect to the torsion (or contortion) tensor.

The spin connection $\omega^{ab}_{\phantom{ab}i}$ appears in $\mathfrak{L}_\textrm{m}$ only through covariant derivatives of $\phi$, in a combination $-(\partial\mathfrak{L}/\partial\phi_{,i})\Gamma_i\phi$, where
\begin{equation}
\Gamma_i=-\frac{1}{2}\omega_{abi}G^{ab}
\end{equation}
is the connection in the covariant derivative of $\phi$:
\begin{equation}
\phi_{|i}=\phi_{,i}-\Gamma_i\phi.
\end{equation}
Consequently, the spin density $\mathfrak{S}_{ab}^{\phantom{ab}i}$ is identical with 
\begin{equation}
\Sigma_{ab}^{\phantom{ab}i}=-\Sigma_{ba}^{\phantom{ba}i}=\frac{\partial\mathfrak{L}_\textrm{m}}{\partial\phi_{,i}}G_{ab}\phi,
\label{spden11}
\end{equation}
referred to as the {\em canonical spin density}.
The {\em spin tensor} is defined as
\begin{equation}
s_{ijk}=\frac{\mathfrak{S}_{ijk}}{\mathfrak{e}}.
\label{spden12}
\end{equation}

\subsubsection{Belinfante--Rosenfeld relation}
The total variation of the matter action with respect to geometrical variables is either
\begin{equation}
\delta S_\textrm{m}=\frac{1}{c}\int d\Omega\mathfrak{T}^{\phantom{i}a}_i\delta e^i_a+\frac{1}{2c}\int d\Omega\mathfrak{S}_{ab}^{\phantom{ab}i}\delta\omega^{ab}_{\phantom{ab}i}
\label{BelRos1}
\end{equation}
or
\begin{equation}
\delta S_\textrm{m}=\frac{1}{2c}\int d\Omega{\cal T}_{ik}\delta g^{ik}+\frac{1}{2c}\int d\Omega\tau_j^{\phantom{j}ik}\delta S^j_{\phantom{j}ik}.
\label{BelRos2}
\end{equation}
The relation (\ref{tet5}) gives
\begin{equation}
\frac{1}{2}\int d\Omega{\cal T}_{ik}\delta g^{ik}=\frac{1}{2}\int d\Omega{\cal T}_{ik}(\delta e_a^i e_b^k+e_a^i \delta e_b^k)\eta^{ab})=\int d\Omega{\cal T}_{ik}e^{ka}\delta e_a^i,
\end{equation}
and (\ref{spcon9}) gives
\begin{eqnarray}
& & \frac{1}{2}\int d\Omega\tau_j^{\phantom{j}ik}\delta S^j_{\phantom{j}ik}=\frac{1}{2}\int d\Omega\tau_j^{\phantom{j}ik}\Bigl(\delta(e^j_a e_{ib}\omega^{ab}_{\phantom{ab}k})+\delta e^a_{i,k}e^j_a+e^a_{i,k}\delta e^j_a\Bigr) \nonumber \\
& & =\frac{1}{2}\int d\Omega\Bigl(\tau_j^{\phantom{j}li}\delta(e^j_a e_{lb})\omega^{ab}_{\phantom{ab}i}+\tau_{ab}^{\phantom{ab}i}\delta\omega^{ab}_{\phantom{ab}i}+(\tau_j^{\phantom{j}ik}e^j_a\delta e^a_i)_{,k}-(\tau_j^{\phantom{j}ik}e^j_a)_{,k}\delta e^a_i+\tau_j^{\phantom{j}ik}e^a_{i,k}\delta e^j_a\Bigr) \nonumber \\
& & =\frac{1}{2}\int d\Omega\Bigl(\tau_j^{\phantom{j}lk}\omega^{cb}_{\phantom{cb}k}e_{lb}\delta e^j_c+\tau_j^{\phantom{j}li}\omega^{ab}_{\phantom{ab}i}e^j_a\delta e_{lb}+\tau_{ab}^{\phantom{ab}i}\delta\omega^{ab}_{\phantom{ab}i}-(\tau_j^{\phantom{j}ik}e^j_a)_{,k}\delta e^a_i+\tau_j^{\phantom{j}lm}e^b_{l,m}\delta e^j_b\Bigr) \nonumber \\
& & +\frac{1}{2}\int dS_k \tau_j^{\phantom{j}ik}e^j_a\delta e^a_i=\frac{1}{2}\int d\Omega\Bigl(-\tau_j^{\phantom{j}lk}\omega^{cb}_{\phantom{cb}k}e_{lb}e^i_c e^j_a\delta e^a_i+\tau_j^{\phantom{j}il}\omega^b_{\phantom{b}al}e^j_b\delta e^a_i+\tau_{ab}^{\phantom{ab}i}\delta\omega^{ab}_{\phantom{ab}i} \nonumber \\
& & -(\tau_{a\phantom{ik}|k}^{\phantom{a}ik}-S^i_{\phantom{i}jk}\tau_a^{\phantom{a}jk}-2S_j\tau_a^{\phantom{a}ij}+\omega^b_{\phantom{b}ak}\tau_b^{\phantom{b}ik})\delta e^a_i-\tau_j^{\phantom{j}lm}e^b_{l,m}e^i_b e^j_a \delta e^a_i\Bigr) \nonumber \\
& & =\frac{1}{2}\int d\Omega\Bigl(\tau_{ab}^{\phantom{ab}i}\delta\omega^{ab}_{\phantom{ab}i}-\tau_{j\phantom{ik};k}^{\phantom{a}ik}e^j_a\delta e^a_i+2S_j\tau_a^{\phantom{a}ij}\delta e^a_i\Bigr).
\end{eqnarray}
Comparing (\ref{BelRos1}) with (\ref{BelRos2}) leads to
\begin{eqnarray}
& & \int d\Omega\mathfrak{T}^{\phantom{i}a}_i\delta e^i_a+\frac{1}{2}\int d\Omega\mathfrak{S}_{ab}^{\phantom{ab}i}\delta\omega^{ab}_{\phantom{ab}i}=\frac{1}{2}\int d\Omega{\cal T}_{ik}\delta g^{ik}+\frac{1}{2}\int d\Omega\Bigl(\tau_{ab}^{\phantom{ab}i}\delta\omega^{ab}_{\phantom{ab}i}-\tau_{j\phantom{ik};k}^{\phantom{j}ik}e^j_a\delta e^a_i \nonumber \\
& & +2S_j\tau_a^{\phantom{a}ij}\delta e^a_i\Bigr)=\int d\Omega{\cal T}_{ik}e^{ka}\delta e^i_a+\frac{1}{2}\int d\Omega\tau_{ab}^{\phantom{ab}i}\delta\omega^{ab}_{\phantom{ab}i}+\frac{1}{2}\int d\Omega\tau_{i\phantom{jk};k}^{\phantom{i}jk}e^a_j\delta e^i_a \nonumber \\
& & -\int d\Omega S_j\tau_b^{\phantom{b}kj}e^a_k e^b_i\delta e^i_a.
\label{BelRos5}
\end{eqnarray}
The terms in (\ref{BelRos5}) with $\delta\omega^{ab}_{\phantom{ab}i}$ give (\ref{spden7}), while the terms with $\delta e^i_a$ give
\begin{equation}
\mathfrak{T}^{\phantom{i}a}_i={\cal T}_{ik}e^{ka}+\frac{1}{2}\tau_{i\phantom{jk};k}^{\phantom{i}jk}e^a_j-S_j\tau_i^{\phantom{i}aj}
\end{equation}
or
\begin{equation}
{\cal T}_{ik}=\mathfrak{T}_{ik}-\frac{1}{2}\nabla_j(\mathfrak{S}_{ik}^{\phantom{ik}j}-\mathfrak{S}_{k\phantom{j}i}^{\phantom{k}j}+\mathfrak{S}^j_{\phantom{j}ik})+S_j(\mathfrak{S}_{ik}^{\phantom{ik}j}-\mathfrak{S}_{k\phantom{j}i}^{\phantom{k}j}+\mathfrak{S}^j_{\phantom{j}ik}).
\label{BelRos7}
\end{equation}
Equation (\ref{BelRos7}) is referred to as the {\em Belinfante--Rosenfeld relation} between the metric and tetrad energy--momentum densites.
The Belinfante--Rosenfeld relation can be written, after dividing by $\mathfrak{e}$, as a tensor equation:
\begin{equation}
T_{ik}=t_{ik}-\frac{1}{2}(\nabla_j-2S_j)(s_{ik}^{\phantom{ik}j}-s_{k\phantom{j}i}^{\phantom{k}j}+s^j_{\phantom{j}ik})=t_{ik}-\frac{1}{2}\nabla_j^\ast(s_{ik}^{\phantom{ik}j}-s_{k\phantom{j}i}^{\phantom{k}j}+s^j_{\phantom{j}ik}),
\label{BelRos8}
\end{equation}
where $\nabla_j^\ast$ is given by (\ref{parint5}).
In the absence of spin, (\ref{BelRos7}) and (\ref{BelRos8}) reduce to
\begin{equation}
\mathfrak{T}_{ik}={\cal T}_{ik},\quad t_{ik}=T_{ik}.
\label{BelRos9}
\end{equation}
\newline
References: \cite{LL2,Lord,Hehl1,KS,Hehl2}

\subsection{Symmetries and conservation laws}
\setcounter{equation}{0}
\subsubsection{Noether theorem}
Let us consider a physical system, described by a Lagrangian density $\mathfrak{L}$ that depends on matter fields $\phi$, their first derivatives $\phi_{,i}$, and the coordinates $x^i$.
The change of the Lagrangian density $\delta\mathfrak{L}$ under an infinitesimal coordinate transformation $\delta x^i=\xi^i$ (\ref{infcor1}) is thus
\begin{equation}
\delta\mathfrak{L}=\frac{\partial\mathfrak{L}}{\partial\phi}\delta\phi+\frac{\partial\mathfrak{L}}{\partial\phi_{,i}}\delta(\phi_{,i})+\frac{\bar{\partial}\mathfrak{L}}{\partial x^i}\xi^i,
\label{Noeth1}
\end{equation}
where the changes $\delta\phi$ and $\delta(\phi_{,i})$ are brought about by the transformation (\ref{infcor1}) and $\bar{\partial}$ denotes partial differentiation with respect to $x^i$ at constant $\phi$ and $\phi_{,i}$.
The variation $\delta\mathfrak{L}$ brought about by this transformation is also given by (\ref{infcor10}):
\begin{equation}
\delta\mathfrak{L}=-\xi^i_{\phantom{i},i}\mathfrak{L}.
\label{Noeth2}
\end{equation}
Using the Lagrange equations (\ref{Lageq8}) and the identities
\begin{eqnarray}
& & \mathfrak{L}_{,i}=\frac{\bar{\partial}\mathfrak{L}}{\partial x^i}+\frac{\partial\mathfrak{L}}{\partial\phi}\phi_{,i}+\frac{\partial\mathfrak{L}}{\partial\phi_{,j}}\phi_{,ji}=\frac{\bar{\partial}\mathfrak{L}}{\partial x^i}+\partial_j\Bigl(\frac{\partial\mathfrak{L}}{\partial\phi_{,j}}\Bigr)\phi_{,i}+\frac{\partial\mathfrak{L}}{\partial\phi_{,j}}\phi_{,ij} \nonumber \\
& & =\frac{\bar{\partial}\mathfrak{L}}{\partial x^i}+\partial_j\Bigl(\frac{\partial\mathfrak{L}}{\partial\phi_{,j}}\phi_{,i}\Bigr), \\
& & \delta(\phi_{,i})=(\delta\phi)_{,i}-\xi^j_{\phantom{j},i}\phi_{,j},
\end{eqnarray}
we bring (\ref{Noeth1}) to
\begin{eqnarray}
& & \delta\mathfrak{L}=\partial_i\Bigl(\frac{\partial\mathfrak{L}}{\partial\phi_{,i}}\Bigr)\delta\phi+\frac{\partial\mathfrak{L}}{\partial\phi_{,i}}\Bigl[(\delta\phi)_{,i}-\xi^j_{\phantom{j},i}\phi_{,j}\Bigr]+\xi^i\frac{\bar{\partial}\mathfrak{L}}{\partial x^i} \nonumber \\
& & =\partial_i\Bigl(\frac{\partial\mathfrak{L}}{\partial\phi_{,i}}\delta\phi\Bigr)-\xi^j_{\phantom{j},i}\frac{\partial\mathfrak{L}}{\partial\phi_{,i}}\phi_{,j}+\xi^i\Bigl[\mathfrak{L}_{,i}-\partial_j\Bigl(\frac{\partial\mathfrak{L}}{\partial\phi_{,j}}\phi_{,i}\Bigr)\Bigr] \nonumber \\
& & =\xi^i\mathfrak{L}_{,i}+\partial_i\Bigl(\frac{\partial\mathfrak{L}}{\partial\phi_{,i}}\delta\phi\Bigr)-\xi^j_{\phantom{j},i}\frac{\partial\mathfrak{L}}{\partial\phi_{,i}}\phi_{,j}-\xi^j\partial_i\Bigl(\frac{\partial\mathfrak{L}}{\partial\phi_{,i}}\phi_{,j}\Bigr).
\label{Noeth5}
\end{eqnarray}
Combining (\ref{Noeth2}) and (\ref{Noeth5}) gives 
\begin{equation}
\xi^i_{\phantom{i},i}\mathfrak{L}+\xi^i\mathfrak{L}_{,i}+\partial_i\Bigl(\frac{\partial\mathfrak{L}}{\partial\phi_{,i}}\delta\phi\Bigr)-\partial_i\Bigl(\xi^j\frac{\partial\mathfrak{L}}{\partial\phi_{,i}}\phi_{,j}\Bigr)=0,
\end{equation}
which is equivalent to the conservation law:
\begin{equation}
\mathfrak{J}^i_{\phantom{i},i}=0,
\label{Noeth6}
\end{equation}
for the {\em Noether current}:
\begin{equation}
\mathfrak{J}^i=\xi^i\mathfrak{L}+\frac{\partial\mathfrak{L}}{\partial\phi_{,i}}(\delta\phi-\xi^j\phi_{,j})=\xi^i\mathfrak{L}+\frac{\partial\mathfrak{L}}{\partial\phi_{,i}}\bar{\delta}\phi.
\label{Noeth7}
\end{equation}
Equation (\ref{Noeth6}) represents the (first) {\em Noether theorem}, which states that each continuous symmetry of a Lagrangian density is associated with a conservation law of some quantity.

The Noether current (\ref{Noeth7}) is a contravariant vector density because $\xi^i$ is a contravariant vector, $\mathfrak{L}$ is a scalar density, $\partial\mathfrak{L}/\partial(\phi_{,i})=\partial\mathfrak{L}/\partial(\phi_{:i})$ according to (\ref{Lageq11}), and $\bar{\delta}\phi$ is a tensor according to (\ref{infcor4}), (\ref{infcor6}), and (\ref{infcor9}).
This current can be written in an explicitly covariant form:
\begin{equation}
\mathfrak{J}^i=\xi^i\mathfrak{L}+\frac{\partial\mathfrak{L}}{\partial\phi_{:i}}\bar{\delta}\phi.
\label{Noeth8}
\end{equation}
The conservation law (\ref{Noeth6}), using (\ref{Chrsym234}), can be written in an explicitly covariant form:
\begin{equation}
\mathfrak{J}^i_{\phantom{i}:i}=\nabla_i^\ast\mathfrak{J}^i=0.
\label{Noeth9}
\end{equation}

\noindent
{\bf Noether theorem without torsion}.\\
The Noether current (\ref{Noeth7}) satisfies the conservation law (\ref{Noeth9}) for any vector $\xi^i$.
In order to obtain a covariant law not involving $\xi^i$, we choose this vector so that it satisfies a covariant constraint $\xi^i_{\phantom{i}:k}=0$ at some point (which is possible for the Levi-Civita connection).
The covariant derivative for any tensor $\phi$ can be written as
\begin{equation}
\phi_{:k}=\phi_{,k}+\mathring{\Gamma}^i_{jk}G^j_i\phi,
\end{equation}
where $G^j_i$ are constant matrices representing the operator in (\ref{infcor14}) at that point.
The transformation law for $\phi$ under (\ref{infcor1}) is given by (\ref{infcor17}):
\begin{equation}
\delta\phi=\xi^k_{\phantom{k},i}G^i_k\phi,
\end{equation}
leading to
\begin{equation}
\bar{\delta}\phi=\delta\phi-\xi^j\phi_{,j}=(\xi^k_{\phantom{k}:i}-\mathring{\Gamma}^k_{ji}\xi^j)G^i_k\phi-\xi^j(\phi_{:j}-\mathring{\Gamma}^k_{ij}G^i_k\phi)=-\xi^j\phi_{:j}.
\end{equation}
The Noether current (\ref{Noeth8}) is therefore
\begin{equation}
\mathfrak{J}^i=\xi^i\mathfrak{L}-\frac{\partial\mathfrak{L}}{\partial\phi_{:i}}\xi^j\phi_{:j}=-\xi^j\Bigl(\frac{\partial\mathfrak{L}}{\partial\phi_{:i}}\phi_{:j}-\delta^i_j\mathfrak{L}\Bigr)=-\xi^j\Theta^{\phantom{j}i}_j,
\end{equation}
where $\Theta^{\phantom{j}i}_j$ is the torsionless canonical energy--momentum density (\ref{cemd3}).
Then the conservation law (\ref{Noeth9}) gives
\begin{equation}
\Theta^{\phantom{j}i}_{j\phantom{i}:i}=0.
\label{Noeth15}
\end{equation}

\subsubsection{Conservation of spin}
A matter Lagrangian density $\mathfrak{L}_\textrm{m}(\phi,\phi_{,i})$ is invariant under tetrad rotations because such rotations do not change the metric tensor:
\begin{equation}
\delta\mathfrak{L}_\textrm{m}=\frac{\partial\mathfrak{L}_\textrm{m}}{\partial\phi}\delta\phi+\frac{\partial\mathfrak{L}_\textrm{m}}{\partial\phi_{,i}}\delta(\phi_{,i})+\mathfrak{T}^{\phantom{i}a}_i\delta e^i_a+\frac{1}{2}\mathfrak{S}_{ab}^{\phantom{ab}i}\delta\omega^{ab}_{\phantom{ab}i}=0,
\label{cam1}
\end{equation}
where the changes $\delta$ are brought about by a tetrad rotation.
The Lorentz group is a group of tetrad rotations.
Consequently, $\mathfrak{L}_\textrm{m}$ is invariant under local, proper Lorentz transformations.
Upon integration of (\ref{cam1}) over spacetime, the first two terms vanish because of the Lagrange equations for $\phi$~(\ref{Lageq8}):
\begin{equation}
\int\Bigl(\mathfrak{T}^{\phantom{i}a}_i\delta e^i_a+\frac{1}{2}\mathfrak{S}_{ab}^{\phantom{ab}i}\delta\omega^{ab}_{\phantom{ab}i}\Bigr)d\Omega=0.
\label{cam2}
\end{equation}

For an infinitesimal Lorentz transformation (\ref{infLor1}), the tetrad $e^a_i$ changes by
\begin{equation}
\delta e^a_i=\tilde{e}^a_i-e^a_i=\Lambda^a_{\phantom{a}b}e^b_i-e^a_i=\epsilon^a_{\phantom{a}i},
\end{equation}
and the tetrad $e_a^i$, because of the identity $\delta(e^a_i e_a^j)=0$, according to
\begin{equation}
\delta e_a^i=-\epsilon_{\phantom{i}a}^i.
\label{cam4}
\end{equation}
The spin connection changes by
\begin{equation}
\delta\omega^{ab}_{\phantom{ab}i}=\delta(e^a_j \omega^{j b}_{\phantom{j b}i})=\epsilon^a_{\phantom{a}j}\omega^{j b}_{\phantom{j b}i}-e^a_j \epsilon^{j b}_{\phantom{j b};i}=\epsilon^a_{\phantom{a}c}\omega^{cb}_{\phantom{cb}i}-e^a_j \epsilon^{j b}_{\phantom{j b}|i}+\epsilon^a_{\phantom{a}c}\omega^{bc}_{\phantom{bc}i}=-\epsilon^{ab}_{\phantom{ab}|i}.
\label{cam5}
\end{equation}
Substituting (\ref{cam4}) and (\ref{cam5}) to (\ref{cam2}), together with partial integration (\ref{parint3}), gives
\begin{eqnarray}
& & -\int\Bigl(\mathfrak{T}^{\phantom{i}a}_i\epsilon^i_{\phantom{i}a}+\frac{1}{2}\mathfrak{S}_{ab}^{\phantom{ab}i}\epsilon^{ab}_{\phantom{ab}|i}\Bigr)d\Omega=-\int\Bigl(\mathfrak{T}_{ij}\epsilon^{ij}+\frac{1}{2}\mathfrak{S}_{ij}^{\phantom{ij}k}\epsilon^{ij}_{\phantom{ij};k}\Bigr)d\Omega \nonumber \\
& & =\int\Bigl(-\mathfrak{T}_{[ij]}\epsilon^{ij}-\frac{1}{2}(\mathfrak{S}_{ij}^{\phantom{ij}k}\epsilon^{ij})_{;k}+\frac{1}{2}\mathfrak{S}_{ij\phantom{k};k}^{\phantom{ij}k}\epsilon^{ij}\Bigr)d\Omega=\int\Bigl(-\mathfrak{T}_{[ij]}-S_k\mathfrak{S}_{ij}^{\phantom{ij}k}+\frac{1}{2}\mathfrak{S}_{ij\phantom{k};k}^{\phantom{ij}k}\Bigr)\epsilon^{ij}d\Omega \nonumber \\
& & =0.
\label{cam6}
\end{eqnarray}
Since the infinitesimal Lorentz rotation $\epsilon^{ij}$ is arbitrary, we obtain the covariant {\em conservation law for the spin density} (6 equations):
\begin{equation}
\mathfrak{S}_{ij\phantom{k};k}^{\phantom{ij}k}-2S_k\mathfrak{S}_{ij}^{\phantom{ij}k}=\mathfrak{T}_{ij}-\mathfrak{T}_{ji}.
\label{cam7}
\end{equation}
Dividing this law by $\mathfrak{e}$ gives the {\em conservation law for the spin tensor}:
\begin{equation}
(\nabla_k-2S_k)s_{ij}^{\phantom{ij}k}=t_{ij}-t_{ji},\quad \nabla_k^\ast s_{ij}^{\phantom{ij}k}=t_{ij}-t_{ji}.
\label{cam8}
\end{equation}

The conservation law (\ref{cam8}) also results from antisymmetrizing the Belinfante--Rosenfeld relation (\ref{BelRos8}) with respect to the indices $i,k$.
If we use the metric-compatible affine connection $\Gamma^{k}_{ij}$, which is invariant under tetrad rotations, instead of the spin connection $\omega^{ab}_{\phantom{ab}i}$ as a variable in $\mathfrak{L}_\textrm{m}$, then we must replace the term with $\delta\omega^{ab}_{\phantom{ab}i}$ in (\ref{cam1}) by a term with $\delta(e^i_{a,j})$.

The conservation law (\ref{cam8}) is an example of the {\em second Noether theorem}: if the variations of two or more quantities satisfy some relation, then the corresponding variational derivatives of the Lagrangian density satisfy some differential equation.
The variations (\ref{cam4}) and (\ref{cam5}) are related through an infinitesimal Lorentz transformation parameter $\epsilon^a_{\phantom{a}b}$ (\ref{infLor1}), resulting in the differential equation (\ref{cam8}) involving the variational derivatives (\ref{dtemd3}) and (\ref{spden2}).

\subsubsection{Conservation of tetrad energy--momentum}
The matter action $S_\textrm{m}$ is invariant under infinitesimal translations of the coordinate system (\ref{infcor1}).
The corresponding changes of the tetrad and spin connection are given by Lie derivatives,
\begin{eqnarray}
& & \bar{\delta}e^i_a=-\mathcal{L}_\xi e^i_a=\xi^i_{\phantom{i},j}e^j_a-\xi^j e^i_{a,j}, \label{ctem1} \\
& & \bar{\delta}\omega^{ab}_{\phantom{ab}i}=-\mathcal{L}_\xi \omega^{ab}_{\phantom{ab}i}=-\xi^j_{\phantom{j},i}\omega^{ab}_{\phantom{ab}j}-\xi^j \omega^{ab}_{\phantom{ab}i,j}.
\label{ctem2}
\end{eqnarray}
Equation (\ref{cam2}) becomes
\begin{equation}
\int\Bigl(\mathfrak{T}^{\phantom{i}a}_i\bar{\delta} e^i_a+\frac{1}{2}\mathfrak{S}_{ab}^{\phantom{ab}i}\bar{\delta}\omega^{ab}_{\phantom{ab}i}\Bigr)d\Omega=0.
\label{ctem3}
\end{equation}
If the variations $\xi^i$ of the coordinates vanish on the boundary of the region of integration, then substituting (\ref{ctem1}) and (\ref{ctem2}) into (\ref{ctem3}) gives
\begin{eqnarray}
& & \int\Bigl(\mathfrak{T}^{\phantom{i}a}_i \xi^i_{\phantom{i},j}e^j_a-\mathfrak{T}^{\phantom{i}a}_i \xi^j e^i_{a,j}-\frac{1}{2}\mathfrak{S}_{ab}^{\phantom{ab}i} \xi^j_{\phantom{j},i}\omega^{ab}_{\phantom{ab}j}-\frac{1}{2}\mathfrak{S}_{ab}^{\phantom{ab}i} \xi^j \omega^{ab}_{\phantom{ab}i,j}\Bigr)d\Omega \nonumber \\
& & =\int\Bigl(-\mathfrak{T}^{\phantom{i}j}_{i\phantom{j},j}-\mathfrak{T}^{\phantom{j}a}_j e^j_{a,i}+\frac{1}{2}(\mathfrak{S}_{ab}^{\phantom{ab}j}\omega^{ab}_{\phantom{ab}i})_{,j}-\frac{1}{2}\mathfrak{S}_{ab}^{\phantom{ab}j}\omega^{ab}_{\phantom{ab}j,i}\Bigr)\xi^i d\Omega=0.
\end{eqnarray}

This equation is satisfied for an arbitrary vector $\xi^i$, thereby we obtain
\begin{eqnarray}
& & \mathfrak{S}_{ab\phantom{j},j}^{\phantom{ab}j}\omega^{ab}_{\phantom{ab}i}+\mathfrak{S}_{ab}^{\phantom{ab}j}(\omega^{ab}_{\phantom{ab}i,j}-\omega^{ab}_{\phantom{ab}j,i})-2\mathfrak{T}^{\phantom{i}j}_{i\phantom{j},j}-2\mathfrak{T}^{\phantom{j}a}_j e^j_{a,i} \nonumber \\
& & =(\mathfrak{S}_{ab\phantom{j}|j}^{\phantom{ab}j}-2S_k \mathfrak{S}_{ab}^{\phantom{ab}k}+\mathfrak{S}_{cb}^{\phantom{cb}j}\omega^c_{\phantom{c}aj}+\mathfrak{S}_{ac}^{\phantom{ac}j}\omega^c_{\phantom{c}bj})\omega^{ab}_{\phantom{ab}i}-2\mathfrak{T}^{\phantom{i}j}_{i\phantom{j},j}-2\mathfrak{T}^{\phantom{j}a}_j e^j_{a,i} \nonumber \\
& & +\mathfrak{S}_{ab}^{\phantom{ab}j}(-R^{ab}_{\phantom{ab}ij}+\omega^a_{\phantom{a}ci}\omega^{cb}_{\phantom{cb}j}-\omega^a_{\phantom{a}cj}\omega^{cb}_{\phantom{cb}i})=0,
\end{eqnarray}
which reduces to
\begin{eqnarray}
& & (\mathfrak{S}_{ab\phantom{j}|j}^{\phantom{ab}j}-2S_k \mathfrak{S}_{ab}^{\phantom{ab}k})\omega^{ab}_{\phantom{ab}i}-R^{ab}_{\phantom{ab}ij}\mathfrak{S}_{ab}^{\phantom{ab}j}-2\mathfrak{T}^{\phantom{i}j}_{i\phantom{j};j}+4S_j \mathfrak{T}^{\phantom{i}j}_i-2\mathfrak{T}_{jk}\omega^{jk}_{\phantom{jk}i}+4S^{jk}_{\phantom{jk}i}\mathfrak{T}_{jk} \nonumber \\
& & =(\mathfrak{S}_{jl\phantom{k};k}^{\phantom{jl}k}-2S_k \mathfrak{S}_{jl}^{\phantom{jl}k})\omega^{jl}_{\phantom{jl}i}-R^{kl}_{\phantom{kl}ij}\mathfrak{S}_{kl}^{\phantom{kl}j}-2\mathfrak{T}^{\phantom{i}j}_{i\phantom{j};j}+4S_j \mathfrak{T}^{\phantom{i}j}_i-2\mathfrak{T}_{jk}\omega^{jk}_{\phantom{jk}i}+4S^{jk}_{\phantom{jk}i}\mathfrak{T}_{jk} \nonumber \\
& & =0.
\label{ctem6}
\end{eqnarray}
The conservation law for the spin density (\ref{cam7}) brings (\ref{ctem6}) to the covariant {\em conservation law for the tetrad energy--momentum density}:
\begin{equation}
\mathfrak{T}^{\phantom{i}j}_{i\phantom{j};j}-2S_j \mathfrak{T}^{\phantom{i}j}_i=2S^j_{\phantom{j}ki}\mathfrak{T}^{\phantom{j}k}_j+\frac{1}{2}R^{kl}_{\phantom{kl}ji}\mathfrak{S}_{kl}^{\phantom{kl}j},
\label{ctem7}
\end{equation}
which is equivalent to
\begin{equation}
\mathfrak{T}^{ij}_{\phantom{ij}:j}=C_{jk}^{\phantom{jk}i}\mathfrak{T}^{jk}+\frac{1}{2}R^{klji}\mathfrak{S}_{klj}.
\label{ctem8}
\end{equation}
This law can be written as the {\em conservation law for the tetrad energy--momentum tensor}:
\begin{equation}
t^{ij}_{\phantom{ij}:j}=C_{jk}^{\phantom{jk}i}t^{jk}+\frac{1}{2}R^{klji}s_{klj}.
\label{ctem9}
\end{equation}
In the absence of spin and torsion, (\ref{ctem8})
reduces to (\ref{Noeth15}) because $\Theta^{ij}=\mathfrak{T}^{ij}$.
Both conservation laws (\ref{cam8}) and (\ref{ctem9}) are independent of the form of the Lagrangian density for the gravitational field.
They would still be satisfied if the Ricci scalar in (\ref{Laggr2}) were replaced with a different scalar function of the curvature tensor.

\subsubsection{Conservation of metric energy--momentum}
The metric and torsion tensors can be taken, instead of the tetrad and spin connection, as the dynamical variables describing spacetime.
Under an infinitesimal coordinate transformation (\ref{infcor1}), the matter Lagrangian density $\mathfrak{L}_\textrm{m}(\phi,\phi_{,i})$ changes according to
\begin{eqnarray}
& & \delta\mathfrak{L}_\textrm{m}=\frac{\partial\mathfrak{L}_\textrm{m}}{\partial\phi}\delta\phi+\frac{\partial\mathfrak{L}_\textrm{m}}{\partial\phi_{,i}}\delta(\phi_{,i})+\frac{\partial\mathfrak{L}_\textrm{m}}{\partial g^{ik}}\delta g^{ik}+\frac{\partial\mathfrak{L}_\textrm{m}}{\partial g^{ik}_{\phantom{ik},l}}\delta(g^{ik}_{\phantom{ik},l}) \nonumber \\
& & +\frac{\partial\mathfrak{L}_\textrm{m}}{\partial S^j_{\phantom{j}ik}}\delta S^j_{\phantom{j}ik}+\frac{\partial\mathfrak{L}_\textrm{m}}{\partial S^j_{\phantom{j}ik,l}}\delta(S^j_{\phantom{j}ik,l}).
\end{eqnarray}
The matter action $S_\textrm{m}=(1/c)\int\mathfrak{L}_\textrm{m}(\phi,\phi_{,i})d\Omega$ is a scalar, thereby it does not change under this transformation:
\begin{eqnarray}
& & \delta S_\textrm{m}=\frac{1}{c}\int\biggl(\frac{\partial\mathfrak{L}_\textrm{m}}{\partial\phi}\delta\phi+\frac{\partial\mathfrak{L}_\textrm{m}}{\partial\phi_{,i}}\delta(\phi_{,i})+\frac{\partial\mathfrak{L}_\textrm{m}}{\partial g^{ik}}\delta g^{ik}+\frac{\partial\mathfrak{L}_\textrm{m}}{\partial g^{ik}_{\phantom{ik},l}}\delta(g^{ik}_{\phantom{ik},l}) \nonumber \\
& & +\frac{\partial\mathfrak{L}_\textrm{m}}{\partial S^j_{\phantom{j}ik}}\delta S^j_{\phantom{j}ik}+\frac{\partial\mathfrak{L}_\textrm{m}}{\partial S^j_{\phantom{j}ik,l}}\delta(S^j_{\phantom{j}ik,l})\biggr)d\Omega=0.
\label{cmem2}
\end{eqnarray}
The first two terms in (\ref{cmem2}) vanish because of the Lagrange equations for $\phi$~(\ref{Lageq8}).
If the variations $\delta g^{ik}$ and $S^j_{\phantom{j}ik}$ vanish on the boundary of the region of integration, then
\begin{eqnarray}
& & \delta S_\textrm{m}=\frac{1}{c}\int\biggl(\frac{\partial\mathfrak{L}_\textrm{m}}{\partial g^{ik}}-\partial_l\frac{\partial\mathfrak{L}_\textrm{m}}{\partial g^{ik}_{\phantom{ik},l}}\biggr)\delta g^{ik}d\Omega+\frac{1}{c}\int\biggl(\frac{\partial\mathfrak{L}_\textrm{m}}{\partial S^j_{\phantom{j}ik}}-\partial_l\frac{\partial\mathfrak{L}_\textrm{m}}{\partial S^j_{\phantom{j}ik,l}}\biggr)\delta S^j_{\phantom{j}ik}d\Omega \nonumber \\
& & =\frac{1}{c}\int\frac{\delta\mathfrak{L}_\textrm{m}}{\delta g^{ik}}\delta g^{ik}d\Omega+\frac{1}{c}\int\frac{\delta\mathfrak{L}_\textrm{m}}{\delta S^j_{\phantom{j}ik}}\delta S^j_{\phantom{j}ik}d\Omega=\frac{1}{2c}\int{\cal T}_{ik}\delta g^{ik}d\Omega+\frac{1}{2c}\int\tau_j^{\phantom{j}ik}\delta S^j_{\phantom{j}ik}d\Omega \nonumber \\
& & =-\frac{1}{2c}\int{\cal T}^{ik}\delta g_{ik}d\Omega+\frac{1}{2c}\int\tau_j^{\phantom{j}ik}\delta S^j_{\phantom{j}ik}d\Omega=0.
\end{eqnarray}

The components of the metric tensor change because of an infinitesimal coordinate transformation (\ref{infcor1}), so the corresponding variation of the metric tensor is given by (\ref{Chrsym22}):
\begin{equation}
\delta g_{ik}=\bar{\delta}g_{ik}=-2\xi_{(i:k)}.
\end{equation}
The variation of the torsion tensor is given by (\ref{infcor13}):
\begin{equation}
\delta S^j_{\phantom{j}ik}=\bar{\delta}S^j_{\phantom{j}ik}=-\mathcal{L}_\xi S^j_{\phantom{j}ik}=\xi^j_{\phantom{j},l}S^l_{\phantom{l}ik}-\xi^l_{\phantom{l},i}S^j_{\phantom{j}lk}-\xi^l_{\phantom{l},k}S^j_{\phantom{j}il}-\xi^l S^j_{\phantom{j}ik,l}.
\end{equation}
The variation of the matter action under (\ref{infcor1}) is therefore equal to
\begin{equation}
\delta S_\textrm{m}=\bar{\delta}S_\textrm{m}=-\frac{1}{2c}\int{\cal T}^{ik}\bar{\delta}g_{ik}d\Omega+\frac{1}{2c}\int\tau_j^{\phantom{j}ik}\bar{\delta}S^j_{\phantom{j}ik}d\Omega=0.
\label{cmem6}
\end{equation}
The first integral in (\ref{cmem6}) is
\begin{eqnarray}
& & -\frac{1}{2c}\int{\cal T}^{ik}\bar{\delta}g_{ik}d\Omega=\frac{1}{c}\int{\cal T}^{ik}\xi_{i:k}d\Omega=\frac{1}{c}\int({\cal T}^{ik}\xi_i)_{:k}d\Omega-\frac{1}{c}\int{\cal T}^{ik}_{\phantom{ik}:k}\xi_i d\Omega \nonumber \\
& & =\frac{1}{c}\int({\cal T}^{ik}\xi_i)_{,k}d\Omega-\frac{1}{c}\int{\cal T}^{ik}_{\phantom{ik}:k}\xi_i d\Omega=\frac{1}{c}\int{\cal T}^{ik}\xi_i dS_k-\frac{1}{c}\int{\cal T}_{l\phantom{k}:k}^{\phantom{l}k}\xi^l d\Omega,
\end{eqnarray}
using (\ref{Chrsym234}) for the contravariant vector density ${\cal T}^{ik}\xi_i$.
The second integral in (\ref{cmem6}) is
\begin{eqnarray}
& & \frac{1}{2c}\int\tau_j^{\phantom{j}ik}\bar{\delta}S^j_{\phantom{j}ik}d\Omega=\frac{1}{2c}\int\bigl((\tau_j^{\phantom{j}ik}\xi^j S^l_{\phantom{l}ik})_{,l}-(\tau_j^{\phantom{j}ik}\xi^l S^j_{\phantom{j}lk})_{,i}-(\tau_j^{\phantom{j}ik}\xi^l S^j_{\phantom{j}il})_{,k}\bigr)d\Omega \nonumber \\
& & +\frac{1}{2c}\int\bigl(-(\tau_j^{\phantom{j}ik}S^l_{\phantom{l}ik})_{,l}\xi^j+(\tau_j^{\phantom{j}ik}S^j_{\phantom{j}lk})_{,i}\xi^l+(\tau_j^{\phantom{j}ik}S^j_{\phantom{j}il})_{,k}\xi^l-\tau_j^{\phantom{j}ik}S^j_{\phantom{j}ik,l}\xi^l\bigr)d\Omega \nonumber \\
& & =\frac{1}{2c}\int\tau_j^{\phantom{j}ik}\xi^j S^l_{\phantom{l}ik}dS_l-\frac{1}{2c}\int\tau_j^{\phantom{j}ik}\xi^l S^j_{\phantom{j}lk}dS_i-\frac{1}{2c}\int\tau_j^{\phantom{j}ik}\xi^l S^j_{\phantom{j}il}dS_k \nonumber \\
& & +\frac{1}{2c}\int\bigl(-(2\tau_j^{\phantom{j}ik}S^j_{\phantom{j}li})_{,k}-(\tau_l^{\phantom{j}ij}S^k_{\phantom{k}ij})_{,k}-\tau_j^{\phantom{j}ik}S^j_{\phantom{j}ik,l}\bigr)\xi^l d\Omega.
\end{eqnarray}

If the variations $\xi^i$ of the coordinates vanish on the boundary of the region of integration, then (\ref{cmem6}) becomes
\begin{equation}
\delta S_\textrm{m}=-\frac{1}{2c}\int\bigl(2{\cal T}_{l\phantom{k}:k}^{\phantom{l}k}+(2\tau_j^{\phantom{j}ik}S^j_{\phantom{j}li}+\tau_l^{\phantom{j}ij}S^k_{\phantom{k}ij})_{,k}+\tau_j^{\phantom{j}ik}S^j_{\phantom{j}ik,l}\bigr)\xi^l d\Omega=0.
\label{cmem9}
\end{equation}
Since the variations $\xi^i$ are arbitrary, (\ref{cmem9}) gives the covariant {\em conservation law for the metric energy--momentum density} (4 equations):
\begin{equation}
{\cal T}_{l\phantom{k}:k}^{\phantom{l}k}+\Bigl(\tau_j^{\phantom{j}ik}S^j_{\phantom{j}li}+\frac{1}{2}\tau_l^{\phantom{j}ij}S^k_{\phantom{k}ij}\Bigr)_{,k}+\frac{1}{2}\tau_j^{\phantom{j}ik}S^j_{\phantom{j}ik,l}=0.
\label{cmem10}
\end{equation}
Dividing the conservation law (\ref{cmem10}) by $\sqrt{-\mathfrak{g}}$ gives
\begin{eqnarray}
& & T_{l\phantom{k}:k}^{\phantom{l}k}+\Bigl(t_j^{\phantom{j}ik}S^j_{\phantom{j}li}+\frac{1}{2}t_l^{\phantom{j}ij}S^k_{\phantom{k}ij}\Bigr)_{,k}+\Bigl(t_j^{\phantom{j}ik}S^j_{\phantom{j}li}+\frac{1}{2}t_l^{\phantom{j}ij}S^k_{\phantom{k}ij}\Bigr)\frac{\sqrt{-\mathfrak{g}}_{,k}}{\sqrt{-\mathfrak{g}}}+\frac{1}{2}t_j^{\phantom{j}ik}S^j_{\phantom{j}ik,l} \nonumber \\
& & =T_{l\phantom{k}:k}^{\phantom{l}k}+\Bigl(t_j^{\phantom{j}ik}S^j_{\phantom{j}li}+\frac{1}{2}t_l^{\phantom{j}ij}S^k_{\phantom{k}ij}\Bigr)_{:k}-t_j^{\phantom{j}im}S^j_{\phantom{j}li}\mathring{\Gamma}^{k}_{mk}+t_j^{\phantom{j}ik}S^j_{\phantom{j}mi}\mathring{\Gamma}^{m}_{lk}+\frac{1}{2}t_m^{\phantom{m}ij}S^k_{\phantom{k}ij}\mathring{\Gamma}^{m}_{lk} \nonumber \\
& & -\frac{1}{2}t_l^{\phantom{l}ij}S^m_{\phantom{m}i\,j}\mathring{\Gamma}^{k}_{mk}+\Bigl(t_j^{\phantom{j}ik}S^j_{\phantom{j}li}+\frac{1}{2}t_l^{\phantom{j}ij}S^k_{\phantom{k}ij}\Bigr)\mathring{\Gamma}^{m}_{mk}+\frac{1}{2}t_j^{\phantom{j}ik}S^j_{\phantom{j}ik:l}-\frac{1}{2}t_j^{\phantom{j}ik}S^m_{\phantom{m}ik}\mathring{\Gamma}^{j}_{ml} \nonumber \\
& & +\frac{1}{2}t_j^{\phantom{j}ik}S^j_{\phantom{j}mk}\mathring{\Gamma}^{m}_{il}+\frac{1}{2}t_j^{\phantom{j}ik}S^j_{\phantom{j}im}\mathring{\Gamma}^{m}_{kl}=0,
\end{eqnarray}
where
\begin{equation}
t_{ijk}=\frac{\tau_{ijk}}{\sqrt{-\mathfrak{g}}}.
\end{equation}
We therefore obtain the {\em conservation law for the metric energy--momentum tensor}:
\begin{equation}
\Bigl(T_l^{\phantom{l}k}+t_j^{\phantom{j}ik}S^j_{\phantom{j}li}+\frac{1}{2}t_l^{\phantom{j}ij}S^k_{\phantom{k}ij}\Bigr)_{:k}+\frac{1}{2}t_j^{\phantom{j}ik}S^j_{\phantom{j}ik:l}=0.
\label{cmem13}
\end{equation}
Equations (\ref{cmem10}) and (\ref{cmem13}) are equivalent to (\ref{ctem7}) and (\ref{ctem9}).

If the matter Lagrangian density does not depend on the torsion tensor, then $t_{ijk}=0$ and (\ref{cmem10}) reduces to
\begin{equation}
{\cal T}^{ik}_{\phantom{ik}:k}=0.
\label{cmem14}
\end{equation}
Equivalently, (\ref{cmem13}) reduces to
\begin{equation}
T^{ik}_{\phantom{ik}:k}=0.
\label{cmem15}
\end{equation}
In this case, vanishing of $\int{\cal T}^{ij}\bar{\delta}g_{ij}d\Omega$ in (\ref{cmem6}) does not imply ${\cal T}^{ij}=0$, because 10 variations $\bar{\delta}g_{ij}$ are not all independent; they are functions of 4 independent variations $\xi^i$.

\subsubsection{Conservation laws in flat spacetime}
{\bf Conservation of energy--momentum from Lagrange equations}.\\
Let us consider a physical system in a locally flat spacetime, in which the torsion and curvature of the gravitational field can be neglected.
Such a system is described by a matter Lagrangian density $\mathfrak{L}_\textrm{m}$, which depends on the coordinates only through matter fields $\phi$ and their first derivatives $\phi_{,i}$.
Differentiating $\mathfrak{L}_\textrm{m}$ gives, using the Lagrange equations (\ref{Lageq8}),
\begin{equation}
\partial_i\mathfrak{L}_\textrm{m}=\frac{\partial\mathfrak{L}_\textrm{m}}{\partial\phi}\phi_{,i}+\frac{\partial\mathfrak{L}_\textrm{m}}{\partial\phi_{,j}}\phi_{,ji}=\partial_j\biggl(\frac{\partial\mathfrak{L}_\textrm{m}}{\partial\phi_{,j}}\biggr)\phi_{,i}+\frac{\partial\mathfrak{L}_\textrm{m}}{\partial\phi_{,j}}\phi_{,ji}=\partial_j\biggl(\frac{\partial\mathfrak{L}_\textrm{m}}{\partial\phi_{,j}}\phi_{,i}\biggr).
\end{equation}
This equation can be written as a conservation law:
\begin{equation}
\theta^{\phantom{i}j}_{i\phantom{j},j}=0,
\label{clLg2}
\end{equation}
for a quantity
\begin{equation}
\theta^{\phantom{i}j}_i=\frac{\partial\mathfrak{L}_\textrm{m}}{\partial\phi_{,j}}\phi_{,i}-\delta^j_i\mathfrak{L}_\textrm{m}.
\label{clLg3}
\end{equation}
The quantity (\ref{clLg3}) is a special case of the canonical energy--momentum density (\ref{cemd3}) in the Galilean and geodesic frame.
The conservation law (\ref{clLg2}) is a special case of (\ref{ctem7}) in that frame.
Since $\mathfrak{e}=1$, (\ref{clLg3}) is equal to the canonical energy--momentum tensor (\ref{cemd5}).\\

\noindent
{\bf Conservation of energy--momentum from Noether theorem for translations}.\\
The Noether current (\ref{Noeth7}) can be written as
\begin{equation}
\mathfrak{J}^i=\frac{\partial\mathfrak{L}_\textrm{m}}{\partial\phi_{,i}}\delta\phi-\theta_j^{\phantom{j}i}\xi^j.
\end{equation}
The Poincar\'{e} group contains translations and Lorentz rotations.
If $x^i$ are the Cartesian coordinates, then for translations, $\xi^i=\epsilon^i=$ const and $\delta\phi=0$, the current (\ref{Noeth7}) is
\begin{equation}
\mathfrak{J}^i=\epsilon^i\mathfrak{L}_\textrm{m}-\frac{\partial\mathfrak{L}_\textrm{m}}{\partial\phi_{,i}}\epsilon^j\phi_{,j}.
\end{equation}
The conservation law (\ref{Noeth6}) for this current is
\begin{equation}
\epsilon^j\theta^{\phantom{j}i}_{j\phantom{i},i}=0,
\label{clLg5}
\end{equation}
which gives (\ref{clLg2}) because $\epsilon^i$ are arbitrary. \\

\noindent
{\bf Conservation of angular momentum from Noether theorem for rotations}.\\
For Lorentz rotations, $\xi^i=\epsilon^i_{\phantom{i}j}x^j$ and $\delta\phi=(1/2)\epsilon_{ij}G^{ij}\phi$, where $G^{ij}$ are the generators of the Lorentz group, the Noether current (\ref{Noeth7}) is
\begin{equation}
\mathfrak{J}^i=\epsilon^{ij}x_j\mathfrak{L}_\textrm{m}+\frac{\partial\mathfrak{L}_\textrm{m}}{\partial\phi_{,i}}\biggl(\frac{1}{2}\epsilon^{kl}G_{kl}\phi-\epsilon^{jk}x_k\phi_{,j}\biggr)=\epsilon^{kl}\biggl(x_k\frac{\partial\mathfrak{L}_\textrm{m}}{\partial\phi_{,i}}\phi_{,l}-x_k\delta^i_l\mathfrak{L}_\textrm{m}+\frac{1}{2}\frac{\partial\mathfrak{L}_\textrm{m}}{\partial\phi_{,i}}G_{kl}\phi\biggr).
\end{equation}
The conservation law (\ref{Noeth6}) for this current is
\begin{equation}
\epsilon^{kl}\biggl(\frac{\partial\mathfrak{L}_\textrm{m}}{\partial\phi_{,i}}\phi_{,[l}x_{k]}-\delta^i_{[l}x_{k]}\mathfrak{L}_\textrm{m}+\frac{1}{2}\frac{\partial\mathfrak{L}_\textrm{m}}{\partial\phi_{,i}}G_{kl}\phi\biggr)_{,i}=0.
\label{clLg8}
\end{equation}
Because $\epsilon^{kl}$ are arbitrary, this equation gives the conservation law,
\begin{equation}
\mathfrak{M}_{kl\phantom{i},i}^{\phantom{kl}i}=0,
\label{clLg9}
\end{equation}
for the {\em angular momentum density}:
\begin{equation}
\mathfrak{M}_{kl}^{\phantom{kl}i}=x_k\theta^{\phantom{l}i}_l-x_l\theta^{\phantom{k}i}_k+\frac{\partial\mathfrak{L}_\textrm{m}}{\partial\phi_{,i}}G_{kl}\phi=x_k\theta^{\phantom{l}i}_l-x_l\theta^{\phantom{k}i}_k+\Sigma_{kl}^{\phantom{kl}i}.
\label{clLg10}
\end{equation}
The angular momentum density is antisymmetric in the first two indices:
\begin{equation}
\mathfrak{M}_{ij}^{\phantom{ij}k}=-\mathfrak{M}_{ji}^{\phantom{ji}k}.
\label{clLg11}
\end{equation}

\noindent
{\bf Orbital and intrinsic angular momentum density}.\\
The angular momentum density is the sum,
\begin{equation}
\mathfrak{M}_{ij}^{\phantom{ij}k}=\Lambda_{ij}^{\phantom{ij}k}+\Sigma_{ij}^{\phantom{ij}k},
\label{clLg12}
\end{equation}
of two tensor densities: the {\em orbital angular momentum density},
\begin{equation}
\Lambda_{kl}^{\phantom{kl}i}=x_k\theta^{\phantom{l}i}_l-x_l\theta^{\phantom{k}i}_k,
\label{clLg13}
\end{equation}
and the {\em intrinsic angular momentum density} (canonical spin density) (\ref{spden11}).
The conservation law (\ref{clLg9}) gives
\begin{equation}
\mathfrak{M}^{kli}_{\phantom{kli},i}=\delta^k_i\theta^{li}+x^k\theta^{li}_{\phantom{li},i}-\delta^l_i\theta^{ki}-x^l\theta^{ki}_{\phantom{ki},i}+\Sigma^{kli}_{\phantom{kli},i}=0,
\end{equation}
which reduces, by means of (\ref{clLg2}), to
\begin{equation}
\theta_{kl}-\theta_{lk}-\Sigma_{kl\phantom{i},i}^{\phantom{kl}i}=0.
\label{clLg15}
\end{equation}
This equation is a special case of the conservation law for the spin density (\ref{cam7}) in the Galilean and geodesic frame.\\

\noindent
{\bf Symmetrization of canonical tensor}.\\
The canonical energy--momentum density $\theta_{ik}$ is not symmetric.
However, the quantity
\begin{equation}
\tau_{ik}=\theta_{ik}+\partial_j\psi_{ik}^{\phantom{ik}j},
\label{clLg16}
\end{equation}
where
\begin{equation}
\psi_{ik}^{\phantom{ik}j}=-\frac{1}{2}(\Sigma_{ik}^{\phantom{ik}j}-\Sigma_{k\phantom{j}i}^{\phantom{k}j}+\Sigma^j_{\phantom{j}ik}),
\label{clLg17}
\end{equation}
is symmetric:
\begin{equation}
\tau_{ik}-\tau_{ki}=\theta_{ik}-\theta_{ki}+\partial_j(\psi_{ik}^{\phantom{ik}j}-\psi_{ki}^{\phantom{ki}j})=\Sigma_{ik\phantom{j},j}^{\phantom{ik}j}+\partial_j(\psi_{ik}^{\phantom{ik}j}-\psi_{ki}^{\phantom{ki}j})=0.
\end{equation}
Since (\ref{clLg17}) is antisymmetric in the last two indices,
\begin{equation}
\psi^{ikj}=-\psi^{ijk},
\label{clLg19}
\end{equation}
the quantity (\ref{clLg16}) is also conserved:
\begin{equation}
\tau^{ik}_{\phantom{ik},k}=\theta^{ik}_{\phantom{ik},k}+\psi^{ikj}_{\phantom{ikj},jk}=\theta^{ik}_{\phantom{ik},k}=0.
\label{clLg20}
\end{equation}
The symmetric energy--momentum density $\tau_{ik}$ is equal to the metric energy--momentum density (\ref{dmemd4}) in the Galilean and geodesic frame.
Equation (\ref{clLg16}) is a special case of the Belinfante--Rosenfeld relation (\ref{BelRos8}) in that frame.

\subsubsection{Momentum four-vector}
We define the {\em momentum four-vector} or {\em four-momentum} of matter on a hypersurface as
\begin{equation}
P_i=\frac{1}{c}\int\mathfrak{T}_i^{\phantom{i}k}dS_k=\frac{1}{c}\int\Theta_i^{\phantom{i}k}dS_k,
\label{fmam1}
\end{equation}
where $dS_k$ is the element of hypersurface.
Either the tetrad $\mathfrak{T}_i^{\phantom{i}k}$ or canonical energy--momentum density $\Theta_i^{\phantom{i}k}$ can be used because they are equal according to (\ref{cemd4}).
The quantity $P_i$ is a four-vector under general coordinate transformations if the hypersurface is infinitesimal (so all the points on it transform the same way).
If the hypersurface is finite, then $P_i$ is a four-vector only under Lorentz transformations.
A hypersurface can be taken as a hyperplane (volume) perpendicular to the $x^0$ axis: $dS_k=\delta_k^0 dV$, where $dV$ is the element of volume.
For this hypersurface of constant time, the four-momentum (\ref{fmam1}) is given by the volume integral:
\begin{equation}
P_i=\frac{1}{c}\int\mathfrak{T}_i^{\phantom{i}0}dV.
\label{fmam2}
\end{equation}

In a flat spacetime, the four-momentum (\ref{fmam1}) reduces to
\begin{equation}
P_i=\frac{1}{c}\int\theta_i^{\phantom{i}k}dS_k=\frac{1}{c}\int\theta_i^{\phantom{i}0}dV,
\label{fmam3}
\end{equation}
where $\theta_i^{\phantom{i}k}$ is the given by (\ref{clLg3}).
A closed hypersurface is a volume surrounding the four-volume between two hyperplanes at times $t_1$ and $t_2$.
Integrating the conservation law (\ref{clLg2}) for the canonical energy--momentum density (\ref{clLg3}) in a flat spacetime over the four-volume, using the Gau\ss--Stokes theorem (\ref{covint7}), gives
\begin{equation}
\int\theta_{i\phantom{k},k}^{\phantom{i}k}d\Omega=\oint\theta_i^{\phantom{i}k}dS_k=\int\theta_i^{\phantom{i}0}dV\Big|_{t_1}^{t_2}=cP_i|_{t_2}-cP_i|_{t_1}=0.
\label{fmam4}
\end{equation}
Consequently, the four-momentum (\ref{fmam3}) in a flat spacetime is conserved:
\begin{equation}
P_i=\mbox{const}.
\label{fmam5}
\end{equation}

If spacetime is not flat, one can construct an energy--momentum pseudotensor for the gravitational field such that the total four-momentum of the gravitational field and matter is conserved, as shown in Section \ref{Gravfieldeqs}.\\

\noindent
{\bf Energy and momentum density}.\\
The integral (\ref{fmam3}) implies that the components $\theta_i^{\phantom{i}0}/c$ form the four-momentum per volume, that is, the {\em four-momentum density}.
The component $\theta_0^{\phantom{0}0}$ is referred to as the {\em energy density}:
\begin{equation}
W=\theta_0^{\phantom{0}0}=\dot{\phi}\frac{\partial\mathfrak{L}_\textrm{m}}{\partial\dot{\phi}}-\mathfrak{L}_\textrm{m},
\label{fmam6}
\end{equation}
which follows from (\ref{clLg3}).
Hereinafter, a dot above a quantity $\phi$ denotes the derivative of $\phi$ with respect to time, $\dot{\phi}=d\phi/dt$, and two dots above $\phi$ denote second derivative of $\phi$ with respect to time, $\ddot{\phi}=d^2\phi/dt^2$.
The spatial components
\begin{equation}
\Pi_\alpha=\frac{1}{c}\theta_\alpha^{\phantom{\alpha}0}
\label{fmam7}
\end{equation}
form the {\em momentum density} ${\bm\Pi}$.\\

\noindent
{\bf Energy and momentum}.\\
We define the {\em energy} as the time component of the four-vector $cP_i$:
\begin{equation}
E=cP_0.
\label{fmam8}
\end{equation}
The energy of matter in some volume is given by the volume integral of the energy density (\ref{fmam6}):
\begin{equation}
E=\int\theta_0^{\phantom{0}0}dV.
\label{fmam9}
\end{equation}
We define the {\em momentum} vector ${\bf P}$ as a spatial vector formed by the spatial components $P^\alpha$ of the four-vector $P^i$:
\begin{equation}
P^i=(P^0,P^\alpha)=(P^0,{\bf P}).
\end{equation}
The momentum of matter in some volume is given by the volume integral of the momentum density (\ref{fmam7}):
\begin{equation}
P_\alpha=\frac{1}{c}\int\theta_\alpha^{\phantom{\alpha}0}dV.
\label{fmam11}
\end{equation}

\noindent
{\bf Energy and momentum flux}.\\
The conservation law (\ref{clLg2}) can be written as
\begin{eqnarray}
& & \frac{\partial\theta_0^{\phantom{0}j}}{\partial x^j}=0,\quad \frac{1}{c}\frac{\partial\theta_0^{\phantom{0}0}}{\partial t}+\frac{\partial\theta_0^{\phantom{0}\beta}}{\partial x^\beta}=0,
\label{fmam51} \\
& & \frac{\partial\theta_\alpha^{\phantom{\alpha}j}}{\partial x^j}=0,\quad \frac{1}{c}\frac{\partial\theta_\alpha^{\phantom{\alpha}0}}{\partial t}+\frac{\partial\theta_\alpha^{\phantom{\alpha}\beta}}{\partial x^\beta}=0.
\label{fmam52}
\end{eqnarray}
Integrating these equations over a volume and using the Gau\ss\, theorem (\ref{spvec33}) gives
\begin{eqnarray}
& & \frac{\partial}{\partial t}\int\theta_0^{\phantom{0}0}dV=-c\oint\theta_0^{\phantom{0}\beta}df_\beta, \label{fmam16} \\
& & \frac{\partial}{\partial t}\int\frac{1}{c}\theta_\alpha^{\phantom{\alpha}0}dV=-\oint\theta_\alpha^{\phantom{\alpha}\beta}df_\beta.
\label{fmam17}
\end{eqnarray}
Integrating the components of the {\em energy flux density} vector {\bf S}:
\begin{equation}
S^\beta=c\theta_0^{\phantom{0}\beta},
\label{fmam18}
\end{equation}
over the element of surface $df_\beta$ gives the {\em energy flux}:
\begin{equation}
\oint S^\beta df_\beta=-\frac{dE}{dt}.
\label{fmam19}
\end{equation}
This relation results from (\ref{fmam9}), (\ref{fmam16}), and (\ref{fmam18}).
The quantity opposite to the energy flux is the rate of the energy leaving the region of integration, referred to as the {\em power}.

Integrating the components $\theta_\alpha^{\phantom{\alpha}\beta}$, which represent the {\em momentum flux density} tensor, over $df_\alpha$ gives the {\em momentum flux} vector.
The {\em stress tensor} is defined as
\begin{equation}
\sigma_\alpha^{\phantom{\alpha}\beta}=-\theta_\alpha^{\phantom{\alpha}\beta}.
\label{fmam20}
\end{equation}
Its integral taken over $df_\beta$ gives the vector opposite to the momentum flux, referred to as the {\em surface force} ${\bf F}$:
\begin{equation}
\oint\sigma_\alpha^{\phantom{\alpha}\beta}df_\beta=F_\alpha.
\label{fmam21}
\end{equation}
The relations (\ref{fmam11}), (\ref{fmam17}), (\ref{fmam20}), and (\ref{fmam21}) equal the time derivative of the momentum $P_\alpha$ to the surface force $F_\alpha$:
\begin{equation}
F_\alpha=\frac{dP_\alpha}{dt},\quad {\bf F}=\frac{d{\bf P}}{dt}.
\label{fmam22}
\end{equation}
The components of the energy--momentum tensor form the following matrix:
\begin{equation}
\theta_i^{\phantom{i}k}=\left( \begin{array}{cc}
W & {\bf S}/c \\
c{\bm\Pi} & -\sigma_\alpha^{\phantom{\alpha}\beta} \end{array} \right).
\label{fmam23}
\end{equation}

\noindent
{\bf Closed system}.\\
If the integration extends beyond the region in which matter is present, then the surface integrals vanish.
Such a system is referred to as {\em closed}.
Equation (\ref{fmam19}) gives the conservation of energy of the matter in a closed system:
\begin{equation}
E=\mbox{const},
\label{fmam53}
\end{equation}
which is the time component of (\ref{fmam5}).
Equation (\ref{fmam22}) gives the conservation of momentum of the matter in a closed system:
\begin{equation}
{\bf P}=\mbox{const},
\label{fmam54}
\end{equation}
corresponding to the spatial components of (\ref{fmam5}).\\

\noindent
{\bf Four-momentum for symmetrized canonical tensor}.\\
Adding the quantity $\psi^{ikj}_{\phantom{ikj},j}$, where the condition (\ref{clLg19}) is satisfied, brings the canonical energy--momentum tensor $\theta^{ik}$ to a symmetric form, $\tau^{ik}$, and preserves the conservation law (\ref{clLg2}).
Using the Gau\ss--Stokes theorem (\ref{covint6}) gives
\begin{equation}
\int\tau_i^{\phantom{i}k}dS_k=\int\theta_i^{\phantom{i}k}dS_k+\int\psi_{i\phantom{kj},j}^{\phantom{i}kj}dS_k=cP_i+\frac{1}{2}\int(\psi_{i\phantom{kj},j}^{\phantom{i}kj}dS_k-\psi_{i\phantom{kj},k}^{\phantom{i}kj}dS_j)=cP_i+\frac{1}{2}\oint\psi_i^{\phantom{i}kj}df^\ast_{kj},
\end{equation}
where the last integral is taken over the surface which bounds the hypersurface.
If this surface is located in a region, where matter is absent, then the surface integral vanishes.
Consequently, replacing $\theta^{ik}$ by $\tau^{ik}$ does not change the four-momentum (\ref{fmam3}):
\begin{equation}
P_i=\frac{1}{c}\int\tau_i^{\phantom{i}k}dS_k=\frac{1}{c}\int\tau_i^{\phantom{i}0}dV.
\label{fmam25}
\end{equation}

\subsubsection{Angular momentum four-tensor}
We define the {\em angular momentum four-tensor} of matter on a hypersurface in a flat spacetime as
\begin{equation}
M_{ik}=\frac{1}{c}\int\mathfrak{M}_{ik}^{\phantom{ik}j}dS_j,
\label{fmam28}
\end{equation}
where $\mathfrak{M}_{ik}^{\phantom{ik}j}$ is the angular momentum density (\ref{clLg10}) and $dS_j$ is the element of hypersurface.
The quantity $M_{ik}$ is a tensor under Lorentz transformations.
The angular momentum four-tensor, following (\ref{clLg11}), is antisymmetric:
\begin{equation}
M_{ik}=-M_{ki}.
\end{equation}
If the hypersurface is taken as a hyperplane perpendicular to the $x^0$ axis, then the angular momentum four-tensor is given by the volume integral:
\begin{equation}
M_{ik}=\frac{1}{c}\int\mathfrak{M}_{ik}^{\phantom{ik}0}dV.
\label{fmam29}
\end{equation}
A closed hypersurface surrounds the four-volume between two hyperplanes at times $t_1$ and $t_2$.
Integrating the conservation law (\ref{clLg9}) for the angular momentum density (\ref{clLg10}) over the four-volume, using the Gau\ss--Stokes theorem (\ref{covint7}), gives
\begin{equation}
\int\mathfrak{M}_{ik\phantom{j},j}^{\phantom{ik}j}d\Omega=\oint\mathfrak{M}_{ik}^{\phantom{ik}j}dS_j=\int\mathfrak{M}_{ik}^{\phantom{ik}0}dV\Big|_{t_1}^{t_2}=cM_{ik}|_{t_2}-cM_{ik}|_{t_1}=0.
\end{equation}
Consequently, the angular momentum four-tensor (\ref{fmam29}) in a flat spacetime is conserved:
\begin{equation}
M_{ik}=\mbox{const}.
\label{fmam37}
\end{equation}

If spacetime is not flat, one can construct an energy--momentum pseudotensor for the gravitational field such that the total angular momentum of the gravitational field and matter is conserved, as shown in Section \ref{Gravfieldeqs}.\\

\noindent
{\bf Orbital and intrinsic angular momentum}.\\
Integrating the angular momentum density (\ref{clLg12}) over a hypersurface gives
\begin{equation}
M_{ik}=L_{ik}+S_{ik},
\label{fmam34}
\end{equation}
where
\begin{equation}
L_{ik}=\frac{1}{c}\int\Lambda_{ik}^{\phantom{ik}j}dS_j=\frac{1}{c}\int(x_i\theta_k^{\phantom{k}j}-x_k\theta_i^{\phantom{i}j})dS_j=\frac{1}{c}\int\Lambda_{ik}^{\phantom{ik}0}dV=\frac{1}{c}\int(x_i\theta_k^{\phantom{k}0}-x_k\theta_i^{\phantom{i}0})dV
\label{fmam35}
\end{equation}
is the {\em orbital angular momentum four-tensor} and
\begin{equation}
S_{ik}=\frac{1}{c}\int\Sigma_{ik}^{\phantom{ik}j}dS_j=\frac{1}{c}\int\Sigma_{ik}^{\phantom{ik}0}dV
\label{fmam36}
\end{equation}
is the {\em intrinsic angular momentum four-tensor}.
These tensors are also antisymmetric:
\begin{equation}
L_{ik}=-L_{ki},\quad S_{ik}=-S_{ki}.
\end{equation}
Unlike $M_{ik}$, they are not separately conserved.
The angular momentum four-tensor can also be written, using (\ref{fmam3}), in terms of $P_i$:
\begin{equation}
M_{ik}=\int(x_i dP_k-x_k dP_i)+S_{ik},\quad L_{ik}=\int(x_i dP_k-x_k dP_i).
\label{fmam45}
\end{equation}
In the absence of the intrinsic angular momentum, (\ref{fmam45}) reduces to
\begin{equation}
M_{ik}=\int(x_i dP_k-x_k dP_i).
\label{fmam46}
\end{equation}

\noindent
{\bf Angular momentum tensor and pseudovector}.\\
The components $\mathfrak{M}_{\alpha\beta}^{\phantom{\alpha\beta}0}/c$ form the spatial {\em angular momentum density}.
Integrating them over the volume gives the components of the spatial {\em angular momentum tensor}:
\begin{equation}
M_{\alpha\beta}=\frac{1}{c}\int\mathfrak{M}_{\alpha\beta}^{\phantom{\alpha\beta}0}dV,
\label{fmam56}
\end{equation}
which is a part of the conserved angular momentum four-tensor (\ref{fmam37}) and therefore also conserved.
Since the angular momentum tensor is antisymmetric,
\begin{equation}
M_{\alpha\beta}=-M_{\beta\alpha},
\end{equation}
we can define the {\em angular momentum pseudovector} ${\bf M}$ dual to $M_{\alpha\beta}$:
\begin{equation}
M^\alpha=\frac{1}{2}e^{\alpha\beta\gamma}M_{\beta\gamma},\quad M_{\alpha\beta}=e_{\alpha\beta\gamma}M^\gamma.
\label{fmam33}
\end{equation}
The conservation of (\ref{fmam56}) gives therefore the conservation of angular momentum of the matter in a closed system:
\begin{equation}
{\bf M}=\mbox{const}.
\label{fmam55}
\end{equation}
Similarly to (\ref{fmam33}), we can define the {\em orbital angular momentum pseudovector} ${\bf L}$ and the {\em intrinsic angular momentum pseudovector} ${\bf S}$:
\begin{eqnarray}
& & L^\alpha=\frac{1}{2}e^{\alpha\beta\gamma}L_{\beta\gamma},\quad L_{\alpha\beta}=e_{\alpha\beta\gamma}L^\gamma, \\
& & S^\alpha=\frac{1}{2}e^{\alpha\beta\gamma}S_{\beta\gamma},\quad S_{\alpha\beta}=e_{\alpha\beta\gamma}S^\gamma, \\
& & M^\alpha=L^\alpha+S^\alpha,\quad {\bf M}={\bf L}+{\bf S}.
\end{eqnarray}
The spatial vectors ${\bf L}$ and ${\bf S}$ are not separately conserved.\\

\noindent
{\bf Angular momentum for symmetrized canonical tensor}.\\
The symmetry of the tensor $\tau^{ik}$ (\ref{clLg16}) can be written, using its conservation law (\ref{clLg20}), as
\begin{equation}
\tau^{ki}-\tau^{ik}=\partial_l(x^i\tau^{kl}-x^k\tau^{il})=0.
\end{equation}
Integrating this equation over a four-volume and using the Gau\ss--Stokes theorem (\ref{covint7}) gives
\begin{equation}
\oint(x^i\tau^{kl}-x^k\tau^{il})dS_l=\int(x^i\tau^{kl}-x^k\tau^{il})dS_l\Big|_{t_1}^{t_2}=\int(x^i\tau^{k0}-x^k\tau^{i0})dV\Big|_{t_1}^{t_2}=0,
\end{equation}
which shows the conservation of the quantity
\begin{equation}
\tilde{L}^{ik}=\frac{1}{c}\int(x^i\tau^{kl}-x^k\tau^{il})dS_l=\frac{1}{c}\int(x^i\tau^{k0}-x^k\tau^{i0})dV=\mbox{const}.
\label{fmam43}
\end{equation}
The symmetry of a second-rank tensor whose ordinary divergence is zero is therefore related to the local conservation of the orbital angular momentum four-tensor constructed from that tensor.
Using (\ref{spden11}), (\ref{clLg16}), (\ref{clLg17}), and the Gau\ss--Stokes theorem (\ref{covint6}) leads to
\begin{eqnarray}
& & c\tilde{L}^{ik}=\int\Lambda^{ikl}dS_l+\int(x^i\psi^{klj}_{\phantom{klj},j}-x^k\psi^{ilj}_{\phantom{ilj},j})dS_l=cL^{ik}+\int\Bigl((x^i\psi^{klj})_{,j}-(x^k\psi^{ilj})_{,j}\Bigr)dS_l \nonumber \\
& & -\int(\psi^{kli}-\psi^{ilk})dS_l=cL^{ik}+\frac{1}{2}\Bigl((x^i\psi^{klj})_{,j}dS_l-(x^i\psi^{klj})_{,l}dS_j-(x^k\psi^{ilj})_{,j}dS_l+(x^k\psi^{ilj})_{,l}dS_j\Bigr) \nonumber \\
& & +\int\Sigma^{ikj}dS_j=cM^{ik}+\frac{1}{2}\oint(x^i\psi^{klj}-x^k\psi^{ilj})df^\ast_{lj}.
\end{eqnarray}
If the integration surface is located in a region, where matter is absent, then the surface integral vanishes.
Consequently, the quantity (\ref{fmam43}) is equal to the angular momentum four-tensor (\ref{fmam28}).
This equality shows that replacing $\theta^{ik}$ in (\ref{clLg13}) by $\tau^{ik}$ changes the values of (\ref{fmam35}) and thereby (\ref{fmam28}).\\

\noindent
{\bf Center of inertia}.\\
The components $M_{\alpha 0}$ of the angular momentum four-tensor (\ref{fmam34}) are conserved in a flat spacetime.
Using (\ref{fmam35}), this conservation gives
\begin{equation}
M_{\alpha 0}=\frac{1}{c}\biggl(\int x_\alpha\theta_0^{\phantom{0}0}dV-x_0\int\theta_\alpha^{\phantom{\alpha}0}dV\biggr)+S_{\alpha 0}=\frac{1}{c}\int x_\alpha\theta_0^{\phantom{0}0}dV-ctP_\alpha+S_{\alpha 0}=\mbox{const},
\end{equation}
divided by the conserved $P_0$ gives
\begin{equation}
X_\alpha=V_\alpha t-\frac{S_{\alpha 0}}{P_0}+\mbox{const},
\label{fmam48}
\end{equation}
where
\begin{equation}
V_\alpha=\frac{cP_\alpha}{P_0}
\label{fmam49}
\end{equation}
and
\begin{equation}
X_\alpha=\frac{\int x_\alpha\theta_0^{\phantom{0}0}dV}{\int\theta_0^{\phantom{0}0}dV}.
\label{fmam50}
\end{equation}
If the intrinsic angular momentum is constant, then the relation (\ref{fmam48}) describes a uniform motion of the {\em center of inertia}, whose coordinates are $X^\alpha$, with velocity $V^\alpha$.
The coordinates of the center of inertia (\ref{fmam50}) are not the spatial components of a four-dimensional vector.

If spacetime is not flat, one can construct an energy--momentum pseudotensor for the gravitational field such that the total four-momentum and angular momentum of the gravitational field and matter are conserved, as shown in Section \ref{Gravfieldeqs}.
The center of inertia corresponding to the gravitational field and matter moves uniformly.

\subsubsection{Meaning of energy, momentum, and angular momentum}
\label{Meaning}
\noindent
{\bf Homogeneity of time}.\\
The conservation law (\ref{clLg5}) for a time translation $\epsilon^0$ gives
\begin{equation}
\epsilon^0\theta^{\phantom{0}i}_{0\phantom{i},i}=0,\quad \theta^{\phantom{0}i}_{0\phantom{i},i}=0,
\end{equation}
which is (\ref{fmam51}).
The conservation of energy therefore follows from the {\em homogeneity of time}: all instants in time are physically equivalent for a closed system.\\

\noindent
{\bf Homogeneity of space}.\\
The conservation law (\ref{clLg5}) for a space translation $\epsilon^\alpha$ gives
\begin{equation}
\epsilon^\alpha\theta^{\phantom{\alpha}i}_{\alpha\phantom{i},i}=0,\quad \theta^{\phantom{\alpha}i}_{\alpha\phantom{i},i}=0,
\end{equation}
which is (\ref{fmam52}).
The conservation of momentum therefore follows from the {\em homogeneity of space}: all points in space are physically equivalent for a closed system.\\

\noindent
{\bf Isotropy of space}.\\
The conservation law (\ref{clLg8}) for a space rotation $\epsilon^{\alpha\beta}$ gives
\begin{equation}
\epsilon^{\alpha\beta}\mathfrak{M}_{\alpha\beta\phantom{i},i}^{\phantom{\alpha\beta}i}=0,\quad \mathfrak{M}_{\alpha\beta\phantom{i},i}^{\phantom{\alpha\beta}i}=0,
\end{equation}
which is a part of (\ref{clLg9}).
The conservation of angular momentum therefore follows from the {\em isotropy of space}: all directions in space are physically equivalent for a closed system.

\subsubsection{Mass}
We define the {\em mass} of matter on a hypersurface as
\begin{equation}
m=\frac{1}{c^2}\int\mathfrak{T}^{\phantom{i}k}_i u^i dS_k=\frac{1}{c^2}\int\Theta^{\phantom{i}k}_i u^i dS_k,
\label{mass1}
\end{equation}
where $u^i$ is the four-velocity of the matter and $dS_k$ is the element of hypersurface.
Either the tetrad $\mathfrak{T}^{\phantom{i}k}_i$ or canonical energy--momentum density $\Theta^{\phantom{i}k}_i$ can be used because they are equal according to (\ref{cemd4}).
Because $\mathfrak{T}^{\phantom{i}k}_i$ and $\Theta^{\phantom{i}k}_i$ are tensor densities, the mass is a scalar.

In a flat spacetime, the mass (\ref{mass1}) is equal to
\begin{equation}
m=\frac{1}{c^2}\int\theta^{ik}u_i dS_k.
\label{mass2}
\end{equation}
Replacing the tensor $\theta^{ik}$ with the symmetrized tensor $\tau^{ik}$ according to (\ref{clLg16}) and using the Gau\ss--Stokes theorem (\ref{covint6}) gives
\begin{eqnarray}
& & \tilde{m}=\frac{1}{c^2}\int\tau^{ik}u_i dS_k=m+\frac{1}{c^2}\int\psi^{ikj}_{\phantom{ikj},j}u_i dS_k=m+\frac{1}{c^2}\int(\psi^{ikj}u_i)_{,j}dS_k-\frac{1}{c^2}\int\psi^{ikj}u_{i,j}dS_k \nonumber \\
& & =m+\frac{1}{2c^2}\int\Bigl((\psi^{ikj}u_i)_{,j}dS_k-(\psi^{ikj}u_i)_{,k}dS_j\Bigr)-\frac{1}{c^2}\int\psi^{ikj}u_{i,j}dS_k \nonumber \\
& & =m+\frac{1}{2c^2}\oint\psi^{ikj}u_i df^\ast_{kj}-\frac{1}{c^2}\int\psi^{ikj}u_{i,j}dS_k.
\end{eqnarray}
If the integration surface is located in a region, where matter is absent, then the surface integral vanishes.
The last term, however, shows that if the four-velocity has a four-gradient then $\tilde{m}$ is different from the mass (\ref{mass2}).

\subsection{Particle limit of field}
\setcounter{equation}{0}
\subsubsection{Spin four-tensor of particle}
\label{SpinParticles}
{\bf Particle}.\\
Let us consider a field which is distributed over a small region in space and consists of points with the coordinates $x^i$, forming an extended body whose motion is represented by a world tube in spacetime.
The motion of the body as a whole is represented by an arbitrary timelike world line $\gamma$ inside the world tube, which consists of points with the coordinates $X^i(\tau)$, where $\tau$ is the proper time on $\gamma$.
We define
\begin{eqnarray}
& & \delta x^i=x^i-X^i, \label{muex1} \\
& & u^i=\frac{dX^i}{ds},
\label{muex3}
\end{eqnarray}
where $ds^2=g_{ij}dX^i dX^j$.
We also define the following integrals:
\begin{eqnarray}
& & M^{ik}=u^0\int\mathfrak{T}^{ik}dV, \label{muex4} \\
& & N^{ijk}=u^0\int\mathfrak{S}^{ijk}dV,
\label{muex6}
\end{eqnarray}
where $dV$ is the element of volume and the integration is taken over the volume of the body.

A {\em particle} is a field that is not spatially extended.
In this case, the quantity (\ref{muex1}) satisfies
\begin{equation}
\delta x^i=0,
\label{muex2}
\end{equation}
and the coordinates $X^i$ represent the spacetime location of the particle.
Since the integration domain in (\ref{muex4}) and (\ref{muex6}) is not spatially extended, these quantities are tensors, and can be represented as covariant hypersurface integrals:
\begin{eqnarray}
& & M^{ik}=u^l\int\mathfrak{T}^{ik}dS_l, \\
& & N^{ijk}=u^l\int\mathfrak{S}^{ijk}dS_l.
\end{eqnarray}

\noindent
{\bf First multipole of spin density}.\\
We can use the conservation laws to determine the spin tensor and the energy--momentum tensor for a particle.
The conservation law for the spin density (\ref{cam7}) is equivalent to
\begin{equation}
\mathfrak{S}^{ijk}_{\phantom{ijk},k}-\Gamma^{i}_{lk}\mathfrak{S}^{jlk}+\Gamma^{j}_{lk}\mathfrak{S}^{ilk}-2\mathfrak{T}^{[ij]}=0.
\label{muex8}
\end{equation}
Integrating (\ref{muex8}) over the volume of the body at a constant time $X^0$ and using Gau\ss' theorem to eliminate surface integrals gives
\begin{equation}
\int\mathfrak{S}^{ij0}_{\phantom{ij0},0}dV-\int\Gamma^{i}_{lk}\mathfrak{S}^{jlk}dV+\int\Gamma^{j}_{lk}\mathfrak{S}^{ilk}dV-2\int\mathfrak{T}^{[ij]}dV=0.
\label{muex9}
\end{equation}
For a particle, the affine connection $\Gamma^{i}_{jk}$ in the integrands in (\ref{muex9}) is equal to its value at the point $X^i$, where the particle is located.
Consequently, we obtain
\begin{equation}
\int\mathfrak{S}^{ij0}_{\phantom{ij0},0}dV-\Gamma^{i}_{lk}\int\mathfrak{S}^{jlk}dV+\Gamma^{j}_{lk}\int\mathfrak{S}^{ilk}dV-2\int\mathfrak{T}^{[ij]}dV=0.
\label{muex99}
\end{equation}
The first term in this equation can be written as
\begin{equation}
\int\mathfrak{S}^{ij0}_{\phantom{ij0},0}dV=\biggl(\int\mathfrak{S}^{ij0}dV\biggr)_{,0}=\frac{1}{u^0}\frac{d}{ds}\int\mathfrak{S}^{ij0}dV.
\end{equation}
Using the integrals (\ref{muex4}) and (\ref{muex6}) turns (\ref{muex99}) into an equation of motion:
\begin{equation}
\frac{d}{ds}\biggl(\frac{N^{ij0}}{u^0}\biggr)-\Gamma^{i}_{lk}N^{jlk}+\Gamma^{j}_{lk}N^{ilk}-2M^{[ij]}=0.
\label{muex13}
\end{equation}

\noindent
{\bf Second multipole of spin density}.\\
The conservation law (\ref{muex8}) gives
\begin{equation}
(x^l\mathfrak{S}^{ijk})_{,k}=\mathfrak{S}^{ijl}+x^l\Gamma^{i}_{lk}\mathfrak{S}^{jlk}-x^l\Gamma^{j}_{lk}\mathfrak{S}^{ilk}+2x^l\mathfrak{T}^{[ij]}.
\label{muex14}
\end{equation}
Integrating (\ref{muex14}) over the volume of the body and using Gau\ss' theorem to eliminate surface integrals gives
\begin{equation}
\int(x^l\mathfrak{S}^{ij0})_{,0}dV=\int\mathfrak{S}^{ijl}dV+\int x^l\Gamma^{i}_{mk}\mathfrak{S}^{jmk}dV-\int x^l\Gamma^{j}_{mk}\mathfrak{S}^{imk}dV+2\int x^l\mathfrak{T}^{[ij]}dV.
\label{muex15}
\end{equation}
In this relation, we use $x^i=X^i$, which follows from (\ref{muex1}) and (\ref{muex2}), and $X^l_{\phantom{l},0}=u^l/u^0$.
Substituting (\ref{muex3}) into (\ref{muex15}) gives
\begin{eqnarray}
& & \frac{u^l}{u^0}\int\mathfrak{S}^{ij0}dV+X^l\int\mathfrak{S}^{ij0}_{\phantom{ij0},0}dV=\int\mathfrak{S}^{ijl}dV+X^l\Bigl(\int\Gamma^{i}_{mk}\mathfrak{S}^{jmk}dV-\int\Gamma^{j}_{mk}\mathfrak{S}^{imk}dV \nonumber \\
& & +2\int\mathfrak{T}^{[ij]}dV\Bigr).
\label{muex156}
\end{eqnarray}
This relation reduces, by means of (\ref{muex9}), to
\begin{equation}
\frac{u^l}{u^0}\int\mathfrak{S}^{ij0}dV=\int\mathfrak{S}^{ijl}dV.
\label{muex88}
\end{equation}

Using the definition (\ref{muex6}) brings (\ref{muex88}) to
\begin{equation}
\frac{u^l}{u^0}N^{ij0}=N^{ijl}.
\label{muex18}
\end{equation}
Putting $l=0$ in (\ref{muex18}) gives the identity.
Contracting (\ref{muex18}) with $u_l$ gives
\begin{eqnarray}
& & \frac{N^{ij0}}{u^0}=N^{ijl}u_l=cS^{ij},
\label{muex180} \\
& & N^{ijl}=cS^{ij}u^l,
\label{muex181}
\end{eqnarray}
where $S^{ij}$ is the intrinsic angular momentum (spin) four-tensor (\ref{fmam36}) of the particle in curved spacetime.
Because $N^{ijl}$ is a tensor, $S^{ij}$ is a tensor.
This tensor has a nonzero value at the location of the particle.
For a system of particles, the spin four-tensor $S^{ij}$ has nonzero values at the locations of the particles.
The quantities analogous to (\ref{muex14}) with higher multiples of $x^i$ do not introduce new relations.

If the spin density is completely antisymmetric, then (\ref{muex18}) gives $N^{il0}=-(u^l/u^0)N^{i00}$ and thus
\begin{equation}
N^{ijk}=0.
\label{Pap73}
\end{equation}
Therefore, such a field cannot be represented as a particle or a system of particles.

\subsubsection{Momentum four-vector of particle}
\label{MomentumParticles}
{\bf First multipole of energy--momentum density}.\\
The conservation law for the tetrad energy--momentum density (\ref{ctem8}) is equivalent to
\begin{equation}
\mathfrak{T}^{ji}_{\phantom{ji},i}+\mathring{\Gamma}^{j}_{ik}\mathfrak{T}^{ik}-C_{ik}^{\phantom{ik}j}\mathfrak{T}^{ik}-\frac{1}{2}R_{ikl}^{\phantom{ikl}j}\mathfrak{S}^{ikl}=0.
\label{muex19}
\end{equation}
Integrating (\ref{muex19}) over the volume of the body at a constant time $X^0$ and using Gau\ss' theorem to eliminate surface integrals gives
\begin{equation}
\int\mathfrak{T}^{j0}_{\phantom{j0},0}dV+\int\mathring{\Gamma}^{j}_{ik}\mathfrak{T}^{ik}dV-\int C_{ik}^{\phantom{ik}j}\mathfrak{T}^{ik}dV-\frac{1}{2}\int R_{ikl}^{\phantom{ikl}j}\mathfrak{S}^{ikl}dV=0,
\label{muex20}
\end{equation}
where $dV$ is the element of volume.
For a particle, the affine connection $\Gamma^{i}_{jk}$, and thus $\mathring{\Gamma}^{i}_{jk}$, $C^i_{\phantom{i}jk}$, and the curvature tensor in the integrands in (\ref{muex20}), are equal to their respective values at the point $X^i$, where the particle is located.
Consequently, we obtain
\begin{equation}
\int\mathfrak{T}^{j0}_{\phantom{j0},0}dV+\mathring{\Gamma}^{j}_{ik}\int\mathfrak{T}^{ik}dV-C_{ik}^{\phantom{ik}j}\int\mathfrak{T}^{ik}dV-\frac{1}{2}R_{ikl}^{\phantom{ikl}j}\int\mathfrak{S}^{ikl}dV=0.
\label{muex21}
\end{equation}
The first term in this equation can be written as
\begin{equation}
\int\mathfrak{T}^{j0}_{\phantom{j0},0}dV=\biggl(\int\mathfrak{T}^{j0}dV\biggr)_{,0}=\frac{1}{u^0}\frac{d}{ds}\int\mathfrak{T}^{j0}dV.
\end{equation}
Using the integrals (\ref{muex4}) and (\ref{muex6}) turns (\ref{muex21}) into an equation of motion:
\begin{equation}
\frac{d}{ds}\biggl(\frac{M^{j0}}{u^0}\biggr)+\mathring{\Gamma}^{j}_{ik}M^{(ik)}-C_{ik}^{\phantom{ik}j}M^{[ik]}-\frac{1}{2}R_{ikl}^{\phantom{ikl}j}N^{ikl}=0.
\label{muex220}
\end{equation}

\noindent
{\bf Second multipole of energy--momentum density}.\\
The conservation law (\ref{muex19}) gives
\begin{eqnarray}
& & (x^l\mathfrak{T}^{ji})_{,i}=\mathfrak{T}^{jl}-x^l\mathring{\Gamma}^{j}_{ik}\mathfrak{T}^{ik}+x^l C_{ik}^{\phantom{ik}j}\mathfrak{T}^{ik}+\frac{1}{2}x^l R_{ikm}^{\phantom{ikm}j}\mathfrak{S}^{ikm}, \label{muex23} \\
& & (x^l x^m\mathfrak{T}^{ji})_{,i}=x^m\mathfrak{T}^{jl}+x^l\mathfrak{T}^{jm}-x^l x^m\mathring{\Gamma}^{j}_{ik}\mathfrak{T}^{ik}+x^l x^m C_{ik}^{\phantom{ik}j}\mathfrak{T}^{ik} \nonumber \\
& & +\frac{1}{2}x^l x^m R_{ikn}^{\phantom{ikn}j}\mathfrak{S}^{ikn}.
\label{muex24}
\end{eqnarray}
Integrating (\ref{muex23}) over the volume of the body and using Gau\ss' theorem to eliminate surface integrals gives
\begin{equation}
\int(x^l\mathfrak{T}^{j0})_{,0}dV=\int\mathfrak{T}^{jl}dV-\int x^l\mathring{\Gamma}^{j}_{ik}\mathfrak{T}^{ik}dV+\int x^l C_{ik}^{\phantom{ik}j}\mathfrak{T}^{ik}dV+\frac{1}{2}\int x^l R_{ikm}^{\phantom{ikm}j}\mathfrak{S}^{ikm}dV.
\label{muex25}
\end{equation}
In this relation, we use $x^i=X^i$ and $X^l_{\phantom{l},0}=u^l/u^0$.
Substituting (\ref{muex3}) into (\ref{muex25}) gives
\begin{eqnarray}
& & \frac{u^l}{u^0}\int\mathfrak{T}^{j0}dV+X^l\int\mathfrak{T}^{j0}_{\phantom{j0},0}dV=\int\mathfrak{T}^{jl}dV-X^l\int\mathring{\Gamma}^{j}_{ik}\mathfrak{T}^{ik}dV+X^l\int C_{ik}^{\phantom{ik}j}\mathfrak{T}^{ik}dV \nonumber \\
& & +\frac{1}{2}X^l\int R_{ikm}^{\phantom{ikm}j}\mathfrak{S}^{ikm}dV.
\end{eqnarray}
This equation reduces, by means of (\ref{muex20}), to
\begin{equation}
\frac{u^l}{u^0}\int\mathfrak{T}^{j0}dV=\int\mathfrak{T}^{jl}dV.
\label{muex27}
\end{equation}

Using the definition (\ref{muex4}) brings (\ref{muex27}) to
\begin{equation}
\frac{u^l}{u^0}M^{j0}=M^{jl}.
\label{muex280}
\end{equation}
Putting $l=0$ in (\ref{muex280}) gives the identity.
Contracting (\ref{muex280}) with $u_l$ gives
\begin{eqnarray}
& & \frac{M^{j0}}{u^0}=M^{jl}u_l=cP^{j},
\label{muex182} \\
& & M^{jl}=cP^{j}u^l,
\label{muex183}
\end{eqnarray}
where $P^{j}$ is the momentum four-vector (\ref{fmam2}) of the particle in curved spacetime.
Because $M^{jl}$ is a tensor, $P^{j}$ is a vector.
This vector has a nonzero value at the location of the particle.
For a system of particles, the momentum four-vector $P^{j}$ has nonzero values at the locations of the particles.
Integrating (\ref{muex24}) over the volume of the body does not introduce new relations.
The quantities analogous to (\ref{muex23}) and (\ref{muex24}) with higher multiples of $x^i$ do not introduce new relations as well.

\subsubsection{Mathisson--Papapetrou equations of motion}
\noindent
{\bf Change of four-spin}.\\
Substituting (\ref{muex180}), (\ref{muex181}), and (\ref{muex183}) into the equation of motion (\ref{muex13}) gives
\begin{equation}
\frac{dS^{ij}}{ds}-\Gamma^{i}_{lk}S^{jl}u^k+\Gamma^{j}_{lk}S^{il}u^k-P^i u^j+P^j u^i=0.
\label{MP1}
\end{equation}
Using
\begin{equation}
\frac{dS^{ij}}{ds}=S^{ij}_{\phantom{ij},k}u^k=S^{ij}_{\phantom{ij};k}u^k-\Gamma^{i}_{lk}S^{lj}u^k-\Gamma^{j}_{lk}S^{il}u^k=\frac{DS^{ij}}{ds}+\Gamma^{i}_{lk}S^{jl}u^k-\Gamma^{j}_{lk}S^{il}u^k
\end{equation}
turns (\ref{MP1}) into a covariant equation:
\begin{equation}
\frac{DS^{ij}}{ds}-P^i u^j+P^j u^i=0.
\label{MP3}
\end{equation}

\noindent
{\bf Change of four-momentum}.\\
Substituting (\ref{muex181}), (\ref{muex182}), and (\ref{muex183}) into the equation of motion (\ref{muex220}) gives
\begin{equation}
\frac{dP^j}{ds}+\mathring{\Gamma}^{j}_{ik}P^i u^k-C_{ik}^{\phantom{ik}j}P^i u^k-\frac{1}{2}R_{ikl}^{\phantom{ikl}j}S^{ik}u^l=0.
\label{MP4}
\end{equation}
Using the relations (\ref{Chrsym6}) and (\ref{Chrsym7}) gives
\begin{equation}
\mathring{\Gamma}^{j}_{ik}-C_{ik}^{\phantom{ik}j}=\Gamma^{j}_{ik}-C^j_{\phantom{j}ik}-C_{ik}^{\phantom{ik}j}=\Gamma^{j}_{ik}-2S_{ik}^{\phantom{ik}j}.
\end{equation}
Using this relation and
\begin{equation}
\frac{dP^j}{ds}=P^j_{\phantom{j},k}u^k=P^j_{\phantom{j};k}u^k-\Gamma^{j}_{ik}P^i u^k=\frac{DP^j}{ds}-\Gamma^{j}_{ik}P^i u^k
\end{equation}
turns (\ref{MP4}) into a covariant equation:
\begin{equation}
\frac{DP^j}{ds}-2S_{ik}^{\phantom{ik}j}P^i u^k-\frac{1}{2}R_{ikl}^{\phantom{ikl}j}S^{ik}u^l=0.
\label{MP7}
\end{equation}
The relations (\ref{MP3}) and (\ref{MP7}) are the {\em Mathisson--Papapetrou equations}, which determine the motion of a particle.\\

\noindent
{\bf Motion of particle in flat spacetime}.\\
In the absence of the external gravitational field and neglecting the gravitational field of a particle, the torsion and curvature tensors vanish.
The spacetime is therefore flat and the coordinates can be taken such that the affine connection vanishes.
The relation (\ref{MP7}) reduces to
\begin{equation}
\frac{dP^j}{ds}=0,
\label{MP8}
\end{equation}
whose integration gives the conservation of the momentum four-vector of the particle along a world line: $P^j=\mbox{const}$, in agreement with (\ref{fmam5}).
The relation (\ref{MP3}) in a flat spacetime can be written, using (\ref{muex3}), as
\begin{equation}
\frac{dX^i}{ds}P^j-\frac{dX^j}{ds}P^i+\frac{dS^{ij}}{ds}=0.
\end{equation}
Its integration, using the constancy of $P^j$ and (\ref{fmam45}), gives the conservation of the angular momentum four-tensor of the particle along a world line: 
\begin{equation}
X^i P^j-X^j P^i+S^{ij}=L^{ij}+S^{ij}=M^{ij}=\mbox{const},
\end{equation}
in agreement with (\ref{fmam37}).
The tensor $L^{ij}=X^i P^j-X^j P^i$ describes the orbital angular momentum associated with the actual motion of the particle, whereas the tensor $S^{ij}$ describes the intrinsic angular momentum of the particle not associated with its motion.

\subsubsection{Mass of particle}
\noindent
{\bf Mass as contraction of four-velocity and four-momentum}.\\
For a particle, the four-velocity $u^i$ in the integrand of the mass (\ref{mass1}) is equal to its value at the point, where the particle is located.
Consequently, we obtain
\begin{equation}
m=\frac{1}{c^2}u^i\int\mathfrak{T}^{\phantom{i}k}_i dS_k=\frac{1}{c}u^i P_i,
\label{mapar1}
\end{equation}
where $P_i$ is the four-momentum (\ref{fmam1}) of the particle.\\

\noindent
{\bf Relation between four-momentum and four-velocity}.\\
The equations of motion (\ref{MP3}) and (\ref{MP7}) are 10 equations for 13 components: $u^i$, $P^i$, and $S^{ik}$, which are functions of $s$.
Contracting (\ref{MP3}) with $u_j$ and using (\ref{mapar1}) gives
\begin{equation}
P^i=mcu^i+\frac{DS^{ij}}{ds}u_j.
\label{mapar2}
\end{equation}
Contracting (\ref{MP3}) with itself gives
\begin{equation}
P^i P_i=m^2 c^2+\frac{1}{2}\frac{DS^{ij}}{ds}\frac{DS_{ij}}{ds}.
\label{mapar3}
\end{equation}
The relation (\ref{mapar2}) gives 10 independent components: $m$, $u^i$, and $S^{ik}$, which are functions of $s$ and can be determined from the equations of motion.\\

\noindent
{\bf Change of mass}.\\
Contracting the equation (\ref{MP7}) with $u_j$ gives
\begin{equation}
\frac{DP^j}{ds}u_j=0
\label{mapar4}
\end{equation}
because the symmetric indices in the four-velocities multiply the antisymmetric indices in the torsion and curvature tensors.
Using (\ref{mapar1}) and (\ref{mapar2}), this relation leads to
\begin{equation}
\frac{dm}{ds}=\frac{1}{c}P^j\frac{Du_j}{ds}=\frac{1}{c}\frac{DS^{jk}}{ds}u_k\frac{Du_j}{ds}.
\label{mapar5}
\end{equation}
If the spin four-tensor is orthogonal to the four-velocity:
\begin{equation}
S^{jk}u_k=0,
\label{mapar17}
\end{equation}
or it vanishes, then the mass is constant:
\begin{equation}
\frac{dm}{ds}=-\frac{1}{c}S^{jk}\frac{Du_k}{ds}\frac{Du_j}{ds}=0.
\label{mapar18}
\end{equation}
Consequently, the four-momentum (\ref{mapar2}) is
\begin{equation}
P^i=mcu^i-S^{ij}\frac{Du_j}{ds}=mcu^i-S^{ij}w_j,
\end{equation}
where $w_j$ is the four-acceleration (\ref{foac4}).\\

\noindent
{\bf Fermi--Walker transport for spin}.\\
Substituting (\ref{mapar2}) into (\ref{MP3}) gives
\begin{equation}
\frac{DS^{ij}}{ds}=\frac{DS^{ik}}{ds}u_k u^j-\frac{DS^{jk}}{ds}u_k u^i.
\label{mapar6}
\end{equation}
Analogously to the Pauli--Luba\'{n}ski pseudovector (\ref{Cas5}), we define a pseudovector:
\begin{equation}
S^i=\frac{1}{2}E^{ijkl}u_j S_{kl},
\label{mapar7}
\end{equation}
where $E^{ijkl}$ is the completely antisymmetric unit pseudotensor (\ref{metten15}).
This pseudovector is orthogonal to the four-velocity:
\begin{equation}
S^i u_i=0.
\label{mapar8}
\end{equation}
If the spin four-tensor satisfies (\ref{mapar17}), then (\ref{KLC8}) gives
\begin{equation}
S_{kl}=E_{klmn}u^m S^n,
\label{mapar7}
\end{equation}
where $E_{klmn}$ is the completely antisymmetric unit pseudotensor (\ref{metten14}).
Using (\ref{mapar6}) and (\ref{mapar7}) gives
\begin{equation}
\frac{DS^i}{ds}=\frac{Du^i}{ds}u_j S^j-u^i\frac{Du_j}{ds}S^j,
\label{mapar9}
\end{equation}
thereby the covariant (with respect to the Levi-Civita connection) change of this pseudovector along a world line is equal to the corresponding Fermi--Walker transport (\ref{FW2}).
Multiplying (\ref{mapar9}) by $S_i$ and using (\ref{mapar8}) gives
\begin{equation}
\frac{d(S^i S_i)}{ds}=0,
\end{equation}
thereby the change of the pseudovector $S^i$ along the world line is a four-rotation with a constant value of the magnitude (precession).\\

\subsubsection{Spinless particles}
\noindent
{\bf Four-momentum of spinless particle}.\\
If the spin density vanishes, $\mathfrak{S}^{ijk}=0$, the matter is {\em spinless}.
The conservation law for the spin density (\ref{cam7}) gives in this case the symmetry of the energy--momentum density, $\mathfrak{T}^{ik}=\mathfrak{T}^{ki}$.
The tensor density $\mathfrak{T}^{ij}$ is also equal to the metric energy--momentum density ${\cal T}^{ij}$ according to (\ref{BelRos9}).
For a spinless particle, the quantity (\ref{muex4}) is then symmetric: $M^{ik}=M^{ki}$, and the quantity (\ref{muex6}) vanishes: $N^{ijk}=0$.
Because of (\ref{muex180}), the spin four-tensor therefore vanishes: $S^{ij}=0$.
Consequently, the four-momentum (\ref{mapar2}) of a spinless particle is proportional to its four-velocity:
\begin{equation}
p^i=mcu^i.
\label{sple1}
\end{equation}
The quantity (\ref{muex4}), following (\ref{muex183}), is therefore
\begin{equation}
M^{ik}=mc^2 u^i u^k.
\label{sple2}
\end{equation}
This quantity for a spinless particle is proportional to the product of the components of the four-velocity.
Accordingly, the energy--momentum density and therefore the energy--momentum tensor for a spinless particle are proportional to this product.
The mass of the particle is constant, $m=\mbox{const}$, according to (\ref{mapar5}).
The relation (\ref{mapar3}) reduces to
\begin{equation}
p^i p_i=m^2 c^2,
\label{sple3}
\end{equation}
which also follows from (\ref{sple1}).
It is analogous to the mass operator (\ref{Cas15}).
If $p^i p_i>0$, then $m>0$.
If $p^i p_i=0$, then $m=0$.

If the spin density vanishes, the torsion tensor also vanishes according to the Cartan equations (\ref{EC16}) and the affine connection is given by the Christoffel symbols.
Consequently, the equation of motion (\ref{MP7}) reduces to
\begin{equation}
\frac{Dp^j}{ds}=mc\frac{Du^j}{ds}=0,
\end{equation}
which is equivalent to the metric geodesic equation (\ref{metgeo3}).
A spinless particle moves in a gravitational field along a metric geodesic, regardless of its mass.
This phenomenon is referred to as the {\em universality of free fall} or {\em weak equivalence principle}.
In a flat spacetime, (\ref{MP8}) gives $du^j/ds=0$, which is equivalent to $u^i=\mbox{const}$.
Accordingly, the coordinates $x^i$ of the particle are linear functions of the proper time $\tau$.\\

\noindent
{\bf Energy--momentum tensor for spinless particle}.\\
The relation (\ref{sple2}) gives
\begin{equation}
\int\mathfrak{T}^{ik}dV=mc^2\frac{u^i u^k}{u^0}
\end{equation}
or
\begin{equation}
\mathfrak{T}^{ik}({\bf x})=mc^2{\bm \delta}({\bf x}-{\bf x}_0)\frac{u^i u^k}{u^0},
\label{emtp5}
\end{equation}
where ${\bm \delta}({\bf x}-{\bf x}_0)$ is the spatial Dirac delta representing a point mass located at ${\bf x}_0$.
We define the {\em mass density} $\mu$ such that
\begin{equation}
\mu\sqrt{\mathfrak{s}}dV=dm,
\label{emtp6}
\end{equation}
where $\mathfrak{s}$ is given by (\ref{spvec9}).
The mass density for a particle located at ${\bf x}_a$ is
\begin{equation}
\mu({\bf x})=\frac{m}{\sqrt{\mathfrak{s}}}{\bm \delta}({\bf x}-{\bf x}_a),
\label{emtp7}
\end{equation}
so (\ref{emtp5}) turns into
\begin{equation}
\mathfrak{T}^{ik}=\mu c^2\sqrt{\mathfrak{s}}\frac{u^i u^k}{u^0}.
\label{emtp8}
\end{equation}
Therefore, the metric energy--momentum tensor for a spinless particle is given by
\begin{eqnarray}
& & T^{ik}(x)=\mu({\bf x})c^2\frac{u^i u^k}{\sqrt{g_{00}}u^0}=\frac{\mu({\bf x})c}{\sqrt{g_{00}}}\frac{dx^i}{ds}\frac{dx^k}{dt}=mc^2{\bm \delta}({\bf x}-{\bf x}_a)\frac{u^i u^k}{\sqrt{-\mathfrak{g}}u^0} \nonumber \\
& & =mc^2\int\frac{u^i u^k}{\sqrt{-\mathfrak{g}}}\delta(x-x_a(\tau))d\tau,
\label{emtp9}
\end{eqnarray}
where $x_a(\tau)$ is the particle's world line as a function of its proper time $\tau$.
For a system of particles, this tensor is equal to
\begin{equation}
T^{ik}({\bf x})=\sum_a m_a c^2{\bm \delta}({\bf x}-{\bf x}_a)\frac{u^i u^k}{\sqrt{-\mathfrak{g}}u^0}.
\label{emtp10}
\end{equation}

\noindent
{\bf Energy and momentum of spinless particle}.\\
In a locally inertial frame of reference, (\ref{foac3}) and (\ref{sple1}) give
\begin{equation}
p^i=(mc\gamma,m\gamma{\bf v})=\biggl(\frac{E}{c},{\bf p}\biggr).
\label{emtp22}
\end{equation}
The energy and momentum of a particle are therefore
\begin{eqnarray}
& & E=mc^2\gamma=\frac{mc^2}{\sqrt{1-v^2/c^2}}, \label{emtp23} \\
& & {\bf p}=m{\bf v}\gamma=\frac{m{\bf v}}{\sqrt{1-v^2/c^2}}.
\label{emtp24}
\end{eqnarray}
Accordingly, (\ref{sple3}) gives
\begin{equation}
E^2=({\bf p}c)^2+(mc^2)^2.
\label{emtp25}
\end{equation}
The formulae (\ref{emtp23}) and (\ref{emtp24}) give
\begin{equation}
{\bf v}=\frac{{\bf p}c^2}{E}.
\label{emtp27}
\end{equation}
Taking the differential of (\ref{emtp25}) gives $EdE=c^2{\bf p}\cdot d{\bf p}$.
This relation, using (\ref{emtp27}), gives
\begin{eqnarray}
dE={\bf v}\cdot d{\bf p}.
\label{emtp28}
\end{eqnarray}
If a particle is massless, $m=0$, then (\ref{emtp25}) and (\ref{emtp27}) give
\begin{equation}
E=pc,\quad v=c.
\end{equation}

\noindent
{\bf Einstein formula for mass--energy equivalence}.\\
In the rest frame of the particle, ${\bf v}=0$, the relation (\ref{emtp25}) reduces to {\em Einstein's formula} for the {\em rest energy}:
\begin{equation}
E=mc^2.
\label{emtp26}
\end{equation}
This relation shows the mass--energy equivalence.\\

\noindent
{\bf Virial theorem}.\\
In the absence of spin and therefore torsion, the conservation law for the metric energy--momentum tensor in the locally Galilean frame of reference follows from the conservation law (\ref{clLg2}) for the canonical energy--momentum density:
\begin{equation}
T^{\phantom{\alpha}i}_{\alpha\phantom{i},i}=0.
\label{emtp12}
\end{equation}
It also follows from the conservation law (\ref{cmem15}).
Let us consider a closed system of particles which carry out a finite motion, in which all quantities vary over finite ranges.
We define the average over a certain time interval $\tau$ of a function $f$ of these quantities as $\overline{f}=(1/\tau)\int_0^\tau fdt$.
The average of the time derivative of a bounded quantity $\overline{\dot{f}}=(1/\tau)\bigl(f(\tau)-f(0)\bigr)\rightarrow0$ as $\tau\rightarrow\infty$.
Therefore, averaging (\ref{emtp12}) over the time gives
\begin{equation}
\overline{T}^{\phantom{\alpha}\beta}_{\alpha\phantom{\beta},\beta}=0.
\label{emtp13}
\end{equation}
Multiplying (\ref{emtp13}) by $x^\alpha$ and integrating over the volume gives, omitting surface integrals,
\begin{equation}
\int x^\alpha\overline{T}^{\phantom{\alpha}\beta}_{\alpha\phantom{\beta},\beta}dV=-\int\overline{T}^{\phantom{\alpha}\alpha}_\alpha dV=0.
\end{equation}
The average energy of the system (\ref{fmam9}) is thus
\begin{equation}
\overline{E}=\int\overline{T}^{\phantom{0}0}_0 dV=\int\overline{T}^{\phantom{i}i}_i dV.
\label{emtp15}
\end{equation}
Contracting the indices in (\ref{emtp9}) and (\ref{emtp10}) and substituting there (\ref{foac3}) gives
\begin{equation}
T^{\phantom{i}i}_i({\bf x})=\mu({\bf x})c^2\frac{1}{u^0}=\sum_a m_a c^2{\bm \delta}({\bf x}-{\bf x}_a)\frac{1}{u^0}=\sum_a m_a c^2{\bm \delta}({\bf x}-{\bf x}_a)\Bigl(1-\frac{v^2_a}{c^2}\Bigr)^{1/2},
\label{emtp16}
\end{equation}
so $T^{\phantom{i}i}_i\geq0$.
Putting (\ref{emtp16}) into (\ref{emtp15}) gives
\begin{equation}
\overline{E}=\sum_a m_a c^2\overline{\Bigl(1-\frac{v^2_a}{c^2}\Bigr)^{1/2}},
\label{emtp17}
\end{equation}
which is referred to as the {\em virial theorem}.

\subsubsection{Fluids and pressure}
\noindent
{\bf Spin tensor for fluid}.\\
A system of particles can be represented as a {\em fluid}, which is characterized by particular forms of the spin and energy--momentum tensors.
For an isotropic (without a preferred direction in its rest frame) fluid, the spin tensor is analogous to that for a particle (\ref{muex181}):
\begin{equation}
s_{ijl}=s_{ij}u_l,
\label{stp8}
\end{equation}
where
\begin{equation}
s_{ij}=s_{ijl}u^l=-s_{ji}.
\label{stp9}
\end{equation}
If this tensor is orthogonal to $u^j$, similarly to (\ref{mapar17}):
\begin{equation}
s_{ij}u^j=0,
\label{stp10}
\end{equation}
then it has 3 independent components.
All contractions of the spin tensor (\ref{stp8}) with the metric tensor vanish.
In a locally Galilean, rest frame of reference, the relation (\ref{stp10}) becomes
\begin{equation}
s_{0\alpha}=0.
\end{equation}
In this frame, the 3 components of $s_{ij}$ are spatial, $s_{\alpha\beta}$, and are equivalent to 3 components of a spatial pseudovector:
\begin{equation}
s^\alpha=\frac{1}{2}e^{\alpha\beta\gamma}s_{\beta\gamma}.
\label{stp14}
\end{equation}

\noindent
{\bf Tetrad energy--momentum tensor for fluid}.\\
For an isotropic fluid, the tetrad energy--momentum tensor is analogous to that for a particle (\ref{muex183}):
\begin{equation}
t_{ik}=c\pi_i u_k,
\label{cemt430}
\end{equation}
where
\begin{equation}
\pi_i=\frac{1}{c}t_{ik}u^k
\label{cemt44}
\end{equation}
is the proper (in the rest frame, in which $u^\alpha=0$) four-momentum density vector.
Its contraction with the four-velocity gives the proper energy density, analogously to the mass (\ref{mapar1}) and using the mass--energy equivalence (\ref{emtp26}):
\begin{equation}
\epsilon=c\pi_i u^i.
\label{cemt46}
\end{equation}
This quantity is a scalar.\\

\noindent
{\bf Pressure in fluid}.\\
The four-momentum density (\ref{cemt44}) preserves its form if a term orthogonal to $u^i$ is added to the tensor (\ref{cemt430}).
Such a term is proportional to the projection tensor (\ref{hyp5}),
\begin{equation}
h_{ik}=g_{ik}-u_i u_k,
\end{equation}
which is orthogonal to $u^k$:
\begin{equation}
h_{ik}u^k=0.
\end{equation}
Consequently, the general form of the tetrad energy--momentum tensor for a fluid is given by
\begin{equation}
t_{ik}=c\pi_i u_k-ph_{ik}=(c\pi_i+pu_i)u_k-pg_{ik}.
\label{cemt43}
\end{equation}
A scalar $p$ is referred to as the {\em pressure}.
The quantities $\epsilon$ and $p$ can also be written as
\begin{eqnarray}
& & \epsilon=t_{ik}u^i u^k, \\
& & p=-\frac{1}{3}t_{ik}h^{ik}.
\end{eqnarray}

Matter described by the tensors (\ref{stp8}) and (\ref{cemt43}) represents an {\em ideal fluid}.
The relation between $\epsilon$ and $p$ is referred to as an {\em equation of state}.
The tensor $t_{ik}$ could also contain terms with covariant derivatives of $u^k$.
In this case, matter would represent a {\em viscous fluid}.\\

\noindent
{\bf Metric energy--momentum tensor for fluid}.\\
The conservation law for the spin density (\ref{cam8}) of an ideal fluid gives
\begin{equation}
c(\pi_i u_j-\pi_j u_i)=\nabla_k^\ast(s_{ij}u^k).
\label{cemt45}
\end{equation}
Contracting (\ref{cemt45}) with $u^j$ gives
\begin{equation}
\pi_i=\frac{1}{c}\epsilon u_i+\frac{1}{c}\nabla_k^\ast(s_{ij}u^k)u^j.
\label{cemt47}
\end{equation}
Therefore, the tetrad energy--momentum tensor can be written as
\begin{equation}
t_{ij}=\epsilon u_i u_j-ph_{ij}+\nabla_k^\ast (s_{il}u^k)u^l u_j.
\end{equation}
Using the Belinfante--Rosenfeld relation (\ref{BelRos8}), gives the metric energy--momentum tensor for isotropic matter with spin:
\begin{equation}
T_{ij}=\epsilon u_i u_j-ph_{ij}+\nabla_k^\ast(s_{il}u^k)u^l u_j-\frac{1}{2}\nabla_k^\ast(s_{ij}u^k+s^k_{\phantom{k}i}u_j+s^k_{\phantom{k}j}u_i).
\label{cemt49}
\end{equation}
Substituting (\ref{cemt47}) into (\ref{cemt45}) gives an equation for the spin tensor:
\begin{equation}
\nabla_k^\ast(s_{ij}u^k)-\nabla_k^\ast(s_{il}u^k)u^l u_j+\nabla_k^\ast(s_{jl}u^k)u^l u_i=0,
\label{cemt50}
\end{equation}
which is analogous to (\ref{mapar6}).
Contracting (\ref{cemt50}) with $u^j$ gives the identity, therefore any 3 components of (\ref{cemt50}) are linear combinations of the other components.
Consequently, we can impose 3 constraints on $s_{ij}$, such as (\ref{stp10}).\\

\noindent
{\bf Spinless fluid}.\\
If the spin density vanishes, $s_{ijl}=0$, then a fluid is spinless.
Substituting $s_{ij}=0$, which follows from (\ref{stp9}), into (\ref{cemt49}) gives the metric (tetrad) energy--momentum tensor for a spinless fluid:
\begin{equation}
T_{ik}=t_{ik}=\epsilon u_i u_k-ph_{ik}=(\epsilon+p)u_i u_k-pg_{ik}.
\label{cemt30}
\end{equation}
The proper four-momentum density vector (\ref{cemt47}) is proportional to the four-velocity:
\begin{equation}
\pi_i=\frac{1}{c}\epsilon u_i.
\end{equation}
In the Galilean frame of reference, combining (\ref{foac3}), (\ref{fmam23}), and (\ref{cemt30}) gives
\begin{eqnarray}
& & W=\frac{\epsilon+pv^2/c^2}{1-v^2/c^2}, \\
& & {\bf S}=\frac{(\epsilon+p){\bf v}}{1-v^2/c^2}, \\
& & \sigma_{\alpha\beta}=-\frac{(\epsilon+p)v_\alpha v_\beta}{c^2-v^2}-p\delta_{\alpha\beta}.
\end{eqnarray}
Contracting the tensor (\ref{cemt30}) gives
\begin{equation}
T=T_i^{\phantom{i}i}=\epsilon-3p.
\label{cemt41}
\end{equation}
The component $T_{00}=\epsilon u_0^2+p(u_0^2-g_{00})$ is, by means of $u_0=(g_{00}dx^0+g_{0\alpha}dx^\alpha)/ds$, (\ref{intprop11}) and (\ref{intprop12}), equal to
\begin{equation}
T_{00}=\epsilon u_0^2+pg_{00}\biggl(\frac{dl}{ds}\biggr)^2,
\end{equation}
so it is positive under physical conditions $\epsilon>0$, $p>0$, and $g_{00}>0$.\\

\noindent
{\bf Pascal law}.\\
In the rest frame, $T_{ik}=\mbox{diag}(\epsilon,p,p,p)$, the energy density $W$ is equal to $\epsilon$, the energy flux density ${\bf S}$ vanishes, and the stress tensor $\sigma_{\alpha\beta}=-p\delta_{\alpha\beta}$.
Consequently, the force (\ref{fmam21}) gives
\begin{equation}
F^\alpha=-\oint p\,df^\alpha=-\oint p\,n^\alpha df.
\label{cemt31}
\end{equation}
This equation, referred to as {\em Pascal's law}, states that the force per unit surface $df$ acting on a surface is parallel, with the opposite sign, to the outward normal vector of this surface $n^\alpha$: $dF^\alpha/df=-pn^\alpha$.\\

\noindent
{\bf Kinetic formulae for ideal gas}.\\
Comparing $T$ in (\ref{cemt41}) with (\ref{emtp16}) gives
\begin{equation}
\epsilon-3p=\mu c^2\frac{1}{u^0}=\sum_a m_a c^2\Bigl(1-\frac{v^2_a}{c^2}\Bigr)^{1/2},
\label{emtp18}
\end{equation}
where $u^0=\gamma=(1-v^2/c^2)^{-1/2}$ and the summation is over all particles in unit volume.
Consequently, the pressure has an upper limit: $p\leq\epsilon/3$.
In the nonrelativistic limit, $p\approx0$, whereas in the ultrarelativistic limit ($v\to c$), $p\to\epsilon/3$.
Comparing the component $T^{00}$ in (\ref{cemt30}) in the locally Galilean rest frame with that in (\ref{emtp9}) and (\ref{emtp10}) gives the energy density:
\begin{equation}
\epsilon=T^{00}=\mu c^2 u^0=\sum_a m_a c^2\Bigl(1-\frac{v^2_a}{c^2}\Bigr)^{-1/2}.
\label{emtp19}
\end{equation}
The relations (\ref{emtp18}) and (\ref{emtp19}) give the pressure:
\begin{equation}
p=\frac{1}{3}\mu c^2(u^0-1/u^0)=\frac{1}{3}\sum_a m_a v^2_a\Bigl(1-\frac{v^2_a}{c^2}\Bigr)^{-1/2}.
\label{emtp20}
\end{equation}
Let us consider a system of noninteracting identical particles of mass $m$, which we refer to as an {\em ideal gas}.
The {\em number density} or {\em concentration} is defined as the number of particles $n$ in unit volume, giving
\begin{equation}
\mu=nm.
\end{equation}
Consequently, we obtain the {\em kinetic formulae} for ideal gases involving the averaged functions of the velocity:
\begin{eqnarray}
& & \epsilon=nmc^2\overline{\gamma}, \\
& & p=\frac{1}{3}nm\overline{v^2\gamma }.
\end{eqnarray}

The number of particles $dN$ in a volume element $dV$ in the rest frame of reference is equal to
\begin{equation}
dN=n\,dV,
\end{equation}
where $n$ is the proper number density.
In a frame of reference moving relative to the rest frame with velocity ${\bf v}$, the same volume element is given by $dV'=dV\sqrt{1-v^2/c^2}$ (\ref{relkin23}), and the number density is $n'$.
Since $dN'=n'\,dV'$ and $dN'=dN$ is an invariant, we have
\begin{equation}
n'=\frac{n}{\sqrt{1-v^2/c^2}}.
\label{rh11}
\end{equation}

\subsubsection{Continuity and Euler equations}
For a spinless fluid, the covariant conservation (\ref{cmem15}) of the metric energy--momentum tensor (\ref{cemt30}) gives
\begin{equation}
\bigl((\epsilon+p)u^k\bigr)_{:k}u^i+(\epsilon+p)u^k u^i_{\phantom{i}:k}=p_{,k}g^{ik}.
\label{rh1}
\end{equation}
Multiplying (\ref{rh1}) by $u_i$ and contracting gives the {\em equation of continuity}:
\begin{equation}
\bigl((\epsilon+p)u^k\bigr)_{:k}=p_{,k}u^k.
\label{rh2}
\end{equation}
Substituting this equation into (\ref{rh1}) gives the {\em Euler equation}:
\begin{equation}
(\epsilon+p)\frac{\mathring{D}u^i}{ds}=p_{,k}h^{ik}.
\label{rh3}
\end{equation}
If the pressure is constant, then (\ref{rh3}) reduces to the metric geodesic equation (\ref{metgeo3}).
Defining a quantity $n$ such that
\begin{equation}
\frac{dn}{n}=\frac{d\epsilon}{\epsilon+p}
\end{equation}
brings (\ref{rh2}) to the conservation law
\begin{equation}
(nu^i)_{:i}=0.
\label{rh5}
\end{equation}
The quantity $n$ thus represents the proper number density of particles composing the fluid.\\

\noindent
{\bf Nonrelativistic limit}.\\
In the nonrelativistic limit, $c\rightarrow\infty$, $\gamma\approx1$, $u^0\sim1$, and $u^\alpha\approx v^\alpha/c$.
For an ideal fluid with mass density $\mu$, the energy density $\epsilon$ and pressure $p$ satisfy $\epsilon\approx\mu c^2$ and $p\ll\epsilon$.
Consequently, the equation of continuity (\ref{rh2}) reduces to
\begin{equation}
(\mu u^k)_{:k}=\frac{1}{\sqrt{-\mathfrak{g}}}(\sqrt{-\mathfrak{g}}\mu u^k)_{,k}=0,
\label{rh41}
\end{equation}
using (\ref{Chrsym16}).
If $g_{00}\approx 1$, then (\ref{spvec6}) gives $-\mathfrak{g}\approx \mathfrak{s}$ and (\ref{rh41}) leads to
\begin{equation}
(\sqrt{\mathfrak{s}}\mu c u^k)_{,k}=(\sqrt{\mathfrak{s}}\mu c u^0)_{,0}+(\sqrt{\mathfrak{s}}\mu c u^\alpha)_{,\alpha}=\frac{\partial}{\partial t}(\sqrt{\mathfrak{s}}\mu)+(\sqrt{\mathfrak{s}}\mu v^\alpha)_{,\alpha}=\frac{\partial}{\partial t}(\sqrt{\mathfrak{s}}\mu)+\sqrt{\mathfrak{s}}\,\mbox{div}(\mu{\bf v})=0,
\end{equation}
using (\ref{spvec28}).
This equation of continuity can be written as
\begin{equation}
\frac{\partial}{\partial t}(\sqrt{\mathfrak{s}}\mu)+\sqrt{\mathfrak{s}}\,\mbox{div}\,{\bf j}=0,
\label{rh6}
\end{equation}
where
\begin{equation}
{\bf j}=\mu {\bf v}
\end{equation}
is the {\em mass flux density} vector.
Integrating (\ref{rh6}) over a volume, using Gau\ss' theorem (\ref{covint29}) and the mass density (\ref{emtp6}), gives
\begin{equation}
\frac{\partial}{\partial t}\int\mu\sqrt{\mathfrak{s}}dV+\oint{\bf j}\cdot \sqrt{\mathfrak{s}}d{\bf f}=\frac{\partial m}{\partial t}+\oint{\bf j}\cdot \sqrt{\mathfrak{s}}d{\bf f}=0.
\end{equation}
This relation represents the conservation of mass of a fluid: the rate of change of the mass of the fluid in a given volume is related to the mass flowing out from this volume through its surface (the mass flux).
An element $d{\bf f}$ of the surface bounding the volume is directed along the outward normal.
The mass flux density vector is parallel to the motion of the fluid and its magnitude is equal to the mass of the fluid flowing through unit area perpendicular to the velocity in unit time.

The Euler equation (\ref{rh3}) gives
\begin{equation}
(\epsilon+p)u^j u^i_{\phantom{i},j}+(\epsilon+p)\mathring{\Gamma}^{i}_{kj}u^j u^k=p_{,k}(g^{ik}-u^i u^k).
\end{equation}
In the nonrelativistic limit of the spatial components $i=\alpha$ of this equation, the dominant terms are
\begin{equation}
(\mu c^2)u^0 u^\alpha_{\phantom{\alpha},0}+(\mu c^2)u^\beta u^\alpha_{\phantom{\alpha},\beta}+(\mu c^2)\mathring{\Gamma}^{\alpha}_{00}(u^0)^2=p_{,\beta}\eta^{\alpha\beta}.
\end{equation}
Using (\ref{grapot2}), this relation gives
\begin{equation}
\mu\biggl(\frac{\partial v^\alpha}{\partial t}+v^\alpha_{\phantom{\alpha},\beta}v^\beta\biggr)+\mu \phi_{,\alpha}=-p_{,\alpha},
\label{rh23}
\end{equation}
where $\phi$ is the gravitational potential (\ref{grapot1}).
This equation in the spatial-vector notation is
\begin{equation}
\mu\frac{d{\bf v}}{dt}=\mu\biggl(\frac{\partial{\bf v}}{\partial t}+({\bf v}\cdot\mbox{{\bf grad}}){\bf v}\biggr)=-\mu\,\mbox{{\bf grad}}\,\phi-\mbox{{\bf grad}}\,p,
\end{equation}
which is equivalent to
\begin{equation}
\frac{d{\bf v}}{dt}=\frac{\partial{\bf v}}{\partial t}+({\bf v}\cdot\mbox{{\bf grad}}){\bf v}={\bf g}-\frac{\mbox{{\bf grad}}\,p}{\mu},
\label{rh8}
\end{equation}
where ${\bf g}$ is the gravitational acceleration (\ref{grapot3}).
The temporal component $i=0$ of the Euler equation (\ref{rh3}) leads to the equation of continuity.
Without pressure gradients, (\ref{rh8}) reduces to (\ref{grapot3}).

In the nonrelativistic limit, the total momentum (\ref{emtp24}) of a fluid is
\begin{equation}
{\bf P}=\int{\bf v}dm=\int\mu{\bf v}\sqrt{\mathfrak{s}}dV.
\end{equation}
Its rate of change follows from $dm/dt=0$, valid for a spinless fluid because of (\ref{mapar18}), and (\ref{rh8}):
\begin{equation}
\frac{d{\bf P}}{dt}=\int\mu\frac{d{\bf v}}{dt}\sqrt{\mathfrak{s}}dV=\int\mu{\bf g}\sqrt{\mathfrak{s}}dV-\oint p\sqrt{\mathfrak{s}}d{\bf f}=\int{\bf g}dm-\oint p\sqrt{\mathfrak{s}}d{\bf f}.
\label{rh9}
\end{equation}
The rate of change of the momentum of a fluid in a given volume is equal to the gravitational force on the fluid within the volume plus the pressure force (directed along the inward normal) on the fluid within the surface bounding the volume.

\subsubsection{Action for spinless particles}
{\bf Lagrangian}.\\
The energy (\ref{fmam9}) is the volume integral of the energy density (\ref{fmam6}):
\begin{equation}
E=\int\biggl(\frac{\partial\mathfrak{L}_\textrm{m}}{\partial\dot{\phi}}\dot{\phi}-\mathfrak{L}_\textrm{m}\biggr)dV.
\end{equation}
For a particle, the time derivative $\dot{\phi}$ of a field $\phi$ in the integrand is equal to its value at the point, where the particle is located.
The spatial coordinates of the particle can be taken as three fields, which are functions of time.
Consequently, the energy is
\begin{equation}
E=\frac{\partial L}{\partial\dot{\phi}}\dot{\phi}-L,
\end{equation}
where
\begin{equation}
L=\int\mathfrak{L}_\textrm{m}dV
\label{fmam8}
\end{equation}
is the Lagrange function or {\em Lagrangian} of the matter.
The time integral of the Lagrangian gives the action (\ref{Lageq1}):
\begin{equation}
S=\int L\,dt=\int\mathfrak{L}_\textrm{m}dV\,dt=\frac{1}{c}\int\mathfrak{L}_\textrm{m}d\Omega.
\label{fmam10}
\end{equation}

\noindent
{\bf Relation between action and interval}.\\
The energy--momentum tensor for a point particle of mass $m$ located at the radius vector ${\bf r}_0$ is given by (\ref{emtp10}):
\begin{equation}
T^{ik}({\bf r})=mc^2{\bm \delta}({\bf r}-{\bf r}_0)\frac{u^i u^k}{\sqrt{-\mathfrak{g}}u^0},
\end{equation}
so the variation of the action with respect to the metric tensor (\ref{dmemd2}) gives
\begin{eqnarray}
& & \delta S=-\frac{1}{2c}\int T^{ik}\delta g_{ik}\sqrt{-\mathfrak{g}}d\Omega=-\frac{mc}{2}\int\frac{u^i u^k}{u^0}\delta g_{ik}dx^0=-\frac{mc}{2}\int u^i u^k\delta g_{ik}ds \nonumber \\
& & =-\frac{mc}{2}\int\frac{\delta g_{ik}dx^i dx^k}{ds}=-mc\int\delta\sqrt{g_{ik}dx^i dx^k}=-mc\,\delta\int ds.
\end{eqnarray}
Therefore, the action for a {\em free} (interacting only with the gravitational field) particle is
\begin{equation}
S=-mc\int_1^2 ds,
\label{acp4}
\end{equation}
where $1$ and $2$ denote the world points corresponding to the initial and final spacetime position of the particle.
The action for a system of noninteracting particles is the sum of the actions corresponding to each particle:
\begin{equation}
S=-\sum_a m_a c\int ds_a.
\label{acp5}
\end{equation}

\noindent
{\bf Variation of action}.\\
The variation of the action (\ref{acp4}) for a particle with respect to the coordinates $x^i$ gives, following (\ref{metgeo2}):
\begin{equation}
\delta S=mc\int_1^2\biggl[g_{ik}\frac{\mathring{D}u^i}{ds}\biggr]\delta x^k ds-mc(u_k\delta x^k)\Big|_1^2.
\label{acp14}
\end{equation}
The principle of stationary action $\delta S=0$ (\ref{Lageq4}), applied to all world lines with fixed endpoints ($\delta x^k=0$ at the initial and final world points) and for an arbitrary variation $\delta x^k$ in the integrand, gives the metric geodesic equation (\ref{metgeo3}) for the motion of the particle.

A particle is a special case of a field existing in spacetime.
The action for a particle (\ref{acp4}) determines the Lagrangian (\ref{fmam10}) for the particle, which satisfies the Lagrange equations that are analogous to (\ref{Lageq8}).
The physics of particles and their systems, such as rigid bodies and ideal fluids, based on the Lagrangian, is referred to as {\em mechanics} and constitutes Chapter \ref{Particles}.
\newline
References: \cite{LL2,Lord}.

\subsection{Gravitational field equations}
\setcounter{equation}{0}
\label{Gravfieldeqs}
\subsubsection{Einstein--Cartan action and equations}
The metric and torsion tensors are two independent, fundamental variables describing a gravitational field.
The action for the gravitational field and matter is equal, following (\ref{Laggr3}), to
\begin{equation}
S=S_\textrm{g}+S_\textrm{m}=-\frac{1}{2\kappa c}\int R\sqrt{-\mathfrak{g}}d\Omega+S_\textrm{m}.
\label{EC1}
\end{equation}
The action (\ref{EC1}) subjected to varying the metric and torsion tensors is called the {\em Einstein--Cartan action} for the gravitational field and matter.
Using (\ref{Riem11}) and applying partial integration (\ref{Chrsym19}) gives
\begin{eqnarray}
& & S_\textrm{g}=-\frac{1}{2\kappa c}\int\Bigl(\mathring{R}-g^{ik}(2C^l_{\phantom{l}il:k}+C^j_{\phantom{j}ij}C^l_{\phantom{l}kl}-C^l_{\phantom{l}im}C^m_{\phantom{m}kl})\Bigr)\sqrt{-\mathfrak{g}}d\Omega \nonumber \\
& & =-\frac{1}{2\kappa c}\int\Bigl(\mathring{R}-g^{ik}(C^j_{\phantom{j}ij}C^l_{\phantom{l}kl}-C^l_{\phantom{l}im}C^m_{\phantom{m}kl})\Bigr)\sqrt{-\mathfrak{g}}d\Omega+\frac{1}{\kappa c}\oint C^{lk}_{\phantom{lk}l}\sqrt{-\mathfrak{g}}dS_k,
\label{EC2}
\end{eqnarray}
where $dS_i$ is the element of the closed hypersurface surrounding the integration four-volume.
The stationarity of action (\ref{Lageq4}), which is a part of the principle of least action, is applied with a condition that the variations of the variables at the boundary of integration four-volume vanish.
Accordingly, the variation of the hypersurface integral taken over this boundary in (\ref{EC2}) vanishes.
This integral therefore does not contribute to the field equations and can be omitted, which reduces (\ref{EC1}) to
\begin{equation}
S=-\frac{1}{2\kappa c}\int\Bigl(\mathring{R}-g^{ik}(C^j_{\phantom{j}ij}C^l_{\phantom{l}kl}-C^l_{\phantom{l}im}C^m_{\phantom{m}kl})\Bigr)\sqrt{-\mathfrak{g}}d\Omega+S_\textrm{m}.
\label{EC3}
\end{equation}

\noindent
{\bf Variation of action over metric}.\\
Firstly, we vary (\ref{EC3}) with respect to the metric tensor.
Using (\ref{dmemd2}) and the identity $\delta\sqrt{-\mathfrak{g}}=-(1/2)\sqrt{-\mathfrak{g}}g_{ik}\delta g^{ik}$, which results from (\ref{metten18}), gives
\begin{eqnarray}
& & \delta_\textrm{g}S=-\frac{1}{2\kappa c}\int\biggl(\delta \mathring{R}_{ik}g^{ik}\sqrt{-\mathfrak{g}}+\mathring{R}_{ik}\delta g^{ik}\sqrt{-\mathfrak{g}}-\frac{1}{2}\mathring{R}\sqrt{-\mathfrak{g}}g_{ik}\delta g^{ik}\biggr)d\Omega \nonumber \\
& & -\frac{1}{2\kappa c}\int\biggl(-C^j_{\phantom{j}ij}C^l_{\phantom{l}kl}+C^l_{\phantom{l}im}C^m_{\phantom{m}kl}+\frac{1}{2}g_{ik}(C^{jm}_{\phantom{jm}j}C^l_{\phantom{l}ml}-C^{lj}_{\phantom{lj}m}C^m_{\phantom{m}jl})\biggr)\sqrt{-\mathfrak{g}}\delta g^{ik}d\Omega \nonumber \\
& & +\frac{1}{2c}\int T_{ik}\sqrt{-\mathfrak{g}}\delta g^{ik}d\Omega.
\label{EC4}
\end{eqnarray}
We define the contravariant {\em metric density},
\begin{equation}
{\sf g}^{ik}=\sqrt{-\mathfrak{g}}g^{ik},
\end{equation}
whose covariant derivative with respect to the Christoffel symbols vanishes: ${\sf g}^{ik}_{\phantom{ik}:l}=0$.
In the first term on the right-hand side of (\ref{EC4}), using (\ref{Riem14}) and applying partial integration (\ref{Chrsym19}) brings this term to zero:
\begin{equation}
\int\delta \mathring{R}_{ik}{\sf g}^{ik}d\Omega=\int\Big((\delta\mathring{\Gamma}^{l}_{ik})_{:l}-(\delta\mathring{\Gamma}^{l}_{il})_{:k}\Bigr){\sf g}^{ik}d\Omega=\oint({\sf g}^{ik}\delta\mathring{\Gamma}^{l}_{ik}dS_l-{\sf g}^{ik}\delta\mathring{\Gamma}^{l}_{il}dS_k)=0.
\end{equation}
We therefore obtain
\begin{eqnarray}
& & \delta_\textrm{g}S=-\frac{1}{2\kappa c}\int G_{ik}\sqrt{-\mathfrak{g}}\delta g^{ik}d\Omega \nonumber \\
& & -\frac{1}{2\kappa c}\int\biggl(-C^j_{\phantom{j}ij}C^l_{\phantom{l}kl}+C^l_{\phantom{l}im}C^m_{\phantom{m}kl}+\frac{1}{2}g_{ik}(C^{jm}_{\phantom{jm}j}C^l_{\phantom{l}ml}-C^{lj}_{\phantom{lj}m}C^m_{\phantom{m}jl})\biggr)\sqrt{-\mathfrak{g}}\delta g^{ik}d\Omega \nonumber \\
& & +\frac{1}{2c}\int T_{ik}\sqrt{-\mathfrak{g}}\delta g^{ik}d\Omega,
\label{EC7}
\end{eqnarray}
where $G_{ik}$ is the Einstein tensor (\ref{Riem23}).\\

\noindent
{\bf Variation of action over torsion}.\\
Secondly, we vary (\ref{EC3}) with respect to the contortion tensor, which is equivalent to varying with respect to the torsion tensor.
Using (\ref{spden4}) and (\ref{spden12}) gives
\begin{eqnarray}
& & \delta_\textrm{C}S=-\frac{1}{\kappa c}\int(C^{kj}_{\phantom{kj}i}-C^{lj}_{\phantom{lj}l}\delta^k_i)\sqrt{-\mathfrak{g}}\delta C^i_{\phantom{i}jk}d\Omega+\frac{1}{2c}\int T_{ik}\sqrt{-\mathfrak{g}}\delta g^{ik}d\Omega \nonumber \\
& & +\frac{1}{2c}\int s^{\phantom{j}ik}_j\sqrt{-\mathfrak{g}}\delta C^j_{\phantom{j}ik}d\Omega.
\label{EC8}
\end{eqnarray}
The total variation of $S$ is then
\begin{equation}
\delta S=\delta_\textrm{g}S+\delta_\textrm{C}S.
\end{equation}

\noindent
{\bf Einstein equations}.\\
Because the variations $\delta g^{ik}$ and $\delta C^j_{\phantom{j}ik}$ are independent, the stationarity of action (\ref{Lageq4}) yields $\delta_\textrm{g}S=\delta_\textrm{C}S=0$.
The condition $\delta_\textrm{g}S=0$ for an arbitrary $\delta g^{ik}$ gives the {\em Einstein equations}:
\begin{equation}
G_{ik}=\mathring{R}_{ik}-\frac{1}{2}\mathring{R}g_{ik}=\kappa(T_{ik}+U_{ik}),
\label{EC10}
\end{equation}
where
\begin{equation}
U_{ik}=\frac{1}{\kappa}\biggl(C^j_{\phantom{j}ij}C^l_{\phantom{l}kl}-C^l_{\phantom{l}ij}C^j_{\phantom{j}kl}-\frac{1}{2}g_{ik}(C^{jm}_{\phantom{jm}j}C^l_{\phantom{l}ml}-C^{mjl}C_{ljm})\biggr)
\label{EC11}
\end{equation}
or
\begin{eqnarray}
& & U_{ik}=\frac{1}{\kappa}\biggl(4S_i S_k-2S^j_{\phantom{j}il}S_{jk}^{\phantom{jk}l}-2S^j_{\phantom{j}il}S^l_{\phantom{l}kj}+S_{ijl}S_k^{\phantom{k}jl} \nonumber \\
& & -\frac{1}{2}g_{ik}(4S^j S_j-2S^l_{\phantom{l}mn}S^{nm}_{\phantom{nm}l}-S^l_{\phantom{l}mn}S_l^{\phantom{l}mn})\biggr).
\label{EC12}
\end{eqnarray}
The Einstein equations (\ref{EC10}) can be written as
\begin{equation}
\mathring{R}_{ik}=\kappa\biggl(T_{ik}+U_{ik}-\frac{1}{2}(T+U)g_{ik}\biggr),
\label{EC13}
\end{equation}
where
\begin{equation}
T=T^i_{\phantom{i}i},\quad U=U^i_{\phantom{i}i}.
\end{equation}

\noindent
{\bf Cartan equations}.\\
The condition $\delta_\textrm{C}S=0$ for an arbitrary $\delta C^j_{\phantom{j}ik}$ gives
\begin{equation}
C^k_{\phantom{k}[ji]}-\delta^k_{[i}C^l_{\phantom{l}j]l}=\frac{1}{2}\kappa s^{\phantom{ij}k}_{ij}.
\label{EC15}
\end{equation}
This equation can be written as the {\em Cartan equations}:
\begin{equation}
S^j_{\phantom{j}ik}-S_i \delta^j_k+S_k \delta^j_i=-\frac{1}{2}\kappa s^{\phantom{ik}j}_{ik}.
\label{EC16}
\end{equation}
The relation (\ref{EC16}) is equivalent to
\begin{eqnarray}
& & S^k_{\phantom{k}ij}=-\frac{1}{2}\kappa(s_{ij}^{\phantom{ij}k}+\delta^k_{[i}s_{j]l}^{\phantom{j]l}l}), \label{EC17} \\
& & C^k_{\phantom{k}ij}=\frac{1}{2}\kappa(s^k_{\phantom{k}ij}-s_{ij}^{\phantom{ij}k}-s_{j\phantom{k}i}^{\phantom{j}k}-g_{ij}s^{kl}_{\phantom{kl}l}+\delta^k_j s_{il}^{\phantom{il}l}).
\label{EC18}
\end{eqnarray}

\noindent
{\bf Combined energy--momentum tensor}.\\
Combining (\ref{EC11}) and (\ref{EC18}) gives
\begin{equation}
U^{ik}=\kappa\biggl(-s^{ij}_{\phantom{ij}[l}s^{kl}_{\phantom{kl}j]}-\frac{1}{2}s^{ijl}s^k_{\phantom{k}jl}+\frac{1}{4}s^{jli}s_{jl}^{\phantom{jl}k}+\frac{1}{8}g^{ik}(-4s^l_{\phantom{l}j[m}s^{jm}_{\phantom{jm}l]}+s^{jlm}s_{jlm})\biggr).
\label{EC19}
\end{equation}
The tensor (\ref{EC19}) represents a correction to the metric energy--momentum tensor from the spin contributions to the geometry of spacetime.
It is quadratic in the spin tensor, thereby representing a spin-spin contact interaction.
Accordingly, changing the signs of all the components of the spin tensor does not affect this correction.
The spin tensor can also appear in $T_{ik}$ because $\mathfrak{L}_\textrm{m}$ can depend on torsion.
The Einstein equations (\ref{EC10}), in which $U_{ik}$ is given by (\ref{EC19}), are field equations with the {\em combined energy--momentum tensor} $\tilde{T}_{ik}$ as a source of the curvature:
\begin{equation}
\mathring{R}_{ik}-\frac{1}{2}\mathring{R}g_{ik}=\kappa\tilde{T}_{ik},\quad \tilde{T}_{ik}=T_{ik}+U_{ik}.
\label{EC21}
\end{equation}
The conservation law (\ref{cmem13}) for the metric energy--momentum tensor is, upon substituting (\ref{EC10}) and (\ref{EC15}), equivalent to the the contracted Bianchi identity (\ref{Riem22}) for the Einstein tensor.
This identity, applied to (\ref{EC10}), gives the Riemannian conservation law for the combined energy--momentum tensor:
\begin{equation}
\tilde{T}^{ik}_{\phantom{ik}:k}=0.
\label{EC20}
\end{equation}
This law is equivalent to (\ref{cmem13}) with (\ref{EC18}).

The relation (\ref{EC17}) between the torsion and spin tensors is algebraic.
Torsion at a given point in spacetime does not vanish only if matter is present at this point, represented in the Lagrangian density by a function which depends on torsion.
If the matter Lagrangian density does not depend on torsion, then the spin tensor vanishes, and so does the torsion tensor.
In {\em vacuum}, which is defined as the absence of matter, $T_{ik}=0$ and $s_{ijk}=0$, the Riemannian Ricci tensor in (\ref{EC13}) also vanishes:
\begin{equation}
\mathring{R}_{ik}=0.
\end{equation}
Unlike the metric, which is related to matter through a differential field equation, torsion does not propagate in vacuum.
The vanishing of $\mathring{R}_{ik}$ and $S_{ijk}$ at a given point in spacetime is a covariant criterion for the absence of matter at this point.
The Riemann tensor $\mathring{R}^i_{\phantom{i}jkl}$ at such a point, however, can be different from zero.

\subsubsection{Sciama--Kibble action}
The tetrad and spin connection, instead of the metric tensor and affine connection, can be regarded as dynamical variables.
The action (\ref{EC1}) subjected to varying the tetrad and spin connection is called the {\em Sciama--Kibble action}.
Using (\ref{BelRos1}) gives
\begin{equation}
\delta S=-\frac{1}{2\kappa c}\int\delta(\mathfrak{e}R)d\Omega+\frac{1}{c}\int\mathfrak{T}^{\phantom{i}a}_i\delta e^i_a d\Omega+\frac{1}{2c}\int\mathfrak{S}_{ab}^{\phantom{ab}i}\delta\omega^{ab}_{\phantom{ab}i}d\Omega.
\label{SK1}
\end{equation}
The Lagrangian density for the gravitational field is given by (\ref{Laggr2}), with the curvature scalar $R$ given by (\ref{tetrep3}) and (\ref{tetrep5}):
\begin{equation}
\mathfrak{e}R=\mathfrak{e}e^i_a e^{jb}(\omega^a_{\phantom{a}bj,i}-\omega^a_{\phantom{a}bi,j}+\omega^a_{\phantom{a}ci}\omega^c_{\phantom{c}bj}-\omega^a_{\phantom{a}cj}\omega^c_{\phantom{c}bi})=2\mathfrak{e}^{ij}_{ab}(\omega^{ab}_{\phantom{ab}j,i}+\omega^a_{\phantom{a}ci}\omega^{cb}_{\phantom{cb}j}),
\end{equation}
where
\begin{equation}
\mathfrak{e}^{ij}_{ab}=\mathfrak{e}e^{[i}_a e^{j]}_b.
\end{equation}
This quantity satisfies $\mathfrak{e}^{ij}_{ab|j}=\mathfrak{e}^{ij}_{ab,j}-\omega^c_{\phantom{c}aj}\mathfrak{e}^{ij}_{cb}-\omega^c_{\phantom{c}bj}\mathfrak{e}^{ij}_{ac}+\Gamma^{i}_{kj}\mathfrak{e}^{kj}_{ab}+\Gamma^{j}_{kj}\mathfrak{e}^{ik}_{ab}-\Gamma^{k}_{kj}\mathfrak{e}^{ij}_{ab}=0$, which results from (\ref{spcon7}).
Varying $\mathfrak{e}R$ and omitting total derivatives which lead to hypersurface integrals gives, using (\ref{tet13}),
\begin{eqnarray}
& & \delta(\mathfrak{e}R)=(2R^a_{\phantom{a}i}-Re^a_i)\mathfrak{e}\delta e^i_a+2\mathfrak{e}^{ij}_{ab}\delta(\omega^{ab}_{\phantom{ab}j,i}+\omega^a_{\phantom{a}ci}\omega^{cb}_{\phantom{cb}j}) \nonumber \\
& & =(2R^a_{\phantom{a}i}-Re^a_i)\mathfrak{e}\delta e^i_a+2(\mathfrak{e}^{ij}_{ab,j}-\omega^c_{\phantom{c}aj}\mathfrak{e}^{ij}_{cb}-\omega^c_{\phantom{c}bj}\mathfrak{e}^{ij}_{ac})\delta\omega^{ab}_{\phantom{ab}i} \nonumber \\
& & =(2R^a_{\phantom{a}i}-Re^a_i)\mathfrak{e}\delta e^i_a-2(S^i_{\phantom{i}kj}\mathfrak{e}^{kj}_{ab}+2S_j\mathfrak{e}^{ij}_{ab})\delta\omega^{ab}_{\phantom{ab}i}.
\end{eqnarray}
The variation (\ref{SK1}) is therefore equal to
\begin{eqnarray}
& & \delta S=-\frac{1}{\kappa c}\int\biggl(R^a_{\phantom{a}i}-\frac{1}{2}Re^a_i\biggr)\mathfrak{e}\delta e^i_a d\Omega+\frac{1}{\kappa c}\int(S^i_{\phantom{i}kj}\mathfrak{e}^{kj}_{ab}+2S_j\mathfrak{e}^{ij}_{ab})\delta\omega^{ab}_{\phantom{ab}i}d\Omega \nonumber \\
& & +\frac{1}{c}\int\mathfrak{T}^{\phantom{i}a}_i\delta e^i_a d\Omega+\frac{1}{2c}\int\mathfrak{S}_{ab}^{\phantom{ab}i}\delta\omega^{ab}_{\phantom{ab}i}d\Omega.
\end{eqnarray}

The condition $\delta S=0$ for an arbitrary $\delta\omega^{ab}_{\phantom{ab}i}$ gives
\begin{equation}
S^i_{\phantom{i}ab}-S_a e^i_b+S_b e^i_a=-\frac{\kappa}{2\mathfrak{e}}\mathfrak{S}^{\phantom{ab}i}_{ab},
\label{SK6}
\end{equation}
which is equivalent to the Cartan equations (\ref{EC16}).
The condition $\delta S=0$ for an arbitrary $\delta e^i_a$ gives
\begin{equation}
R^a_{\phantom{a}i}-\frac{1}{2}Re^a_i=\frac{\kappa}{\mathfrak{e}}\mathfrak{T}^{\phantom{i}a}_i,
\label{SK7}
\end{equation}
which is equivalent to
\begin{equation}
R_{ki}-\frac{1}{2}Rg_{ik}=\kappa t_{ik}.
\label{SK8}
\end{equation}

\noindent
{\bf Equivalence of Einstein and Cartan equations to conservation laws}.\\
Substituting (\ref{EC16}) and (\ref{SK8}) into the conservation law for the spin tensor (\ref{cam8}) gives
\begin{equation}
-2(S^k_{\phantom{k}ij;k}-S_{i;j}+S_{j;i})=R_{ji}-R_{ij}-4S_k(S^k_{\phantom{k}ij}-S_i\delta^k_j+S_j\delta^k_i),
\end{equation}
which is equivalent to the contracted cyclic identity (\ref{Riem16}).
Therefore, the contracted cyclic identity imposes the conservation law for the spin density.
Substituting (\ref{EC16}) and (\ref{SK8}) into the conservation law for the tetrad energy--momentum tensor (\ref{ctem9}) gives
\begin{equation}
R^j_{\phantom{j}i;j}-\frac{1}{2}R_{;i}=2S_j\biggl(R^j_{\phantom{j}i}-\frac{1}{2}R\delta^j_i\biggr)+2S^j_{\phantom{j}ki}\biggl(R^k_{\phantom{k}j}-\frac{1}{2}R\delta^k_j\biggr)-(S^j_{\phantom{j}kl}-S_k\delta^j_l+S_l\delta^j_k)R^{kl}_{\phantom{kl}ji},
\end{equation}
which is equivalent to the contracted Bianchi identity (\ref{Riem17}).
Therefore, the contracted Bianchi identity imposes the conservation law for the energy--momentum density.
The gravitational field equations therefore contain the equations of motion of matter.

Substituting (\ref{EC16}) and (\ref{SK8}) into the Belinfante--Rosenfeld relation (\ref{BelRos8}) gives
\begin{eqnarray}
& & \kappa T_{ik}=R_{ki}-\frac{1}{2}Rg_{ik}+\nabla_j^\ast(S^j_{\phantom{j}ik}+2S_k\delta^j_i-2S_{(ik)}^{\phantom{(ik)}j}-2S^j g_{ik})=R_{ki}-\frac{1}{2}Rg_{ik} \nonumber \\
& & +\nabla_j^\ast(-C^j_{\phantom{j}ki}+C^l_{\phantom{l}kl}\delta^j_i-C^{lj}_{\phantom{lj}l}g_{ik}).
\label{SK11}
\end{eqnarray}
Combining (\ref{Riem9}), (\ref{Riem11}) and (\ref{SK11}) gives
\begin{eqnarray}
& & \kappa T_{ik}=\mathring{R}_{ik}-\frac{1}{2}\mathring{R}g_{ik}+C^l_{\phantom{l}ki:l}-C^l_{\phantom{l}kl:i}+C^j_{\phantom{j}ki}C^l_{\phantom{l}jl}-C^j_{\phantom{j}kl}C^l_{\phantom{l}ji}-\frac{1}{2}g_{ik}(-2C^{lj}_{\phantom{lj}l:j} \nonumber \\
& & -C^{lj}_{\phantom{lj}l}C^m_{\phantom{m}jm}+C^{mjl}C_{ljm})-C^j_{\phantom{j}ki:j}-C^j_{\phantom{j}lj}C^l_{\phantom{l}ki}+C^l_{\phantom{l}kj}C^j_{\phantom{j}li}+C^l_{\phantom{l}ij}C^j_{\phantom{j}kl}+C^j_{\phantom{j}kj:i} \nonumber \\
& & -C^l_{\phantom{l}ki}C^j_{\phantom{j}lj}-g_{ik}(C^{lj}_{\phantom{lj}l:j}+C^j_{\phantom{j}lj}C^{ml}_{\phantom{ml}m})-C_j(-C^j_{\phantom{j}ki}+C^l_{\phantom{l}kl}\delta^j_i-C^{lj}_{\phantom{lj}l}g_{ik}),
\end{eqnarray}
which is equivalent to the Einstein equations (\ref{EC10}).
Therefore, the relation between the Ricci tensor and the Riemannian Ricci tensor is equivalent to the Belinfante--Rosenfeld relation, whereas (\ref{SK8}) is another form of the Einstein equations.
Varying the action for the gravitational field and matter with respect to the metric tensor and the tensorial, antisymmetric part of the affine connection (torsion tensor) constitutes the {\em metric--affine variational principle of stationary action}.

\subsubsection{Einstein--Hilbert action and Einstein equations}
\noindent
{\bf General relativity}.\\
In almost all physical situations, the Cartan equations give a torsion tensor whose squares of the leading components are negligibly small in magnitude relative to the leading components of the Riemann tensor (as in (\ref{EC11})).
In those situations, we can approximate the torsion tensor as zero.
In this approximation, the affine connection is equal to the Levi-Civita connection.
Varying the action for the gravitational field and matter with respect to the metric tensor, with the affine connection constrained to be equal to the Levi-Civita connection, constitutes the {\em metric variational principle of stationary action}.
If the Lagrangian density for matter does not depend on the affine connection, then the spin density vanishes, and so does the torsion tensor.
In this case, the metric-affine field equations reduce to the metric field equations and we can use the metric principle of stationary action.

If the torsion tensor vanishes, the Einstein--Cartan action (\ref{EC1}) reduces to
\begin{equation}
\mathring{S}=-\frac{1}{2\kappa c}\int \mathring{R}\sqrt{-\mathfrak{g}}d\Omega+S_\textrm{m},
\label{EH1}
\end{equation}
which corresponds to the Lagrangian density (\ref{Laggr4}).
The action (\ref{EH1}) subjected to varying the metric tensor is called the {\em Einstein--Hilbert action} for the gravitational field and matter.
The Einstein--Hilbert action is a special case of the Einstein--Cartan action, where the affine connection is constrained to be symmetric and thus equal to the Levi-Civita connection.
Varying (\ref{EH1}) with respect to the metric tensor gives, similarly to (\ref{EC7}),
\begin{equation}
\delta_\textrm{g}\mathring{S}=-\frac{1}{2\kappa c}\int\biggl(\mathring{R}_{ik}-\frac{1}{2}\mathring{R}g_{ik}\biggr)\sqrt{-\mathfrak{g}}\delta g^{ik}d\Omega+\frac{1}{2c}\int T_{ik}\sqrt{-\mathfrak{g}}\delta g^{ik}d\Omega.
\label{EH2}
\end{equation}
Applying the stationarity of action $\delta_\textrm{g}\mathring{S}=0$ to (\ref{EH2}) for an arbitrary $\delta g^{ik}$ gives the {\em Einstein equations} of the {\em general theory of relativity}:
\begin{equation}
G_{ik}=\mathring{R}_{ik}-\frac{1}{2}\mathring{R}g_{ik}=\kappa T_{ik}
\label{EH3}
\end{equation}
or
\begin{equation}
\mathring{R}_{ik}=\kappa\biggl(T_{ik}-\frac{1}{2}Tg_{ik}\biggr).
\label{EH4}
\end{equation}
These equations follow from (\ref{EC10}) for spinless matter: $U_{ik}=0$.

Because $\delta\int \mathring{R}\sqrt{-\mathfrak{g}}d\Omega=\delta\int{\sf G}\sqrt{-\mathfrak{g}}d\Omega$, where ${\sf G}$ is the noncovariant quantity (\ref{Laggr7}), the left-hand side of the Einstein equations is
\begin{equation}
G_{ik}=\frac{1}{\sqrt{-\mathfrak{g}}}\frac{\delta(\sqrt{-\mathfrak{g}}{\sf G})}{\delta g^{ik}}=\frac{1}{\sqrt{-\mathfrak{g}}}\biggl(\frac{\partial(\sqrt{-\mathfrak{g}}{\sf G})}{\partial g^{ik}}-\partial_l\frac{\partial(\sqrt{-\mathfrak{g}}{\sf G})}{\partial(\partial_l g^{ik})}\biggr).
\end{equation}
The covariant conservation of the Einstein tensor (\ref{Riem22}) imposes the conservation of the metric energy--momentum tensor (\ref{cmem15}).
The gravitational field equations therefore contain the equations of motion of matter, like for the Einstein--Cartan action.\\

\noindent
{\bf Structure of Einstein equations}.\\
The Einstein equations (\ref{EH3}) are 10 second-order partial differential equations for: $10-4=6$ independent components of the metric tensor $g_{ik}$ (the factor 4 is the number of the coordinates which can be chosen arbitrarily), 3 independent components of the four-velocity $u^i$, and either $\epsilon$ or $p$ (which are related to each other by the equation of state).
The contracted Bianchi identity (\ref{Riem22}) gives the equations of motion of matter.
In vacuum, the Einstein equations are $10-4=6$ independent equations (the factor 4 is the number of constraints from the contracted Bianchi identity) for 6 independent components of the metric tensor $g_{ik}$.

In the Einstein equations (\ref{EH3}), the only second time-derivatives of $g_{ik}$ are the derivatives of the spatial components of the metric tensor, $\ddot{g}_{\alpha\beta}$, and they appear only in the $\alpha\beta$ components of the equations.
Therefore, the initial values (at $t=0$) for $g_{\alpha\beta}$ and $\dot{g}_{\alpha\beta}$ can be chosen arbitrarily.
The first time-derivatives $\dot{g}_{0\alpha}$ and $\dot{g}_{00}$ appear only in the $\alpha\beta$ components of the field equations (\ref{EH3}).
The $0\alpha$ and $00$ components of the field equations (\ref{EH3}) give the initial values for $g_{0\alpha}$ and $g_{00}$.
The undetermined initial values for $\dot{g}_{0\alpha}$ and $\dot{g}_{00}$ correspond to 4 degrees of freedom for a free gravitational field.
A general gravitational field has 8 degrees of freedom: 4 degrees of freedom for a free gravitational field, 3 related to the four-velocity, and 1 related to the matter ($\epsilon$ or $p$).
The above analysis regarding $g_{ik}$ also applies to the Einstein equations (\ref{EC10}) because the torsion tensor is algebraically related to the matter.

The Einstein equations (\ref{EH3}) are a special case of the Einstein equations (\ref{EC10}).
They are valid when the matter fields do not depend on the affine connection, for which the spin density vanishes, so $U_{ik}=0$.
They are also an accurate approximation of (\ref{EC10}) when the matter fields depend on the connection but the tensor $U_{ik}$ can be neglected relative to $T_{ik}$.
In the metric-affine variational principle of stationary action, in which the variations $\delta\omega^{ab}_{\phantom{ab}i}$ are independent of $\delta e^i_a$, the spin density is independent of the energy--momentum density.
The Einstein and Cartan equations contain the covariant conservation laws for the energy--momentum and spin tensors, which generalize the special-relativistic conservation laws (\ref{clLg2}) and (\ref{clLg9}).
The angular momentum density in (\ref{clLg9}) contains both the orbital and intrinsic parts.

In the metric variational principle, in which the variations $\delta\omega^{ab}_{\phantom{ab}i}=\delta\mathring{\omega}^{ab}_{\phantom{ab}i}$ are functions of the variations $\delta e^i_a$ and their derivatives according to (\ref{spcon15}), the spin density is a function of the energy--momentum density.
The Einstein equations contain only the covariant conservation law for the energy--momentum tensor, which generalizes the special-relativistic conservation law (\ref{clLg2}) with a symmetric energy--momentum tensor.
The resulting conservation law (\ref{clLg9}) contains the angular momentum density only with the orbital part that depends on the energy--momentum density.
Accordingly, the metric variational principle does not account for the intrinsic angular momentum (spin) of matter.
Consequently, the existence of spin (which does not depend on energy and momentum) requires the metric-affine variational principle.\\

\noindent
{\bf Poisson equation}.\\
In the nonrelativistic limit, the leading component of the energy--momentum tensor (\ref{emtp9}) is
\begin{equation}
T_{00}=\mu c^2.
\end{equation}
The leading component of the Riemannian Ricci tensor is given by (\ref{grapot5}).
Therefore, the Einstein equations in the nonrelativistic limit reduce to the {\em Poisson equation}:
\begin{equation}
\triangle\phi=4\pi G\mu,
\label{grapot7}
\end{equation}
where
\begin{equation}
G=\frac{c^4\kappa}{8\pi}
\end{equation}
is {\em Newton's gravitational constant}.
In vacuum, where $\mu=0$, the Poisson equation reduces to the {\em Laplace equation}:
\begin{equation}
\triangle\phi=0.
\end{equation}
If $g_{ik}=\eta_{ik}$, then the general theory of relativity is said to reduce to the {\em special theory of relativity}.

\subsubsection{Utiyama action}
The action (\ref{EH1}) subjected to varying the tetrad is called the {\em Utiyama action}.
The Utiyama action is a special case of the Sciama--Kibble action, where the torsion tensor is approximated as zero and the spin connection is constrained to be equal to the Levi-Civita spin connection (\ref{spcon15}) which depends on the tetrad.
Using (\ref{dtemd1}) gives
\begin{equation}
\delta\mathring{S}=-\frac{1}{2\kappa c}\int\delta(\mathfrak{e}\mathring{R})d\Omega+\frac{1}{c}\int\mathfrak{T}^{\phantom{i}a}_i\delta e^i_a d\Omega.
\end{equation}
The Lagrangian density for the gravitational field is given by (\ref{Laggr4}), with the Riemann scalar $\mathring{R}$ given by (\ref{tetrep6}) and (\ref{tetrep8}):
\begin{equation}
\mathfrak{e}\mathring{R}=\mathfrak{e}e^i_a e^{jb}(\mathring{\omega}^a_{\phantom{a}bj,i}-\mathring{\omega}^a_{\phantom{a}bi,j}+\mathring{\omega}^a_{\phantom{a}ci}\mathring{\omega}^c_{\phantom{c}bj}-\mathring{\omega}^a_{\phantom{a}cj}\mathring{\omega}^c_{\phantom{c}bi})=2\mathfrak{e}^{ij}_{ab}(\mathring{\omega}^{ab}_{\phantom{ab}j,i}+\mathring{\omega}^a_{\phantom{a}ci}\mathring{\omega}^{cb}_{\phantom{cb}j}).
\end{equation}
Varying $\mathfrak{e}\mathring{R}$ and omitting total derivatives gives in the absence of torsion, using $\delta\mathfrak{e}=\mathfrak{e}e^i_a\delta e^a_i$ and $\mathfrak{e}^{ij}_{ab|j}=\mathfrak{e}^{ij}_{ab,j}-\mathring{\omega}^c_{\phantom{c}aj}\mathfrak{e}^{ij}_{cb}-\mathring{\omega}^c_{\phantom{c}bj}\mathfrak{e}^{ij}_{ac}=0$ (which results from (\ref{spcon7})),
\begin{eqnarray}
& & \delta(\mathfrak{e}\mathring{R})=(2\mathring{R}^a_{\phantom{a}i}-\mathring{R}e^a_i)\mathfrak{e}\delta e^i_a+2\mathfrak{e}^{ij}_{ab}\delta(\mathring{\omega}^{ab}_{\phantom{ab}j,i}+\mathring{\omega}^a_{\phantom{a}ci}\mathring{\omega}^{cb}_{\phantom{cb}j})=(2\mathring{R}^a_{\phantom{a}i}-\mathring{R}e^a_i)\mathfrak{e}\delta e^i_a \nonumber \\
& & +2(\mathfrak{e}^{ij}_{ab,j}-\mathring{\omega}^c_{\phantom{c}aj}\mathfrak{e}^{ij}_{cb}-\mathring{\omega}^c_{\phantom{c}bj}\mathfrak{e}^{ij}_{ac})\delta\mathring{\omega}^{ab}_{\phantom{ab}i}=(2\mathring{R}^a_{\phantom{a}i}-\mathring{R}e^a_i)\mathfrak{e}\delta e^i_a.
\end{eqnarray}
Equaling $\delta\mathring{S}=0$ gives
\begin{equation}
\mathring{R}^a_{\phantom{a}i}-\frac{1}{2}\mathring{R}e^a_i=\kappa t^{\phantom{i}a}_i,
\end{equation}
which is equivalent to the Einstein equations (\ref{EH3}) because of (\ref{dmemd5}) and (\ref{BelRos8}) (in the absence of torsion).

\subsubsection{Einstein pseudotensor and principle of equivalence}
\noindent
{\bf Energy--momentum pseudotensor for gravitational field}.\\
Because the noncovariant quantity $\mathcal{G}$ (\ref{Laggr6}) differs from $\sqrt{-\mathfrak{g}}\mathring{R}$ by a total divergence, we can use the Gau\ss--Stokes theorem (\ref{covint7}) to write the action for the gravitational field as
\begin{equation}
S_g=-\frac{1}{2\kappa c}\int\Bigl(\mathcal{G}-{\sf g}^{np}(C^i_{\phantom{i}ni}C^m_{\phantom{m}pm}-C^i_{\phantom{i}nm}C^m_{\phantom{m}pi})\Bigr)d\Omega-\frac{1}{2\kappa c}\oint\mathfrak{g}^{ik}\mathring{\Gamma}^{l}_{ik}dS_l+\frac{1}{2\kappa c}\oint\mathfrak{g}^{ik}\mathring{\Gamma}^{l}_{il}dS_k.
\label{Eps1}
\end{equation}
The hypersurface integrals in (\ref{Eps1}) do not contribute to the field equations and can be omitted.
The action for the gravitational field and matter (\ref{EC3}) thus reduces to
\begin{equation}
S=\frac{1}{c}\int\biggl(-\frac{1}{2\kappa}\Bigl(\mathcal{G}-{\sf g}^{np}(C^i_{\phantom{i}ni}C^m_{\phantom{m}pm}-C^i_{\phantom{i}nm}C^m_{\phantom{m}pi})\Bigr)+\mathfrak{L}_\textrm{m}\biggr)d\Omega,
\end{equation}
corresponding to the Lagrangian density:
\begin{equation}
\mathfrak{L}=-\frac{1}{2\kappa}\Bigl(\mathcal{G}-{\sf g}^{np}(C^i_{\phantom{i}ni}C^m_{\phantom{m}pm}-C^i_{\phantom{i}nm}C^m_{\phantom{m}pi})\Bigr)+\mathfrak{L}_\textrm{m}.
\label{Eps2}
\end{equation}
The condition $\delta_\textrm{g}S=0$, which is equivalent to
\begin{equation}
\frac{\delta}{\delta g^{jl}}\biggl(-\frac{1}{2\kappa}\Bigl(\mathcal{G}-{\sf g}^{np}(C^i_{\phantom{i}ni}C^m_{\phantom{m}pm}-C^i_{\phantom{i}nm}C^m_{\phantom{m}pi})\Bigr)+\mathfrak{L}_\textrm{m}\biggr)=0,
\label{Eps3}
\end{equation}
gives the Einstein equations (\ref{EC10}).
Because $\mathcal{G}$ depends on $g^{ij}$ and its first derivatives $g^{ij}_{\phantom{ij},k}$, we can construct a canonical energy--momentum density (\ref{clLg3}) corresponding to the gravitational field, treating $-\mathcal{G}/(2\kappa)$ like $\mathfrak{L}_\textrm{m}$ and $g^{ik}$ like a matter field $\phi$:
\begin{equation}
\mathfrak{t}^{\phantom{i}k}_i=-\frac{1}{2\kappa}\biggl(\frac{\partial\mathcal{G}}{\partial g^{jl}_{\phantom{jl},k}}g^{jl}_{\phantom{jl},i}-\delta^k_i\mathcal{G}\biggr).
\label{Eps4}
\end{equation}
This quantity is not a tensor density because $\mathcal{G}$ is not a scalar density.
Its division by $\sqrt{-\mathfrak{g}}$ defines the {\em Einstein energy--momentum pseudotensor} for the gravitational field:
\begin{equation}
\frac{\mathfrak{t}^{\phantom{i}k}_i}{\sqrt{-\mathfrak{g}}}=-\frac{1}{2\kappa\sqrt{-\mathfrak{g}}}\biggl(\frac{\partial(\sqrt{-\mathfrak{g}}{\sf G})}{\partial g^{jl}_{\phantom{jl},k}}g^{jl}_{\phantom{jl},i}-\delta^k_i\sqrt{-\mathfrak{g}}{\sf G}\biggr),
\label{Eps5}
\end{equation}
where ${\sf G}$ is the quantity (\ref{Laggr7}).\\

\noindent
{\bf Explicit form of pseudotensor}.\\
Using (\ref{Chrsym4}) and (\ref{Chrsym14}) gives
\begin{eqnarray}
& & \frac{\partial\mathring{\Gamma}^{m}_{il}}{\partial g^{rs}_{\phantom{rs},n}}=-\frac{1}{2}(g_{l(r}\delta^m_{s)}\delta^n_i+g_{i(r}\delta^m_{s)}\delta^n_l-g^{mn}g_{i(r}g_{s)l}), \\
& & \frac{\partial\mathring{\Gamma}^{l}_{ml}}{\partial g^{rs}_{\phantom{rs},n}}=-\frac{1}{2}g_{rs}\delta^n_m.
\end{eqnarray}
Consequently, we obtain
\begin{eqnarray}
& & \frac{\partial{\sf G}}{\partial g^{rs}_{\phantom{rs},n}}=2g^{ik}\mathring{\Gamma}^{m}_{il}\frac{\partial\mathring{\Gamma}^{l}_{mk}}{\partial g^{rs}_{\phantom{rs},n}}-g^{ik}\mathring{\Gamma}^{l}_{ml}\frac{\partial\mathring{\Gamma}^{m}_{ik}}{\partial g^{rs}_{\phantom{rs},n}}-g^{ik}\mathring{\Gamma}^{m}_{ik}\frac{\partial\mathring{\Gamma}^{l}_{ml}}{\partial g^{rs}_{\phantom{rs},n}} \nonumber \\
& & =-\mathring{\Gamma}^{n}_{rs}+\frac{1}{2}\Bigl(\mathring{\Gamma}^{l}_{sl}\delta^n_r+\mathring{\Gamma}^{l}_{rl}\delta^n_s-\mathring{\Gamma}^{l}_{ml}g^{mn}g_{rs}\Bigr)+\frac{1}{2}\mathring{\Gamma}^{n}_{jl}g^{jl}g_{rs},
\end{eqnarray}
which leads to
\begin{equation}
\frac{\partial\mathcal{G}}{\partial g^{rs}_{\phantom{rs},k}}g^{rs}_{\phantom{rs},i}=-\sqrt{-\mathfrak{g}}\mathring{\Gamma}^{k}_{rs}g^{rs}_{\phantom{rs},i}+\sqrt{-\mathfrak{g}}\mathring{\Gamma}^{l}_{rl}g^{rk}_{\phantom{rk},i}+\mathring{\Gamma}^{l}_{ml}g^{mk}(\sqrt{-\mathfrak{g}})_{,i}-\mathring{\Gamma}^{k}_{jl}g^{jl}(\sqrt{-\mathfrak{g}})_{,i}.
\end{equation}
The Einstein pseudotensor (\ref{Eps5}) can thus be written as
\begin{equation}
\frac{\mathfrak{t}^{\phantom{i}k}_i}{\sqrt{-\mathfrak{g}}}=\frac{1}{2\kappa\sqrt{-\mathfrak{g}}}(\mathring{\Gamma}^{k}_{lm}{\sf g}^{lm}_{\phantom{lm},i}-\mathring{\Gamma}^{l}_{ml}{\sf g}^{mk}_{\phantom{mk},i}+\delta^k_i\mathcal{G}).
\label{Eps17}
\end{equation}
Accordingly, $\mathfrak{t}_{ik}$ is not symmetric in the indices $i,k$.\\

\noindent
{\bf Principle of equivalence}.\\
Since the derivatives ${\sf g}^{ik}_{\phantom{ik},j}$ are homogeneous linear functions of the Christoffel symbols, the Einstein pseudotensor (\ref{Eps17}) is a homogeneous quadratic function of the Christoffel symbols, so it vanishes in the locally Galilean and geodesic frame of reference.
It can also differ from zero in the Minkowski spacetime (in the absence of the gravitational field) if we choose the coordinates such that the Christoffel symbols do not vanish.
Therefore, the energy of the gravitational field is not absolutely localized in spacetime; it depends on the choice of the coordinates.
A physically meaningful energy--momentum pseudotensor can be constructed if the coordinates are asymptotically (far from the sources of the field) Cartesian, so the Christoffel symbols tend asymptotically to zero.
If we neglect torsion, then the gravitational field can be always eliminated locally by transforming the coordinate system to the locally Galilean and geodesic frame of reference in which the Einstein pseudotensor vanishes.
This property of the gravitational field is referred to as the {\em principle of equivalence}.\\

\noindent
{\bf Conservation of total energy--momentum of field and matter}.\\
Differentiating (\ref{Eps4}) gives
\begin{eqnarray}
& & 2\kappa\mathfrak{t}^{\phantom{i}k}_{i\phantom{k},k}=-\partial_k\frac{\partial\mathcal{G}}{\partial g^{jl}_{\phantom{jl},k}}g^{jl}_{\phantom{jl},i}-\frac{\partial\mathcal{G}}{\partial g^{jl}_{\phantom{jl},k}}g^{jl}_{\phantom{jl},ik}+\mathcal{G}_{,i}=-\partial_k\frac{\partial\mathcal{G}}{\partial g^{jl}_{\phantom{jl},k}}g^{jl}_{\phantom{jl},i}-\frac{\partial\mathcal{G}}{\partial g^{jl}_{\phantom{jl},k}}g^{jl}_{\phantom{jl},ik}+\frac{\partial\mathcal{G}}{\partial g^{jl}}g^{jl}_{\phantom{jl},i} \nonumber \\
& & +\frac{\partial\mathcal{G}}{\partial g^{jl}_{\phantom{jl},k}}g^{jl}_{\phantom{jl},ki}=\biggl(\frac{\partial\mathcal{G}}{\partial g^{jl}}-\partial_k\frac{\partial\mathcal{G}}{\partial g^{jl}_{\phantom{jl},k}}\biggr)g^{jl}_{\phantom{jl},i}=\frac{\delta\mathcal{G}}{\delta g^{jl}}g^{jl}_{\phantom{jl},i}.
\end{eqnarray}
Using the metric energy--momentum density (\ref{dmemd4}), the quantity (\ref{EC11}), and the variation (\ref{Eps3}) leads to
\begin{equation}
\mathfrak{t}^{\phantom{i}k}_{i\phantom{k},k}=\frac{\delta\bigl(\mathfrak{L}_\textrm{m}+\sqrt{-\mathfrak{g}}g^{np}(C^k_{\phantom{k}nk}C^m_{\phantom{m}pm}-C^k_{\phantom{k}nm}C^m_{\phantom{m}pk})/(2\kappa)\bigr)}{\delta g^{jl}}g^{jl}_{\phantom{jl},i}=\frac{1}{2}({\cal T}_{jl}+\sqrt{-\mathfrak{g}}U_{jl})g^{jl}_{\phantom{jl},i}.
\end{equation}
Using (\ref{dmemd5}) and (\ref{EC21}), this divergence can be written in terms of the combined energy--momentum tensor:
\begin{equation}
\mathfrak{t}^{\phantom{i}k}_{i\phantom{k},k}=\frac{1}{2}\sqrt{-\mathfrak{g}}\tilde{T}_{jl}g^{jl}_{\phantom{jl},i}.
\label{Eps7}
\end{equation}
The Riemannian conservation law (\ref{EC20}) gives
\begin{equation}
(\sqrt{-\mathfrak{g}}\tilde{T}^{\phantom{i}k}_i)_{,k}=\mathring{\Gamma}^{l}_{ik}\sqrt{-\mathfrak{g}}\tilde{T}^{\phantom{l}k}_l=\frac{1}{2}g^{lm}g_{km,i}\sqrt{-\mathfrak{g}}\tilde{T}^{\phantom{l}k}_l=-\frac{1}{2}g^{lm}_{\phantom{lm},i}\sqrt{-\mathfrak{g}}\tilde{T}_{lm}.
\label{Eps8}
\end{equation}
The total energy--momentum density for the gravitational field and matter is given by
\begin{equation}
\mathfrak{t}^{\phantom{i}k}_i+\sqrt{-\mathfrak{g}}\tilde{T}^{\phantom{i}k}_i=\mathfrak{t}^{\phantom{i}k}_i+\frac{\sqrt{-\mathfrak{g}}}{\kappa}G^{\phantom{i}k}_i.
\label{Eps12}
\end{equation}
As a result of adding (\ref{Eps7}) and (\ref{Eps8}), the ordinary divergence of this quantity vanishes, as that in (\ref{clLg20}):
\begin{equation}
(\mathfrak{t}^{\phantom{i}k}_i+\sqrt{-\mathfrak{g}}\tilde{T}^{\phantom{i}k}_i)_{,k}=\biggl(\mathfrak{t}^{\phantom{i}k}_i+\frac{\sqrt{-\mathfrak{g}}}{\kappa}G^{\phantom{i}k}_i\biggr)_{,k}=0.
\label{Eps9}
\end{equation}

The conservation law (\ref{Eps9}) infers that the quantity (\ref{Eps12}) can be written as
\begin{equation}
\mathfrak{t}^{\phantom{i}k}_i+\frac{\sqrt{-\mathfrak{g}}}{\kappa}G^{\phantom{i}k}_i=\eta^{\phantom{i}kl}_{i\phantom{kl},l},
\label{Eps19}
\end{equation}
where
$\eta^{\phantom{i}kl}_i$ satisfies
\begin{equation}
\eta^{\phantom{i}kl}_i=-\eta^{\phantom{i}lk}_i,
\end{equation}
analogously to (\ref{clLg19}).
Equations (\ref{Chrsym11}), (\ref{Chrsym14}), (\ref{Laggr6}) and (\ref{Eps17}) infer that
\begin{equation}
\eta^{ikl}=\frac{1}{2\kappa\sqrt{-\mathfrak{g}}}\bigl((-\mathfrak{g})(g^{ik}g^{lm}-g^{il}g^{km})\bigr)_{,m}
\label{Eps28}
\end{equation}
satisfies (\ref{Eps19}).\\

\noindent
{\bf Four-momentum and angular momentum of field and matter}.\\
Integrating (\ref{Eps9}) over the four-volume and using Gau\ss' theorem (\ref{covint7}) gives
\begin{equation}
\oint(\mathfrak{t}^{\phantom{i}k}_i+\sqrt{-\mathfrak{g}}\tilde{T}^{\phantom{i}k}_i)dS_k=0.
\end{equation}
The corresponding four-momentum (\ref{fmam1}) of the gravitational field and matter, which is not a vector (it transforms like a vector only for Lorentz transformations), is therefore conserved:
\begin{equation}
P_i=\frac{1}{c}\int(\mathfrak{t}^{\phantom{i}k}_i+\sqrt{-\mathfrak{g}}\tilde{T}^{\phantom{i}k}_i)dS_k=\mbox{const}.
\label{Eps11}
\end{equation}
Putting (\ref{Eps17}) in this relation gives
\begin{equation}
P_i=\frac{1}{2\kappa c}\int\Bigl(\mathring{\Gamma}^{k}_{lm}(\sqrt{-\mathfrak{g}}g^{lm})_{,i}-\mathring{\Gamma}^{l}_{ml}
(\sqrt{-\mathfrak{g}}g^{mk})_{,i}+\delta^k_i\sqrt{-\mathfrak{g}}{\sf G}\Bigr)dS_k+\frac{1}{c}\int\sqrt{-\mathfrak{g}}\tilde{T}^{\phantom{i}k}_i dS_k.
\label{Eps23}
\end{equation}
This formula also follows from the relation analogous to the canonical energy--momentum density (\ref{cemd2}):
\begin{equation}
P_i=\frac{1}{c}\int\biggl(\sum_A\frac{\partial\mathfrak{L}}{\partial\phi_{A,k}}\phi_{A,i}-\delta^k_i\mathfrak{L}\biggr)dS_k,
\end{equation}
where the Lagrangian density is given by (\ref{Eps2}).
If the summation is taken over all matter fields $\phi_A$, then the matter Lagrangian density gives the second integral in (\ref{Eps23}).
If the contravariant components $g^{lm}$ of the metric tensor are taken as matter fields, then the gravitational Lagrangian density gives the first integral in (\ref{Eps23}).

Substituting (\ref{Eps19}) into (\ref{Eps11}) gives, using the Gau\ss--Stokes theorem (\ref{covint6}) and (\ref{Eps12}),
\begin{equation}
P_i=\frac{1}{c}\int\eta^{\phantom{i}kl}_{i\phantom{kl},l}dS_k=\frac{1}{2c}\int(\eta^{\phantom{i}kl}_{i\phantom{kl},l}dS_k-\eta^{\phantom{i}kl}_{i\phantom{kl},k}dS_l)=\frac{1}{2c}\oint\eta^{\phantom{i}kl}_{i\phantom{kl}}df^\ast_{kl},
\label{Eps21}
\end{equation}
where $df^\star_{ik}$ is the element of the closed surface which bounds the hypersurface.
If the hypersurface is a volume hypersurface with the element $dS_k=\delta_k^0 dV$, then the four-momentum of the gravitational field and matter (\ref{Eps21}) in a given volume can be written as a surface integral:
\begin{equation}
P_i=\frac{1}{2c}\oint\eta^{\phantom{i}0\alpha}_{i\phantom{0\alpha}}df^\ast_{0\alpha}=\frac{1}{c}\oint\eta^{\phantom{i}0\alpha}_{i\phantom{0\alpha}}df_\alpha,
\label{Eps22}
\end{equation}
where $df_\alpha$ is the element of the closed surface which bounds the volume.
If we neglect torsion, then $U_{ik}=0$ and the left-hand side of (\ref{Eps12}) reduces to $\mathfrak{t}^{\phantom{i}k}_i+\sqrt{-\mathfrak{g}}T^{\phantom{i}k}_i$.
The conservation law (\ref{Eps9}) reduces to $(\mathfrak{t}^{\phantom{i}k}_i+\sqrt{-\mathfrak{g}}T^{\phantom{i}k}_i)_{,k}=0$.
This quantity can also be written as (\ref{Eps19}), thereby the corresponding four-momentum of the gravitational field and matter is also given by (\ref{Eps22}).
We can construct from $P^i$ (\ref{Eps11}) the angular momentum four-tensor (\ref{fmam46}):
\begin{equation}
M^{ik}=\frac{1}{c}\int\bigl(x^i(\mathfrak{t}^{kl}+\sqrt{-\mathfrak{g}}\tilde{T}^{kl})-x^k(\mathfrak{t}^{il}+\sqrt{-\mathfrak{g}}\tilde{T}^{il})\bigr)dS_l,
\label{Eps18}
\end{equation}
which is not a tensor (it transforms like a tensor only for Lorentz transformations).
Because the quantity $\mathfrak{t}^{ik}$ is not symmetric, the four-tensor (\ref{Eps18}) is not conserved.

The construction of a conserved four-momentum for the gravitational field and matter is possible because the Lagrangian density for the gravitational field $\mathfrak{L}_\textrm{g}$ (\ref{Laggr2}) is linear in the curvature tensor.
Accordingly, it is linear in second derivatives of the metric tensor, so we can use the noncovariant quantity $\mathcal{G}$ (\ref{Laggr6}).
Another scalar density which is linear in curvature is $\epsilon^{ijkl}R_{ijkl}$.
Using (\ref{Chrsym20}), (\ref{Riem8}) and (\ref{Riem19}), and omitting a total derivative, this parity-violating quantity reduces to $-2\epsilon^{iklm}C^j_{\phantom{j}km}C_{jil}$, which does not depend on the derivatives of the metric tensor and thus does not describe a gravitational field.

\subsubsection{M{\o}ller pseudotensor}
{\bf Lagrangian of gravitational field without derivatives of spin connection}.\\
The Riemann scalar $\mathring{R}$ (\ref{tetrep8}) is linear in first derivatives of the Levi-Civita spin connection $\mathring{\omega}^a_{\phantom{a}bi}$:
\begin{eqnarray}
& & \mathfrak{e}\mathring{R}=(\mathfrak{e}e^i_a e^j_b \mathring{\omega}^{ab}_{\phantom{ab}j})_{,i}-(\mathfrak{e}e^i_a e^j_b)_{,i}\mathring{\omega}^{ab}_{\phantom{ab}j}-(\mathfrak{e}e^i_a e^j_b \mathring{\omega}^{ab}_{\phantom{ab}i})_{,j}+(\mathfrak{e}e^i_a e^j_b)_{,j}\mathring{\omega}^{ab}_{\phantom{ab}i}+\mathfrak{e}e^i_a e^j_b \mathring{\omega}^{ac}_{\phantom{ac}i}\mathring{\omega}^{\phantom{c}b}_{c\phantom{b}j} \nonumber \\
& & -\mathfrak{e}e^i_a e^j_b \mathring{\omega}^{ac}_{\phantom{ac}j}\mathring{\omega}^{\phantom{c}b}_{c\phantom{b}i}=2(\mathfrak{e}e^i_a e^j_b \mathring{\omega}^{ab}_{\phantom{ab}j})_{,i}-2(\mathfrak{e}e^i_a e^j_b)_{,i}\mathring{\omega}^{ab}_{\phantom{ab}j}+\mathfrak{e}e^i_a e^j_b \mathring{\omega}^{ac}_{\phantom{ac}i}\mathring{\omega}^{\phantom{c}b}_{c\phantom{b}j} \nonumber \\
& & -\mathfrak{e}e^i_a e^j_b \mathring{\omega}^{ac}_{\phantom{ac}j}\mathring{\omega}^{\phantom{c}b}_{c\phantom{b}i},
\end{eqnarray}
where we used (\ref{tetrep6}).
We can therefore subtract from $\mathfrak{e}\mathring{R}$ total derivatives without altering the field equations, replacing it by a noncovariant quantity $\mathcal{M}$ that does not contain first derivatives of the Levi-Civita spin connection:
\begin{eqnarray}
& & \mathcal{M}=-2(\mathfrak{e}e^i_a e^j_b)_{,i}\mathring{\omega}^{ab}_{\phantom{ab}j}+\mathfrak{e}e^i_a e^j_b \mathring{\omega}^{ac}_{\phantom{ac}i}\mathring{\omega}^{\phantom{c}b}_{c\phantom{b}j}-\mathfrak{e}e^i_a e^j_b \mathring{\omega}^{ac}_{\phantom{ac}j}\mathring{\omega}^{\phantom{c}b}_{c\phantom{b}i} \nonumber \\
& & =-2\mathfrak{e}(\mathring{\Gamma}^k_{ki}\mathring{\omega}^{ij}_{\phantom{ij}j}+\mathring{\omega}^i_{\phantom{i}ai}\mathring{\omega}^{aj}_{\phantom{aj}j}-\mathring{\Gamma}^i_{ki}\mathring{\omega}^{kj}_{\phantom{kj}j}+\mathring{\omega}^j_{\phantom{j}bi}\mathring{\omega}^{ib}_{\phantom{ib}j}-\mathring{\Gamma}^j_{ki}\mathring{\omega}^{ik}_{\phantom{ik}j}) \nonumber \\
& & +\mathfrak{e}(\mathring{\omega}^{ic}_{\phantom{ic}i}\mathring{\omega}^{\phantom{c}j}_{c\phantom{j}j}-\mathring{\omega}^{ic}_{\phantom{ic}j}\mathring{\omega}^{\phantom{c}j}_{c\phantom{j}i})=\mathfrak{e}(\mathring{\omega}^{ia}_{\phantom{ia}i}\mathring{\omega}^j_{\phantom{j}aj}-\mathring{\omega}^{ia}_{\phantom{ia}j}\mathring{\omega}^j_{\phantom{j}ai}),
\label{Mol2}
\end{eqnarray}
using (\ref{Chrsym14}) and (\ref{spcon15}).
We also define
\begin{equation}
{\sf M}=\frac{\mathcal{M}}{\mathfrak{e}}=\mathring{\omega}^{ia}_{\phantom{ia}i}\mathring{\omega}^j_{\phantom{j}aj}-\mathring{\omega}^{ia}_{\phantom{ia}j}\mathring{\omega}^j_{\phantom{j}ai}.
\end{equation}
The Riemannian part (\ref{Laggr4}) of the Lagrangian density for the gravitational field reduces accordingly to
\begin{equation}
\mathring{\mathfrak{L}}_\textrm{g}=-\frac{1}{2\kappa}\mathcal{M}=-\frac{1}{2\kappa}\mathfrak{e}{\sf M},
\end{equation}
similarly to (\ref{Laggr8}).\\

\noindent
{\bf Energy--momentum pseudotensor for gravitational field}.\\
Because the noncovariant quantity $\mathcal{M}$ (\ref{Mol2}) differs from $\mathfrak{e}\mathring{R}$ by a total divergence, the action for the gravitational field and matter (\ref{EC3}) is equivalent to
\begin{equation}
S=\frac{1}{c}\int\biggl(-\frac{1}{2\kappa}\Bigl(\mathcal{M}-\mathfrak{e}(C^{ia}_{\phantom{ia}i}C^j_{\phantom{j}aj}-C^{ia}_{\phantom{ia}j}C^j_{\phantom{j}ai})\Bigr)+\mathfrak{L}_\textrm{m}\biggr)d\Omega.
\end{equation}
Using (\ref{spcon17}), this action can be written as
\begin{equation}
S=\frac{1}{c}\int\biggl(\frac{\mathfrak{e}}{2\kappa}(\omega^{ia}_{\phantom{ia}i}\omega^j_{\phantom{j}aj}-\omega^{ia}_{\phantom{ia}j}\omega^j_{\phantom{j}ai}-2\omega^{ia}_{\phantom{ia}i}\mathring{\omega}^j_{\phantom{j}aj}+2\omega^{ia}_{\phantom{ia}j}\mathring{\omega}^j_{\phantom{j}ai})+\mathfrak{L}_\textrm{m}\biggr)d\Omega.
\end{equation}
The condition $\delta_\textrm{e}S=0$, which is equivalent to
\begin{equation}
\frac{\delta}{\delta e^j_a}\biggl(-\frac{1}{2\kappa}\Bigl(\mathcal{M}-\mathfrak{e}g^{np}(C^i_{\phantom{i}ni}C^m_{\phantom{m}pm}-C^i_{\phantom{i}nm}C^m_{\phantom{m}pi})\Bigr)+\mathfrak{L}_\textrm{m}\biggr)=0,
\label{Mol7}
\end{equation}
gives the Einstein equations (\ref{SK7}).
Because of (\ref{spcon1}), $\mathcal{M}$ depends on the tetrad $e^i_a$ and its first derivatives $e^i_{a,j}$.
Therefore, analogously to (\ref{Eps4}), we can construct a canonical energy--momentum density corresponding to the gravitational field, treating $-\mathcal{M}/(2\kappa)$ like $\mathfrak{L}_\textrm{m}$ and $e^i_a$ like a matter field $\phi$:
\begin{equation}
\mathfrak{m}^{\phantom{i}k}_i=-\frac{1}{2\kappa}\biggl(\frac{\partial\mathcal{M}}{\partial e^j_{a,k}}e^j_{a,i}-\delta^k_i\mathcal{M}\biggr).
\label{Mol8}
\end{equation}
This quantity is not a tensor density because $\mathcal{M}$ is not a scalar density.
Its division by $\mathfrak{e}$ defines the {\em M{\o}ller energy--momentum pseudotensor} for the gravitational field:
\begin{equation}
\frac{\mathfrak{m}^{\phantom{i}k}_i}{\mathfrak{e}}=-\frac{1}{2\kappa\mathfrak{e}}\biggl(\frac{\partial(\mathfrak{e}{\sf M})}{\partial e^j_{a,k}}e^j_{a,i}-\delta^k_i\mathfrak{e}{\sf M}\biggr).
\label{Mol9}
\end{equation}

\noindent
{\bf Explicit form of pseudotensor}.\\
Using (\ref{Chrsym4}), (\ref{Chrsym14}), (\ref{tet5}), (\ref{tet10}), and (\ref{spcon15}) gives
\begin{eqnarray}
& & \frac{\partial\mathring{\omega}^i_{\phantom{i}aj}}{\partial e^m_{b,l}}=\frac{1}{2}(\delta^i_m\delta^b_a\delta^l_j-\delta^i_m e^l_a e^b_j-e^{ib}e^l_a g_{jm}-e^{ib}e_{am}\delta^l_j+g^{il}e_{am}e^b_j+g^{il}\delta^b_a g_{jm}), \\
& & \frac{\partial\mathring{\omega}^i_{\phantom{i}ai}}{\partial e^m_{b,l}}=\delta^b_a\delta^l_m-e^l_a e^b_m.
\end{eqnarray}
Consequently, we obtain
\begin{equation}
\frac{\partial{\sf M}}{\partial e^m_{b,l}}=2(\delta^l_m\mathring{\omega}^{jb}_{\phantom{jb}j}-e^b_m\mathring{\omega}^{jl}_{\phantom{jl}j}-\mathring{\omega}^{lb}_{\phantom{lb}m}+\mathring{\omega}_m^{\phantom{m}bl}+\mathring{\omega}_m^{\phantom{m}lb}).
\end{equation}
The M{\o}ller energy--momentum pseudotensor (\ref{Mol9}) can thus be written, using (\ref{spcon15}), as
\begin{equation}
\frac{\mathfrak{m}^{\phantom{i}k}_i}{\mathfrak{e}}=\frac{1}{\kappa}\biggl(-\mathring{\omega}^k_{\phantom{k}ai}\mathring{\omega}^{ja}_{\phantom{ja}j}-\mathring{\omega}^j_{\phantom{j}ai}\mathring{\omega}_j^{\phantom{j}ak}+\mathring{\Gamma}^k_{li}\mathring{\omega}^{jl}_{\phantom{jl}j}-\mathring{\Gamma}^l_{li}\mathring{\omega}^{jk}_{\phantom{jk}j}+\mathring{\omega}^{ljk}g_{il,j}-\mathring{\omega}^{kjl}g_{jl,i}+\frac{1}{2}\delta^k_i{\sf M}\biggr).
\label{Mol21}
\end{equation}
Accordingly, $\mathfrak{m}_{ik}$ is not symmetric in the indices $i,k$.\\

\noindent
{\bf Conservation of total energy--momentum of field and matter}.\\
Differentiating (\ref{Mol8}) gives
\begin{eqnarray}
& & 2\kappa\mathfrak{m}^{\phantom{i}k}_{i\phantom{k},k}=-\partial_k\frac{\partial\mathcal{M}}{\partial e^j_{a,k}}e^j_{a,i}-\frac{\partial\mathcal{M}}{\partial e^j_{a,k}}e^j_{a,ik}+\mathcal{M}_{,i}=-\partial_k\frac{\partial\mathcal{M}}{\partial e^j_{a,k}}e^j_{a,i}-\frac{\partial\mathcal{M}}{\partial e^j_{a,k}}e^j_{a,ik}+\frac{\partial\mathcal{M}}{\partial e^j_a}e^j_{a,i} \nonumber \\
& & +\frac{\partial\mathcal{M}}{\partial e^j_{a,k}}e^j_{a,ik}=\biggl(\frac{\partial\mathcal{M}}{\partial e^j_a}-\partial_k\frac{\partial\mathcal{M}}{\partial e^j_{a,k}}\biggr)e^j_{a,i}=\frac{\delta\mathcal{M}}{\delta e^j_a}e^j_{a,i}.
\end{eqnarray}
Using the tetrad energy--momentum density (\ref{dtemd3}), the quantity (\ref{EC11}), and the variation (\ref{Mol7}) leads to
\begin{equation}
\mathfrak{m}^{\phantom{i}k}_{i\phantom{k},k}=\frac{\delta\bigl(\mathfrak{L}_\textrm{m}+\mathfrak{e}g^{np}(C^k_{\phantom{k}nk}C^m_{\phantom{m}pm}-C^k_{\phantom{k}nm}C^m_{\phantom{m}pk})/(2\kappa)\bigr)}{\delta e^j_a}e^j_{a,i}=(\mathfrak{T}^{\phantom{j}a}_j+\mathfrak{e}U^{\phantom{j}a}_j)e^j_{a,i}.
\label{Mol11}
\end{equation}
The conservation law (\ref{Eps8}) gives, using (\ref{tet5}),
\begin{equation}
(\mathfrak{e}\tilde{T}^{\phantom{i}k}_i)_{,k}=-\frac{1}{2}g^{lm}_{\phantom{lm},i}\mathfrak{e}\tilde{T}_{lm}=-e^j_{a,i}\mathfrak{e}\tilde{T}^{\phantom{j}a}_j.
\label{Mol12}
\end{equation}
Adding (\ref{Mol11}) and (\ref{Mol12}) leads, using (\ref{SK7}) and (\ref{SK11}), to
\begin{equation}
(\mathfrak{m}^{\phantom{i}k}_i+\mathfrak{e}\tilde{T}^{\phantom{i}k}_i)_{,k}=e^j_{a,i}(\mathfrak{T}^{\phantom{j}a}_j-{\cal T}^{\phantom{j}a}_j)=e^j_{a,i}\frac{\mathfrak{e}}{\kappa}\nabla_k^\ast(C^{ka}_{\phantom{ka}j}-2S^a\delta^k_j+2S^k e^a_j).
\end{equation}

In the absence of torsion, $U_{ik}=0$ and the total energy--momentum density for the gravitational field and matter is given by
\begin{equation}
\mathfrak{m}^{\phantom{i}k}_i+\mathfrak{e}T^{\phantom{i}k}_i=\mathfrak{m}^{\phantom{i}k}_i+\frac{\mathfrak{e}}{\kappa}G^{\phantom{i}k}_i.
\end{equation}
The ordinary divergence of this quantity vanishes, as that in (\ref{Eps9}):
\begin{equation}
(\mathfrak{m}^{\phantom{i}k}_i+\mathfrak{e}T^{\phantom{i}k}_i)_{,k}=0.
\label{Mol14}
\end{equation}

\noindent
{\bf Four-momentum and angular momentum of field and matter}.\\
Integrating (\ref{Mol14}) over the four-volume and using Gau\ss' theorem gives
\begin{equation}
\oint(\mathfrak{m}^{\phantom{i}k}_i+\mathfrak{e}T^{\phantom{i}k}_i)dS_k=0.
\end{equation}
The corresponding four-momentum (\ref{fmam1}) of the gravitational field and matter, which is not a vector (it transforms like a vector only for Lorentz transformations), is therefore conserved:
\begin{equation}
P_i=\frac{1}{c}\int(\mathfrak{m}^{\phantom{i}k}_i+\mathfrak{e}T^{\phantom{i}k}_i)dS_k=\mbox{const}.
\label{Mol16}
\end{equation}
We can construct from $P^i$ (\ref{Mol16}) the angular momentum four-tensor (\ref{fmam46}):
\begin{equation}
M^{ik}=\frac{1}{c}\int\bigl(x^i(\mathfrak{m}^{kl}+\mathfrak{e}T^{kl})-x^k(\mathfrak{m}^{il}+\mathfrak{e}T^{il})\bigr)dS_l,
\label{Mol22}
\end{equation}
which is not a tensor (it transforms like a tensor only for Lorentz transformations).
Because the quantity $\mathfrak{m}^{ik}$ is not symmetric, the four-tensor (\ref{Mol22}) is not conserved.

Since the Levi-Civita spin connection and the Christoffel symbols are homogeneous linear functions of the derivatives $e^i_{a,k}$, these derivatives are homogeneous linear functions of the Christoffel symbols.
Consequently, the M{\o}ller energy--momentum pseudotensor (\ref{Mol21}) is a homogeneous quadratic function of the Christoffel symbols, so it vanishes in the locally Galilean and geodesic frame of reference.
This pseudotensor depends on the choice of both the coordinates and tetrad.
To fix the tetrad, one can impose on it 6 constraints which are covariant under constant Lorentz transformations but not under general Lorentz transformations (otherwise such constraints would not fix the tetrad since Lorentz transformations are tetrad rotations).

\subsubsection{Landau--Lifshitz energy--momentum pseudotensor}
We define
\begin{eqnarray}
& & \lambda^{iklm}=\frac{1}{2\kappa}(-\mathfrak{g})(g^{ik}g^{lm}-g^{il}g^{km}), \label{Eps25} \\
& & h^{ikl}=\lambda^{iklm}_{\phantom{iklm},m}=-h^{ilk}.
\label{Eps26}
\end{eqnarray}
The quantity (\ref{Eps25}) satisfies the identities:
\begin{equation}
\lambda^{iklm}-\lambda^{lkim}=\lambda^{ikml},\quad \lambda^{iklm}=-\lambda^{ilkm}.
\label{Eps27}
\end{equation}
The quantity (\ref{Eps26}) is equal, by means of (\ref{Chrsym11}) and (\ref{Chrsym14}), to
\begin{eqnarray}
& & h^{ikl}=\frac{1}{2\kappa}(-\mathfrak{g})(\mathring{\Gamma}^{l}_{jm}g^{km}g^{ij}+\mathring{\Gamma}^{k}_{mn}g^{mn}g^{il}-\mathring{\Gamma}^{m}_{mn}g^{kn}g^{il} \nonumber \\
& & -\mathring{\Gamma}^{k}_{jm}g^{lm}g^{ij}-\mathring{\Gamma}^{l}_{mn}g^{mn}g^{ik}+\mathring{\Gamma}^{m}_{mn}g^{ln}g^{ik}).
\end{eqnarray}

In the locally Galilean and geodesic system of coordinates, the Einstein equations (\ref{EC21}) are
\begin{equation}
\mathring{R}^{ik}-\frac{1}{2}\mathring{R}g^{ik}=\kappa\tilde{T}^{ik},
\end{equation}
with the Ricci tensor obtained from the Riemann tensor (\ref{Riem3}) with the Christoffel symbols equal to zero.
\begin{equation}
\mathring{R}^{np}=\frac{1}{2}g^{kn}g^{mp}g^{il}(g_{im,kl}+g_{kl,im}-g_{il,km}-g_{km,il}).
\end{equation}
These equations can be written as
\begin{equation}
\tilde{T}^{ik}=\biggl(\frac{1}{2\kappa(-\mathfrak{g})}\bigl((-\mathfrak{g})(g^{ik}g^{lm}-g^{il}g^{km})\bigr)_{,m}\biggr)_{,l}.
\label{LLps12}
\end{equation}
Because the first derivatives of the metric tensor in this system are zero, the factor $1/(-\mathfrak{g})$ can be carried outside the derivative, yielding
\begin{equation}
(-\mathfrak{g})
\tilde{T}^{ik}=h^{ikl}_{\phantom{ikl},l}.
\label{LLps1}
\end{equation}
Using $(-\mathfrak{g})=1$, the Einstein equations simplify to
\begin{equation}
\tilde{T}^{ik}=h^{ikl}_{\phantom{ikl},l},
\label{LLps2}
\end{equation}
and the Riemannian conservation law (\ref{EC20}) reduces to $\tilde{T}^{ik}_{\phantom{ik},k}=0$, which is consistent with (\ref{LLps2}).

In an arbitrary system of coordinates, (\ref{LLps1}) is not valid.
We define a quantity ${\sf t}^{ik}$ such that
\begin{equation}
(-\mathfrak{g})({\sf t}^{ik}+\tilde{T}^{ik})=h^{ikl}_{\phantom{ikl},l}.
\label{LLps3}
\end{equation}
Therefore, the following divergence vanishes:
\begin{equation}
\bigl((-\mathfrak{g})({\sf t}^{ik}+\tilde{T}^{ik})\bigr)_{,k}=0.
\label{LLps4}
\end{equation}
The quantity ${\sf t}^{ik}$ is referred to as the {\em Landau--Lifshitz energy--momentum pseudotensor} for the gravitational field.\\

\noindent
{\bf Explicit form of pseudotensor}.\\
The explicit expression for the Landau--Lifshitz pseudotensor is
\begin{eqnarray}
& & {\sf t}^{ik}=\frac{1}{2\kappa}\Bigl((g^{il}g^{km}-g^{ik}g^{lm})(2\mathring{\Gamma}^{n}_{lm}\mathring{\Gamma}^{p}_{np}-\mathring{\Gamma}^{n}_{lp}\mathring{\Gamma}^{p}_{mn}-\mathring{\Gamma}^{n}_{ln}\mathring{\Gamma}^{p}_{mp}) \nonumber \\
& & +g^{il}g^{mn}(\mathring{\Gamma}^{k}_{lp}\mathring{\Gamma}^{p}_{mn}+\mathring{\Gamma}^{k}_{mn}\mathring{\Gamma}^{p}_{lp}-\mathring{\Gamma}^{k}_{np}\mathring{\Gamma}^{p}_{lm}-\mathring{\Gamma}^{k}_{lm}\mathring{\Gamma}^{p}_{np}) \nonumber \\
& & +g^{kl}g^{mn}(\mathring{\Gamma}^{i}_{lp}\mathring{\Gamma}^{p}_{mn}+\mathring{\Gamma}^{i}_{mn}\mathring{\Gamma}^{p}_{lp}-\mathring{\Gamma}^{i}_{np}\mathring{\Gamma}^{p}_{lm}-\mathring{\Gamma}^{i}_{lm}\mathring{\Gamma}^{p}_{np}) \nonumber \\
& & +g^{lm}g^{np}(\mathring{\Gamma}^{i}_{ln}\mathring{\Gamma}^{k}_{mp}-\mathring{\Gamma}^{i}_{lm}\mathring{\Gamma}^{k}_{np})\Bigr)
\label{LLps7}
\end{eqnarray}
or
\begin{eqnarray}
& & (-\mathfrak{g}){\sf t}^{ik}=\frac{1}{2\kappa}\biggl({\sf g}^{ik}_{\phantom{ik},l}{\sf g}^{lm}_{\phantom{lm},m}-{\sf g}^{il}_{\phantom{il},l}{\sf g}^{km}_{\phantom{km},m}+\frac{1}{2}g^{ik}g_{lm}{\sf g}^{ln}_{\phantom{ln},p}{\sf g}^{pm}_{\phantom{pm},n} \nonumber \\
& & -(g^{il}g_{mn}{\sf g}^{kn}_{\phantom{kn},p}{\sf g}^{mp}_{\phantom{mp},l}+g^{kl}g_{mn}{\sf g}^{in}_{\phantom{in},p}{\sf g}^{mp}_{\phantom{mp},l})+g_{lm}g^{np}{\sf g}^{il}_{\phantom{il},n}{\sf g}^{km}_{\phantom{km},p} \nonumber \\
& & +\frac{1}{8}(2g^{il}g^{km}-g^{ik}g^{lm})(2g_{np}g_{qr}-g_{pq}g_{nr}){\sf g}^{nr}_{\phantom{nr},l}{\sf g}^{pq}_{\phantom{pq},m}\biggr).
\end{eqnarray}
This pseudotensor is symmetric in the indices $i,k$.
Without carrying the factor $1/(-\mathfrak{g})$ outside the derivative in (\ref{LLps12}), ${\sf t}^{ik}$ would not be symmetric.\\

\noindent
{\bf Four-momentum and angular momentum of field and matter}.\\
Integrating (\ref{LLps4}) over the four-volume and using Gau\ss' theorem gives
\begin{equation}
\oint(-\mathfrak{g})({\sf t}^{ik}+\tilde{T}^{ik})dS_k=0.
\end{equation}
The corresponding four-momentum (\ref{fmam1}) of the gravitational field and matter is therefore conserved:
\begin{equation}
P^i=\frac{1}{c}\int(-\mathfrak{g})({\sf t}^{ik}+\tilde{T}^{ik})dS_k=\mbox{const}.
\label{LLps5}
\end{equation}
The quantity ${\sf t}^{ik}$ is not a tensor density, thereby the conserved four-momentum $P^i$ (\ref{LLps5}) is not a vector.
The four-momentum $P^i$ is not a vector even for Lorentz transformations, because of the factor $(-\mathfrak{g})$ instead of the weight-1 density $\sqrt{-\mathfrak{g}}$.
Dividing $P^i$ by $\sqrt{-\mathfrak{g}}$ at some fixed point (a natural choice is infinity) turns it into a vector with respect to Lorentz transformations.
For the volume hypersurface $x^0=\mbox{const}$ with the element $dS_0=dV$, the four-momentum is
\begin{equation}
P^i=\frac{1}{c}\int(-\mathfrak{g})({\sf t}^{i0}+\tilde{T}^{i0})dV.
\end{equation}
The quantity $(-\mathfrak{g})({\sf t}^{00}+\tilde{T}^{00})$ is the total energy density of the field and matter, $(-\mathfrak{g})({\sf t}^{\alpha 0}+\tilde{T}^{\alpha 0})/c$ are the components of the total momentum density, $(-\mathfrak{g})({\sf t}^{0\beta}+\tilde{T}^{0\beta})c$ are the components of the total energy flux density, and $(-\mathfrak{g})({\sf t}^{\alpha\beta}+\tilde{T}^{\alpha\beta})$ are the components of the total momentum flux density.
Without matter, these components contain only ${\sf t}^{ik}$ and represent the gravitational field alone.
Using (\ref{LLps3}) turns (\ref{LLps5}), for the volume hypersurface bounded by the closed spatial surface $df_\alpha$, into
\begin{equation}
P^i=\frac{1}{c}\int h^{ikl}_{\phantom{ikl},l}dS_k=\frac{1}{2c}\int(h^{ikl}_{\phantom{ikl},l}dS_k-h^{ikl}_{\phantom{ikl},k}dS_l)=\frac{1}{2c}\oint h^{ikl}df^\ast_{kl}=\frac{1}{c}\oint h^{i0\alpha}df_\alpha.
\label{LLps6}
\end{equation}

Because the quantity ${\sf t}^{ik}$ is symmetric, the corresponding angular momentum four-tensor of the gravitational field and matter, constructed from $P^i$ as that in (\ref{fmam46}), is conserved:
\begin{equation}
M^{ik}=\frac{1}{c}\int(-\mathfrak{g})\bigl(x^i({\sf t}^{kl}+\tilde{T}^{kl})-x^k({\sf t}^{il}+\tilde{T}^{il})\bigr)dS_l=\mbox{const}.
\label{LLps9}
\end{equation}
Dividing $M^{ik}$ by $\sqrt{-\mathfrak{g}}$ at some fixed point turns it into an antisymmetric tensor with respect to Lorentz transformations.
For the volume hypersurface, the angular momentum four-tensor is
\begin{equation}
M^{ik}=\frac{1}{c}\int(-\mathfrak{g})\bigl(x^i({\sf t}^{k0}+\tilde{T}^{k0})-x^k({\sf t}^{i0}+\tilde{T}^{i0})\bigr)dV.
\label{LLps11}
\end{equation}
Using (\ref{Eps26}), (\ref{Eps27}), and (\ref{LLps3}) for the volume hypersurface turns (\ref{LLps9}) into
\begin{eqnarray}
& & M^{ik}=\frac{1}{c}\int(x^i h^{klm}_{\phantom{klm},m}-x^k h^{ilm}_{\phantom{ilm},m})dS_l=\frac{1}{c}\int(x^i\lambda^{klmn}_{\phantom{klmn},nm}-x^k\lambda^{ilmn}_{\phantom{ilmn},nm})dS_l \nonumber \\
& & =\frac{1}{c}\int(x^i\lambda^{klmn}_{\phantom{klmn},n}-x^k\lambda^{ilmn}_{\phantom{ilmn},n})_{,m}dS_l-\frac{1}{c}\oint(\delta^i_m\lambda^{klmn}-\delta^k_m\lambda^{ilmn})_{,n}dS_l \nonumber \\
& & =\frac{1}{2c}\oint(x^i\lambda^{klmn}_{\phantom{klmn},n}-x^k\lambda^{ilmn}_{\phantom{ilmn},n})df^\ast_{lm}-\frac{1}{2c}\oint(\lambda^{klin}-\lambda^{ilkn})df^\ast_{ln} \nonumber \\
& & =\frac{1}{c}\oint(x^i h^{k0\alpha}-x^k h^{i0\alpha}+\lambda^{i0\alpha k})df_\alpha.
\label{LLps10}
\end{eqnarray}
The conservation of $M^{0\alpha}$ in (\ref{LLps11}) divided by the conserved $P^0$ in (\ref{LLps5}) gives a uniform motion (\ref{fmam48}) (without the intrinsic angular momentum) of the center of inertia for the gravitational field and matter.
The velocity of this motion is given by (\ref{fmam49}) and (\ref{LLps5}), and the coordinates of the center of inertia are
\begin{equation}
X^\alpha=\frac{\int x^\alpha(-\mathfrak{g})({\sf t}^{00}+\tilde{T}^{00})dV}{\int(-\mathfrak{g})({\sf t}^{00}+\tilde{T}^{00})dV}.
\end{equation}
These coordinates, like (\ref{fmam50}), are not the spatial components of a four-dimensional vector.\\

\noindent
{\bf Relation to Einstein pseudotensor}.\\
The Einstein and Landau--Lifshitz pseudotensors are examples of quantities which in the absence of the gravitational field reduce to $\tilde{T}^{ik}$, and which upon integration over $dS_k$ give a conservation of some quantity. 
There exists an infinite number of such pseudotensors, but the Landau--Lifshitz pseudotensor is the only one which contains only the first derivatives of $g_{ik}$ and is also symmetric.
The quantity (\ref{Eps26}) is related to $\eta^{ikl}$ (\ref{Eps28}) by
\begin{equation}
h^{ikl}=\sqrt{-\mathfrak{g}}\eta^{ikl},
\label{LLps16}
\end{equation}
which relates the Einstein and Landau--Lifshitz pseudotensors.
Accordingly, the Landau--Lifshitz four-momentum (\ref{LLps6}) differs from the Einstein four-momentum (\ref{Eps22}) by an additional factor $\sqrt{-\mathfrak{g}}$ in the integrand.

The quantity $\eta^{\phantom{i}kl}_i$ is not unique because the relation (\ref{Eps19}) is invariant under a transformation
\begin{equation}
\eta^{\phantom{ik}l}_{ik}\rightarrow\eta^{\phantom{ik}l}_{ik}+\zeta^{\phantom{i}klm}_{i\phantom{klm},m},
\end{equation}
where $\zeta^{\phantom{i}klm}_i$ is a quantity satisfying
\begin{equation}
\zeta^{\phantom{i}klm}_i=-\zeta^{\phantom{i}kml}_i.
\end{equation}
Taking $\eta^{\phantom{i}kl}_i$, which is related to $h^{\phantom{i}kl}_i$ in a different way than (\ref{LLps16}), leads to a different energy--momentum pseudotensor.
For example, choosing
\begin{equation}
\eta^{\phantom{i}kl}_i=\frac{1}{\sqrt{-\mathfrak{g}}}(2h^{\phantom{i}kl}_i-\delta^k_i h^{\phantom{j}jl}_j+\delta^l_i h^{\phantom{j}jk}_j)
\end{equation}
simplifies the right-hand side of (\ref{Eps19}):
\begin{equation}
\eta^{\phantom{i}kl}_{i\phantom{kl},l}=\frac{\sqrt{-\mathfrak{g}}}{\kappa}(g_{in,m}-g_{im,n})g^{km}g^{ln}.
\end{equation}

\begin{footnotesize}
\subsubsection{Palatini variation}
Instead of varying the action for the gravitational field and matter with respect to the torsion tensor, we can vary it with respect to the affine connection ($\delta\Gamma^{j}_{ik}$ is a tensor) and use the metric compatibility of the connection (\ref{metten5}).
Varying $S_\textrm{g}$ in (\ref{EC1}) with respect to $\Gamma^{k}_{ij}$ gives, by means of (\ref{Ric7}),
\begin{equation}
\delta_{\Gamma}S_\textrm{g}=-\frac{1}{2\kappa c}\int\delta R_{ik}{\sf g}^{ik}d\Omega=-\frac{1}{2\kappa c}\int\bigl((\delta\Gamma^{l}_{ik})_{;l}-(\delta\Gamma^{l}_{il})_{;k}-2S^j_{\phantom{j}lk}\delta\Gamma^{l}_{ij}\bigl)g^{ik}\sqrt{-\mathfrak{g}}d\Omega.
\label{Pal1}
\end{equation}
Partial integration and omitting total derivatives in (\ref{Pal1}) gives, using (\ref{parint3}),
\begin{equation}
\delta_{\Gamma}S_\textrm{g}=\frac{1}{2\kappa c}\int(\delta\Gamma^{l}_{ik}{\sf g}^{ik}_{\phantom{ik};l}-2S_l\delta\Gamma^{l}_{ik}{\sf g}^{ik}-\delta\Gamma^{l}_{il}{\sf g}^{ik}_{\phantom{ik};k}+2S_k\delta\Gamma^{l}_{il}{\sf g}^{ik}+2S^j_{\phantom{j}lk}\delta\Gamma^{l}_{ij}{\sf g}^{ik})d\Omega.
\end{equation}
The variation of the action is thus
\begin{eqnarray}
& & \delta_{\Gamma}S=\frac{1}{2\kappa c}\int(\delta\Gamma^{l}_{ik}{\sf g}^{ik}_{\phantom{ik};l}-2S_l\delta\Gamma^{l}_{ik}{\sf g}^{ik}-\delta\Gamma^{l}_{il}{\sf g}^{ik}_{\phantom{ik};k}+2S_k\delta\Gamma^{l}_{il}{\sf g}^{ik}+2S^j_{\phantom{j}lk}\delta\Gamma^{l}_{ij}{\sf g}^{ik})d\Omega \nonumber \\
& & +\frac{1}{2c}\int\Pi^{i\phantom{j}k}_{\phantom{i}j}\delta\Gamma^{j}_{ik}d\Omega,
\end{eqnarray}
where
\begin{equation}
\Pi^{i\phantom{j}k}_{\phantom{i}j}=2\frac{\delta\mathfrak{L}_\textrm{m}}{\delta\Gamma^{j}_{ik}}.
\end{equation}

Since the connection is metric-compatible, the condition $\delta S=0$ gives
\begin{equation}
g^{ik}S_j-\delta^k_j S^i+S^{ki}_{\phantom{ki}j}=\frac{\kappa}{2\sqrt{-\mathfrak{g}}}\Pi^{i\phantom{j}k}_{\phantom{i}j}.
\label{Pal5}
\end{equation}
Comparing (\ref{Pal5}) with the Cartan equations (\ref{SK6}) shows that
\begin{equation}
\Pi^{i\phantom{j}k}_{\phantom{i}j}=-\mathfrak{S}^{i\phantom{j}k}_{\phantom{i}j}=-\Pi_j^{\phantom{j}ik}.
\label{Pal6}
\end{equation}
Contracting the indices $i,j$ gives
\begin{equation}
\Pi^{i\phantom{i}k}_{\phantom{i}i}=0,
\end{equation}
which also results from the invariance of the Lagrangian density under a projective transformation (\ref{affgeo10}) (the symmetric part of the Ricci tensor is invariant under this transformation):
\begin{equation}
\delta\mathfrak{L}=\delta\mathfrak{L}_\textrm{m}=\frac{1}{2}\Pi^{i\phantom{j}k}_{\phantom{i}j}\delta\Gamma^{j}_{ik}=\frac{1}{2}\Pi^{i\phantom{j}k}_{\phantom{i}j}\delta^j_i\delta A_k=0.
\end{equation}
The antisymmetry relation in (\ref{Pal6}) algebraically constrains possible forms of matter Lagrangians.
Thereby, it is not a conservation law.
If the matter Lagrangian density $\mathfrak{L}_\textrm{m}$ does not depend on the affine connection, then the variation of the action with respect to the connection is referred to as the {\em Palatini variation}.
In this case, (\ref{Pal5}) turns the torsion tensor into zero, so the connection is formed by the Christoffel symbols and the field equations are the Einstein equations (\ref{EH3}).
\end{footnotesize}

\subsubsection{Raychaudhuri equation}
\label{Ray}
Let us consider a congruence of particles with four-velocity $u^i$.
We define the {\em expansion scalar} $\theta$, the traceless {\em shear tensor} $\sigma_{ik}$, and the antisymmetric {\em vorticity tensor} $\omega_{ik}$ according to
\begin{eqnarray}
& & \theta=u^i_{\phantom{i}:i}, \\
& & \sigma_{ik}=u_{(i:k)}-\frac{1}{3}\theta h_{ik}-w_{(i}u_{k)}, \\
& & \omega_{ik}=u_{[i:k]}-w_{[i}u_{k]},
\end{eqnarray}
where $w^i$ is the four-acceleration (\ref{foac4}).
Expansion has $\theta>0$ and contraction has $\theta<0$.
These definitions give
\begin{equation}
u_{i:j}h^j_k=\sigma_{ik}+\omega_{ik}+\frac{1}{3}\theta h_{ik}.
\end{equation}
Contracting $u^i_{\phantom{i}:jk}-u^i_{\phantom{i}:kj}=\mathring{R}^i_{\phantom{i}lkj}u^l$ with respect to the indices $i,j$ gives $\theta_{:k}u^k-u^j_{\phantom{j}:kj}u^k=-\mathring{R}_{kl}u^k u^l$ or
\begin{equation}
-\mathring{R}_{kl}u^k u^l=\theta_{:k}u^k-w^i_{\phantom{i}:i}+u_{i:k}u^{k:i}.
\label{Ray5}
\end{equation}
Defining
\begin{equation}
\sigma^2=\frac{1}{2}\sigma_{ik}\sigma^{ik},\quad \omega^2=\frac{1}{2}\omega_{ik}\omega^{ik}
\end{equation}
gives
\begin{equation}
u_{i:k}u^{k:i}=2(\sigma^2-\omega^2)+\frac{1}{3}\theta^2,
\end{equation}
which brings (\ref{Ray5}) to the {\em Raychaudhuri equation}:
\begin{equation}
\frac{d\theta}{ds}=-2(\sigma^2-\omega^2)-\frac{1}{3}\theta^2+w^i_{\phantom{i}:i}-\mathring{R}_{ik}u^i u^k.
\label{Ray8}
\end{equation}

\noindent
{\bf Energy conditions}.\\
For an ideal fluid, the Einstein equations give
\begin{equation}
\mathring{R}_{ik}u^i u^k=\frac{\kappa}{2}(\epsilon+3p).
\end{equation}
We define four energy conditions.
The {\em null energy condition} is satisfied if
\begin{equation}
T_{ij}k^i k^j\ge0
\end{equation}
for any null, future-pointing vector $k^i$.
For an ideal fluid, this condition gives
\begin{equation}
\epsilon+p\ge0.
\end{equation}
The {\em weak energy condition} is satisfied if
\begin{equation}
T_{ij}u^i u^j\ge0
\end{equation}
for any causal (null or timelike), future-pointing vector $u^i$.
For a perfect fluid, this condition gives
\begin{equation}
\epsilon+p\ge0,\quad \epsilon\ge0.
\end{equation}

The {\em strong energy condition} is satisfied if
\begin{equation}
\biggl(T_{ij}-\frac{1}{2}Tg_{ij}\biggr)u^i u^j\ge0
\end{equation}
for any causal, future-pointing vector $u^i$.
For a perfect fluid, this condition gives
\begin{equation}
\epsilon+p\ge0,\quad \epsilon+3p\ge0.
\label{strong}
\end{equation}
The {\em dominant energy condition} is satisfied if the weak condition is satisfied and $T_{ij}u^j$ is a causal, future-pointing vector.
For a perfect fluid, this condition gives
\begin{equation}
\epsilon+p\ge0,\quad \epsilon\ge|p|.
\end{equation}
This condition guarantees that particles in a congruence do not move faster than light.
The dominant condition infers the weak condition.
The weak condition infers the null condition.
The strong condition infers the null condition.\\

\noindent
{\bf Singularity}.\\
If the strong condition is satisfied and particles in a congruence move without rotation and acceleration ($\omega_{ik}=w^i=0$), then (\ref{Ray8}) and the Einstein equations give
\begin{equation}
\frac{d\theta}{d\tau}\le-\frac{c}{3}\theta^2.
\end{equation}
If the initial value of $\theta$ at $\tau=0$ is $\theta_0$, then
\begin{equation}
\theta^{-1}\ge\theta^{-1}_0+\frac{c\tau}{3}.
\end{equation}
Therefore, if $\theta_0<0$ (initial contraction) then $\theta$ diverges to a curvature {\em singularity}, which is a point (or a system of points) in spacetime where the density of matter and curvature are infinite, as $\tau$ increases: $\theta\rightarrow-\infty$.
Singularities are unphysical and their appearance in a system indicates that a theory describing such a system is incomplete and must include a violation of the strong energy condition.

\subsubsection{Spin fluid}
A fluid with spin, which is a source of torsion according to the Einstein--Cartan theory of gravity, is referred to as a {\em spin fluid}.
For a spin fluid, substituting the spin tensor (\ref{stp8}) into the Cartan equations (\ref{EC17}) gives the torsion tensor:
\begin{equation}
S^j_{\phantom{j}ik}=-\frac{1}{2}\kappa s_{ik}u^j,
\label{rs3}
\end{equation}
and the contortion tensor (\ref{EC18}):
\begin{equation}
C^k_{\phantom{k}ij}=\frac{1}{2}\kappa(s^k_{\phantom{k}i}u_j-s_{ij}u^k-s_j^{\phantom{j}k}u_i).
\end{equation}
Using (\ref{stp10}) gives
\begin{equation}
S_i=0,\quad \nabla_i^\ast=\nabla_i.
\end{equation}
Therefore, the corresponding metric energy--momentum tensor (\ref{cemt49}) reduces to
\begin{equation}
T_{ij}=\epsilon u_i u_j-ph_{ij}+\nabla_k(s_{il}u^k)u^l u_j-\frac{1}{2}\nabla_k(s_{ij}u^k+s^k_{\phantom{k}i}u_j+s^k_{\phantom{k}j}u_i).
\label{rs2}
\end{equation}
Two terms in (\ref{rs2}) can be written, using (\ref{stp10}), (\ref{cemt45}), and (\ref{cemt47}), as
\begin{eqnarray}
& & \nabla_k(s_{il}u^k)u^l u_j-\frac{1}{2}\nabla_k(s_{ij}u^k)=c(\pi_i u_l-\pi_l u_i)u^l u_j-\frac{1}{2}c(\pi_i u_j-\pi_j u_i) \nonumber \\
& & =\frac{1}{2}c(\pi_i u_j+\pi_j u_i)-\epsilon u_i u_j=\frac{1}{2}[\nabla_k(s_{il}u^k)u^l u_j+\nabla_k(s_{jl}u^k)u^l u_i]=-\nabla_l(s_{k(i}u^l)u^k u_{j)} \nonumber \\
& & =-\nabla_l s_{k(i}u^l u^k u_{j)}=-\nabla_l(s_{k(i}u_{j)})u^l u^k=-\nabla_l(s^k_{\phantom{k}(i}u_{j)})u_k u^l.
\end{eqnarray}
The metric energy--momentum tensor (\ref{rs2}) is then
\begin{equation}
T^{ij}=\epsilon u^i u^j-ph^{ij}-(\delta^l_k+u_k u^l)\nabla_l(s^{k(i}u^{j)}).
\label{rs5}
\end{equation}

The last term on the right of (\ref{rs5}) can be decomposed according to (\ref{Chrsym6}) into
\begin{eqnarray}
& & -(\delta^l_k+u_k u^l)\nabla_l(s^{k(i}u^{j)})=-(\delta^l_k+u_k u^l)\mathring{\nabla}_l(s^{k(i}u^{j)})-(\delta^l_k+u_k u^l)(C^k_{\phantom{k}ml}s^{m(i}u^{j)} \nonumber \\
& & +C^i_{\phantom{i}ml}s^{k(m}u^{j)}+C^j_{\phantom{j}ml}s^{k(i}u^{m)}).
\end{eqnarray}
This term reduces, by means of (\ref{stp10}) and (\ref{EC15}), to
\begin{eqnarray}
& & -(\delta^l_k+u_k u^l)\mathring{\nabla}_l(s^{k(i}u^{j)})-\delta^l_k(C^i_{\phantom{i}ml}s^{k(m}u^{j)}+C^j_{\phantom{j}ml}s^{k(i}u^{m)})-u_k u^l C^k_{\phantom{k}ml}s^{m(i}u^{j)} \nonumber \\
& & =-(\delta^l_k+u_k u^l)\mathring{\nabla}_l(s^{k(i}u^{j)})-C^i_{\phantom{i}mk}s^{k(m}u^{j)}-C^j_{\phantom{j}mk}s^{k(i}u^{m)} \nonumber \\
& & =-(\delta^l_k+u_k u^l)\mathring{\nabla}_l(s^{k(i}u^{j)})+\frac{1}{2}\kappa(s_{mk}u^i+s_k^{\phantom{k}i}u_m+s_m^{\phantom{m}i}u_k)s^{k(m}u^{j)} \nonumber \\
& & +\frac{1}{2}\kappa(s_{mk}u^j+s_k^{\phantom{k}j}u_m+s_m^{\phantom{m}j}u_k)s^{k(m}u^{i)} \nonumber \\
& & =-(\delta^l_k+u_k u^l)\mathring{\nabla}_l(s^{k(i}u^{j)})-\frac{1}{2}\kappa(s_{kl}s^{kl}u^i u^j-s^{ik}s^j_{\phantom{j}k}).
\end{eqnarray}
Therefore, the tensor (\ref{rs5}) becomes
\begin{equation}
T^{ij}=\epsilon u^i u^j-ph^{ij}-(\delta^l_k+u_k u^l)\mathring{\nabla}_l(s^{k(i}u^{j)})-\kappa s^2 u^i u^j+\frac{1}{2}\kappa s^{ik}s^j_{\phantom{j}k},
\label{rs8}
\end{equation}
where
\begin{equation}
s^2=\frac{1}{2}s^{ij}s_{ij}>0.
\label{rs9}
\end{equation}

\noindent
{\bf Effective energy density and pressure}.\\
Substituting (\ref{stp8}) into (\ref{EC19}) and using (\ref{stp10}) gives
\begin{equation}
U^{ij}=\frac{1}{2}\kappa s^2 u^i u^j+\frac{1}{4}\kappa s^2 g^{ij}-\frac{1}{2}\kappa s^{ik}s^j_{\phantom{j}k}.
\label{rs10}
\end{equation}
Adding (\ref{rs8}) and (\ref{rs10}) brings the combined energy--momentum tensor $\tilde{T}^{ij}$ in the Einstein equations (\ref{EC21}) to
\begin{equation}
\tilde{T}^{ij}=\Bigl(\epsilon-\frac{1}{4}\kappa s^2\Bigr)u^i u^j-\Bigl(p-\frac{1}{4}\kappa s^2\Bigr)h^{ij}-(\delta^l_k+u_k u^l)\mathring{\nabla}_l(s^{k(i}u^{j)}).
\label{rs11}
\end{equation}
If the spin orientation of particles in a spin fluid is random, then the macroscopic spacetime average of $s_{ij}$ and its gradients, such as of the last term on the right of (\ref{rs11}), vanish.
On the contrary, the terms that are quadratic in the spin tensor do not vanish after averaging.
Therefore, the combined energy--momentum tensor of a macroscopic spin fluid describes a perfect fluid with the effective energy density
\begin{equation}
\tilde{\epsilon}=\tilde{T}_{ij}u^i u^j=\epsilon-\frac{1}{4}\kappa s^2
\label{rs12}
\end{equation}
and the effective pressure
\begin{equation}
\tilde{p}=-\frac{1}{3}\tilde{T}_{ij}h^{ij}=p-\frac{1}{4}\kappa s^2.
\label{rs13}
\end{equation}

If the spin orientation of particles in a spin fluid is not random, then the combined energy density of a macroscopic spin fluid is
\begin{eqnarray}
& & \tilde{\epsilon}=\epsilon-\frac{1}{4}\kappa s^2-(\delta^l_k+u_k u^l)u^i\mathring{\nabla}_l s^k_{\phantom{k}i}=\epsilon-\frac{1}{4}\kappa s^2-\mathring{\nabla}_k s^k_{\phantom{k}i}u^i \nonumber \\
& & =\epsilon-\frac{1}{4}\kappa s^2+s^{ki}\mathring{\nabla}_{[k}u_{i]}=\epsilon-\frac{1}{4}\kappa s^2+s^{ki}\partial_{[k}u_{i]}.
\label{rs14}
\end{eqnarray}
In a locally Galilean frame of reference which is also a rest frame, (\ref{rs14}) becomes
\begin{equation}
\tilde{\epsilon}=\epsilon-\frac{1}{4}\kappa s^2+\frac{1}{2}{\bf s}\cdot\mbox{{\bf curl}}\,{\bf v}.
\end{equation}
where ${\bf s}$ is the spatial spin-density pseudovector (\ref{stp14}).\\

\noindent
{\bf Avoidance of singularity}.\\
The effective energy density (\ref{rs12}) and pressure (\ref{rs13}) can be negative if the quantity $s^2$ is sufficiently large.
Consequently, a spin fluid could violate the strong energy condition (\ref{strong}) and thus prevent a singularity.
\newline
References: \cite{Schr,LL2,Lord,Hehl1,Uti,KS,Hehl2,Mol}.

\subsection{Spinor fields}
\setcounter{equation}{0}
\subsubsection{Dirac matrices}
{\bf Pauli matrices}.\\
The Dirac matrices $\gamma^a$ defined by (\ref{spin1}) are complex.
A particular solution of (\ref{spin1}) is given by the {\em Dirac representation}:
\begin{equation}
\gamma^0=\left( \begin{array}{cc}
I_2 & 0 \\
0 & -I_2 \end{array} \right),\quad 
\gamma^\alpha=\left( \begin{array}{cc}
0 & \sigma^\alpha \\
-\sigma^\alpha & 0 \end{array} \right),
\label{Dirmat1}
\end{equation}
where $I_2$ is the two-dimensional unit matrix, and
\begin{equation}
\sigma^1=\left( \begin{array}{cc}
0 & 1 \\
1 & 0 \end{array} \right),\quad 
\sigma^2=\left( \begin{array}{cc}
0 & -i\\
i & 0 \end{array} \right),\quad 
\sigma^3=\left( \begin{array}{cc}
1 & 0\\
0 & -1 \end{array} \right)
\label{Dirmat2}
\end{equation}
are the {\em Pauli matrices} (all indices are coordinate invariant).
The Pauli matrices are traceless, $\mbox{tr}(\sigma^\alpha)=0$, and Hermitian, $\sigma^{\alpha\dag}=\sigma^\alpha$ (the Hermitian conjugation of a matrix $A$ is the combination of the complex conjugation and transposition, $A^\dag=A^{\ast T}$).
They also satisfy
\begin{equation}
\sigma_\alpha\sigma_\beta=\delta_{\alpha\beta}I_2+i\varepsilon_{\alpha\beta\gamma}\sigma_\gamma,
\label{Dirmat3}
\end{equation}
and their squares are $(\sigma_\alpha)^2=I_2$.
The identity (\ref{Dirmat3}) gives the commutation relation
\begin{equation}
\Bigl[\frac{\sigma_\alpha}{2},\frac{\sigma_\beta}{2}\Bigr]=i\varepsilon_{\alpha\beta\gamma}\frac{\sigma_\gamma}{2},
\end{equation}
so $\sigma_\alpha/2$ form the lowest, two-dimensional representation of the angular momentum operator $M_\alpha$ (\ref{Cas11}).\\

\noindent
{\bf Properties of Dirac matrices}.\\
The properties of $\sigma^\alpha$ infer that the Dirac matrices are traceless, $\mbox{tr}(\gamma^a)=0$, and satisfy
\begin{equation}
\quad \gamma^{a\dag}=\gamma^0\gamma^a\gamma^0,\quad \gamma^{a\ast}=\gamma^2\gamma^a\gamma^2,
\label{Dirmat5}
\end{equation}
which gives $\gamma^{0\dag}=\gamma^0$ and $\gamma^{\alpha\dag}=-\gamma^\alpha$.
Hereinafter, $\gamma^0$, $\gamma^1$, $\gamma^2$ and $\gamma^3$ refer to the Dirac matrices with a Lorentz index, $\gamma^a$.
The relation (\ref{spin1}) yields the total antisymmetry of $\gamma^0\gamma^1\gamma^2\gamma^3$:
\begin{equation}
\gamma^0\gamma^1\gamma^2\gamma^3=\gamma^{[0}\gamma^1\gamma^2\gamma^{3]}.
\label{Dirmat6}
\end{equation}
We define
\begin{equation}
\gamma^5=-\frac{i}{24}e_{abcd}\gamma^a\gamma^b\gamma^c\gamma^d=i\gamma^0\gamma^1\gamma^2\gamma^3,
\label{Dirmat7}
\end{equation}
which is traceless, $\mbox{tr}(\gamma^5)=0$, and Hermitian, $\gamma^{5\dag}=\gamma^5$.
It also satisfies
\begin{equation}
\{\gamma^a,\gamma^5\}=0,\quad (\gamma^5)^2=I_4,\quad \gamma^5_{\phantom{5}|i}=0,
\label{Dirmat8}
\end{equation}
where the last relation results from (\ref{metten20}) and (\ref{spinco13}).
In the Dirac representation,
\begin{equation}
\gamma^5=\gamma^{5\ast}=\left( \begin{array}{cc}
0 & I_2 \\
I_2 & 0 \end{array} \right).
\label{Dirmat9}
\end{equation}

The anticommutation relation (\ref{spin1}) gives
\begin{eqnarray}
& & \gamma^a\gamma_a=4I_4, \\
& & \gamma^a\gamma^b\gamma_a=-2\gamma^b, \\
& & \gamma^a\gamma^b\gamma^c\gamma_a=4\eta^{bc}I_4, \\
& & \gamma^a\gamma^b\gamma^c\gamma^d\gamma_a=-2\gamma^d\gamma^c\gamma^b, \\
& & \gamma^a\gamma^b\gamma^c=\eta^{ab}\gamma^c+\eta^{bc}\gamma^a-\eta^{ac}\gamma^b+ie^{abcd}\gamma_d\gamma^5, \label{Dirmat14} \\
& & \{\gamma^a,\gamma^{[b}\gamma^{c]}\}=2\gamma^{[a}\gamma^b\gamma^{c]}.
\label{Dirmat15}
\end{eqnarray}
The Dirac representation is not unique; the relation (\ref{spin1}) is invariant under a similarity transformation $\gamma^a\rightarrow S\gamma^a S^{-1}$, where $S$ is a nondegenerate ($\mbox{det}\,S\neq0$) matrix.
Accordingly, $\psi\rightarrow S\psi$ and $\bar{\psi}\rightarrow\bar{\psi}S^{-1}$.
Taking $S=(1/\sqrt{2})\left( \begin{array}{cc}
I_2 & -I_2 \\
I_2 & I_2 \end{array} \right)$ turns the Dirac representation into the {\em chiral} or {\em Weyl representation}, in which
\begin{equation}
\gamma^0=\left( \begin{array}{cc}
0 & I_2 \\
I_2 & 0 \end{array} \right),\quad 
\gamma^\alpha=\left( \begin{array}{cc}
0 & \sigma^\alpha \\
-\sigma^\alpha & 0 \end{array} \right),\quad 
\gamma^5=\left( \begin{array}{cc}
-I_2 & 0 \\
0 & I_2 \end{array} \right).
\end{equation}

\noindent
{\bf Adjoint spinor}.\\
For an infinitesimal Lorentz transformation (\ref{infLor1}), the relations (\ref{spin5}) and (\ref{spin6}) give $L=I_4+(1/8)\epsilon_{ab}(\gamma^a\gamma^b-\gamma^b\gamma^a)$.
Using $(AB)^\dag=B^\dag A^\dag$ gives
\begin{equation}
L^\dag=I_4+\frac{1}{8}\epsilon_{ab}(\gamma^{b\dag}\gamma^{a\dag}-\gamma^{a\dag}\gamma^{b\dag}),
\end{equation}
which is equal to $L^{-1}$ (so $L$ is unitary) for rotations and equal to $L$ for boosts.
The relation (\ref{Dirmat5}) gives
\begin{equation}
(\gamma^{b\dag}\gamma^{a\dag}-\gamma^{a\dag}\gamma^{b\dag})\gamma^0=-\gamma^0(\gamma^a\gamma^b-\gamma^b\gamma^a),\quad (G^{ab})^\dagger\gamma^0=-\gamma^0 G^{ab}.
\label{Dirmat20}
\end{equation}
and thus
\begin{equation}
L^\dag\gamma^0=\gamma^0+\frac{1}{8}\epsilon_{ab}(\gamma^{b\dag}\gamma^{a\dag}-\gamma^{a\dag}\gamma^{b\dag})\gamma^0=\gamma^0-\frac{1}{8}\epsilon_{ab}\gamma^0(\gamma^a\gamma^b-\gamma^b\gamma^a)=\gamma^0 L^{-1}.
\end{equation}
Therefore, the quantity $\psi^\dag\gamma^0$ transforms under (\ref{spin7}) like an adjoint spinor:
\begin{equation}
\psi^\dag\gamma^0\rightarrow\psi^\dag L^\dag\gamma^0=\psi^\dag\gamma^0 L^{-1}.
\end{equation}
Accordingly, we can associate these two quantities:
\begin{equation}
\bar{\psi}=\psi^\dag\gamma^0.
\label{Dirmat23}
\end{equation}

\noindent
{\bf Bilinear pseudoforms}.\\
A spinor $\psi$ and its adjoint $\psi^\dag\gamma^0$ can be used to construct tensors, as in (\ref{spin11}) and (\ref{spin12}): $\psi^\dag\gamma^0\psi$ transforms like a scalar, $\psi^\dag\gamma^0\gamma^i\psi$ is a vector, and $\psi^\dag\gamma^0\gamma^{[i}\gamma^{j]}\psi$ is an antisymmetric tensor.
Because $\gamma^5$ (\ref{Dirmat7}) is constructed from the completely antisymmetric unit pseudotensor, which is a tensor density, $\psi^\dag\gamma^0\gamma^5\psi$ is a pseudoscalar, and $\psi^\dag\gamma^0\gamma^i\gamma^5\psi$ is a pseudovector.
Higher-rank tensors constructed from $\psi$ and $\psi^\dag\gamma^0$ reduce to the above 5 kinds of tensors because of (\ref{Dirmat14}).
To show that $\bar{\psi}\gamma^5\psi$ transforms like a pseudoscalar, we substitute (\ref{spin4}) into (\ref{Dirmat7}) and use (\ref{Dirmat6}), which gives
\begin{equation}
\gamma^5=i\Lambda^0_{\phantom{0}a}\Lambda^1_{\phantom{1}b}\Lambda^2_{\phantom{2}c}\Lambda^3_{\phantom{3}d}L\gamma^a \gamma^b \gamma^c \gamma^d L^{-1}=i\Lambda^0_{\phantom{0}a}\Lambda^1_{\phantom{1}b}\Lambda^2_{\phantom{2}c}\Lambda^3_{\phantom{3}d}L\gamma^{[a}\gamma^b \gamma^c \gamma^{d]}L^{-1}.
\end{equation}
Using (\ref{KLC3}) and
\begin{equation}
\gamma^{[a}\gamma^b \gamma^c \gamma^{d]}=-ie^{abcd}\gamma^5,
\end{equation}
which results from (\ref{Dirmat7}), gives
\begin{equation}
\gamma^5=e^{abcd}\Lambda^0_{\phantom{0}a}\Lambda^1_{\phantom{1}b}\Lambda^2_{\phantom{2}c}\Lambda^3_{\phantom{3}d}L\gamma^5 L^{-1}=\mbox{det}(\Lambda^a_{\phantom{a}b})L\gamma^5 L^{-1}.
\end{equation}
Therefore, we have
\begin{equation}
\bar{\psi}\gamma^5\psi\rightarrow\bar{\psi}L^{-1}\mbox{det}(\Lambda^a_{\phantom{a}b})L\gamma^5 L^{-1}L\psi=\mbox{det}(\Lambda^a_{\phantom{a}b})\bar{\psi}\gamma^5\psi,
\end{equation}
which is the transformation law for a Lorentz scalar density, and thus a pseudoscalar.
Similarly,
\begin{equation}
\bar{\psi}\gamma^c\gamma^5\psi\rightarrow\bar{\psi}L^{-1}\Lambda^c_{\phantom{c}d}\mbox{det}(\Lambda^a_{\phantom{a}b})L\gamma^d \gamma^5 L^{-1}L\psi=\mbox{det}(\Lambda^a_{\phantom{a}b})\Lambda^c_{\phantom{c}d}\bar{\psi}\gamma^d \gamma^5\psi,
\end{equation}
which is the transformation law for a Lorentz vector density, and thus a pseudovector.\\

\noindent
{\bf Chirality}.\\
We define the {\em chirality projection operators}
\begin{equation}
P_\pm=\frac{I_4\pm\gamma^5}{2},\quad P_++P_-=I_4,\quad P_\pm^2=I_4,\quad P_+P_-=P_-P_+=0.
\end{equation}
They project a spinor $\psi$ into the {\em right-handed} spinor $\psi_R$ and {\em left-handed} spinor $\psi_L$,
\begin{equation}
\psi_R=P_+\psi,\quad \psi_L=P_-\psi,\quad \psi=\psi_R+\psi_L.
\label{Dirmat27}
\end{equation}
A spinor $\psi$ can be decomposed into two {\em two-dimensional spinors} $u$ and $v$:
\begin{equation}
\psi=\left( \begin{array}{c}
u \\
v \end{array} \right),
\label{Dirmat28}
\end{equation}
which are columns with two components.
In the chiral representation:
\begin{equation}
\psi_L=\left( \begin{array}{c}
u \\
0 \end{array} \right),\quad \psi_R=\left( \begin{array}{c}
0 \\
v \end{array} \right).
\end{equation}

\noindent
{\bf Spinor representation of rotation}.\\
An infinitesimal rotation is described by a Lorentz matrix (\ref{infLor1}) with $\epsilon_{0\alpha}=0$.
The corresponding spinor transformation matrix (\ref{spin5}) in the Dirac representation is, using (\ref{rotbo5}),
\begin{equation}
L=I_4+\frac{1}{4}\epsilon_{\alpha\beta}\gamma^\alpha\gamma^\beta=I_4+\frac{1}{4}e_{\alpha\beta\gamma}\gamma^\alpha\gamma^\beta\vartheta_\gamma=I_4-\frac{i}{2}\vartheta_\alpha\left( \begin{array}{cc}
\sigma^\alpha & 0 \\
0 & \sigma^\alpha \end{array} \right).
\end{equation}
Therefore, the (unitary) spinor transformation matrix for a finite rotation by an angle $\vartheta$ about an axis parallel to a unit vector ${\bf n}$, ${\bm\vartheta}=\vartheta{\bf n}$, is
\begin{equation}
L=\exp\Biggl[-\frac{i}{2}{\bm\vartheta}\cdot\left( \begin{array}{cc}
{\bm\sigma} & 0 \\
0 & {\bm\sigma} \end{array} \right)\Biggr]=\cos\frac{\vartheta}{2}I_4-i\,\sin\frac{\vartheta}{2}\,{\bf n}\cdot\left( \begin{array}{cc}
{\bm\sigma} & 0 \\
0 & {\bm\sigma} \end{array} \right),
\end{equation}
where $\bm\sigma$ is the spatial vector composed from the Pauli matrices $\sigma^\alpha$.
If we split a spinor $\psi$ into two parts $u$ and $v$ (\ref{Dirmat28}), then both $u$ and $v$ transform under rotations according to
\begin{equation}
u\rightarrow Su,\quad v\rightarrow Sv,
\end{equation}
where
\begin{equation}
S=\cos\frac{\vartheta}{2}I_2-i\,\sin\frac{\vartheta}{2}\,{\bf n}\cdot{\bm\sigma}.
\end{equation}
A rotation by a full angle $2\pi$ changes the sign of a spinor.
A rotation by $4\pi$ brings a spinor to its original position.\\

\noindent
{\bf Spinor representation of boost}.\\
An infinitesimal boost is described by a Lorentz matrix (\ref{infLor1}) with $\epsilon_{\alpha\beta}=0$.
The corresponding spinor transformation matrix (\ref{spin5}) in the Dirac representation is, using (\ref{rotbo6}),
\begin{equation}
L=I_4+\frac{1}{2}\epsilon_{0\alpha}\gamma^0\gamma^\alpha=I_4+\frac{1}{2}\eta_\alpha\gamma^0\gamma^\alpha=I_4+\frac{1}{2}\eta_\alpha\left( \begin{array}{cc}
0 & \sigma^\alpha \\
\sigma^\alpha & 0 \end{array} \right).
\end{equation}
Therefore, the spinor transformation matrix for a finite boost with a rapidity $\eta$ along an axis parallel to a unit vector ${\bf n}$, ${\bm\eta}=\eta{\bf n}$, is
\begin{equation}
L=\exp\Biggl[\frac{1}{2}{\bm\eta}\cdot\left( \begin{array}{cc}
0 & {\bm\sigma} \\
{\bm\sigma} & 0 \end{array} \right)\Biggr]=\cosh\frac{\eta}{2}I_4+\sinh\frac{\eta}{2}\,{\bf n}\cdot\left( \begin{array}{cc}
0 & {\bm\sigma} \\
{\bm\sigma} & 0 \end{array} \right).
\end{equation}
Using (\ref{relkin6}) gives
\begin{equation}
L=\frac{1}{\sqrt{2(1+\gamma)}}\left( \begin{array}{cc}
(1+\gamma)I_2 & \gamma{\bm\beta}\cdot{\bm\sigma} \\
\gamma{\bm\beta}\cdot{\bm\sigma} & (1+\gamma)I_2 \end{array} \right).
\label{Dirmat36}
\end{equation}
If this boost transforms a particle of mass $m$ from rest to a motion with momentum ${\bf p}$ and energy $E$, then (\ref{Dirmat36}) is equivalent, because of (\ref{emtp23}) and (\ref{emtp24}), to
\begin{equation}
L=\frac{1}{\sqrt{2mc^2(E+mc^2)}}\left( \begin{array}{cc}
(E+mc^2)I_2 & {\bm\sigma}\cdot{\bf p}c \\
{\bm\sigma}\cdot{\bf p}c & (E+mc^2)I_2 \end{array} \right).
\label{Dirmat37}
\end{equation}

\subsubsection{Lagrangian density for spinor field}
A Lagrangian density for dynamical spinor fields must contain first derivatives of spinors and be real.
The simplest scalar containing derivatives of spinors is quadratic in $\psi$, $\bar{\psi}\gamma^i\psi_{;i}$, where $\psi_{;i}$ is the covariant derivative of $\psi$ (\ref{spinco3}).
Its kinetic part, containing derivatives of spinors, is equal to $\bar{\psi}\gamma^i\psi_{,i}$ and is complex.
Since the complex conjugate of this quantity is
\begin{equation}
(\bar{\psi}\gamma^i\psi_{,i})^\ast=(\bar{\psi}\gamma^i\psi_{,i})^\dag=\psi^\dag_{,i}\gamma^{i\dag}\bar{\psi}^\dag=\bar{\psi}_{,i}\gamma^0\gamma^{i\dag}\gamma^0\psi=\bar{\psi}_{,i}\gamma^i\psi,
\end{equation}
both scalars $\bar{\psi}\gamma^i\psi_{,i}+\bar{\psi}_{,i}\gamma^i\psi$ and $i(\bar{\psi}\gamma^i\psi_{,i}-\bar{\psi}_{,i}\gamma^i\psi)$ are real.
The former scalar is, however, equal to a total divergence $(\bar{\psi}\gamma^i\psi)_{,i}$, thereby a Lagrangian density proportional to such term does not contribute to field equations.
Therefore, the simplest kinetic part of a spinor Lagrangian density is proportional to $i(\bar{\psi}\gamma^i\psi_{,i}-\bar{\psi}_{,i}\gamma^i\psi)$.
Another scalar that can be used in a spinor Lagrangian density is proportional to $\bar{\psi}\psi$, which is real:
\begin{equation}
(\bar{\psi}\psi)^\ast=(\bar{\psi}\psi)^\dag=(\psi^\dag\gamma^0\psi)^\dag=\psi^\dag\gamma^0\psi^{\dag\dag}=\bar{\psi}\psi.
\end{equation}
The simplest Lagrangian density for a spinor field is thus given by
\begin{equation}
\mathfrak{L}_\psi=\frac{i\mathfrak{e}}{2}(\bar{\psi}\gamma^i\psi_{;i}-\bar{\psi}_{;i}\gamma^i\psi)-mc\mathfrak{e}\bar{\psi}\psi=\frac{i\mathfrak{e}}{2}e^i_a(\bar{\psi}\gamma^a\psi_{;i}-\bar{\psi}_{;i}\gamma^a\psi)-mc\mathfrak{e}\bar{\psi}\psi,
\label{Lagsp3}
\end{equation}
where $m$ is a real scalar constant, called the {\em spinor mass}.
It is referred to as the {\em Dirac Lagrangian density}.
Putting the definition of the covariant derivative of a spinor (\ref{spinco3}) into (\ref{Lagsp3}) gives
\begin{equation}
\mathfrak{L}_\psi=\frac{i\mathfrak{e}}{2}(\bar{\psi}\gamma^i\psi_{,i}-\bar{\psi}_{,i}\gamma^i\psi)-\frac{i\mathfrak{e}}{2}\bar{\psi}\{\gamma^i,\Gamma_i\}\psi-mc\mathfrak{e}\bar{\psi}\psi.
\label{Lagsp4}
\end{equation}
Using the Fock--Ivanenko coefficients (\ref{spinco17}) as the spinor connection $\Gamma_i$ turns (\ref{Lagsp4}) into
\begin{eqnarray}
& & \mathfrak{L}_\psi=\frac{i\mathfrak{e}}{2}(\bar{\psi}\gamma^i\psi_{,i}-\bar{\psi}_{,i}\gamma^i\psi)+\frac{i\mathfrak{e}}{8}\omega_{abi}\bar{\psi}\{\gamma^i,\gamma^a \gamma^b\}\psi-mc\mathfrak{e}\bar{\psi}\psi \nonumber \\
& & =\frac{i\mathfrak{e}}{2}(\bar{\psi}\gamma^i\psi_{,i}-\bar{\psi}_{,i}\gamma^i\psi)+\frac{i\mathfrak{e}}{8}\omega_{abi}\bar{\psi}\{\gamma^i,\gamma^{[a}\gamma^{b]}\}\psi-mc\mathfrak{e}\bar{\psi}\psi.
\label{Lagsp5}
\end{eqnarray}

\subsubsection{Spin tensor for spinor field}
\label{SpinSpinors}
The spin density (\ref{spden2}) corresponding to the Dirac Lagrangian density (\ref{Lagsp5}) is
\begin{equation}
\mathfrak{S}_{\mu\nu}^{\phantom{\mu\nu}i}=2\frac{\delta\mathfrak{L}_\psi}{\delta\omega^{\mu\nu}_{\phantom{\mu\nu}i}}=\frac{1}{4}i\mathfrak{e}\bar{\psi}\{\gamma^i,\gamma_{[\mu}\gamma_{\nu]}\}\psi=\frac{1}{8}i\mathfrak{e}\bar{\psi}\{\gamma^i,[\gamma_\mu,\gamma_\nu]\}\psi=\frac{1}{2}i\mathfrak{e}\bar{\psi}\{\gamma^i,G_{\mu\nu}\}\psi,
\label{Lagsp60}
\end{equation}
where $G_{\mu\nu}$ are the generators (\ref{spin6}) of the spinor representation of the Lorentz group.
By means of (\ref{Dirmat15}), this spin density can be written as
\begin{equation}
\mathfrak{S}^{ijk}=\frac{i\mathfrak{e}}{2}\bar{\psi}\gamma^{[i}\gamma^j\gamma^{k]}\psi.
\label{Lagsp6}
\end{equation}
The quantity (\ref{Lagsp6}) is completely antisymmetric,
\begin{equation}
\mathfrak{S}^{ijk}=\mathfrak{S}^{[ijk]},
\label{Lagsp7}
\end{equation}
and independent of the spinor mass.
The corresponding spin tensor is also completely antisymmetric:
\begin{equation}
s^{ijk}=\frac{i}{2}\bar{\psi}\gamma^{[i}\gamma^j\gamma^{k]}\psi=s^{[ijk]}=-e^{ijkl}s_l,
\label{Lagsp8}
\end{equation}
where
\begin{equation}
s^i=\frac{1}{2}\bar{\psi}\gamma^i\gamma^5\psi
\label{Lagsp9}
\end{equation}
is the {\em spin-density pseudovector}.
The Cartan equations (\ref{EC17}) with the spin tensor (\ref{Lagsp8}) give a completely antisymmetric torsion tensor:
\begin{equation}
S_{ijk}=-\frac{i\kappa}{4}\bar{\psi}\gamma_{[i}\gamma_j\gamma_{k]}\psi,
\end{equation}
so the torsion vector vanishes, $S_i=0$.
Therefore, the contortion tensor is, using (\ref{Dirmat14}),
\begin{equation}
C_{ijk}=S_{ijk}=\frac{\kappa}{4}e_{ijkl}\bar{\psi}\gamma^l\gamma^5\psi=\frac{\kappa}{2}e_{ijkl}s^l.
\label{Lagsp11}
\end{equation}
The torsion tensor is dual to a pseudovector that is proportional to the Dirac spin-density pseudovector (\ref{Lagsp9}).
The pseudovector density
\begin{equation}
{\sf j}^i_A=\mathfrak{e}\bar{\psi}\gamma^i\gamma^5\psi=2\mathfrak{e}s^i
\end{equation}
is the {\em Dirac axial current density}.
This quantity and the pseudovector (\ref{Lagsp9}) are real because
\begin{equation}
(\bar{\psi}\gamma^i\gamma^5\psi)^\ast=(\psi^\dag\gamma^0\gamma^i\gamma^5\psi)^\dag=\psi^\dag\gamma^5\gamma^{i\dag}\gamma^0\psi=\psi^\dag\gamma^5\gamma^0\gamma^i\psi=\psi^\dag\gamma^0\gamma^i\gamma^5\psi=\bar{\psi}\gamma^i\gamma^5\psi.
\label{Lagsp13}
\end{equation}
Therefore, the spin tensor (\ref{Lagsp8}) is also real, in accordance with the reality of (\ref{Lagsp3}).

For a system of spinor fields, the Lagrangian density is equal to the sum of the Lagrangian densities (\ref{Lagsp3}) for each spinor and a Lagrangian density representing fields that carry the interaction between the spinors.
The Lagrangian density for the interaction fields does not depend on derivatives of spinors.
Since $C_{ijk}$ appears only in the additive, kinetic term in the Lagrangian density, $(i/2)\mathfrak{e}e^i_a(\bar{\psi}\gamma^a\psi_{;i}-\bar{\psi}_{;i}\gamma^a\psi)$, the spin tensor for a system of spinor fields is additive.
Accordingly, the spin tensor for such a system is also completely antisymmetric.

The complete antisymmetry of the spin density (\ref{Lagsp7}) leads to (\ref{Pap73}), which is consistent with (\ref{Lagsp8}) only if $\psi=0$.
Therefore, a spinor field cannot be a point particle or a system of point particles.

\subsubsection{Dirac equation}
{\bf Dirac equation in terms of general connection}.\\
Varying (\ref{Lagsp4}) with respect to $\bar{\psi}$ and omitting total divergences gives
\begin{equation}
\delta\mathfrak{L}_\psi=\frac{i}{2}\delta\bar{\psi}\bigl(\mathfrak{e}\gamma^k\psi_{,k}+(\mathfrak{e}\gamma^k\psi)_{,k}-\mathfrak{e}\{\Gamma_k,\gamma^k\}\psi\bigr)-\mathfrak{e}mc\delta\bar{\psi}\psi.
\end{equation}
The stationarity of the action $\delta S=0$ under $\delta\bar{\psi}$ therefore gives
\begin{equation}
\frac{i}{2}\bigl(\mathfrak{e}\gamma^k\psi_{,k}+(\mathfrak{e}\gamma^k\psi)_{,k}-\mathfrak{e}\{\Gamma_k,\gamma^k\}\psi\bigr)-\mathfrak{e}mc\psi=0.
\label{Direq2}
\end{equation}
Substituting
\begin{equation}
(\mathfrak{e}\gamma^k\psi)_{,k}=\mathfrak{e}\gamma^k\psi_{,k}+\mathfrak{e}\gamma^k_{\phantom{k};k}\psi-2\mathfrak{e}S_k\gamma^k\psi=\mathfrak{e}\gamma^k\psi_{,k}+\mathfrak{e}[\Gamma_k,\gamma^k]\psi
\end{equation}
into (\ref{Direq2}) gives the {\em Dirac equation}:
\begin{equation}
i\gamma^k\psi_{,k}-i\gamma^k\Gamma_k\psi-mc\psi=0,\quad i\gamma^k\psi_{;k}=mc\psi.
\label{Direq4}
\end{equation}
Varying (\ref{Lagsp4}) with respect to $\psi$ and using the stationarity of the action under $\delta\psi$ gives the adjoint Dirac equation:
\begin{equation}
-i\bar{\psi}_{,k}\gamma^k-i\bar{\psi}\Gamma_k\gamma^k-mc\bar{\psi}=0,\quad -i\bar{\psi}_{;k}\gamma^k=mc\bar{\psi}.
\label{Direq7}
\end{equation}

\noindent
{\bf Dirac equation in terms of Levi-Civita connection}.\\
Putting (\ref{Lagsp11}) into (\ref{spinco23}) yields
\begin{eqnarray}
& & \gamma^k\psi_{;k}=\gamma^k\psi_{:k}+\frac{\kappa}{16}e_{ijkl}(\bar{\psi}\gamma^l\gamma^5\psi)\gamma^k\gamma^i\gamma^j\psi=\gamma^k\psi_{:k}+\frac{i\kappa}{16}e_{ijkl}(\bar{\psi}\gamma^l\gamma^5\psi)e^{ijkm}\gamma_m\gamma^5\psi \nonumber \\
& & =\gamma^k\psi_{:k}-\frac{3i\kappa}{8}(\bar{\psi}\gamma^l\gamma^5\psi)\gamma_l\gamma^5\psi.
\end{eqnarray}
Therefore, the Dirac equation (\ref{Direq4}) becomes
\begin{equation}
i\gamma^k\psi_{:k}+\frac{3\kappa}{8}(\bar{\psi}\gamma_k\gamma^5\psi)\gamma^k\gamma^5\psi=mc\psi.
\label{Direq6}
\end{equation}
The Dirac equation (\ref{Direq7}) becomes the adjoint conjugate of (\ref{Direq6}):
\begin{equation}
-i\bar{\psi}_{:k}\gamma^k+\frac{3\kappa}{8}(\bar{\psi}\gamma_k\gamma^5\psi)\bar{\psi}\gamma^k\gamma^5=mc\bar{\psi}.
\label{Direq8}
\end{equation}
Equations (\ref{Direq6}) and (\ref{Direq8}) are cubic in spinor fields.
They are also the field equations corresponding to an effective metric Lagrangian density with a quartic, axial-axial spin-spin interaction:
\begin{equation}
\mathring{\mathfrak{L}}_\psi=\frac{i\mathfrak{e}}{2}(\bar{\psi}\gamma^i\psi_{:i}-\bar{\psi}_{:i}\gamma^i\psi)-mc\mathfrak{e}\bar{\psi}\psi+\frac{3\kappa\mathfrak{e}}{16}(\bar{\psi}\gamma_k\gamma^5\psi)(\bar{\psi}\gamma^k\gamma^5\psi).
\label{Direq9}
\end{equation}

\noindent
{\bf Conservation of Dirac vector current}.\\
Subtracting (\ref{Direq4}) multiplied by $i\bar{\psi}$ from (\ref{Direq7}) multiplied by $i\psi$ gives, using (\ref{spinco7}) and (\ref{cursp4}),
\begin{equation}
\bar{\psi}\gamma^k\psi_{;k}+\bar{\psi}_{;k}\gamma^k\psi=\bar{\psi}\gamma^k\psi_{|k}+\bar{\psi}_{|k}\gamma^k\psi=(\bar{\psi}\gamma^k\psi)_{|k}=(\bar{\psi}\gamma^k\psi)_{;k}=\frac{1}{\mathfrak{e}}(\mathfrak{e}\bar{\psi}\gamma^k\psi)_{;k}=0.
\label{Direq10}
\end{equation}
Applying (\ref{anticd2}) to the contravariant vector density $\mathfrak{e}\bar{\psi}\gamma^k\psi$ gives, using $S_i=0$,
\begin{equation}
(\mathfrak{e}\bar{\psi}\gamma^k\psi)_{,k}=0.
\label{Direq137}
\end{equation}
The vector density
\begin{equation}
{\sf j}^k_V=\mathfrak{e}\bar{\psi}\gamma^k\psi,
\label{Direq11}
\end{equation}
referred to as the {\em Dirac vector current density}, is thus conserved:
\begin{equation}
{\sf j}^k_{V,k}=\frac{\partial{\sf j}^0_V}{c\partial t}+\partial_\alpha{\sf j}^\alpha_V=0.
\label{Direq12}
\end{equation}
The corresponding {\em Dirac current vector} is
\begin{equation}
j^k=\frac{{\sf j}^k_V}{\mathfrak{e}}=\bar{\psi}\gamma^k\psi.
\label{Direq127}
\end{equation}
This quantity and the density (\ref{Direq11}) are real because
\begin{equation}
(\bar{\psi}\gamma^k\psi)^\ast=(\psi^\dag\gamma^0\gamma^k\psi)^\dag=\psi^\dag\gamma^{k\dag}\gamma^0\psi=\psi^\dag\gamma^0\gamma^k\psi=\bar{\psi}\gamma^k\psi.
\end{equation}
The time component
\begin{equation}
{\sf j}^0_V=\mathfrak{e}\bar{\psi}\gamma^0\psi=\mathfrak{e}\psi^\dag\psi
\end{equation}
is thus real and positive.
The conservation of the Dirac vector current density can also be shown by subtracting (\ref{Direq6}) multiplied by $i\bar{\psi}$ from (\ref{Direq8}) multiplied by $i\psi$, giving a relation analogous to (\ref{Direq10}):
\begin{equation}
\bar{\psi}\gamma^k\psi_{:k}+\bar{\psi}_{:k}\gamma^k\psi=(\bar{\psi}\gamma^k\psi)_{:k}=\frac{1}{\mathfrak{e}}(\mathfrak{e}\bar{\psi}\gamma^k\psi)_{,k}=0.
\end{equation}

\noindent
{\bf Klein--Gordon--Fock equation}.\\
The Dirac equation (\ref{Direq4}) gives
\begin{equation}
-\gamma^j(\gamma^k\psi_{;k})_{|j}=imc\gamma^j\psi_{;j}
\end{equation}
or, because of (\ref{cursp4}),
\begin{equation}
-\gamma^j\gamma^k\psi_{;kj}=-\gamma^{(j}\gamma^{k)}\psi_{;kj}-\gamma^j\gamma^k\psi_{|[kj]}=m^2 c^2\psi.
\label{Direq16}
\end{equation}
The relations (\ref{cursp1}) and (\ref{cursp10}) turn (\ref{Direq16}) into
\begin{equation}
\psi_{;i}^{\phantom{;i}i}+m^2 c^2\psi=\frac{1}{2}\gamma^j\gamma^k(K_{kj}\psi+2S^l_{\phantom{l}kj}\psi_{;l})=\frac{1}{8}R_{ilkj}\gamma^j\gamma^k\gamma^i\gamma^l \psi+S^l_{\phantom{l}kj}\gamma^j\gamma^k\psi_{;l},
\label{Direq17}
\end{equation}
where $K_{ij}$ is the curvature spinor.
If a spinor is equal either to its left-handed projection, $\psi=\psi_L$, or right-handed projection, $\psi=\psi_R$, then it is called a {\em Weyl spinor}.
Multiplying (\ref{Direq4}) by $P_\pm$ and using (\ref{Dirmat8}) gives
\begin{equation}
iP_\pm\gamma^i\psi_{;i}=i\gamma^i P_\mp\psi_{;i}=mcP_\pm\psi
\end{equation}
or
\begin{equation}
i\gamma^i\psi^{L(R)}_{;i}=mc\psi^{R(L)}.
\end{equation}
Therefore, if $\psi$ is a Weyl spinor then $m=0$.

The nonlinear, cubic terms in (\ref{Direq6}) and (\ref{Direq8}) represent a spinor self-interaction, corresponding to a torsion-generated spin-spin interaction in the tensor (\ref{EC19}).
At densities satisfying $\kappa\bar{\psi}\psi\ll mc$, the effects of torsion are negligible and these equations can be approximated by linear equations:
\begin{equation}
i\gamma^k\psi_{:k}=mc\psi,\quad -i\bar{\psi}_{:k}\gamma^k=mc\bar{\psi}.
\end{equation}
Equation (\ref{Direq17}) reduces to the {\em Klein--Gordon--Fock equation}:
\begin{equation}
\psi_{:i}^{\phantom{:i}i}+m^2 c^2\psi=\frac{1}{8}\mathring{R}_{ilkj}\gamma^j\gamma^k\gamma^i\gamma^l\psi,
\end{equation}
where $\mathring{R}_{ijkl}$ is the Riemann tensor.

If the mass of a spinor vanishes, the Klein--Gordon--Fock equation in a flat spacetime reduces to the 
{\em d'Alembert equation} or {\em wave equation}:
\begin{equation}
\psi_{,i}^{\phantom{,i}i}=\Box\psi=\frac{1}{c^2}\frac{\partial^2\psi}{\partial t^2}-\triangle\psi=0,
\end{equation}
where $\Box$ is the d'Alembertian (\ref{relkin100}) and $\triangle$
is the Laplacian (\ref{spvec30}).

\subsubsection{Energy--momentum tensor for spinor field}
Varying the Lagrangian density (\ref{Lagsp3}) with respect to $e^i_a$ gives the tetrad energy--momentum density for a Dirac spinor field:
\begin{equation}
\mathfrak{T}^{\phantom{i}a}_i=\frac{i\mathfrak{e}}{2}(\bar{\psi}\gamma^a\psi_{;i}-\bar{\psi}_{;i}\gamma^a\psi-e^a_i\bar{\psi}\gamma^j\psi_{;j}+e^a_i\bar{\psi}_{;j}\gamma^j\psi)+mc\mathfrak{e}e^a_i\bar{\psi}\psi,
\label{emts1}
\end{equation}
which is the same as the canonical energy--momentum density.
The conservation law (\ref{ctem9}) applied to the energy--momentum density (\ref{emts1}) gives the Dirac equations (\ref{Direq4}) and (\ref{Direq7}).
We can obtain the metric energy--momentum tensor for such a field using the Belinfante--Rosenfeld relation (\ref{BelRos8}), the spin tensor (\ref{Lagsp8}), and the torsion tensor in (\ref{Lagsp11}).
We can also derive the metric energy--momentum tensor by varying (\ref{Lagsp3}) with respect to $g^{ik}$:
\begin{equation}
T_{ik}=\frac{2}{\sqrt{-g}}\frac{\delta\mathfrak{L}_\psi}{\delta g^{ik}}.
\end{equation}
Using an identity
\begin{equation}
\frac{\delta\gamma^j}{\delta g^{ik}}=\frac{1}{2}\delta^j_{(i}\gamma_{k)},
\end{equation}
which results from the definition of the Dirac matrices (\ref{spin2}), leads to
\begin{equation}
T_{ik}=\frac{i}{2}(\bar{\psi}\gamma_{(i}\psi_{;k)}-\bar{\psi}_{;(i}\gamma_{k)}\psi-g_{ik}\bar{\psi}\gamma^j\psi_{;j}+g_{ik}\bar{\psi}_{;j}\gamma^j\psi)+mcg_{ik}\bar{\psi}\psi.
\label{emts4}
\end{equation}
The conservation law (\ref{cmem15}) applied to the energy--momentum tensor (\ref{emts4}) gives the Dirac equations (\ref{Direq4}) and (\ref{Direq7}).
Substituting the Dirac equation (\ref{Direq4}) into the (\ref{emts4}) gives
\begin{equation}
T_{ik}=\frac{i}{2}(\bar{\psi}\delta^j_{(i}\gamma_{k)}\psi_{;j}-\bar{\psi}_{;j}\delta^j_{(i}\gamma_{k)}\psi).
\label{emts5}
\end{equation}

Putting (\ref{spinco23}) and (\ref{spinco24}) with (\ref{Lagsp11}) into (\ref{emts5}) yields
\begin{equation}
T_{ik}=\frac{i}{2}(\bar{\psi}\delta^j_{(i}\gamma_{k)}\psi_{:j}-\bar{\psi}_{:j}\delta^j_{(i}\gamma_{k)}\psi)+\frac{\kappa}{2}(-s_i s_k+s^l s_l g_{ik}).
\end{equation}
Substituting the completely antisymmetric spin tensor (\ref{Lagsp8}) into (\ref{EC19}) gives
\begin{equation}
U^{ik}=\frac{\kappa}{4}\biggl(s^{ijl}s^k_{\phantom{k}jl}-\frac{1}{2}g^{ik}s^{jlm}s_{jlm}\biggr)=\frac{\kappa}{4}(2s^i s^k+s^l s_l g^{ik}).
\end{equation}
The combined energy--momentum tensor for a Dirac field is thus
\begin{equation}
T_{ik}+U_{ik}=\frac{i}{2}(\bar{\psi}\delta^j_{(i}\gamma_{k)}\psi_{:j}-\bar{\psi}_{:j}\delta^j_{(i}\gamma_{k)}\psi)+\frac{3\kappa}{4}s^l s_l g_{ik}.
\label{emts8}
\end{equation}
The tensor (\ref{emts8}) is equal to the energy--momentum tensor for the effective metric Lagrangian density (\ref{Direq9}):
\begin{equation}
T_{ik}=\frac{2}{\sqrt{-g}}\frac{\delta\mathring{\mathfrak{L}}_\psi}{\delta g^{ik}}.
\end{equation}

The first term on the right of (\ref{emts8}) is the Riemannian part of the energy--momentum tensor for a Dirac field and can be macroscopically averaged as an ideal fluid with the energy density $\epsilon$ and pressure $p$.
In the comoving frame of reference, in which $s^0=0$ because of $s_i u^i=0$, the second term on the right of (\ref{emts8}) is equal to $-(3/4)\kappa{\bf s}^2 g_{ik}$, where ${\bf s}$ is the spatial spin pseudovector.
The average value of ${\bf s}^2$ is proportional to $n^2$, where $n$ is the concentration of spinor particles.
The averaged second term on the right of (\ref{emts8}) has therefore a negative contribution to the energy density, $0>\tilde{\epsilon}\propto n^2$, and a positive contribution to the pressure, $\tilde{p}=-\tilde{\epsilon}$.

\subsubsection{Discrete symmetries of spinors}
The spinor representation of the parity transformation (\ref{sub3}) is given by
\begin{equation}
L_P=C_P\gamma^0,
\label{disc1}
\end{equation}
where $C_P=\mbox{const}$.
Substituting (\ref{disc1}) into (\ref{spin4}) with $L=L_P$ gives, using (\ref{Dirmat5}),
\begin{equation}
\gamma^a=\Lambda^a_{\phantom{a}b}\gamma^0\gamma^b\gamma^0=\Lambda^a_{\phantom{a}b}\gamma^{b\dag},
\end{equation}
which is satisfied if $\Lambda^a_{\phantom{a}b}=\Lambda^a_{\phantom{a}b}(P)$.
Since the double parity transformation is equivalent to the identity transformation, $L_P^2=I_4$, we have $C_P=\pm1$.
The spinor representation of the time-reversal transformation (\ref{sub4}) is given by
\begin{equation}
L_T=C_T\gamma^0\gamma^5,
\label{disc3}
\end{equation}
where $C_T=\mbox{const}$.
Substituting (\ref{disc3}) into (\ref{spin4}) with $L=L_T$ gives, using (\ref{Dirmat5}) and (\ref{Dirmat8}),
\begin{equation}
\gamma^a=\Lambda^a_{\phantom{a}b}\gamma^0\gamma^5\gamma^b\gamma^5\gamma^0=-\Lambda^a_{\phantom{a}b}\gamma^{b\dag},
\end{equation}
which is satisfied if $\Lambda^a_{\phantom{a}b}=\Lambda^a_{\phantom{a}b}(T)$.
Since the double time-reversal transformation is equivalent to the identity transformation, $L_T^2=I_4$, we have $C_T=\pm i$.

The {\em charge conjugation} of a spinor $\psi$ is defined as
\begin{equation}
\psi^c=-i\gamma^2\psi^\ast,\quad \psi^\ast=-i\gamma^2\psi^c.
\label{disc5}
\end{equation}
The double charge-conjugation transformation is equivalent to the identity transformation:
\begin{equation}
(\psi^c)^c=-i\gamma^2(\psi^c)^\ast=-i\gamma^2(-i\gamma^2\psi^\ast)^\ast=\gamma^2\gamma^{2\ast}\psi=\psi.
\end{equation}
The charge conjugation of the left-handed projection of a spinor is the right-handed projection of the charge conjugation of the spinor and vice versa (as in (\ref{Dirmat27})):
\begin{equation}
\Bigl((I_4\mp\gamma^5)\psi\Bigr)^c=-i\gamma^2\Bigl((I_4\mp\gamma^5)\psi\Bigr)^\ast=-i\gamma^2(I_4\mp\gamma^5)\psi^\ast=-i(I_4\pm\gamma^5)\gamma^2\psi^\ast=(I_4\pm\gamma^5)\gamma^c.
\end{equation}

\subsubsection{Spinor as particle}
A spinor can be associated with a particle, whose four-velocity is given by
\begin{equation}
u^i=\frac{\bar{\psi}\gamma^i \psi}{\bar{\psi}\psi}.
\label{franiko}
\end{equation}
It is equal to the ratio of two real quantities: the Dirac current vector $j^i=\bar{\psi}\gamma^i\psi$ (\ref{Direq127}) and the scalar $\bar{\psi}\psi$.

It is straightforward to demonstrate that the relation (\ref{franiko}) is satisfied for a spinor in flat spacetime (in the absence of torsion).
In the rest frame of reference, a solution of the Dirac equation (\ref{Direq4}) is given by (\ref{Dirmat28}), where $u$ and $v$ are two-component columns with constant elements.
For a particle moving with momentum ${\bf p}$ and energy $E$, the corresponding spinor is given by $L\psi$, where the matrix $L$ is the spinor representation of a boost (\ref{Dirmat37}):
\begin{equation}
\psi({\bf r},t)=\frac{1}{\sqrt{2mc^2(E+mc^2)}}\left( \begin{array}{cc}
    (E+mc^2)I_2 & {\bm\sigma}\cdot{\bf p}c \\
    {\bm\sigma}\cdot{\bf p}c & (E+mc^2)I_2 \end{array} \right)\left( \begin{array}{c}
    u \\
    v \end{array} \right)\exp[i({\bf p}\cdot{\bf r}-Et)].
    \label{plane}
\end{equation}

For a case with $u=\left( \begin{array}{c}
    1 \\
    0 \\ \end{array} \right)$ and $v=\left( \begin{array}{c}
    0 \\
    0 \\ \end{array} \right)$,
the scalar bilinear composed from (\ref{plane}) is
\begin{equation}    \bar{\psi}\psi=\psi^\dagger\gamma^0\psi=\frac{1}{2mc^2(E+mc^2)}\Bigl[(E+mc^2)^2-(1,0)({\bm\sigma}\cdot{\bf p})({\bm\sigma}\cdot{\bf p})c^2
    \left( \begin{array}{c}
    1 \\
    0 \\ \end{array} \right)\Bigr]=\frac{(E+mc^2)^2-{\bf p}^2 c^2}{2mc^2(E+mc^2)}=1.
\end{equation}
The vector bilinear components composed from (\ref{plane}), using the relation $p^i=mcu^i$ for a free particle and an identity $\sigma^\alpha\sigma^\beta+\sigma^\beta\sigma^\alpha=-2\eta^{\alpha\beta}I_2$, which follows from (\ref{Dirmat3}), are
\begin{eqnarray}
    & & \bar{\psi}\gamma^0\psi=\psi^\dagger\psi=\frac{1}{2mc^2(E+mc^2)}\Bigl[(E+mc^2)^2+(1,0)({\bm\sigma}\cdot{\bf p})({\bm\sigma}\cdot{\bf p})c^2
    \left( \begin{array}{c}
    1 \\
    0 \\ \end{array} \right)\Bigr] \nonumber \\
    & & =\frac{(E+mc^2)^2+{\bf p}^2 c^2}{2mc^2(E+mc^2)}=\frac{E}{mc^2}=u^0
\end{eqnarray}
and
\begin{eqnarray}
    & & \bar{\psi}\gamma^\alpha\psi=\psi^\dagger\gamma^0\gamma^\alpha\psi=\frac{E+mc^2}{2mc^2(E+mc^2)}\Bigl[(1,0)\sigma^\alpha({\bm\sigma}\cdot{\bf p})c\left( \begin{array}{c}
    1 \\
    0 \\ \end{array} \right)+(1,0)({\bm\sigma}\cdot{\bf p})c\sigma^\alpha\left( \begin{array}{c}
    1 \\
    0 \\ \end{array} \right)\Bigr]\nonumber \\
    & & =\frac{p^\alpha}{mc}=u^\alpha.
\end{eqnarray}
These bilinears give the four-velocity (\ref{franiko}).
This formula is satisfied for the quantities $u$ and $v$ with other constant components too.
It is also valid in curved spacetime because at the location of the particle one can construct a locally flat system of coordinates, in which (\ref{franiko}) is valid.

The conservation of the momentum four-vector $P_i$ follows from the symmetry of a system under spacetime translations, as stated in Section \ref{Meaning}.
The four-momentum of a spinor is therefore associated with a generator of translation, which in curved spacetime is given by a covariant derivative:
\begin{equation}
P_i=\frac{i(\bar{\psi}\psi_{;i}-\bar{\psi}_{;i}\psi)}{2\bar{\psi}\psi},
\label{franiko1}
\end{equation}
following (\ref{Poin22}) and (\ref{Cas9}).
This quantity is real because $\bar{\psi}\psi_{;i}-\bar{\psi}_{;i}\psi$ is imaginary:
\begin{eqnarray}
& & (\bar{\psi}\psi_{;i}-\bar{\psi}_{;i}\psi)^\ast=(\bar{\psi}\psi_{;i}-\bar{\psi}_{;i}\psi)^\dagger=\bigl(\psi^\dagger\gamma^0\psi_{;i}-(\psi^\dagger\gamma^0)_{|i}\psi\bigr)^\dagger=\bigl(\psi^\dagger(\gamma^0\psi)_{|i}-\psi^\dagger_{;i}\gamma^0\psi\bigr)^\dagger \nonumber \\
& & =(\psi^\dagger\gamma^0)_{|i}\psi-\psi^\dagger\gamma^0\psi_{;i}=-(\bar{\psi}\psi_{;i}-\bar{\psi}_{;i}\psi),
\end{eqnarray}
using (\ref{spinco7}), (\ref{cursp4}), and (\ref{Dirmat23}).

The conservation of the intrinsic angular momentum (spin) four-tensor $S_{ik}$ follows from the symmetry of a system under spacetime rotations, as stated in Section \ref{Meaning}.
The spin four-tensor of a spinor is therefore associated with a generator of rotation $G_{ik}$ (\ref{spin6}):
\begin{equation}
S_{ik}=\frac{i\bar{\psi}G_{ik}\psi}{\bar{\psi}\psi},
\label{franiko2}
\end{equation}
following (\ref{Cas10}).
This quantity is real because $\bar{\psi}G_{ik}\psi$ is imaginary:
\begin{equation}
(\bar{\psi}G_{ik}\psi)^\ast=(\bar{\psi}G_{ik}\psi)^\dagger=(\psi^\dagger\gamma^0 G_{ik}\psi)^\dagger=\psi^\dagger G_{ik}^\dagger\gamma^0\psi=-\psi^\dagger\gamma^0 G_{ik}\psi=-\bar{\psi}G_{ik}\psi,
\end{equation}
using (\ref{Dirmat20}) and (\ref{Dirmat23}).

It can be demonstrated that if a spinor satisfies the Dirac equation, then the four-velocity $u^i$ (\ref{franiko}), four-momentum $P_i$ (\ref{franiko1}), and spin $S_{ik}$ (\ref{franiko2}) of the corresponding particle satisfy the relation (\ref{mapar2}) and the Mathisson--Papapetrou equations of motion (\ref{MP3}) and (\ref{MP7}).
\newline
References: \cite{Lord,Hehl1,KS,Hehl4,FraNiko}.

\subsection{Electromagnetic field}
\setcounter{equation}{0}
\subsubsection{Gauge invariance and electromagnetic potential}
The Lagrangian density (\ref{Lagsp3}) is a real combination of the complex Dirac matrices $\gamma^i$ and spinors $\psi$, $\bar{\psi}$.
We define a {\em gauge transformation of the first type} of the spinor fields,
\begin{equation}
\psi\rightarrow\psi'=e^{ie\alpha}\psi,\quad \bar{\psi}\rightarrow\bar{\psi}'=e^{-ie\alpha}\bar{\psi},
\label{gi1}
\end{equation}
where $e$ is a real scalar constant, called the {\em spinor electric charge}.
If $\alpha$ is a real scalar constant, then (\ref{Lagsp3}) is invariant under (\ref{gi1}).
Such a transformation is called {\em global}.
If $\alpha=\alpha(x^i)$ is a scalar function of the coordinates, then (\ref{Lagsp3}) is not invariant under (\ref{gi1}) because
\begin{equation}
\psi'_{;k}=e^{ie\alpha}(\psi_{;k}+ie\alpha_{,k}\psi).
\label{gi2}
\end{equation}
Such a transformation is called {\em local}.

For a local transformation, we must introduce a compensating vector field $A_k$, referred to as the {\em electromagnetic potential}, such that the Weyl {\em electromagnetic covariant derivative}
\begin{equation}
D_k=\nabla_k-ieA_k
\end{equation}
of a spinor $\psi$,
\begin{equation}
D_k\psi=\psi_{;k}-ieA_k\psi,
\label{gi4}
\end{equation}
transforms under (\ref{gi1}) like $\psi$:
\begin{equation}
D_k\psi'=e^{ie\alpha}D_k\psi.
\end{equation}
This requirement gives
\begin{equation}
\psi'_{;k}-ieA'_k\psi'=e^{ie\alpha}(\psi_{;k}-ieA_k\psi),
\end{equation}
which, with (\ref{gi1}) and (\ref{gi2}), yields the transformation law for the electromagnetic potential,
\begin{equation}
A_k\rightarrow A'_k=A_k+\alpha_{,k}.
\label{gi7}
\end{equation}
This law is called a {\em gauge transformation of the second type}.
The adjoint conjugation of (\ref{gi4}) is
\begin{equation}
D_k\bar{\psi}=\bar{\psi}_{;k}+ieA^\ast_k\bar{\psi}.
\label{gi8}
\end{equation}
The scalar $\bar{\psi}\psi$ is invariant under (\ref{gi1}), thereby
\begin{equation}
D_k\bar{\psi}\psi+\bar{\psi}D_k\psi=D_k(\bar{\psi}\psi)=\partial_k(\bar{\psi}\psi),
\end{equation}
which constrains the electromagnetic potential to be real:
\begin{equation}
A^\ast_k=A_k.
\end{equation}
All laws of physics are {\em gauge-invariant}: they do not change their forms under the transformations (\ref{gi1}) and (\ref{gi7}).\\

\noindent
{\bf Electric and magnetic potentials}.\\
The time component of $A^i$, $\phi=A^0$, is called the {\em electric potential} and the spatial components $A^\alpha$ form the {\em magnetic potential} ${\bf A}$:
\begin{equation}
A^i=(\phi,{\bf A}).
\label{gi11}
\end{equation}
The gauge transformation (\ref{gi7}) can be written as
\begin{equation}
\phi'=\phi+\frac{1}{c}\frac{\partial\alpha}{\partial t},\quad {\bf A}'={\bf A}-\mbox{{\bf grad}}\,\alpha.
\label{gi12}
\end{equation}
In the local Minkowski spacetime, for a boost in the $X$ direction with speed $V=c\beta$, the components $A^i$ transform according to (\ref{relkin33}):
\begin{eqnarray}
& & \phi=\gamma(\phi'+\beta A'_x), \nonumber \\
& & A_x=\gamma(A'_x+\beta \phi'), \nonumber \\
& & A_y=A'_y,\quad  A_z=A'_z.
\label{gi13}
\end{eqnarray}
For a boost in an arbitrary direction with velocity ${\bf V}=c{\bm\beta}$, the components 
$A^i$ transform according to (\ref{relkin15}):
\begin{equation}
\left( \begin{array}{c}
\phi \\
{\bf A} \end{array} \right)
=\left( \begin{array}{cc}
\gamma & \gamma{\bm\beta} \\
\gamma{\bm\beta} & 1+[(\gamma-1){\bm\beta}/\beta^2]{\bm\beta} \end{array} \right)
\left( \begin{array}{cc}
\phi' \\
{\bf A}' \end{array} \right).
\end{equation}

\subsubsection{Electromagnetic field tensor}
\label{Faraday}
The commutator of total covariant derivatives of a spinor is given by (\ref{cursp1}) with the curvature spinor $K_{ij}$ given by (\ref{cursp7}), where the tensor $B_{ij}$ is related to the vector $A_i$ in (\ref{spinco15}) by (\ref{cursp9}).
Therefore, the commutator of the electromagnetic covariant derivatives of a spinor, $[D_i,D_j]\psi$, is given by (\ref{cursp1}) with the curvature spinor
\begin{equation}
K_{ij}=\frac{1}{4}R_{klij}\gamma^k \gamma^l+ieF_{ij}I_4,
\end{equation}
where the antisymmetric tensor
\begin{equation}
F_{ij}=-F_{ji}=A_{j,i}-A_{i,j}=A_{j:i}-A_{i:j}
\label{eft2}
\end{equation}
is referred to as the {\em electromagnetic field tensor}.
The electromagnetic field tensor is analogous to the curvature tensor: it appears in the expression for the commutator of electromagnetic covariant derivatives of a spinor, while the curvature tensor appears in the expression for the commutator of coordinate-covariant derivatives of a tensor.
Substituting (\ref{gi7}) into (\ref{eft2}) gives
\begin{equation}
F'_{ij}=F_{ij},
\end{equation}
so the electromagnetic field tensor is gauge invariant.
Consequently, the laws of physics can include the electromagnetic potential only through the electromagnetic field tensor.
The definition (\ref{eft2}) is equivalent to the {\em first Maxwell--Minkowski equation}:
\begin{equation}
F_{ij,k}+F_{jk,i}+F_{ki,j}=F_{ij:k}+F_{jk:i}+F_{ki:j}=0
\label{eft4},
\end{equation}
which can be written as
\begin{equation}
\epsilon^{ijkl}F_{jk,l}=\epsilon^{ijkl}F_{jk:l}=0.
\label{eft5}
\end{equation}

\noindent
{\bf Electric and magnetic fields}.\\
We define a spatial vector ${\bf E}$ whose covariant components are related to the $0\alpha$ components of the electromagnetic field tensor (\ref{eft2}):
\begin{equation}
E_\alpha=F_{0\alpha}.
\end{equation}
We also define a spatial tensor $B_{\alpha\beta}$ equal to the spatial part of (\ref{eft2}):
\begin{equation}
B_{\alpha\beta}=F_{\alpha\beta},\quad B^\alpha=-\frac{1}{2}\eta^{\alpha\beta\gamma}B_{\beta\gamma},\quad B_{\alpha\beta}=-\eta_{\alpha\beta\gamma}B^\gamma,
\end{equation}
where $\eta^{\alpha\beta\gamma}$ and $\eta_{\alpha\beta\gamma}$ are the components of the spatial unit antisymmetric pseudotensor (\ref{spvec106}).
A spatial pseudovector ${\bf B}$ is defined as dual to this tensor.
The spatial vector ${\bf E}$ is referred to as the {\em electric field} and the spatial pseudovector ${\bf B}$ is referred to as the {\em magnetic field}.
In a locally geodesic and Galilean frame of reference, these fields depend on the components of the electromagnetic potential (\ref{gi11}) according to (\ref{eft2}):
\begin{eqnarray}
& & {\bf E}=-\frac{1}{c}\frac{\partial{\bf A}}{\partial t}-\mbox{{\bf grad}}\,\phi, \label{eft10} \\
& & {\bf B}=\mbox{{\bf curl}}\,{\bf A}.
\label{eft11}
\end{eqnarray}
They are invariant under the gauge transformation (\ref{gi12}).\\

\noindent
{\bf Lorentz transformation of electric and magnetic fields}.\\
In a locally geodesic and Galilean frame of reference, the tensor $F_{ij}$ is given by
\begin{equation}
F_{ij}=\left( \begin{array}{cccc}
0 & E_x & E_y & E_z \\
-E_x & 0 & -B_z & B_y \\
-E_y & B_z & 0 & -B_x \\
-E_z & -B_y & B_x & 0 \end{array} \right),\quad {\bf E}=(F_{01},F_{02},F_{03}),\quad {\bf B}=(F_{32},F_{13},F_{21}).
\end{equation}
The components $F_{ij}$ determine how the electric and magnetic fields transform under the Lorentz transformation.
For a boost in the $X$ direction with speed $V=c\beta$, these components transform according to (\ref{relkin16}) in which the antisymmetry $F_{ij}=-F_{ji}$ is used:
\begin{eqnarray}
& & F_{01}=\gamma(F_{01'}-\beta F_{00'})=\gamma^2(F_{0'1'}-\beta F_{1'1'}-\beta F_{0'0'}+\beta^2 F_{1'0'})=F_{01'}, \nonumber \\
& & F_{0\perp}=\gamma(F_{0'\perp'}-\beta F_{1'\perp'}), \nonumber \\
& & F_{1\perp}=\gamma(F_{1'\perp'}-\beta F_{0'\perp'}).
\label{eft136}
\end{eqnarray}
Consequently, the electric and magnetic fields transform according to
\begin{eqnarray}
& & E_x=E'_x,\quad  E_y=\gamma(E'_y+\beta B'_z),\quad  E_z=\gamma(E'_z-\beta B'_y), \nonumber \\
& & B_x=B'_x,\quad  B_y=\gamma(B'_y-\beta E'_z),\quad  B_z=\gamma(B'_z+\beta E'_y).
\label{eft137}
\end{eqnarray}
For a boost in an arbitrary direction with velocity ${\bf V}=c{\bm\beta}$, the fields transform according to
\begin{eqnarray}
& & {\bf E}=\gamma({\bf E}'-{\bm \beta}\times{\bf B}')+\frac{1-\gamma}{\beta^2}({\bm \beta}\cdot{\bf E}'){\bm \beta}, \label{eft13} \\
& & {\bf B}=\gamma({\bf B}'+{\bm \beta}\times{\bf E}')+\frac{1-\gamma}{\beta^2}({\bm \beta}\cdot{\bf B}'){\bm \beta}.
\label{eft14}
\end{eqnarray}

The simplest invariants (under proper Lorentz transformations) of the electromagnetic field are quadratic in $F_{ij}$:
\begin{equation}
F_{ij}F^{ij}=2(B^2-E^2)=\mbox{const},\quad e^{ijkl}F_{ij}F_{kl}=8{\bf E}\cdot{\bf B}=\mbox{const}.
\label{eft19}
\end{equation}
If the vectors ${\bf E}$ and ${\bf B}$ are mutually perpendicular in frame $K$, ${\bf E}\cdot{\bf B}=0$, then they are mutually perpendicular in other inertial frames.
If ${\bf E}$ and ${\bf B}$ are equal in magnitude in $K$, $B^2-E^2=0$, then they are equal in magnitude in other inertial frames. 
The transformation laws (\ref{eft13}) and (\ref{eft14}) imply that if ${\bf E}'=0$ in the frame $K'$, then in the frame $K$:
\begin{equation}
{\bf E}=-\gamma{\bm \beta}\times{\bf B}'=-{\bm \beta}\times{\bf B},
\end{equation}
and if ${\bf B}'=0$ in the frame $K'$, then in the frame $K$:
\begin{equation}
{\bf B}=\gamma{\bm \beta}\times{\bf E}'={\bm \beta}\times{\bf E}.
\label{eft21}
\end{equation}
If the vectors ${\bf E}$ and ${\bf B}$ are mutually perpendicular in $K$, but not equal in magnitude, then there exists a frame $K'$ in which the field is either electric, ${\bf B}'=0$ (if $E>B$), or magnetic, ${\bf E}'=0$ (if $E<B$).
The velocity of $K'$ relative to $K$ is perpendicular to ${\bf E}$ and ${\bf B}$, and it is equal in magnitude to respectively either $cB/E$ or $cE/B$.
Equivalently, if one of the vectors ${\bf E},{\bf B}$ vanishes in one frame of reference then these vectors are mutually perpendicular in other inertial frames.

Except for the case where the vectors ${\bf E}$ and ${\bf B}$ are mutually perpendicular and equal in magnitude, there exist frames in which these vectors are parallel to each other at a given point.
These frames move relative to one another with velocities parallel to both vectors.
One of such frames, $K'$ (in which ${\bf E}'\parallel{\bf B}'$), has a velocity ${\bf V}$ relative to $K$ which is perpendicular to both vectors ${\bf E}$ and ${\bf B}$.
Substituting the formulae
\begin{eqnarray}
& & {\bf E}'=\gamma({\bf E}+{\bm \beta}\times{\bf B})+\frac{1-\gamma}{\beta^2}({\bm \beta}\cdot{\bf E}){\bm \beta}, \\
& & {\bf B}'=\gamma({\bf B}-{\bm \beta}\times{\bf E})+\frac{1-\gamma}{\beta^2}({\bm \beta}\cdot{\bf B}){\bm \beta},
\end{eqnarray}
which are inverse to (\ref{eft13}) and (\ref{eft14}), into the condition ${\bf E}'\times{\bf B}'=0$ and using ${\bm\beta}=k{\bf E}\times{\bf B}$, where $k$ is a constant of proportionality, gives $\Bigl({\bf E}+k({\bf E}\times{\bf B})\times{\bf B}\Bigr)\times\Bigl({\bf B}-k({\bf E}\times{\bf B})\times{\bf E}\Bigr)=0$ or $k=(1+\beta^2)/(E^2+B^2)$, thereby
\begin{equation}
\frac{{\bf V}/c}{1+V^2/c^2}=\frac{{\bf E}\times{\bf B}}{E^2+B^2}.
\end{equation}

\subsubsection{First pair of Maxwell equations}
The components of (\ref{eft4}) with all spatial indices, $B_{\alpha\beta,\gamma}+B_{\beta\gamma,\alpha}+B_{\gamma\alpha,\beta}=0$, give, using (\ref{spvec28}),
\begin{equation}
\mbox{div}\,{\bf B}=0.
\label{eft8}
\end{equation}
This equation also follows from applying the divergence operator to (\ref{eft11}) gives (\ref{eft8}).
The components of (\ref{eft4}) with one temporal index, $B_{\alpha\beta,0}+E_{\alpha,\beta}-E_{\beta,\alpha}=0$, gives, using (\ref{spvec29}),
\begin{equation}
\mbox{{\bf curl}}\,{\bf E}=-\frac{1}{c\sqrt{\mathfrak{s}}}\frac{\partial(\sqrt{\mathfrak{s}}{\bf B})}{\partial t},
\label{eft9}
\end{equation}
where $\mathfrak{s}$ is given by (\ref{spvec9}).
Applying the divergence operator to (\ref{eft9}) gives (\ref{eft8}).
The relations (\ref{eft8}) and (\ref{eft9}) are referred to as the {\em first pair of the Maxwell equations}.
In a locally geodesic and Galilean frame of reference, in which $\mathfrak{s}=1$, the first pair of the Maxwell equations (\ref{eft8}) and (\ref{eft9}) is given by
\begin{eqnarray}
& & \mbox{div}\,{\bf B}=0, \label{eft15} \\
& & \mbox{{\bf curl}}\,{\bf E}=-\frac{1}{c}\frac{\partial{\bf B}}{\partial t}.
\label{eft16}
\end{eqnarray}
Applying the curl operator to (\ref{eft10}) gives (\ref{eft16}).\\

\noindent
{\bf Faraday law}.\\
Integrating (\ref{eft8}) over a volume, using (\ref{covint29}), and integrating (\ref{eft9}) over a surface, using (\ref{covint28}), respectively give
\begin{eqnarray}
& & \oint{\bf B}\cdot \sqrt{\mathfrak{s}}d{\bf f}=0, \\
& & \oint{\bf E}\cdot d{\bf l}=-\frac{1}{c}\frac{\partial}{\partial t}\biggl(\int{\bf B}\cdot \sqrt{\mathfrak{s}}d{\bf f}\biggr).
\end{eqnarray}
In a locally flat spacetime, these relations reduce to
\begin{eqnarray}
& & \oint{\bf B}\cdot d{\bf f}=0, \\
& & \oint{\bf E}\cdot d{\bf l}=-\frac{1}{c}\frac{\partial}{\partial t}\biggl(\int{\bf B}\cdot d{\bf f}\biggr).
\end{eqnarray}
Therefore, the flux of the magnetic field, which is referred to as the {\em magnetic flux}, through a closed surface vanishes.
The circulation of the electric field along a closed contour, which is referred to as the {\em electromotive force}, is equal to the minus time derivative of the flux of the magnetic field through the surface enclosed by this contour, constituting {\em Faraday's law of electromagnetic induction}.

\subsubsection{Lagrangian density for electromagnetic field}
The simplest gauge-invariant Lagrangian density representing the electromagnetic field is a linear combination of terms quadratic in $F_{ij}$: $\sqrt{-\mathfrak{g}}F_{ij}F^{ij}$ and $\epsilon^{ijkl}F_{ij}F_{kl}$, which a locally Galilean frame of reference reduce to (\ref{eft19}).
The second, parity-violating term is a total divergence because of (\ref{eft5}):
\begin{equation}
\epsilon^{ijkl}F_{ij}F_{kl}=2(\epsilon^{ijkl}F_{ij}A_l)_{,k},
\end{equation}
so it does not contribute to the field equations.
Therefore, the Lagrangian density for the electromagnetic field is given by
\begin{equation}
\mathfrak{L}_{EM}=-\frac{1}{16\pi}\sqrt{-\mathfrak{g}}F_{ij}F^{ij},
\label{Lagem2}
\end{equation}
where the Gau{\ss}ian factor $1/16\pi$ sets the units of $A_i$.
In the locally geodesic and Galilean frame of reference, (\ref{Lagem2}) becomes
\begin{equation}
\mathfrak{L}_{EM}=\frac{1}{8\pi}(E^2-B^2).
\end{equation}
Therefore, in order for the action $S$ to have a minimum, there must be the minus sign in front of the right-hand side of (\ref{Lagem2}).
Otherwise an arbitrarily rapid change of ${\bf A}$ in time would result in an arbitrarily large value of ${\bf E}$, according to (\ref{eft10}), and thus an arbitrarily low value of $S$, thereby the action would have no minimum.\\

\begin{footnotesize}
A generalization of the tensor (\ref{eft2}) to the covariant derivatives with respect to the affine connection $\Gamma^{k}_{ij}$, $\tilde{F}_{ij}=A_{j;i}-A_{i;j}=F_{ij}+2S^k_{\phantom{k}ij}A_k$, is not gauge invariant, thereby the torsion tensor cannot appear in a gauge-invariant Lagrangian density which is quadratic in $F_{ij}$.
Therefore, the electromagnetic field, unlike spinor fields, does not couple to torsion.
Accordingly, the electromagnetic field is not minimally coupled to the affine connection.
\end{footnotesize}

\subsubsection{Electromagnetic current and electric charge}
We define a four-vector density:
\begin{equation}
{\sf j}^i=-c\frac{\delta\mathfrak{L}_\textrm{m}}{\delta A_i},
\label{emc1}
\end{equation}
referred to as the {\em electromagnetic current density}.
We also define a four-vector:
\begin{equation}
j^i=\frac{{\sf j}^i}{\sqrt{-\mathfrak{g}}},
\end{equation}
referred to as the {\em electromagnetic current}.
The invariance of the action under an arbitrary infinitesimal gauge transformation $\delta A_i=A'_i-A_i=\phi_{,i}$ gives, upon partial integration and omitting a total divergence,
\begin{equation}
\delta S=-\frac{1}{c^2}\int{\sf j}^i\delta A_j d\Omega=-\frac{1}{c^2}\int{\sf j}^i\phi_{,i}d\Omega=\frac{1}{c^2}\int{\sf j}^i_{\phantom{i},i}\phi\,d\Omega=0.
\end{equation}
Consequently, the electromagnetic current density and current are conserved:
\begin{equation}
{\sf j}^i_{\phantom{i},i}=0,\quad j^i_{\phantom{i}:i}=0.
\label{emc4}
\end{equation}
The conservation law (\ref{emc4}) has a form of the equation of continuity, as (\ref{Direq12}).

We define the {\em electric charge} of matter on a hypersurface as
\begin{equation}
e=\frac{1}{c}\int{\sf j}^i dS_i,
\end{equation}
where $dS_i$ is the element of hypersurface.
Because ${\sf j}^i$ is a vector density, the electric charge is a scalar.
If the hypersurface is taken as a hyperplane perpendicular to the $x^0$ axis, then the charge is given by the volume integral:
\begin{equation}
e=\frac{1}{c}\int{\sf j}^0 dV,
\label{emc0}
\end{equation}
where $dV$ is the element of volume.

\subsubsection{Charged particle}
{\bf Multipoles of current density}.\\
Let us consider a particle, whose spacetime location is represented by the coordinates $X^i$, as in sections \ref{SpinParticles} and \ref{MomentumParticles}.
Integrating the conservation law for the electromagnetic current density (\ref{emc4}) over the volume of the body at a constant time $X^0$ and using Gau\ss' theorem to eliminate surface integrals gives
\begin{equation}
\int{\sf j}^0_{\phantom{0},0}dV=\biggl(\int{\sf j}^0 dV\biggr)_{,0}=\frac{1}{u^0}\frac{d}{ds}\int{\sf j}^0 dV=\frac{c}{u^0}\frac{de}{ds}=0,
\label{emc6}
\end{equation}
where $u^i=dX^i/ds$ (\ref{muex3}).
The relation (\ref{emc6}) shows the constancy of the charge of a particle along a world line.

The conservation law (\ref{emc4}) also gives
\begin{eqnarray}
(x^k{\sf j}^i)_{,i}=x^k_{\phantom{k},i}{\sf j}^i+x^k{\sf j}^i_{\phantom{i},i}=\delta^k_i{\sf j}^i={\sf j}^k.
\label{emc7}
\end{eqnarray}
Integrating this relation over the volume of the body and using Gau\ss' theorem to eliminate surface integrals gives
\begin{equation}
\biggl(\int x^k{\sf j}^0 dV\biggr)_{,0}=\int{\sf j}^k dV.
\label{emc8}
\end{equation}
Using $x^i=X^i$, $X^l_{\phantom{l},0}=u^l/u^0$, and (\ref{emc6}) brings (\ref{emc8}) to
\begin{equation}
\frac{u^k}{u^0}\int{\sf j}^0 dV=\int{\sf j}^k dV.
\end{equation}
For a particle located at ${\bf x}_a$, the density ${\sf j}^i({\bf x})$ is proportional to ${\bm \delta}({\bf x}-{\bf x}_a)$, giving
\begin{equation}
{\sf j}^k=\frac{u^k}{u^0}{\sf j}^0,\quad j^k=\frac{u^k}{u^0}j^0.
\label{emc10}
\end{equation}
Putting $k=0$ in these relations gives the identity.
The electromagnetic current density of a particle is proportional to its four-velocity, analogously to (\ref{emtp8}).
The quantities analogous to (\ref{emc7}) with higher multiples of $x^i$ do not introduce new relations.\\

\noindent
{\bf Electromagnetic current for particle}.\\
We define the {\em electric charge density} $\rho$ such that
\begin{equation}
j^0=\frac{c\rho}{\sqrt{g_{00}}}.
\end{equation}
Consequently, the electromagnetic current vector is equal to
\begin{equation}
j^k=\frac{c\rho u^k}{\sqrt{g_{00}}u^0}=\frac{c\rho}{\sqrt{g_{00}}}\frac{dx^k}{dx^0}=\frac{\rho}{\sqrt{g_{00}}}\frac{dx^k}{dt}.
\label{emc137}
\end{equation}
The charge density is equal to the electric charge in unit volume, which is equivalent to
\begin{equation}
\rho\sqrt{\mathfrak{s}}dV=de.
\label{emc12}
\end{equation}
Accordingly, the charge density is a three-dimensional scalar density.
The invariance of the charge also follows from this equality of contravariant vectors:
\begin{equation}
de\,dx^i=\rho\sqrt{\mathfrak{s}}dV\,dx^i=\sqrt{-\mathfrak{g}}dV\,dt\,\frac{\rho}{\sqrt{g_{00}}}\frac{dx^i}{dt}=\frac{1}{c}\sqrt{-\mathfrak{g}}d\Omega j^i,
\end{equation}
in which $\sqrt{-\mathfrak{g}}d\Omega$ is a scalar.

For particles with charges $e_a$ located at ${\bf x}_a$, the charge density is given by
\begin{equation}
\rho({\bf x})=\sum_a\frac{e_a}{\sqrt{\mathfrak{s}}}{\bm \delta}({\bf x}-{\bf x}_a).
\end{equation}
The volume integral of the corresponding electromagnetic current density is
\begin{equation}
\int{\sf j}^0 dV=\sum_a\int\sqrt{-\mathfrak{g}}\frac{ce_a}{\sqrt{g_{00}}\sqrt{\mathfrak{s}}}{\bm \delta}({\bf x}-{\bf x}_a)dV=c\sum_a e_a,
\end{equation}
in agreement with (\ref{emc0}).
Following (\ref{emc10}), the electromagnetic current vector for a system of charged particles is
\begin{equation}
j^k({\bf x})=\sum_a\frac{cu^k}{u^0}\frac{e_a}{\sqrt{-\mathfrak{g}}}{\bm \delta}({\bf x}-{\bf x}_a),
\label{emc15}
\end{equation}
analogously to (\ref{emtp10}).
The relation (\ref{emc6}) represents the conservation of the total electric charge of a physical system.
For a particle moving along a world line $x_a(\tau)$, the current is
\begin{equation}
j^k(x)=\frac{cu^k}{u^0}\frac{e}{\sqrt{-\mathfrak{g}}}{\bm \delta}({\bf x}-{\bf x}_a)=ec\int\frac{u^k}{\sqrt{-\mathfrak{g}}}\delta(x-x_a(\tau))d\tau,
\end{equation}
where $\delta$ is the product of the delta functions for all four coordinates.\\

\noindent
{\bf Continuity equation for charge}.\\
The conservation law (\ref{emc4}) is equivalent, using (\ref{spvec6}) and (\ref{emc137}), to
\begin{eqnarray}
& & (\sqrt{-\mathfrak{g}}j^i)_{,i}=\Bigl(\sqrt{\mathfrak{s}}\rho\frac{dx^i}{dt}\Bigr)_{,i}=\Bigl(\sqrt{\mathfrak{s}}\rho\frac{dx^0}{dt}\Bigr)_{,0}+\Bigl(\sqrt{\mathfrak{s}}\rho\frac{dx^\alpha}{dt}\Bigr)_{,\alpha}=\frac{\partial}{\partial t}(\sqrt{\mathfrak{s}}\rho)+\Bigl(\sqrt{\mathfrak{s}}\rho v^\alpha\Bigr)_{,\alpha} \nonumber \\
& & =\frac{\partial}{\partial t}(\sqrt{\mathfrak{s}}\rho)+\sqrt{\mathfrak{s}}\,\mbox{div}(\rho{\bf v})=0.
\end{eqnarray}
This law has a form of the equation of continuity:
\begin{equation}
\frac{\partial}{\partial t}(\sqrt{\mathfrak{s}}\rho)+\sqrt{\mathfrak{s}}\,\mbox{div}\,{\bf j}=0,
\label{Max29}
\end{equation}
where ${\bf j}$ is the spatial {\em electric current density} vector:
\begin{equation}
{\bf j}=\rho{\bf v}.
\end{equation}
This vector is the spatial part of the electromagnetic current vector:
\begin{equation}
j^i=(c\rho,{\bf j}),
\end{equation}
which follows from $u^i/u^0=(1,{\bf v}/c)$.
The equation of continuity (\ref{Max29}) is analogous to (\ref{rh6}).

Integrating (\ref{Max29}) over a volume, using (\ref{covint29}) and (\ref{emc12}), gives
\begin{equation}
i=-\frac{de}{dt},
\label{emc41}
\end{equation}
where
\begin{equation}
i=\oint{\bf j}\cdot\sqrt{\mathfrak{s}}d{\bf f}
\label{emc42}
\end{equation}
is the {\em electric current} flowing through the closed surface enclosing the volume.
The electric current through an arbitrary surface satisfies
\begin{equation}
i=\frac{de}{dt},\quad i=\int{\bf j}\cdot\sqrt{\mathfrak{s}}d{\bf f}.
\end{equation}
The relations (\ref{emc41}) and (\ref{emc42}) represent the conservation of the total electric charge: the electric current leaving a region in space is equal to the minus time derivative of the electric charge inside this region.

In a locally flat spacetime, in which $-\mathfrak{g}=\mathfrak{s}=1$, the equation of continuity (\ref{Max29}) reduces to
\begin{equation}
j^i_{\phantom{i},i}=\frac{\partial\rho}{\partial t}+\mbox{div}\,{\bf j}=0.
\label{emc19}
\end{equation}
For one particle located at ${\bf x}_0(t)$, $\rho({\bf x})=e{\bm \delta}({\bf x}-{\bf x}_0)$, (\ref{emc19}) is explicitly satisfied because
\begin{eqnarray}
& & \frac{\partial\rho}{\partial t}=e\frac{\partial}{\partial t}{\bm \delta}({\bf x}-{\bf x}_0)=e{\bf v}\cdot\frac{\partial}{\partial {\bf x}_0}{\bm \delta}({\bf x}-{\bf x}_0)=-e{\bf v}\cdot\frac{\partial}{\partial {\bf x}}{\bm \delta}({\bf x}-{\bf x}_0) \nonumber \\
& & =-\frac{\partial}{\partial {\bf x}}\cdot\Bigl(e{\bf v}{\bm \delta}({\bf x}-{\bf x}_0)\Bigr)=-\mbox{div}\,{\bf j},
\end{eqnarray}
where ${\bf v}=d{\bf x}_0/dt$.
For a system of charged particles, we also have
\begin{equation}
\int{\bf j}\,dV=\sum_a e_a{\bf v}_a,
\label{emc21}
\end{equation}
where ${\bf v}_a=d{\bf x}_a/dt$ is the velocity of the particle of charge $e_a$.

\subsubsection{Second pair of Maxwell equations}
The Lagrangian density for the electromagnetic field and charged matter is the sum of (\ref{Lagem2}) and the term $-\sqrt{-\mathfrak{g}}A_i j^i$ arising from (\ref{emc1}):
\begin{equation}
\mathfrak{L}_{\textrm{EM}+\textrm{q}}=-\frac{1}{16\pi}\sqrt{-\mathfrak{g}}F_{ik}F^{ik}-\frac{1}{c}\sqrt{-\mathfrak{g}}A_k j^k,
\label{Max1}
\end{equation}
where we omitted terms corresponding to the gravitational field and matter fields which do not depend on $A_k$.
Varying (\ref{Max1}) with respect to the electromagnetic potential $A_k$ and omitting a total divergence gives
\begin{eqnarray}
& & \delta_\textrm{A}\mathfrak{L}_{\textrm{EM}+\textrm{q}}=-\frac{1}{8\pi}\sqrt{-\mathfrak{g}}F^{ik}\delta F_{ik}-\frac{{\sf j}^k}{c}\delta A_k=-\frac{1}{8\pi}\sqrt{-\mathfrak{g}}F^{ik}(\delta A_{k,i}-\delta A_{i,k})-\frac{{\sf j}^k}{c}\delta A_k \nonumber \\
& & =\frac{1}{4\pi}\sqrt{-\mathfrak{g}}F^{ik}\delta A_{k,i}-\frac{{\sf j}^k}{c}\delta A_k=\frac{1}{4\pi}(\sqrt{-\mathfrak{g}}F^{ik})_{,i}\delta A_k-\frac{1}{c}\sqrt{-\mathfrak{g}}j^k\delta A_k.
\label{Max2}
\end{eqnarray}
Consequently, the principle of least action $\delta S=0$ for arbitrary variations $\delta A_k$ yields the {\em second Maxwell--Minkowski equation}
\begin{equation}
(\sqrt{-\mathfrak{g}}F^{ik})_{,i}=\frac{4\pi}{c}{\sf j}^k,
\label{Max3}
\end{equation}
which is equivalent to
\begin{equation}
F^{ik}_{\phantom{ik}:i}=\frac{4\pi}{c}j^k.
\label{Max4}
\end{equation}

The variation (\ref{Max2}) can also be written as
\begin{equation}
\delta_\textrm{A}\mathfrak{L}_{\textrm{EM}+\textrm{q}}=-\frac{1}{8\pi}\mathfrak{e}F^{ik}(\delta A_{k:i}-\delta A_{i:k})-\frac{{\sf j}^k}{c}\delta A_k=-\frac{1}{4\pi}\mathfrak{e}F^{ik}\delta A_{k:i}-\frac{{\sf j}^k}{c}\delta A_k.
\end{equation}
Accordingly, we have
\begin{equation}
\frac{\partial\mathfrak{L}_{\textrm{EM}+\textrm{q}}}{\partial A_{k:i}}=\frac{\partial\mathfrak{L}_{\textrm{EM}+\textrm{q}}}{\partial A_{k,i}}=-\frac{1}{4\pi}\mathfrak{e}F^{ik}.
\label{Max6}
\end{equation}
The Lagrange equations (\ref{Lageq8}) with $\mathfrak{L}=\mathfrak{L}_{\textrm{EM}+\textrm{q}}$ for the field $A_k$
\begin{equation}
\frac{\partial\mathfrak{L}}{\partial A_k}-\partial_i\biggl(\frac{\partial\mathfrak{L}}{\partial(A_{k,i})}\biggr)=0,
\end{equation}
are equivalent to (\ref{Max3}) with (\ref{emc1}).
The electromagnetic field equation (\ref{Max3}) infers that ${\sf j}^i$ is conserved, ${\sf j}^i_{\phantom{i},i}=0$, which corresponds to the conservation of the total electric charge, but does not constrain the motion of particles.
Therefore, a configuration of charged particles producing the electromagnetic field can be arbitrary, subject only to the condition that the total charge be conserved, unlike a configuration of particles producing the gravitational field which is not arbitrary but constrained by the gravitational field equations.\\

\noindent
{\bf Second pair of Maxwell equations in vector form}.\\
We define
\begin{eqnarray}
& & D^\alpha=-\sqrt{g_{00}}F^{0\alpha},
\label{Max12} \\
& & H^{\alpha\beta}=\sqrt{g_{00}}F^{\alpha\beta},\quad H_\alpha=-\frac{1}{2}\eta_{\alpha\beta\gamma}H^{\beta\gamma},\quad H^{\alpha\beta}=-\eta^{\alpha\beta\gamma}H_\gamma.
\end{eqnarray}
The relations $F_{0\alpha}=g_{0i}g_{\alpha j}F^{ij}$ and $F^{\alpha\beta}=g^{\alpha i}g^{\beta j}F_{ij}$ give then
\begin{equation}
D_\alpha=\frac{E_\alpha}{\sqrt{g_{00}}}+g^\beta H_{\alpha\beta},\quad B^{\alpha\beta}=\frac{H^{\alpha\beta}}{\sqrt{g_{00}}}-g^\alpha E^\beta+g^\beta E^\alpha.
\end{equation}
In the spatial-vector notation, they become
\begin{equation}
{\bf D}=\frac{{\bf E}}{\sqrt{g_{00}}}-{\bf g}\times{\bf H}, \quad {\bf B}=\frac{{\bf H}}{\sqrt{g_{00}}}+{\bf g}\times{\bf E}.
\label{Max13}
\end{equation}
Using (\ref{spvec6}) and (\ref{Max12}) brings the temporal component of (\ref{Max3}) to
\begin{equation}
\mbox{div}\,{\bf D}=\frac{1}{\sqrt{\mathfrak{s}}}(\sqrt{\mathfrak{s}}D^\alpha)_{,\alpha}=4\pi\rho.
\label{Max15}
\end{equation}
The spatial components of (\ref{Max3}) are
\begin{equation}
\frac{1}{\sqrt{\mathfrak{s}}}(\sqrt{\mathfrak{s}}H^{\alpha\beta})_{,\beta}+\frac{1}{\sqrt{\mathfrak{s}}}(\sqrt{\mathfrak{s}}D^\alpha)_{,0}=-4\pi\rho\frac{dx^\alpha}{dx^0},
\end{equation}
which in the spatial-vector notation gives
\begin{equation}
\mbox{{\bf curl}}\,{\bf H}=\frac{1}{c\sqrt{\mathfrak{s}}}\frac{\partial(\sqrt{\mathfrak{s}}{\bf D})}{\partial t}+\frac{4\pi}{c}{\bf j}.
\label{Max17}
\end{equation}
The relations (\ref{Max15}) and (\ref{Max17}) are referred to as the {\em second pair of the Maxwell equations}.

In a locally flat spacetime, in which $g_{00}=\mathfrak{s}=1$ and ${\bf g}=0$, the relations (\ref{Max13}) reduce to
\begin{equation}
{\bf D}={\bf E},\quad {\bf B}={\bf H}.
\end{equation}
Accordingly, the second pair of the Maxwell equations (\ref{Max15}) and (\ref{Max17}) is given by
\begin{eqnarray}
& & \mbox{div}\,{\bf E}=4\pi\rho, \label{Max21} \\
& & \mbox{{\bf curl}}\,{\bf B}=\frac{1}{c}\frac{\partial{\bf E}}{\partial t}+\frac{4\pi}{c}{\bf j}.
\label{Max22}
\end{eqnarray}
Applying the divergence operator to (\ref{Max22}), using (\ref{Max21}), gives (\ref{emc19}).\\

\noindent
{\bf Gau\ss\, and Amp\`{e}re--{\O}rsted--Maxwell laws}.\\
Integrating (\ref{Max15}) over a volume, using (\ref{covint29}), and integrating (\ref{Max17}) over a surface, using (\ref{covint28}), respectively give
\begin{eqnarray}
& & \oint{\bf D}\cdot \sqrt{\mathfrak{s}}d{\bf f}=4\pi e, \label{Max31} \\
& & \oint{\bf H}\cdot d{\bf l}=\frac{1}{c}\frac{\partial}{\partial t}\biggl(\int{\bf D}\cdot \sqrt{\mathfrak{s}}d{\bf f}\biggr)+\frac{4\pi}{c}\int{\bf j}\cdot \sqrt{\mathfrak{s}}d{\bf f}.
\label{Max32}
\end{eqnarray}
In a locally flat spacetime, the relations (\ref{Max31}) and (\ref{Max32}) reduce to
\begin{eqnarray}
& & \oint{\bf E}\cdot d{\bf f}=4\pi e, \\
& & \oint{\bf B}\cdot d{\bf l}=\frac{1}{c}\frac{\partial}{\partial t}\biggl(\int{\bf E}\cdot d{\bf f}\biggr)+\frac{4\pi}{c}\int{\bf j}\cdot d{\bf f}.
\end{eqnarray}
Therefore, the flux of the electric field, which is referred to as the {\em electric flux}, through a closed surface is proportional to the total charge inside the volume enclosed by this surface, constituting {\em Gau\ss' law}.
The circulation of the magnetic field along a closed contour is equal to the sum of two terms, constituting the {\em Amp\`{e}re--{\O}rsted--Maxwell law}.
The first term is proportional to the time derivative of the flux of the electric field through the surface enclosed by this contour, which is referred to as the {\em displacement current}.
The second term is proportional to the current.\\

\noindent
{\bf Principle of superposition}.\\
The two pairs of the Maxwell equations are linear in the fields {\bf E} and {\bf B}.
The sum of any two solutions of the Maxwell equations is also a solution of these equations.
Therefore, the electromagnetic field of a system of sources (particles) is the sum of the fields from each source.
The additivity of the electromagnetic field is referred to as the {\em principle of superposition}.

\subsubsection{Energy--momentum tensor for electromagnetic field}
{\bf Metric energy--momentum tensor}.\\
The variation with respect to the metric tensor (\ref{dmemd4}) of the Lagrangian density (\ref{Lagem2}),
\begin{eqnarray}
& & \delta_\textrm{g}\mathfrak{L}_\textrm{EM}=-\frac{1}{16\pi}F_{ik}F_{lm}\delta(\sqrt{-\mathfrak{g}}g^{il}g^{km})=\frac{1}{32\pi}\sqrt{-\mathfrak{g}}g_{lm}F_{ik}F^{ik}\delta g^{lm}-\frac{1}{8\pi}\sqrt{-\mathfrak{g}}F_{ik}F_{lm}g^{il}\delta g^{km} \nonumber \\
& & =\frac{1}{8\pi}\sqrt{-\mathfrak{g}}\biggl(\frac{1}{4}g_{ik}F_{lm}F^{lm}-F_i^{\phantom{i}j}F_{kj}\biggr)\delta g^{ik},
\end{eqnarray}
gives the metric energy--momentum tensor (\ref{dmemd5}) for the electromagnetic field:
\begin{equation}
T_{ik}=\frac{1}{4\pi}\biggl(\frac{1}{4}g_{ik}F_{lm}F^{lm}-F_i^{\phantom{i}j}F_{kj}\biggr).
\label{emtem2}
\end{equation}
The corresponding energy density $W$, energy flux density vector ${\bf S}$ called the {\em Poynting vector}, and stress tensor $\sigma_{\alpha\beta}$ called the {\em Maxwell stress tensor}, are given in the locally geodesic and Galilean frame of reference, using (\ref{fmam23}), by
\begin{eqnarray}
& & W=\frac{1}{8\pi}(E^2+B^2), \label{emtem3} \\
& & {\bf S}=\frac{c}{4\pi}{\bf E}\times{\bf B}, \\
& & \sigma_{\alpha\beta}=\frac{1}{4\pi}\biggl(E_\alpha E_\beta+B_\alpha B_\beta-\frac{1}{2}\delta_{\alpha\beta}(E^2+B^2)\biggr).
\label{emtem5}
\end{eqnarray}
Under Lorentz transformations, $W$, ${\bf S}$ and $\sigma_{\alpha\beta}$ transform like the corresponding components of a tensor of rank (0,2) (\ref{fmam23}), according to (\ref{relkin16}).

The energy--momentum tensor for the electromagnetic field is traceless:
\begin{equation}
T_{ik}g^{ik}=0,
\label{emtem9}
\end{equation}
so (\ref{emtp16}) and the virial theorem (\ref{emtp17}) remain unchanged if the particles interact electromagnetically.
The condition (\ref{emtem9}) also gives, using (\ref{cemt41}),
\begin{equation}
\epsilon=3p,
\end{equation}
so (\ref{emtp18}) shows that the free electromagnetic field is ultrarelativistic.
In a frame of reference, in which the vectors ${\bf E}$ and ${\bf B}$ are parallel to one another or one of them vanishes (and the $x$ axis is along the direction of these vectors), the nonzero components of the tensor $T_{ik}$ are
\begin{equation}
T_{00}=-T_{11}=T_{22}=T_{33}=W.
\end{equation}
If the vectors ${\bf E}$ (along the $x$ axis) and ${\bf B}$ (along the $y$ axis) are mutually perpendicular and equal in magnitude then
\begin{equation}
T_{00}=T_{03}=T_{33}=W.
\end{equation}

\noindent
{\bf Conservation law for energy}.\\
Multiplying (\ref{eft16}) by ${\bf B}$ and (\ref{Max22}) by ${\bf E}$ and adding these scalar products gives
\begin{equation}
\frac{1}{c}{\bf E}\cdot\frac{\partial{\bf E}}{\partial t}+\frac{1}{c}{\bf B}\cdot\frac{\partial{\bf B}}{\partial t}=-\frac{4\pi}{c}{\bf j}\cdot{\bf E}-({\bf B}\cdot\mbox{{\bf curl}}\,{\bf E}-{\bf E}\cdot\mbox{{\bf curl}}\,{\bf B}).
\end{equation}
Using (\ref{spvec444}), from this relation we obtain
\begin{equation}
\frac{1}{2c}\frac{\partial}{\partial t}(E^2+B^2)=-\frac{4\pi}{c}{\bf j}\cdot{\bf E}-\mbox{div}({\bf E}\times{\bf B}),
\end{equation}
which has a form of the continuity equation with an additional term:
\begin{equation}
\frac{\partial W}{\partial t}+\mbox{div}\,{\bf S}=-{\bf j}\cdot{\bf E}.
\label{emtem8}
\end{equation}
Integrating (\ref{emtem8}) over the volume and using Gau\ss' theorem (\ref{covint29}) give
\begin{equation}
\frac{\partial}{\partial t}\int WdV+\int{\bf j}\cdot{\bf E}dV+\oint{\bf S}\cdot d{\bf f}=0.
\label{emtem336}
\end{equation}
This relation, using (\ref{emc21}), can be written as
\begin{equation}
\frac{\partial}{\partial t}\int WdV+\sum_a e_a{\bf v}_a\cdot{\bf E}({\bf r}_a)+\oint{\bf S}\cdot d{\bf f}=0.
\label{emtem33}
\end{equation}

\noindent
{\bf Tetrad and canonical energy--momentum tensors}.\\
The variation with respect to the tetrad (\ref{dtemd3}) of the Lagrangian density (\ref{Lagem2}),
\begin{equation}
\delta_\textrm{e}\mathfrak{L}_\textrm{EM}=-\frac{1}{16\pi}F_{ij}F_{kl}\eta^{ab}\eta^{cd}\delta(\mathfrak{e}e^i_a e^j_b e^k_c e^l_d)=\frac{1}{16\pi}\mathfrak{e}e^a_i F_{jk}F^{jk}\delta e^i_a-\frac{1}{4\pi}\mathfrak{e}F_{ij}F^{aj}\delta e^i_a,
\end{equation}
gives the tetrad energy--momentum tensor (\ref{dtemd5}) for the electromagnetic field:
\begin{equation}
t_i^{\phantom{i}a}=\frac{1}{4\pi}\biggl(\frac{1}{4}e^a_i F_{lm}F^{lm}-F_{ij}F^{aj}\biggr).
\label{emtem14}
\end{equation}
The corresponding tensor $t_{ik}$ is equal to (\ref{emtem2}).
This equality is a consequence of the reduced Belinfante--Rosenfeld relation (\ref{BelRos9}), which is valid if the spin tensor (\ref{spden12}) is equal to zero.
The spin tensor for the electromagnetic field vanishes because the Lagrangian density (\ref{Lagem2}) does not depend on the torsion tensor.
The canonical energy--momentum density (\ref{cemd3}) for the electromagnetic field is given by
\begin{equation}
\Theta^{\phantom{i}j}_i=\frac{\partial\mathfrak{L}_\textrm{EM}}{\partial A_{k:j}}A_{k:i}-\delta^j_i \mathfrak{L}_\textrm{EM}.
\end{equation}
The relation (\ref{Max6}) gives
\begin{equation}
\Theta^{\phantom{i}j}_i=-\frac{1}{4\pi}\mathfrak{e}F^{jk}A_{k:i}+\frac{1}{16\pi}\mathfrak{e}\delta^j_i F_{kl}F^{kl}.
\end{equation}
The canonical energy--momentum density is not identical with the tetrad energy--momentum density $\mathfrak{e}t_i^{\phantom{i}j}$ corresponding to (\ref{emtem14}) because the relation (\ref{cemd4}) is valid only for fields that are minimally coupled to the affine connection.\\

\begin{footnotesize}
If the Lagrangian density for the electromagnetic field were minimally coupled to the affine connection, $\tilde{\mathfrak{L}}_\textrm{EM}=-\frac{1}{16\pi}\mathfrak{e}\tilde{F}_{ij}\tilde{F}^{ij}$, then $\delta_\textrm{A}\tilde{\mathfrak{L}}_{\textrm{EM}+\textrm{q}}=-\frac{1}{4\pi}\mathfrak{e}\tilde{F}^{ik}\delta A_{k;i}-\frac{{\sf j}^k}{c}\delta A_k=-\frac{1}{4\pi}\mathfrak{e}(2S_i\tilde{F}^{ik}-\tilde{F}^{ik}_{\phantom{ik};i})\delta A_k-\frac{{\sf j}^k}{c}\delta A_k$, where a total divergence was omitted.
The resulting field equation would be $\nabla^\ast_i\tilde{F}^{ik}=\frac{4\pi}{c}j^k$.
The variation $\delta_\textrm{C}\tilde{\mathfrak{L}}_{\textrm{EM}+\textrm{q}}=-\frac{1}{4\pi}\mathfrak{e}\tilde{F}^{ik}\delta S^j_{\phantom{j}ik}A_j=-\frac{1}{4\pi}\mathfrak{e}\tilde{F}^{ik}A_j\delta C_{jik}$ gives the spin tensor, $s_{ijk}=-\frac{1}{2\pi}A_{[i}\tilde{F}_{j]k}$.
The second Einstein-Cartan equation gives the corresponding torsion tensor, $S_{kij}=\frac{\kappa}{4\pi}A_{[i}\tilde{F}_{j]k}+\frac{\kappa}{8\pi}A^l g_{k[j}F_{i]l}$, leading to $\tilde{F}_{ij}=A_{j,i}-A_{i,j}+\frac{\kappa}{4\pi}A^k A_{[i}\tilde{F}_{j]k}$ and $\tilde{F}^{ik}_{\phantom{ik};i}=\frac{4\pi}{c}j^k+\frac{\kappa}{8\pi}A^l F_{il}F^{ik}$.
In the presence of spinors, $S_{kij}$ would have another part (\ref{Lagsp11}), $S_{kij}=\frac{\kappa}{4\pi}A_{[i}\tilde{F}_{j]k}+\frac{\kappa}{8\pi}A^l g_{k[j}F_{i]l}+\frac{\kappa}{2}e_{ijkl}s^l$, leading to $\tilde{F}_{ij}=A_{j,i}-A_{i,j}+\frac{\kappa}{4\pi}A^k A_{[i}\tilde{F}_{j]k}+\kappa e_{ijkl}A^k s^l$.

The metric energy--momentum tensor for $\tilde{\mathfrak{L}}_\textrm{EM}$ is $T_{ik}=\frac{1}{4\pi}(\frac{1}{4}g_{ik}\tilde{F}_{lm}\tilde{F}^{lm}-\tilde{F}_i^{\phantom{i}j}\tilde{F}_{kj})$.
The Belinfante--Rosenfeld relation (\ref{BelRos8}) gives $t_{ik}=T_{ik}-\frac{1}{4\pi}\nabla^\ast_j(A_i\tilde{F}_k^{\phantom{k}j})=T_{ik}-\frac{1}{4\pi}A_{i;j}\tilde{F}_k^{\phantom{k}j}$, where the sourceless field equation $\nabla^\ast_i\tilde{F}^{ik}=0$ was used.
The corresponding canonical energy--momentum density is $\Theta^{\phantom{i}j}_i=\frac{\partial\tilde{\mathfrak{L}}_\textrm{EM}}{\partial A_{k;j}}A_{k;i}-\delta^j_i \tilde{\mathfrak{L}}_\textrm{EM}=-\frac{1}{4\pi}\mathfrak{e}\tilde{F}^{jk}A_{k;i}+\frac{1}{16\pi}\mathfrak{e}\delta^j_i\tilde{F}_{kl}\tilde{F}^{kl}=\mathfrak{e}t_i^{\phantom{i}j}$, showing that the relation (\ref{cemd4}) is valid.
\end{footnotesize}

\subsubsection{Lorentz force}
Let us consider a charged particle interacting with the electromagnetic field.
The total energy--momentum tensor for the particle and electromagnetic field is covariantly conserved, which gives the motion of the particle.
The electromagnetic part yields, by means of (\ref{eft4}) and (\ref{Max4}),
\begin{eqnarray}
& & T_{i\phantom{k}:k}^{\phantom{i}k}=\frac{1}{4\pi}\biggl(\frac{1}{2}F_{lm:i}F^{lm}-F_{il:k}F^{kl}-F_{il}F^{kl}_{\phantom{kl}:k}\biggr)=\frac{1}{4\pi}\biggl(-\frac{1}{2}F_{mi:l}F^{lm}-\frac{1}{2}F_{il:m}F^{lm} \nonumber \\
& & -F_{il:k}F^{kl}-F_{il}F^{kl}_{\phantom{kl}:k}\biggr)=-\frac{1}{4\pi}F_{il}F^{kl}_{\phantom{kl}:k}=-\frac{1}{c}F_{il}j^l.
\label{Lofo1}
\end{eqnarray}
The particle part gives, using (\ref{emtp9}),
\begin{equation}
T_{i\phantom{k}:k}^{\phantom{i}k}=\Bigl(\mu c^2\frac{u_i u^k}{\sqrt{g_{00}}u^0}\Bigr)_{:k}.
\end{equation}
The two parts together thus satisfy
\begin{equation}
\Bigl(\mu c^2\frac{u_i u^k}{\sqrt{g_{00}}u^0}\Bigr)_{:k}-\frac{1}{c}F_{il}j^l=0.
\label{Lofo3}
\end{equation}
Contracting (\ref{Lofo3}) with $u^i$, and using (\ref{emc10}) and $(u_i u^i)_{:k}=0$, gives
\begin{equation}
\Bigl(\mu c^2\frac{u^k}{\sqrt{g_{00}}u^0}\Bigr)_{:k}=0.
\end{equation}
which, after using (\ref{emc137}), turns (\ref{Lofo3}) into
\begin{equation}
\mu c^2\frac{u^k}{\sqrt{g_{00}}u^0}u_{i:k}=F_{il}\rho\frac{u^l}{\sqrt{g_{00}}u^0}.
\end{equation}
This relation, using $\mu/\rho=m/e$, gives
\begin{equation}
mc\frac{\mathring{D}u^i}{ds}=\frac{e}{c}F^{ij}u_j.
\label{Lofo6}
\end{equation}
It is equivalent, using the four-momentum $p^l=mcu^l$ (\ref{sple1}), to
\begin{equation}
\frac{\mathring{D}p^i}{ds}=\frac{e}{c}F^{ij}u_j.
\label{Lofo64}
\end{equation}

This relation is the equation of motion of a particle of mass $m$ and charge $e$ in the electromagnetic field $F_{ij}$.
It has a form allowed by (\ref{metgeo9}).
Multiplying (\ref{Lofo6}) by $u_i$ and contracting gives the identity: the left-hand side vanishes because of the orthogonality of the four-acceleration and the four-velocity (\ref{foac5}), and the right-hand side vanishes because of the antisymmetry of the electromagnetic field tensor.
Consequently, (\ref{Lofo6}) has 3 independent components.
The right-hand side of (\ref{Lofo6}) is referred to as the {\em Lorentz force}.
In the absence of the electromagnetic field, the equation of motion of this particle reduces to the metric geodesic equation (\ref{metgeo3}).\\

\noindent
{\bf Lorentz force in vector form}.\\
In the locally geodesic and Galilean frame of reference, we have $\mathring{D}/ds=d/ds=(u^0/c)d/dt$ and the four-velocity is given by (\ref{foac3}).
The 3 independent spatial components of (\ref{Lofo6}) are, using (\ref{sple1}),
\begin{equation}
\frac{dp^\alpha}{dt}=mc\frac{du^\alpha}{dt}=eF^{\alpha 0}+\frac{e}{c}F^{\alpha\beta}v_\beta.
\end{equation}
In the spatial-vector notation, this equation of motion is given by
\begin{equation}
\frac{d{\bf p}}{dt}={\bf F}=e{\bf E}+\frac{e}{c}{\bf v}\times{\bf B},
\label{Lofo8}
\end{equation}
where ${\bf p}$ is the momentum of the particle (\ref{emtp24}) and ${\bf F}$ is the spatial vector of the Lorentz force.
The temporal component of (\ref{Lofo6}) is
\begin{equation}
\frac{dp^0}{dt}=mc\frac{du_0}{dt}=\frac{e}{c}F_{0\alpha}v^\alpha.
\end{equation}
In the spatial-vector notation, this equation of motion is given by
\begin{equation}
\frac{d\mathfrak{E}}{dt}=e{\bf v}\cdot{\bf E},
\label{Lofo10}
\end{equation}
where $\mathfrak{E}$ is the energy of the particle (\ref{emtp23}), which also results from multiplying (\ref{Lofo8}) by ${\bf v}$ and using (\ref{emtp28}).
The equation of motion (\ref{Lofo8}) can be written for the acceleration of the particle, using (\ref{emtp27}) and (\ref{Lofo10}):
\begin{equation}
{\bf a}=\frac{d}{dt}\Bigl(\frac{{\bf p}c^2}{\mathfrak{E}}\Bigr)=\frac{c^2}{\mathfrak{E}}\Bigl(e{\bf E}+\frac{e}{c}{\bf v}\times{\bf B}\Bigr)-\frac{e({\bf v}\cdot{\bf E}){\bf p}c^2}{\mathfrak{E}^2}=\frac{e}{m\gamma}\Bigl({\bf E}+\frac{1}{c}{\bf v}\times{\bf B}-\frac{1}{c^2}({\bf v}\cdot{\bf E}){\bf v}\Bigr).
\end{equation}
The relations (\ref{emtem33}) and (\ref{Lofo10}) give the conservation law (\ref{fmam16}) for the total energy of the electromagnetic field and particles:
\begin{equation}
\frac{\partial}{\partial t}\Biggl(\int WdV+\sum_a \mathfrak{E}_a\Biggr)+\oint{\bf S}\cdot d{\bf f}=0.
\end{equation}

\subsubsection{Action for charged particles}
The action for a system of noninteracting particles is given by (\ref{acp5}).
A particle interacts with other particles through fields that carry the interaction.
Consequently, if an interaction is present, then we must add to (\ref{acp5}) the action describing fields that carry that interaction.
For the electromagnetic interaction, such a field is described by the electromagnetic potential and couples to the electromagnetic current.
The electromagnetic current vector for a point particle of charge $e$ located at the radius vector ${\bf r}_0$ is given by (\ref{emc15}):
\begin{equation}
j^k({\bf r})=\frac{cu^k}{u^0}\frac{e}{\sqrt{-\mathfrak{g}}}{\bm \delta}({\bf r}-{\bf r}_0).
\label{acp10}
\end{equation}
Substituting (\ref{acp10}) into the second term of (\ref{Max1}) gives the action for this coupling:
\begin{equation}
S_\textrm{e}=-\frac{1}{c^2}\int\sqrt{-\mathfrak{g}}A_k j^k dV\,dx^0=-\frac{e}{c}\int A_k\frac{u^k}{u^0}dx^0=-\frac{e}{c}\int A_k dx^k.
\label{acp11}
\end{equation}
The total action for a particle of mass $m$ and charge $e$ interacting with the electromagnetic potential $A_i$ is therefore the sum of (\ref{acp4}) and (\ref{acp11}):
\begin{equation}
S=-mc\int ds-\frac{e}{c}\int A_i dx^i.
\label{acp12}
\end{equation}
For a system of particles, the total action is the sum of the actions (\ref{acp12}) for each particle.
Under the gauge transformation (\ref{gi7}), the action (\ref{acp12}) changes by the integral of a total differential:
\begin{equation}
S'=-mc\int ds-\frac{e}{c}\int A_i dx^i-\frac{e}{c}\int d\alpha,
\end{equation}
so the conditions $\delta S=0$ and $\delta S'=0$ are equivalent, and the corresponding equations of motion are gauge invariant.

For a particle, the variation of the mass part of the action with respect to the coordinates $x^i$ is given by (\ref{acp14}).
The variation of $\int A_i dx^i$ with respect to the coordinates $x^i$ is
\begin{eqnarray}
& & \delta\int A_i dx^i=\int\delta A_i dx^i+\int A_i\delta dx^i=\int A_{i,j}\delta x^j dx^i+\int A_i d\delta x^i \nonumber \\
& & =\int A_{i,j}\delta x^j dx^i+\int d(A_i\delta x^i)-\int dA_i\delta x^i=\int A_{i,j}\delta x^j dx^i+\int d(A_i\delta x^i) \nonumber \\
& & -\int A_{i,j}dx^j\delta x^i=\int F_{ij}dx^j\delta x^i+\int d(A_i\delta x^i) \nonumber \\
& & =\int F_{ij}u^j\delta x^i ds+\int d(A_i\delta x^i).
\end{eqnarray}
Therefore, the variation of (\ref{acp12}) is, using the four-momentum $p^l=mcu^l$ (\ref{sple1}):
\begin{eqnarray}
& & \delta S=mc\int\frac{\mathring{D}u_i}{ds}\delta x^i ds-\frac{e}{c}\int F_{ij}u^j\delta x^i ds-mc\int d(u_i\delta x^i)-\frac{e}{c}\int d(A_i\delta x^i) \nonumber \\
& & =\int\biggl(\frac{\mathring{D}p_i}{ds}-\frac{e}{c}\int F_{ij}u^j\biggr)\delta x^i ds-\int d\biggl(p_i\delta x^i+\frac{e}{c}A_i\delta x^i\biggr) \nonumber \\
& & =\int_1^2\biggl(\frac{\mathring{D}p_i}{ds}-\frac{e}{c}\int F_{ij}u^j\biggr)\delta x^i ds-\biggl(p_i+\frac{e}{c}A_i\biggr)\delta x^i\biggl|_1^2,
\label{acp16}
\end{eqnarray}
where the limits 1 and 2 denote the endpoints of the particle's world line.
The principle of stationary action $\delta S=0$ for an arbitrary variation $\delta x^i$ that vanishes at the endpoints gives the Lorentz equation of motion (\ref{Lofo64}) of a particle in the electromagnetic field.

\subsubsection{Charged spinors}
The gauge-invariant modification of the Dirac Lagrangian density (\ref{Lagsp3}) is
\begin{equation}
\mathfrak{L}_\psi=\frac{i\mathfrak{e}}{2}e^i_a(\bar{\psi}\gamma^a D_i\psi-D_i\bar{\psi}\gamma^a\psi)-mc\mathfrak{e}\bar{\psi}\psi=\frac{i\mathfrak{e}}{2}e^i_a(\bar{\psi}\gamma^a\psi_{;i}-\bar{\psi}_{;i}\gamma^a\psi)-mc\mathfrak{e}\bar{\psi}\psi-eA_i\mathfrak{e}\bar{\psi}\gamma^i\psi,
\label{gi14}
\end{equation}
which is the Lagrangian density for charged spinor matter: $\mathfrak{L}_\textrm{m}=\mathfrak{L}_\psi$.
The Lagrangian density for the electromagnetic field and a charged spinor is thus
\begin{equation}
\mathfrak{L}=-\frac{\mathfrak{e}}{16\pi}F_{ik}F^{ik}+\frac{i\mathfrak{e}}{2}e^i_a(\bar{\psi}\gamma^a D_i\psi-D_i\bar{\psi}\gamma^a\psi)-mc\mathfrak{e}\bar{\psi}\psi.
\end{equation}
Consequently, the electromagnetic current density 
(\ref{emc1}) for a spinor field is proportional to the conserved Dirac vector current density (\ref{Direq11}):
\begin{equation}
{\sf j}^i=ec\mathfrak{e}\bar{\psi}\gamma^i\psi=ec{\sf j}^i_V.
\label{emc5}
\end{equation}

For an infinitesimal gauge transformation, $\alpha\ll 1$, (\ref{gi1})) and (\ref{gi7}) give
\begin{equation}
\delta\psi=ie\alpha\psi,\quad \delta\bar{\psi}=-ie\alpha\bar{\psi},\quad \delta A_k=\alpha_{,k},\quad \xi^i=0.
\end{equation}
If this transformation is global, $\alpha=\mbox{const}$, then the corresponding Noether current (\ref{Noeth7}) is
\begin{eqnarray}
& & \mathfrak{J}^i=\frac{\partial\mathfrak{L}}{\partial A_{k,i}}\delta A_k+\frac{\partial\mathfrak{L}}{\partial\psi_{,i}}\delta\psi+\delta\bar{\psi}\frac{\partial\mathfrak{L}}{\partial\bar{\psi}_{,i}}=-\frac{\mathfrak{e}}{4\pi}F^{ik}\delta A_k+\frac{i\mathfrak{e}}{2}\bar{\psi}\gamma^i\delta\psi-\frac{i\mathfrak{e}}{2}\delta\bar{\psi}\gamma^i\psi \nonumber \\
& & =-e\mathfrak{e}\bar{\psi}\gamma^i\psi\alpha=-\frac{\alpha}{c}{\sf j}^i,
\end{eqnarray}
using (\ref{emc5}).
Consequently, the conservation law (\ref{Noeth6}) gives the conservation (\ref{emc4}) of the electromagnetic current associated with the spinor.
The relation (\ref{emc5}) gives
\begin{equation}
\rho=e\psi^\dag\psi,\quad {\bf j}=ec\bar{\psi}{\bm\gamma}\psi,
\label{emc23}
\end{equation}
where ${\bm\gamma}$ is a spatial vector composed from the Dirac matrices $\gamma^\alpha$.
Identifying the electric charge $e$ in (\ref{emc12}) with the spinor electric charge $e$ in (\ref{emc23}) gives an integral constraint on a Dirac spinor field:
\begin{equation}
\int\psi^\dag\psi\sqrt{\mathfrak{s}}dV=1.
\end{equation}

\subsubsection{Charge conjugation}
The spin density corresponding to the Lagrangian density (\ref{gi14}) remains equal to (\ref{Lagsp6}); it is independent of the spinor electric charge $e$.
The electromagnetic potential corresponds, up to the multiplication by an arbitrary constant, to the vector multiple of $I_4$ in the formula for the spinor connection (\ref{spinco15}).
The electromagnetic potential is analogous to the affine connection: it modifies a derivative of a spinor so such derivative transforms like a spinor under unitary gauge transformations of the first type, while the connection modifies a derivative of a tensor so such derivative transforms like a tensor under coordinate transformations.

The gauge-invariant modification of the Dirac equation (\ref{Direq4}) is
\begin{equation}
i\gamma^k\psi_{;k}+eA_k\gamma^k\psi=mc\psi,
\end{equation}
whose adjoint conjugate generalizes (\ref{Direq7}):
\begin{equation}
-i\bar{\psi}_{;k}\gamma^k+eA_k\bar{\psi}\gamma^k=mc\bar{\psi}.
\end{equation}
The gauge-invariant modification of the Dirac equation (\ref{Direq6}) is
\begin{equation}
i\gamma^k\psi_{:k}+eA_k\gamma^k\psi=mc\psi-\frac{3\kappa}{8}(\bar{\psi}\gamma^5\gamma_k\psi)\gamma^5\gamma^k\psi,
\label{gi17}
\end{equation}
whose adjoint conjugate generalizes (\ref{Direq8}):
\begin{equation}
-i\bar{\psi}_{:k}\gamma^k+eA_k\bar{\psi}\gamma^k=mc\bar{\psi}-\frac{3\kappa}{8}(\bar{\psi}\gamma^5\gamma_k\psi)\bar{\psi}\gamma^5\gamma^k.
\end{equation}

Taking the complex conjugate of (\ref{gi17}) gives, using (\ref{Lagsp13}),
\begin{equation}
-i\gamma^{k\ast}\psi^\ast_{:k}+eA_k\gamma^{k\ast}\psi^\ast=mc\psi^\ast-\frac{3\kappa}{8}(\bar{\psi}\gamma^5\gamma_k\psi)\gamma^{5\ast}\gamma^{k\ast}\psi^\ast.
\label{gi19}
\end{equation}
The relations (\ref{spinco7}) and (\ref{spinco13}) give
\begin{equation}
\psi^c_{;k}=\psi^c_{|k}=(-i\gamma^2\psi^\ast)_{|k}=-i\gamma^2\psi^\ast_{|k}=-i\gamma^2\psi^\ast_{;k}.
\label{gi20}
\end{equation}
Substituting (\ref{disc5}) and (\ref{gi20}) into (\ref{gi19}) gives, using (\ref{Dirmat5}) and (\ref{Dirmat9}),
\begin{equation}
-i\gamma^2\gamma^k\gamma^2(-i\gamma^2)\psi_{:k}^c+eA_k\gamma^2\gamma^k\gamma^2(-i\gamma^2)\psi^c=mc(-i\gamma^2)\psi^c-\frac{3\kappa}{8}(\bar{\psi}\gamma^5\gamma_k\psi)\gamma^5\gamma^2\gamma^k\gamma^2(-i\gamma^2)\psi^c.
\label{gi21}
\end{equation}
Using (\ref{Dirmat8}), the relation (\ref{gi21}) becomes
\begin{equation}
\gamma^2\gamma^k\psi_{:k}^c+ieA_k\gamma^2\gamma^k\psi^c=-imc\gamma^2\psi^c+\frac{3i\kappa}{8}(\bar{\psi}\gamma^5\gamma_k\psi)\gamma^2\gamma^5\gamma^k\psi^c.
\label{gi22}
\end{equation}
Multiplying (\ref{gi22}) by $-i\gamma^2$ from the left brings this equation to
\begin{equation}
i\gamma^k\psi_{:k}^c-eA_k\gamma^k\psi^c=mc\psi^c-\frac{3\kappa}{8}(\bar{\psi}\gamma^5\gamma_k\psi)\gamma^5\gamma^k\psi^c.
\label{gi23}
\end{equation}

The Hermitian conjugate of (\ref{disc5}) gives
\begin{equation}
\psi^T=\psi^{\ast\dagger}=i\psi^{c\dagger}\gamma^{2\dagger}=-i\psi^{c\dagger}\gamma^2.
\end{equation}
Thus we obtain, using (\ref{Lagsp13}),
\begin{eqnarray}
& & \bar{\psi}\gamma^5\gamma_k\psi=(\psi^\dagger\gamma^0\gamma^5\gamma_k\psi)^\ast=\psi^T\gamma^2\gamma^0\gamma^2\gamma^5\gamma^2\gamma_k\gamma^2\psi^\ast=(-i\psi^{c\dagger}\gamma^2)\gamma^2\gamma^0\gamma^2\gamma^5\gamma^2\gamma_k\gamma^2(-i\gamma^2\psi^c) \nonumber \\
& & =-\psi^{c\dagger}\gamma^0\gamma^2\gamma^5\gamma^2\gamma_k\psi^c=-\overline{\psi^c}\gamma^5\gamma_k\psi^c.
\label{gi25}
\end{eqnarray}
Substituting (\ref{gi25}) into (\ref{gi23}) gives the Dirac equation for the charge-conjugate spinor field $\psi^c$:
\begin{equation}
i\gamma^k\psi_{:k}^c-eA_k\gamma^k\psi^c=mc\psi^c+\frac{3\kappa}{8}(\overline{\psi^c}\gamma^5\gamma_k\psi^c)\gamma^5\gamma^k\psi^c.
\label{gi26}
\end{equation}
Comparing (\ref{gi26}) with (\ref{gi17}) shows that $\psi$ and $\psi^c$ correspond to the same value of the mass $m$ and to the opposite values of the electric charge, $e$ and $-e$.
Accordingly, the charge-conjugation transformation does not change the mass of a spinor, but changes the sign of its electric charge.
The field equations for $\psi$ and $\psi^c$ are asymmetric because of the opposite signs of the corresponding cubic terms relative to the mass terms.
This asymmetry is related to the fact that the scalar $\bar{\psi}\psi$ changes sign under the charge-conjugation transformation:
\begin{equation}
\overline{\psi^c}\psi^c=-\bar{\psi}\psi,
\end{equation}
whereas the Lorentz square of $\bar{\psi}\gamma^5\gamma^k\psi$ does not change sign:
\begin{equation}
(\overline{\psi^c}\gamma^5\gamma^k\psi^c)(\overline{\psi^c}\gamma^5\gamma_k\psi^c)=(\bar{\psi}\gamma^5\gamma^k\psi)(\bar{\psi}\gamma^5\gamma_k\psi).
\end{equation}
The first two terms in the effective metric Lagrangian density (\ref{Direq9}) are thus antisymmetric under charge conjugation, while the last, four-fermion term is symmetric.
Therefore, $(\psi,m,e)$ and $(\psi^c,m,-e)$ are asymmetric under the charge-conjugation transformation and do not satisfy the same field equation.
Torsion generates an asymmetry between a spinor and its charge conjugate.
At densities satisfying $\kappa\bar{\psi}\psi\ll mc$, the cubic terms in (\ref{gi17}) and (\ref{gi26}) can be neglected.
In this approximation, $(\psi,m,e)$ and $(\psi^c,m,-e)$ are symmetric under the charge-conjugation transformation and satisfy the same field equation.
\newline
References: \cite{LL2,Lord}.
\newline
\newline
A particle is a special case of a field existing in spacetime.
The physics of particles and their systems, such as rigid bodies and ideal fluids, is referred to as {\em mechanics} and will constitute future Chapters 3 (Particles) and 4 (Systems).
This material is logically presented in {\em The Course of Theoretical Physics} by L. D. Landau and E. M. Lifshitz \cite{LL2,LL1,LL6}.


\newpage
\begin{thebibliography}{}
\bibitem{Schr} E. Schr\"{o}dinger, {\em Space-Time Structure} (Cambridge University Press, 1950).
\bibitem{LL2} L. D. Landau and E. M. Lifshitz, {\em The Classical Theory of Fields} (Pergamon, 1975).
\bibitem{Lord} E. A. Lord, {\em Tensors, Relativity and Cosmology} (McGraw-Hill, 1976).
\bibitem{Hehl1} F. W. Hehl, P. von der Heyde, and G. D. Kerlick, Phys. Rev. D {\bf 10}, 1066 (1974); F. W. Hehl, P. von der Heyde, G. D. Kerlick, and J. M. Nester, Rev. Mod. Phys. {\bf 48}, 393 (1976); V. de Sabbata and M. Gasperini, {\em Introduction to Gravitation} (World Scientific, 1986); V. de Sabbata and C. Sivaram, {\em Spin and Torsion in Gravitation} (World Scientific, 1994).
\bibitem{Wald} R. M. Wald, {\em General Relativity} (University of Chicago Press, 1984).
\bibitem{Uti} R. Utiyama, Phys. Rev. {\bf 101}, 1597 (1956).
\bibitem{KS} T. W. B. Kibble, J. Math. Phys. {\bf 2}, 212 (1961); D. W. Sciama, in: {\em Recent Developments in General Relativity}, p. 415 (Pergamon, 1962); Rev. Mod. Phys. {\bf 36}, 463 (1964); {\bf 36}, 1103 (1964).
\bibitem{MTW} C. W. Misner, K. S. Thorne, and J. A. Wheeler, {\em Gravitation} (Freeman, 1973).
\bibitem{Hehl2} F. W. Hehl, Phys. Lett. A {\bf 36}, 225 (1971); Gen. Relativ. Gravit. {\bf 4}, 333 (1973); {\bf 5}, 491 (1974).
\bibitem{NSH} A. Papapetrou, Proc. Roy. Soc. London A {\bf 209}, 248 (1951); K. Nomura, T. Shirafuji, and K. Hayashi, Prog. Theor. Phys. {\bf 86}, 1239 (1991); N. J. Pop{\l}awski, Phys. Lett. B {\bf 690}, 73 (2010); {\bf 727}, 575 (2013).
\bibitem{Mol} C. M{\o}ller, Ann. Phys. {\bf 4}, 347 (1958).
\bibitem{Hehl4} F. W. Hehl and B. K. Datta, J. Math. Phys. {\bf 12}, 1334 (1971).
\bibitem{FraNiko} F. R. B. Guedes and N. J. Pop{\l}awski, arXiv:2211.03234 (2022).
\bibitem{LL1} L. D. Landau and E. M. Lifshitz, {\em Mechanics} (Pergamon, 1976).
\bibitem{LL6} L. D. Landau and E. M. Lifshitz, {\em Fluid Mechanics} (Pergamon, 1982); L. D. Landau and E. M. Lifshitz, {\em Theory of Elasticity} (Pergamon, 1986); L. D. Landau and E. M. Lifshitz, {\em Electrodynamics of Continuous Media} (Pergamon, 1984).
\end{thebibliography}
\end{document}